\documentclass{article}

\usepackage[final, nonatbib]{neurips_2022}
\usepackage[autonum, blackhypersetup]{shortex}
\usepackage[numbers, sort&compress]{natbib} 

\hypersetup{
  breaklinks=true,
  colorlinks=true,
  pdfusetitle=true
}

\usepackage[utf8]{inputenc} 
\usepackage[T1]{fontenc}    
\usepackage{url}            
\usepackage{booktabs}       
\usepackage{amsfonts}       
\usepackage{microtype}      

\usepackage{algorithm}
\usepackage{algpseudocode}
\usepackage{mathtools}
\usepackage{tikz}
\usetikzlibrary{arrows}
\usetikzlibrary{bayesnet}
\usetikzlibrary{backgrounds}
\usepackage{caption}
\usepackage{subcaption}
\usepackage{amsthm}
\usepackage{enumerate}
\usepackage{stackengine}

\def\bfx{\mathbf{x}}
\def\bfX{\mathbf{X}}
\def\calT{\mathcal{T}}
\def\calX{\mathcal{X}}
\def\calQ{\mathcal{Q}}
\def\calB{\mathcal{B}}

\def\calF{\mathcal{F}}
\def\KL{\text{KL}}
\def\reals{\mathbb{R}}
\def\E{\mathbb{E}}
\def\Gap{\mathrm{Gap}}
\def\pis{\boldsymbol{\pi}}
\def\dee{\mathrm{d}}
\def\data{\textbf{Y}_m}
\def\MLE{x_{\mathrm{MLE},m}}
\def\iidsim{\stackrel{iid}{\sim}}
\def\deq{\stackrel{d}{=}}

\newcounter{xxx}
\definecolor{nice_blue}{rgb}{0.062, 0.45, 0.75}
\definecolor{nice_red}{rgb}{0.91, 0.42, 0.39}
\definecolor{nice_green}{rgb}{0.0, 0.5, 0.0}

\title{Parallel Tempering With a Variational Reference}

\author{%
  Nikola Surjanovic\\
  Department of Statistics\\
  University of British Columbia\\
  \texttt{nikola.surjanovic@stat.ubc.ca}\\
   \And
  Saifuddin Syed\\
  Department of Statistics\\
  University of Oxford\\
  \texttt{saifuddin.syed@stats.ox.ac.uk}\\
   \And
  Alexandre Bouchard-C\^ot\'e\\
  Department of Statistics\\
  University of British Columbia\\
  \texttt{bouchard@stat.ubc.ca}\\
   \And
  Trevor Campbell\\
  Department of Statistics\\
  University of British Columbia\\
  \texttt{trevor@stat.ubc.ca} \\
}

\begin{document}

\maketitle

\begin{abstract}
  
Sampling from complex target distributions is a challenging task fundamental to
Bayesian inference.  Parallel tempering (PT) addresses this problem by
constructing a Markov chain on the expanded state space of a sequence of
distributions interpolating between the posterior distribution and a fixed
reference distribution, which is typically chosen to be the prior. However, in the typical case
where the prior and posterior are nearly mutually singular, PT methods are
computationally prohibitive. 
In this work we
address this challenge by constructing a generalized
annealing path connecting the posterior to an adaptively tuned variational
reference. The reference distribution is tuned to minimize the forward
(inclusive) KL divergence to the posterior distribution using a simple,
gradient-free moment-matching procedure. We show that our adaptive procedure
converges to the forward KL minimizer, and that the forward KL divergence serves
as a good proxy to a previously developed measure of PT performance.
We also show that in the large-data limit in typical Bayesian models, the proposed 
method improves in performance, while traditional PT deteriorates arbitrarily.
Finally, we introduce PT with two references---one fixed, one  variational---with a novel split annealing
path that ensures stable variational reference adaptation.
The paper concludes with experiments that demonstrate the large empirical
gains achieved by our method in a wide range of realistic Bayesian inference
scenarios.
\end{abstract}


\section{Introduction}

Parallel tempering (PT) is a popular approach to sampling from challenging
probability distributions used in many scientific disciplines
\cite{akiyama2022first,akiyama2021first, earl2005parallel, ballnus2017comprehensive}. 
PT methods involve running Markov chain Monte Carlo (MCMC) on the expanded state
space of a sequence of distributions that 
connect the target distribution of interest, $\pi_1$, to a simple reference distribution, $\pi_0$, 
for which \iid sampling is tractable. The key innovation in PT is that the MCMC chain enables distributions
along the path to swap states (or \emph{communicate}). This communication enables \iid draws from the reference $\pi_0$
to aid in exploration of the challenging target $\pi_1$. Indeed, recent work has shown that the 
effectiveness of a PT method is essentially characterized by the efficiency of this communication
via the \emph{global communication barrier} (GCB) from $\pi_0$ to $\pi_1$ \cite{syed2019nrpt}. Intuitively, 
the GCB is low when the reference $\pi_0$ is similar to the target $\pi_1$; in this case, 
the distributions along the path have substantial overlap and proposed swaps are generally accepted.
The GCB is also inversely related to the \textit{restart rate}, which quantifies how frequently 
\iid samples from $\pi_0$ traverse the path to $\pi_1$ (a \textit{restart}) \cite{syed2019nrpt}.

In the setting of Bayesian posterior inference---a key application of
PT, and the focus of this work---the target $\pi_1$ is the posterior distribution, and the reference
distribution $\pi_0$ is typically set to the prior. From the perspective of 
PT communication efficiency, this is a poor choice in general; the prior is often 
quite different from the posterior, resulting in a high GCB.
As an extreme (but common) example, we show in this work that when the
posterior distribution concentrates in the large-data limit, 
the restart rate with a fixed reference tends to
zero and PT becomes computationally infeasible (\cref{prop:GCB_infty_posterior}). On 
the other hand, the posterior often
exhibits regularities---asymptotic normality in certain parameters, for
example---that motivate the need for a choice of PT reference that can automatically adapt to
the target to obtain computational gains.  

\begin{figure}[t]
    \centering
    \includegraphics[width=\textwidth]{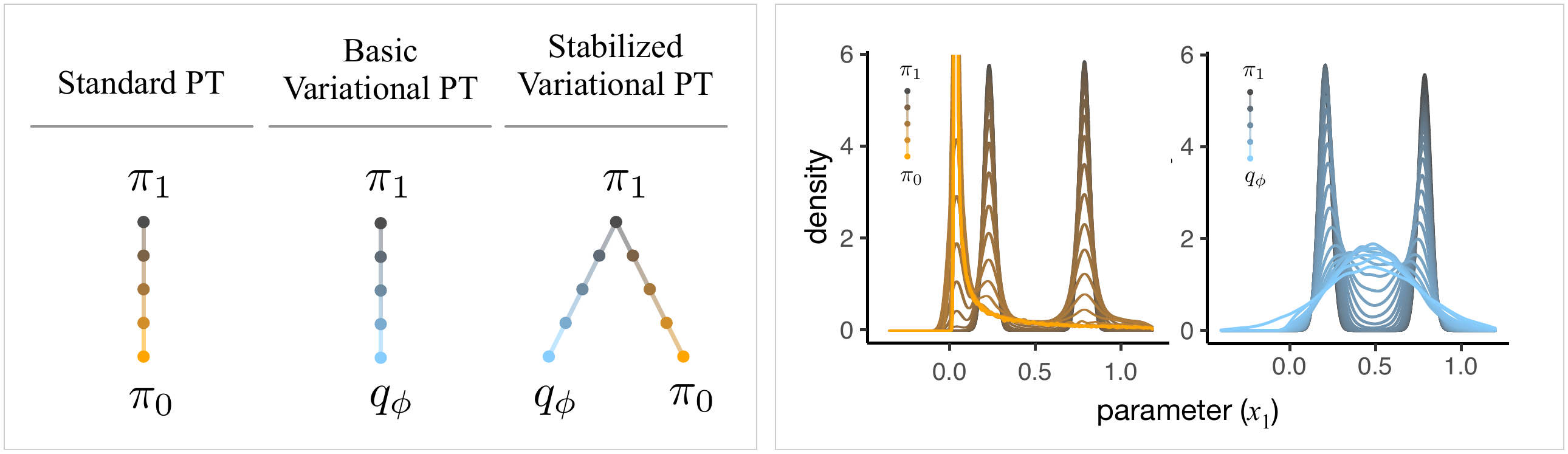}
	\caption{{\footnotesize  \textbf{Left box:} visualization of the three main PT
	algorithms considered in this work. Nodes represent distributions
	interpolating between tractable reference distributions (bottom, either fixed
	$\pi_0$ or variational $q_\phi$), and
	an intractable distribution (top, $\pi_1$, typically a Bayesian posterior). Edges encode the structure of the possible swaps
	performed by the various PT algorithms. \textbf{Right box:} examples of a path
	of marginal distributions obtained on a Bayesian ODE parameter estimation
	problem with more than one latent variable for an mRNA transfection dataset \cite[page
	8]{ballnus2017comprehensive}. The two modes are non-trivial to switch as
	they require changing other parameters (not shown) simultaneously.
	In this example, the marginal of the prior places a small amount of mass on the second mode 
	whereas the marginal of the variational reference places significant mass on both modes.
	Because the variational reference covers both modes, the length of the annealing path
	from the reference to the target is shorter and it is easier to obtain samples from the 
	target distribution using parallel tempering.
  }}
    \label{fig:reference_posterior}
\end{figure}

\begin{figure}[t]
    \centering
    \includegraphics[width=\textwidth]{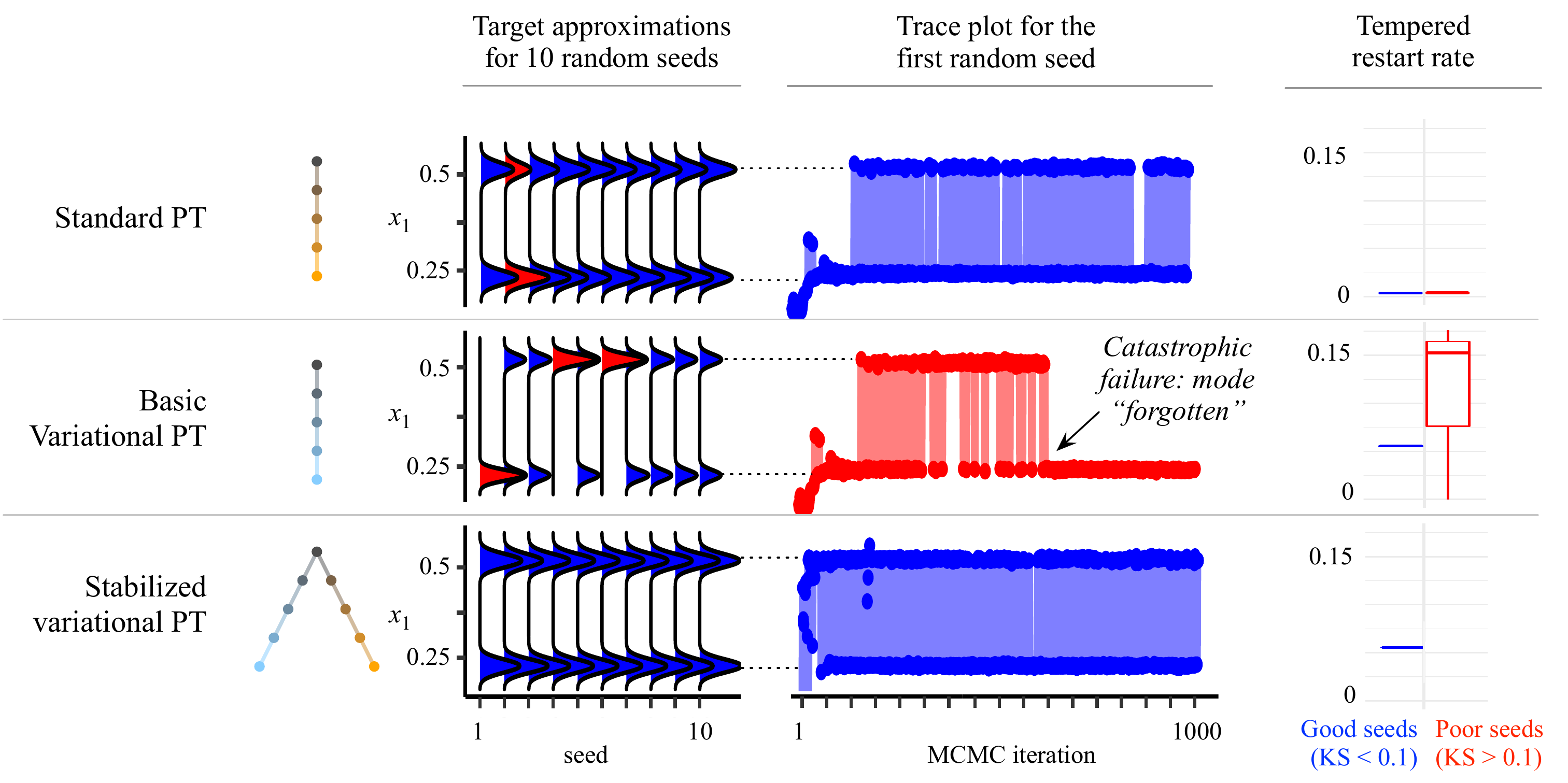}
	\caption{{\footnotesize   Comparison of three PT methods for Bayesian ODE
	parameter estimation with an mRNA transfection dataset \cite[page
	8]{ballnus2017comprehensive}.  \textbf{Top row:} standard PT succeeds in achieving a
	positive but small restart rate (a measure of PT efficiency; higher is
	better). \textbf{Middle row:} basic implementation of only one variational
	reference distribution leads to catastrophic failures in 3 out of 10
	independent runs. The variational distribution sometimes ``forgets'' one of
	the modes (red denotes ``bad'' approximations with a Kolmogorov-Smirnov
	distance for the marginal posterior over the model parameter $x_1$ larger than 0.1) and becomes stuck
	in the other, which leads to an overestimation of the restart rate. 
\textbf{Bottom row:} using both a fixed and an adaptive reference addresses this 
issue and leads to markedly better performance in
terms of restart rate compared to standard PT. All methods have a comparable cost per iteration;
all use the same variational reference family and run the same total
number of chains and iterations.}}
    \label{fig:demo}
\end{figure}

In this work, we develop and analyze a novel PT algorithm
that automatically adapts a variational reference distribution within a 
parametric family, $\calQ=\{q_\phi : \phi\in\Phi\}$. 
This adaptive reference family addresses the shortcomings of using the prior as
a PT reference: we show that in the large-data limit, the restart rate with an
appropriate variational reference improves arbitrarily (\cref{prop:GCB_zero_posterior}).
We find that even when one is not in the large data setting, our method 
can provide large empirical gains compared to fixed-reference PT in a wide range of realistic 
Bayesian inference scenarios.
The method is based on two
major methodological contributions. First, we adapt the parameter $\phi$ to minimize the
forward (inclusive) KL divergence $\text{KL}(\pi_1 \| q_{\phi})$ instead of
directly taking gradients with respect to the GCB itself. 
This approach is particularly advantageous when $\calQ$ 
is an exponential family: \cref{thm:GCB_KL} shows that the forward KL is a good surrogate of the GCB,
and  minimizing the forward KL amounts
to matching moments, which involves no extra tuning effort from the user. 
We perform moment matching in a simple iterative fashion, in rounds of increasingly many PT draws;
\cref{thm:adapt_tune} identifies conditions that guarantee that the variational parameter estimate converges to the optimum. 
Second, we combine two references---one fixed, one variational---by 
``gluing'' two PT algorithms together (each based on one of the references, see \cref{fig:reference_posterior}).
We demonstrate that this ``stabilized'' method is necessary for obtaining a 
reliable PT algorithm: adaptation with just the variational reference
alone can lead to ``forgetting'' the
structure of the posterior distribution (e.g., multi-modality, as shown in \cref{fig:demo}).
Although this requires more computational effort,
we show that under idealized conditions
the restart rate of our adaptive method is no lower than half the restart
rate of standard, fixed reference PT (\cref{thm:restart_rate})
after accounting for the doubled computation time. In practice,
it is often much better.
Finally, the paper presents an
extensive empirical study of the performance of our method in a variety
of real-world Bayesian models, including spatial models (sparse random field  Poisson
regression) and functional data analysis (Bayesian estimation of ODE parameters),
among others. We find that our
method can substantially increase the performance of PT. 

\noindent {\bf Related work.} The idea of adapting the tractable end-point of
a path of distributions has been explored most thoroughly in the context of
estimation of normalization constants
\cite{lefebvre_path_2009,lefebvre_path_2010,vitoratou_thermodynamic_2017,hawryluk_referenced_2021}.
However, the importance of having both a fixed and adaptive reference
has not appeared in this literature.  In the PT literature, the closest related work appears to be \cite[section 5]{paquet_perturbation_2009} and \cite[section
4.1]{cameron_recursive_2014}, which consider PT algorithms with non-prior references.
However, no adaptive algorithm is proposed; the reference distribution is set to a fixed variational approximation in \cite{paquet_perturbation_2009} and to a Gaussian
approximation at an analytically tractable mode in \cite{cameron_recursive_2014}.
Reference adaption has been
considered in passing in the annealed importance sampling and sequential Monte Carlo literature,
but without carefully considering the importance of stabilization discussed in our work, e.g. \cite{heng_gibbs_2021}.
Another related line of work is the design of adaptive independence
Metropolis-Hastings (IMH) proposals
\cite{de_freitas_variational_2001,gasemyr_adaptive_2003,rue_approximating_2004,maire_adaptive_2019}.
In particular, the theoretical work studying convergence of adaptive IMH has
recognized the usefulness of combining the adaptive reference with a fixed reference
\cite[Section 7]{andrieu_ergodicity_2006}. However, compared to the PT context,
this combination is more straightforward to achieve in IMH via alternation of
samplers. Finally, variational inference has previously been used to aid Monte Carlo
methods, e.g., to learn proposals in sequential Monte Carlo \cite{Gu15}, annealed importance sampling \cite{Salimans15},
and MCMC \cite{Levy18}, and to form an initial reference for MCMC \cite{Hoffman17,Ruiz19}.
None of these works connect the KL objective directly to the performance of a PT algorithm as we do in this paper.


\section{Background}
This work builds on a recent state-of-the-art PT method, non-reversible parallel tempering (NRPT) 
\cite{syed2019nrpt}. The goal is to sample 
from a target distribution, $\pi_1$, on a state space $\mcX$. This is achieved by constructing a sequence of $N+1$ 
distributions from $\pi_1$ to a reference distribution $\pi_0$ also on $\mcX$. 
Given an 
\emph{annealing schedule} $\mcB_N = (\beta_n)_{n=0}^N$, where 
$0 = \beta_0 < \beta_1 < \cdots < \beta_N = 1$ with mesh size $\|\calB_N\|=\max_n|\beta_n-\beta_{n-1}|$, 
the distribution for each chain $n=0, \dots, N$ is given by $\pi_{\beta_n}$, where for all $\beta\in[0,1]$, $\pi_{\beta}(x) \propto \pi_0^{1-{\beta}}(x) \cdot \pi_1^{\beta}(x)$.

In parallel tempering, we simulate a Markov chain $\textbf{X}_t=(X^0_t,\dots,X^N_t)$ that 
targets the product distribution $\boldsymbol{\pi}$ on $\mcX^{N+1}$ given by 
$\boldsymbol{\pi}(\bfx) = \pi_{\beta_0}(x^0) \cdot \pi_{\beta_1}(x^1)\cdots \pi_{\beta_N}(x^N)$
as the unique invariant probability distribution of the Markov chain.
Note that the target distribution $\pi_1$ is a marginal of $\boldsymbol{\pi}$, hence Monte Carlo averages based on $X^N_t$ converge to the correct posterior expectations.
Each iteration of PT involves two steps---
\textit{local exploration} and \textit{communication}---as shown in \cref{alg:NRPT}. The local exploration 
step involves applying, for each chain $n$, a $\pi_{\beta_n}$-invariant Markov kernel (e.g.~Hamiltonian Monte Carlo).
The communication step involves Metropolized swaps of states between adjacent chains;
one alternates between swaps of 
chains $n$ and $n+1$ for all $n$ in $S_\text{even}$ followed by $S_\text{odd}$, where $S_\text{even}$ and $S_\text{odd}$ are the 
even and odd subsets of $\{0,\dots,N-1\}$, respectively. 
The deterministic alternation between even and odd swaps enjoys 
remarkable theoretical and empirical properties 
and ensures that the performance of parallel tempering does not degrade 
as $N$ increases \cite{syed2019nrpt}. 
Each proposed swap of component $n$ and $n+1$ of $\bfx = (x^0,\dots,x^N)$ is accepted or rejected independently with probability
\[\label{eq:acceptance_probability}
  \alpha_n(\bfx) = 1 \wedge \frac{\pi_{\beta_n}(x^{n+1})\cdot\pi_{\beta_{n+1}}(x^{n})}
  {\pi_{\beta_n}(x^{n})\cdot\pi_{\beta_{n+1}}(x^{n+1})}.
\]

\begin{algorithm}[t]
	\footnotesize
\caption{Non-reversible parallel tempering (NRPT)}
\label{alg:NRPT}
\begin{algorithmic}
\Require Initial state $\bfx_0$, annealing schedule $\mathcal{B}_N$, \# iterations $T$, 
annealing path $\pi_\beta$
\State $r_n \gets 0$ for $n = 0, 1, \ldots, N-1$
\For{$t=1,2,\ldots,T$}
  \State $\bfx \gets \texttt{LocalExploration}(\bfx_{t-1})$ 
  \Comment{Local exploration kernels (e.g., HMC)}
  \State $S_t \gets $ {\bf if } $t$ is even $S_\text{even}$ {\bf else} $S_\text{odd}$
  \For{$n=0,1,\ldots, N-1$}
    \State $\alpha_n \gets \alpha_n(\bfx)$ \Comment{Acceptance probability from \eqref{eq:acceptance_probability}}
    \State $r_n \gets r_n + (1-\alpha_n)/T$ \Comment{Store chain communication 
    rejection rate estimates}
    \State $U_n \gets \text{Unif}(0,1)$
    \State {\bf if} {$n \in S_t$ and $U_n \leq \alpha_n$} {\bf then}  $x^{n+1}, \, x^n \gets x^{n}, \, x^{n+1}$ \Comment{Swap components $n$ and $n+1$ of $\bfx$}
  \EndFor
  \State $\bfx_t\gets \bfx$
\EndFor
\State \textbf{Return:} $(\bfx_t)_{t=1}^T$, $(r_n)_{n=0}^{N-1}$
\end{algorithmic}
\end{algorithm}

To characterize the performance of a parallel tempering method we study how often 
samples \textit{restart}, i.e. travel from the reference $\pi_0$ to the target $\pi_1$ \cite{syed2019nrpt}. Studying restarts isolates the effect of communication from the 
problem-specific characteristics of local exploration. (An improved local exploration kernel
will of course improve overall performance.) 
When it is possible to obtain i.i.d.~samples from $\pi_0$, the number
of restarts is empirically found to be related to the effective sample size in the target distribution chain \cite{syed2019nrpt}.
Formally, the 
\emph{restart rate} $\tau(\calB_N)$ from $\pi_0$ to $\pi_1$ with schedule $\calB_N$ is the fraction 
of PT iterations that result in a restart. The maximum value of $\tau$ is $\nicefrac{1}{2}$, since a
communication swap is proposed with $\pi_0$ at every other iteration of PT. If the local exploration 
is efficient (\cref{assump:ELE}) then $\tau(\calB_N)=(2+2\sum_{n=0}^{N-1}\frac{r_n}{1-r_n})^{-1}$ 
\cite[Corollary 1]{syed2019nrpt}, where $r_n = 1-\EE[\alpha_n(\bfX)]$ and $\bfX\distas \boldsymbol{\pi}$.
Asymptotically as $N\to\infty$, the round trip rate converges to a constant known as the \emph{asymptotic restart rate} $\tau$: 
$\lim_{\|\calB_N\|\to 0}\tau(\calB_N)=\tau=(2+2\Lambda(\pi_0,\pi_1))^{-1}$ \cite[Theorem 3]{syed2019nrpt},
where $\Lambda(\pi_0,\pi_1)$ is the  
\emph{global communication barrier} (GCB) between $\pi_0$ and $\pi_1$, 
\[ 
  \Lambda(\pi_0, \pi_1) = \frac{1}{2} \int_0^1 \mathbb{E}[|\ell(X_\beta) -
  \ell(X_\beta')|] \, \dee\beta, \qquad \ell(x) = \log \frac{\pi_1(x)}{\pi_0(x)}, \qquad X_\beta, X'_\beta \distiid \pi_\beta.\label{eq:GCB}
\]
The GCB can be estimated using the rejection rates $r_n$ for all adjacent pairs of chains,
$\Lambda(\pi_0, \pi_1) \approx \sum_{n=0}^{N-1} r_n$, where $r_n = 1-\EE[\alpha_n(\bfX)]$ for 
$\bfX\distas \boldsymbol{\pi}$, with the approximation error decreasing to zero at a rate $O(N^{-2})$ 
as the number of chains $N$ increases \cite[Section 5.2]{syed2019nrpt}. GCB values near zero imply that 
swaps are typically accepted and communication is efficient. 


\section{Parallel tempering with a variational reference}
\label{sec:variational_PT}
A key degree of freedom one has when using PT is the reference distribution, $\pi_0$.
Although the standard approach is to set $\pi_0$ to the prior,
\cref{eq:GCB} suggests that the GCB might be quite large---and hence communication 
performance quite poor---when the prior
and posterior differ significantly, which commonly occurs in practice. Indeed, \cref{prop:GCB_infty_posterior} 
motivates the importance of choosing the reference carefully:
in a typical Bayesian model, as one obtains more data,
the restart rate for the prior reference tends to zero.
This result relies on \cref{assump:regularity} in \cref{sec:largedataasymp}, which 
stipulates standard technical conditions sufficient for, e.g., asymptotic consistency of the MLE, 
a Bernstein-von Mises result for asymptotic normality of the posterior \cite{miller2021asymptotic}, along
with PT-specific assumptions such as efficient local exploration.
We emphasize this result is a motivation for our methodology; the proposed algorithms apply much more broadly
and not only in the data limit.

\bprop[Large-data restart rate, fixed reference]
\label{prop:GCB_infty_posterior}
Consider data $\data=\{Y_i\}_{i=1}^m$ generated \iid from a model with likelihood
$L(y | x_0)$, $x_0\in\mcX \subset \reals^d$, satisfying the conditions in
\cref{assump:regularity}. Denote $\pi_{1,m}$ to be the posterior
conditioned on $\data$. Then, in the large-data limit, the asymptotic
restart rate $\tau_m$ associated with the
annealing path from $\pi_0$ to $\pi_{1,m}$ 
degrades arbitrarily, 
i.e., $\tau_m \to 0$ almost surely as $m\to\infty$.
\eprop

In this section, we demonstrate that allowing the reference to be
\emph{tunable} addresses this issue.

\subsection{Annealing paths with a variational reference}
\label{sec:paths_variational_reference}
Let $\calQ=\{q_\phi: \phi \in \Phi\}$ be a parametric family of distributions on $\calX$, and for each $\phi \in \Phi$, denote the annealing path from the reference distribution $q_\phi$ to
the target $\pi_1$ by
\[
 \forall\, \beta\in[0,1], \quad \pi_{\phi,\beta}(x) \propto q_\phi(x)^{1-\beta}\cdot\pi_1(x)^\beta.
\]
Note that for this modified annealing path, the target distribution $\pi_1$ remains the same 
although the reference may change. In the Bayesian framework, this means that
the prior $\pi_0$ and posterior $\pi_1$ remain the same while the variational reference $q_\phi$
is tuned. To ensure the variational
reference family $\mcQ$ is compatible with the asymptotic PT theory developed in \cite{syed2019nrpt}, we will assume 
$\mcQ$ is a \emph{PT-suitable family} for the target $\pi_1$, i.e., each $q_\phi\in\mcQ$ shares the same support at the target and satisfies some mild moment conditions (\cref{def:regular_reference} in \cref{sec:theoretical_conditions}). 
We will also assume throughout that the fixed reference
$\pi_0$ is itself \emph{PT-suitable} for $\pi_1$.
PT-suitability is sufficient to guarantee that the restart rate is inversely related to the GCB and that the schedule-tuning procedure from \cite[Section 5.4]{syed2019nrpt} is justified.

\subsection{Exponential variational reference family}
The variational reference family $\calQ$ should be flexible enough
to match the target $\pi_1$ reasonably well, but also simple enough to enable i.i.d.\ sampling, pointwise evaluation, and tractable optimization.
\cref{prop:GCB_zero_posterior} suggests that in the large-data limit, the family
of multivariate Gaussian distributions often suffices. In particular, unlike the fixed prior reference---whose restart rate 
decays to zero in the large-data limit---there exists a sequence of
multivariate normal reference distributions so that the restart rate tends to 
its maximum value of $\nicefrac{1}{2}$. 
Note again that this large-data setting is just one instance in which a tunable reference helps; our method in this work applies much more broadly, and does not 
require a Gaussian reference or rely on the asymptotic setup in \cref{prop:GCB_zero_posterior}. In particular, our method is advantageous in any setting where the GCB decreases compared to fixed-reference PT. 
\bprop[Large-data restart rate, variational reference]
\label{prop:GCB_zero_posterior}
Consider the setting of \cref{prop:GCB_infty_posterior}, and suppose
$\calQ=\{\distNorm(\mu,\Sigma):\mu\in \reals^d,\Sigma\in\reals^{d\times d}, \Sigma=\Sigma^\top \succ 0\}$ is a
PT-suitable family for all targets $\pi_{1,m}$ almost surely. Then for any fixed $N > 1$, there
exists a random sequence $\mu_m \in \reals^d$, $\Sigma_m \in \reals^{d \times
d}$ such that for any schedule $\calB_N$,
in the large-data limit,
the restart rate $\tau_m(\calB_N)$ associated with $\pi_{1,m}$, $\distNorm(\mu_m, \Sigma_m)$ converges to the maximum possible value. I.e., for any schedule $\calB_N$ we have
$\tau_m(\calB_N) \convp \nicefrac{1}{2}$ as $m\to\infty$.
\eprop
\cref{prop:GCB_zero_posterior} motivates the use of a tunable variational reference that can adapt to the target, as opposed to a fixed reference. In this work, we consider the general class of
exponential reference families of full-rank where the distributions take the form
$q_\phi(x) = h(x) \exp(\phi^\top \eta(x) - A(\phi)),$
for base density $h$, natural parameter $\phi$, sufficient statistic $\eta$, and
log partition function $A$. Aside from being flexible enough to match posteriors
arbitrarily well in the large-data limit, a key advantage of an exponential reference family 
is that it is straightforward to fit: one can obtain the forward (inclusive) KL divergence minimizer $q_{\phi_\text{KL}}$ using a 
simple gradient-free moment matching procedure because $\EE_{\phi_\KL}[\eta] = \EE_{1}[\eta]$, where
$\EE_{\phi}$ and $\EE_1$ denote expectations with respect to $q_{\phi}$ and $\pi_1$, respectively \cite{herbrich2005minimising}.  
Indeed, under slightly more stringent technical assumptions in the setting 
of \cref{prop:GCB_infty_posterior,prop:GCB_zero_posterior}---namely
\cref{assump:posterior_moments}---\cref{prop:GCB_zero_posterior_2} shows that we may use this 
forward KL fit as the reference sequence $\distNorm(\mu_m, \Sigma_m)$ for which the restart rate is asymptotically maximized.
\cref{assump:posterior_moments} stipulates that the differences between the posterior mean and MLE, as well as 
between the inverse Fisher information and scaled posterior variance, are not too large.

\bprop[Large-data restart rate, moment matched reference]
\label{prop:GCB_zero_posterior_2}
Consider the setting of \cref{prop:GCB_zero_posterior} and suppose that \cref{assump:posterior_moments} also holds. Then, the conclusion of \cref{prop:GCB_zero_posterior} holds if $\mu_m, \Sigma_m$ are set to the mean
and variance of $\pi_{1,m}$ conditioned on $\data$, respectively. 
\eprop

\subsection{Tuning the variational reference}
\label{sec:variational_PT_alg}

In practice, we fit the exponential family reference iteratively by running NRPT for multiple
tuning rounds $r=1, \dots, R$; in each tuning round $r$ we run $T_r=2^r$ iterations with
variational parameter $\hphi_r$. Using the generated states
$(\bfX_{t,r})_{t=1}^{T_r}$, we obtain the parameter $\hphi_{r+1}$ for
round $r+1$ by solving
\[\label{eq:update_reference}
    \E_{\hat{\phi}_{r+1}}[\eta] = 
        \frac{1}{T_{r}} \sum_{t=1}^{T_r} \eta(X_{t,r}^N).
\]
Note that by relying on sufficient statistics, we are not required to keep in 
memory the MCMC trace or to loop over MCMC samples when performing variational 
parameter optimization.  
For example, when $\calQ$ is a Gaussian family, \cref{eq:update_reference}
simplifies to setting the mean vector and covariance matrix to the empirical
mean and covariance obtained from the target chain samples $X_{t,r}^N$. 
When a full (non-diagonal) covariance matrix is used, one should start tuning
when $T_r \geq d$. We additionally use the samples from each round to tune
the annealing schedule $\mathcal{B}_N$ using the procedure from \cite{syed2019nrpt}.

\cref{thm:adapt_tune} shows that if the absolute spectral gap $\Gap(\phi)$ \cite{fan2021hoeffding} of the PT Markov
chain with reference $q_\phi$ is bounded away from zero, and the
number of iterations in each round tends to infinity at an appropriate rate, then $\hphi_r$ 
will converge almost surely to the forward KL minimizer $\phi_\KL$. 
Although \cref{thm:adapt_tune} stipulates that $\eta$ is bounded,
this is a technicality that is not required in practice.

\begin{theorem}[Convergence of variational reference tuning]
\label{thm:adapt_tune}
Suppose $\calQ=\{q_\phi:\phi\in\Phi\}$ is a PT-suitable exponential family of full rank
with sufficient statistic $\eta(x)$ bounded in $x$. Further assume that $\phi_\KL$ exists and is unique.
Suppose each round of tuning starts at stationarity and there is $\kappa>0$ such that
$\Gap(\phi)\geq \kappa>0$ for all $\phi$ and $T_r=\Omega(2^r)$ as $r\to\infty$. Then, (1): 
$\hphi_r \to\phi_\KL$ almost surely as $r \to \infty$; and (2): for all
$0<\epsilon<\frac{1}{2}$, almost surely there exists an $R(\epsilon)$ such that
for all $r\geq R(\epsilon)$, 
$\|\E_{\hat\phi_{r}}[\eta] - \E_1[\eta]\| \leq T_r^{-\frac{1}{2}+\epsilon}.$ 
\end{theorem}

\subsection{Forward KL as a surrogate objective}
We now provide a general theoretical justification for the minimization of the forward KL
divergence as opposed to the global communication barrier, 
which is appealing as it enables a simple gradient-free moment matching procedure.
First, note that when $\pi_1 \in \calQ$, minimizing the KL divergence and the
GCB is equivalent since they are both divergences \cite{syed2019nrpt}.
\cref{thm:GCB_KL} generalizes this to the more usual case where 
$\pi_1 \notin \calQ$, demonstrating that the GCB at the forward KL minimum is bounded by quantities
that depend on the flexibility of the variational family. In particular, provided that there
exists a $\phi_0 \in \Phi$ such that the difference between log densities of
the target $\pi_1$ and reference $q_{\phi_0}$ is bounded by a function $g$,
then the GCB evaluated at the forward KL minimizer is bounded by a term
involving expectations of $g$ under the target and distributions $q_\phi$ that are close to $\pi_1$. 

\begin{theorem}[Forward KL proxy for the GCB]
\label{thm:GCB_KL}
Suppose that $\calQ=\{q_\phi : \phi \in \Phi\}$ is a PT-suitable exponential family of full rank.
Let $g$ be any function such that for some $\phi_0 \in \Phi$ and for all $x\in\mcX$, $|\log \pi_1(x) - \log q_{\phi_0}(x)| \leq g(x).$
Then, if $\phi_\text{KL} = \arg \min_\phi \text{KL}(\pi_1 \| q_\phi)$ 
exists and is unique, we have that
$\Lambda(q_{\phi_\text{KL}}, \pi_1) \leq \sqrt{\frac{1}{2}\left(\mathbb{E}_{1}[g] + \sup_{\phi \in \Phi'} \mathbb{E}_{\phi}[g]\right)},$
where $\Phi' = \{\phi : \text{KL}(\pi_1 \| q_\phi) \leq \text{KL}(\pi_1 \| q_{\phi_0})\}$.
\end{theorem}

We consider in \cref{sec:upper_bounds_GCB} two simple examples to verify that the upper bound 
given by \cref{thm:GCB_KL} is small enough for practical purposes.

\subsection{Stabilization with a fixed reference} \label{sec:stabilization}

In \cref{sec:variational_PT_alg} we provided 
a result (\cref{thm:adapt_tune}) guaranteeing convergence of the
adaptive scheme assuming the existence of an absolute spectral gap.  
In practice, the risk is that certain regions of $\Phi$ may
significantly degrade the absolute spectral gap under the basic variational scheme
discussed so far. For example, as shown in \cref{fig:demo} (middle row), if
the posterior is multimodal, the variational reference may quickly center on
one mode; because subsequent rounds of tuning use samples that depend on the
variational reference, these samples may largely come from that one mode,
causing the variational reference to remain trapped there for many tuning
rounds.  

To address this issue, we introduce parallel tempering with two
reference distributions, using both the original (fixed) reference and a
variational reference, which we call ``stabilized variational PT'', illustrated
in \cref{fig:reference_posterior}. 
We create an annealing path that starts at $q_\phi$, proceeds 
along an annealing path to $\pi_1$, and then moves on a new path 
from $\pi_1$ to $\pi_0$, connecting all three distributions. 
This modification adds significant robustness; as long as there are some restarts from 
the fixed reference, the target chain will escape the local optima and provide a more accurate 
estimate of $\pi_1$ used to tune $q_\phi$. In general, a well-tuned variational reference  can provide a
significant reduction in GCB compared to just a fixed path, but keeping the fixed reference ensures 
that the method will never do significantly worse than standard NRPT even if the variational 
reference tuning performs poorly (\cref{thm:restart_rate} below). Our variational PT algorithm
with two references is presented in \cref{alg:variational_PT}, in which \texttt{UpdateSchedule} refers to \cite[Algorithms 2, 3]{syed2019nrpt}.

To formalize the notion of a piecewise path, let $\pi_{\phi,\beta}$ be
the annealing path between $q_\phi$ and $\pi_1$, and let $\pi_\beta$ be the
annealing path between the fixed reference $\pi_0$ and $\pi_1$. We define the
concatenated (piecewise) path $\bar\pi_{\phi,\beta}$,
\[\label{eq:concatenation_path}
\bar{\pi}_{\phi,\beta} = 
\begin{cases}
\pi_{\phi,2\beta} & 0\leq \beta\leq\frac{1}{2},\\
\pi_{2-2\beta} & \frac{1}{2}\leq \beta\leq 1.
\end{cases}
\]
This new annealing path can be used within the NRPT \cref{alg:NRPT} as any other path. 
To tune the annealing schedule within each leg of PT with two 
references, we define the schedules $\calB_{\phi,N_\phi}=(\beta_{\phi,n})_{n=0}^{N_\phi}$ and $\calB_{N}=(\beta_n)_{n=0}^N$
for the legs connecting $q_\phi$ and $\pi_0$ to $\pi_1$, respectively. 
Then, we define the concatenated schedule 
$\bar\calB_{\phi,\bar{N}}=(\bar\beta_{\phi,n})_{n=0}^{\bar{N}}$ where $\barN=N_\phi+N$ and
\[\label{eq:concatenation_schedule}
    \bar{\beta}_{\phi,n}=
    \begin{cases}
    \frac{1}{2} \beta_{\phi,n} & 0\leq n \leq N_\phi\\
    1 - \frac{1}{2}\beta_{\barN-n} &  N_\phi\leq n\leq \bar{N}
    \end{cases}.
\]
This concatenated schedule $\bar\calB_{\phi,\barN}$ and path $\bar\pi_{\phi,\beta}$ are provided as input to the NRPT algorithm.

\begin{algorithm}[t!]
	\footnotesize
\caption{Variational PT}
\label{alg:variational_PT}
\begin{algorithmic}
\Require initial state $\bfx_0$, \# chains $\barN=2N$, \# total tuning rounds $R$, target $\pi_1$, reference family $\calQ=\{q_\phi : \phi \in \Phi\}$, initial reference parameter $\phi$, fixed reference $\pi_0$ 
\State $\calB_{\phi,N},\calB_N \gets (0, 1/N, 2/N, \ldots, 1)$ \Comment{Initialize annealing parameters uniformly with $N_\phi=N$}
\For{$r = 1,2,\ldots,R$}
  \State $T \gets 2^r$ \Comment{Double the number of iterations in the next tuning 
  round}
  \State $\bar{\pi}_{\phi,\beta} \gets \texttt{Concatenate}(\pi_{\phi,\beta},\pi_\beta)$ \Comment{Concatenate paths using \eqref{eq:concatenation_path}}
  \State $\bar{\calB}_{\phi,\barN} \gets \texttt{Concatenate}(\calB_{\phi,N},\calB_N)$ \Comment{Concatenate schedules using \eqref{eq:concatenation_schedule}}
  \State $(\bfx_t)_{t=1}^T, \, (r_n)_{n=0}^{\barN-1} \gets 
  \texttt{NRPT}(\bfx_0, \bar{\calB}_{\barN}, T, \bar\pi_{\phi,\beta})$ \Comment{PT with two references}
  \State $\calB_{\phi,N} \gets \texttt{UpdateSchedule}((r_n)_{n=0}^{N-1}, \calB_{\phi,N} )$ \Comment{Tune 
  annealing parameters for $\pi_\phi$ \cite{syed2019nrpt}}
  \State $\calB_N \gets \texttt{UpdateSchedule}((r_n)_{n=\bar{N}-1}^N, \calB_N )$ \Comment{Tune 
  annealing parameters for $\pi$ \cite{syed2019nrpt}}
  \State $\bfx_0 \gets \bfx_T$ \Comment{Initialization for next round}
  \State 
  	$\phi \gets \texttt{UpdateReference}((\bfx_t)_{t=1}^T)$ \Comment{Tune according to \cref{eq:update_reference} or another procedure}
\EndFor
\State \textbf{Return: }$(\bfx_t)_{t=1}^T$
\end{algorithmic}
\end{algorithm}

Finally, we provide an analysis of the worst-case performance of variational PT with
two reference distributions. We show that the asymptotic restart rate of PT with two references
is always greater than or equal to the restart rate with either one of the two references alone.
Because PT with two references requires twice the amount of computation, this amounts to a 
worst case of half the performance of regular PT with a fixed reference. In practice we 
often find that including a variational reference substantially improves the PT restart rate. 

Let $\bar{\tau}(\bar{\calB}_{\phi,\bar{N}})$ be the restart rate for $\pi_1$
for the concatenated path, i.e.\ the rate at which samples from either reference
$q_\phi$ or $\pi_0$ reach the target $\pi_1$. \cref{thm:restart_rate}
shows that if the Markov chain \emph{efficiently explores locally} (\cref{assump:ELE}), then
the restart rate of multiple-reference PT is the sum of the restart rate for
$\pi_1$ between $q_\phi$ and $\pi_0$ denoted $\tau_\phi(\calB_{\phi,N_\phi})$
and $\tau(\calB_N)$ respectively.

\begin{theorem}[Restart rate of NRPT with two reference distributions]
\label{thm:restart_rate}
Let $q_\phi,\pi_0$ be PT-suitable references for the target $\pi_1$. Suppose the PT chains with 
references $q_\phi,\pi_0$ with schedules $\calB_{\phi,N_\phi},\calB_{N}$ respectively efficiently explore 
locally (see \cref{assump:ELE}). Then 
$\bar\tau_\phi(\bar{\calB}_{\phi,\bar{N}})  
    =\tau_\phi(\calB_{\phi,N_\phi})+\tau(\calB_{N}).$ 
Moreover, if $\|\bar \calB_{\phi,\bar N}\| \to 0$, then
\[
\lim_{\bar{N}\to\infty}\bar\tau_\phi(\bar{\calB}_{\phi,\bar{N}})  
= \frac{1}{2 + 2 \Lambda(q_\phi, \pi_1)} + \frac{1}{2 + 2 \Lambda(\pi_0, \pi_1)}.
\]
\end{theorem}


\newcommand{\VPTF}{\texttt{VPT\_full}\xspace}
\newcommand{\VPTD}{\texttt{VPT\_diag}\xspace}
\newcommand{\NRPT}{\texttt{NRPT}\xspace}
\newcommand{\MSC}{\texttt{MSC}\xspace}

\section{Experiments}  
\label{sec:experiments}

We consider various Bayesian inference problems: 11  
based on real data, and 4 based on synthetic data 
(see \cref{tab:models} in \cref{sec:experiments_details} for the details of each). 
The range of problem settings considered include spatial statistics, 
Bayesian ODE parameter inference, 
phylogenetic inference, and several distinct Bayesian hierarchical models.  
In all examples the variational reference is a multivariate 
normal distribution with either a diagonal estimated covariance matrix (\VPTD)
or a full covariance matrix (\VPTF). 
The code for the experiments is made 
publicly available:
Julia code is available at \url{https://github.com/UBC-Stat-ML/VariationalPT} and 
Blang code is at \url{https://github.com/UBC-Stat-ML/bl-vpt-nextflow}.
A distributed implementation is also under development at 
\url{https://github.com/Julia-Tempering/Pigeons.jl}. 
Experimental details can be found in \cref{sec:experiments_details}.

\subsection{Comparative efficiency of variational PT families and a PT baseline}
\label{sec:VPT_versus_NRPT}

\begin{figure}[t]
    \centering
    \begin{subfigure}{0.325\textwidth}
      \centering
      \stackinset{l}{0.135\textwidth}{t}{0.06125\textwidth}
        {\includegraphics[width=0.4\textwidth]{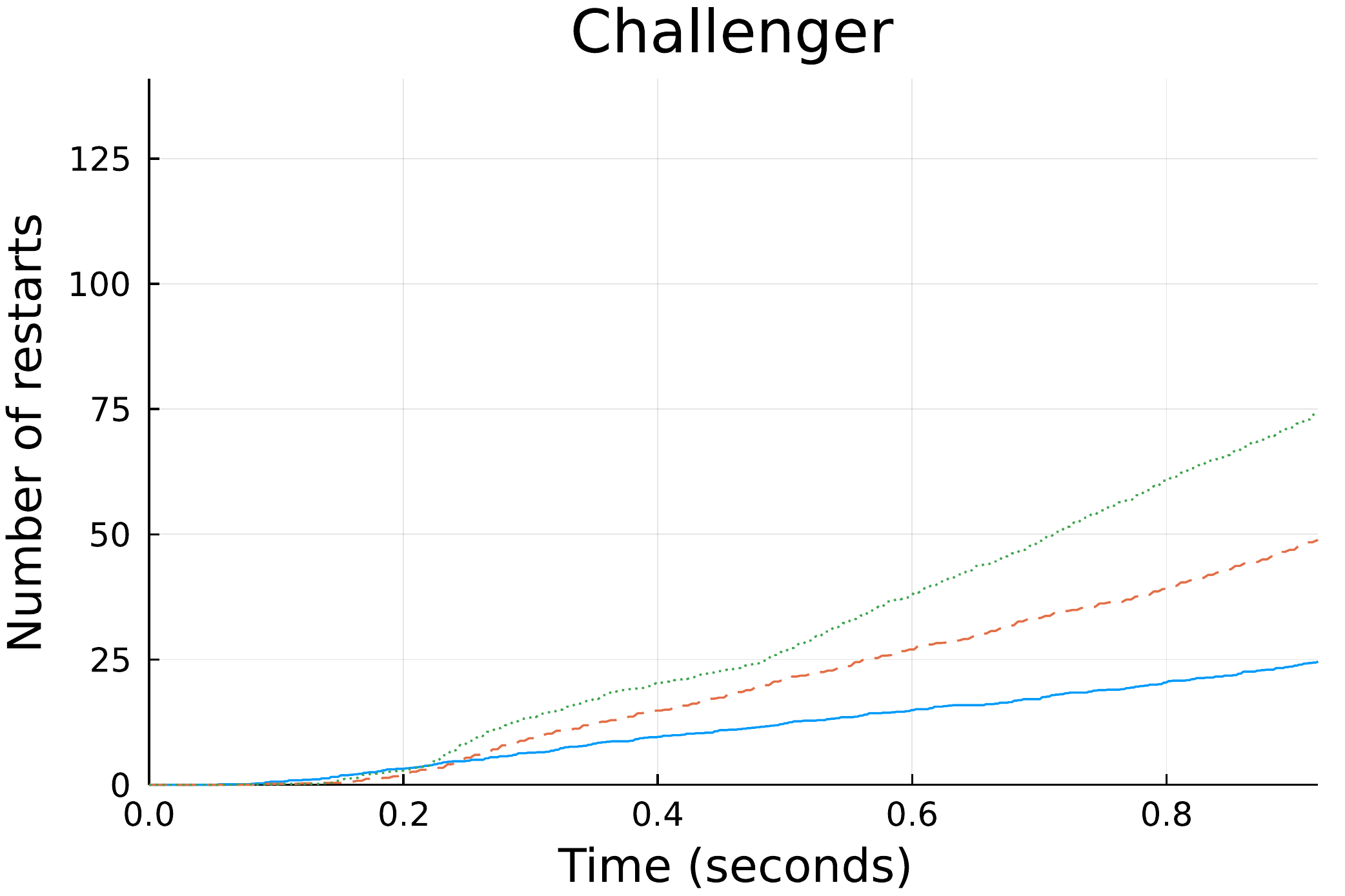}}
        {\includegraphics[width=\textwidth]{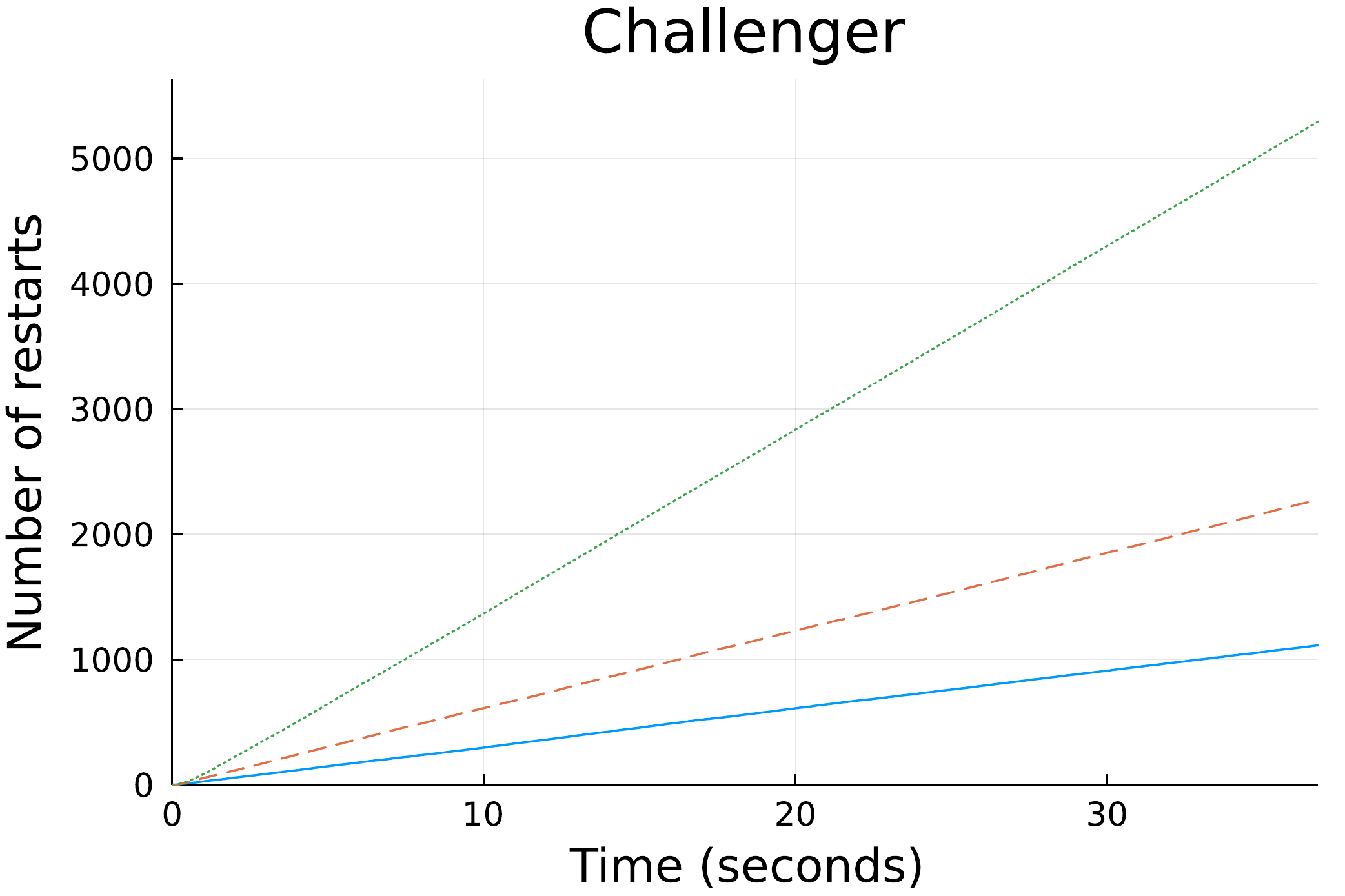}}
    \end{subfigure}
    \begin{subfigure}{0.325\textwidth}
      \centering
      \stackinset{l}{0.135\textwidth}{t}{0.06125\textwidth}
      {\includegraphics[width=0.4\textwidth]{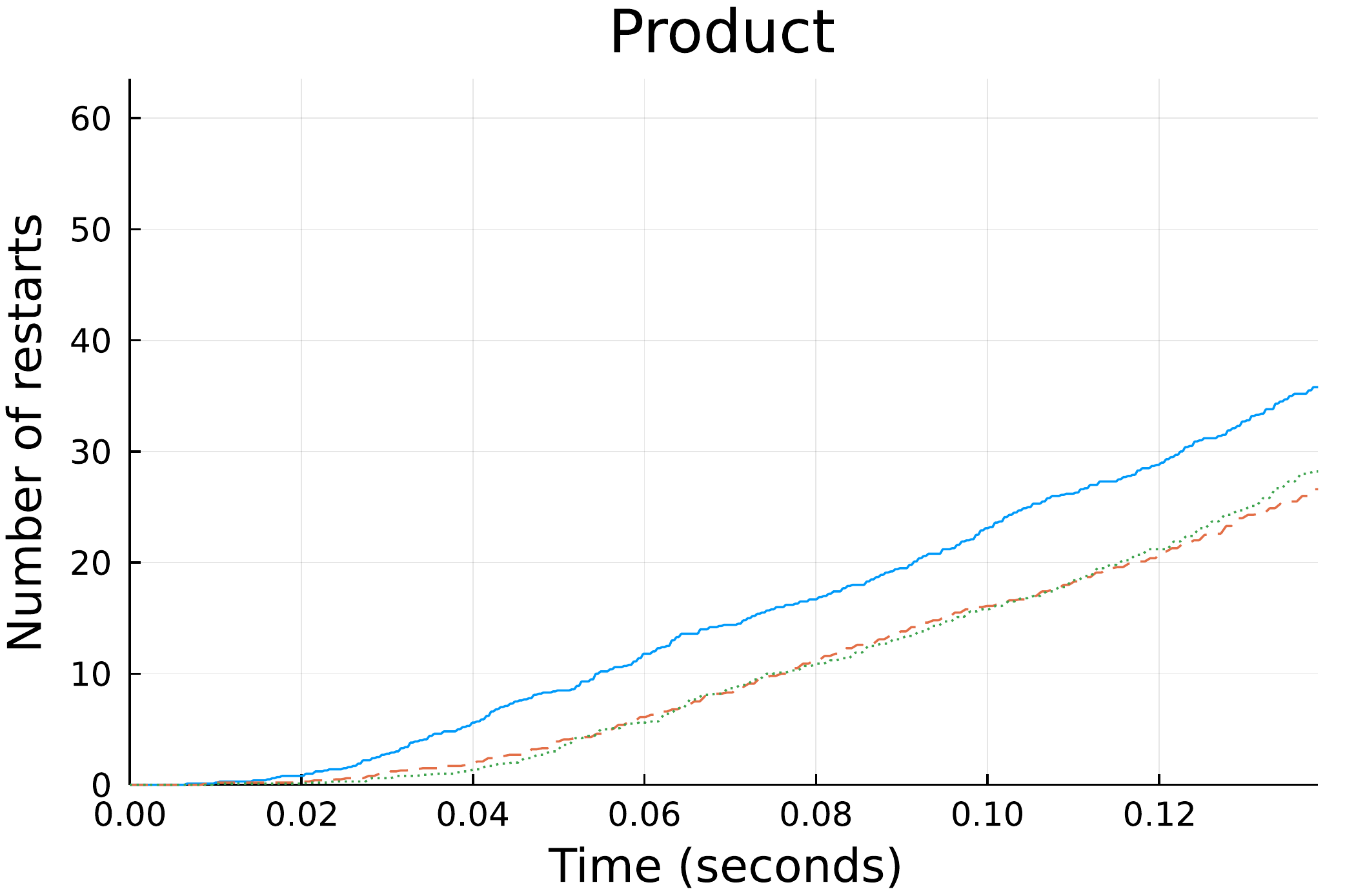}}
      {\includegraphics[width=\textwidth]{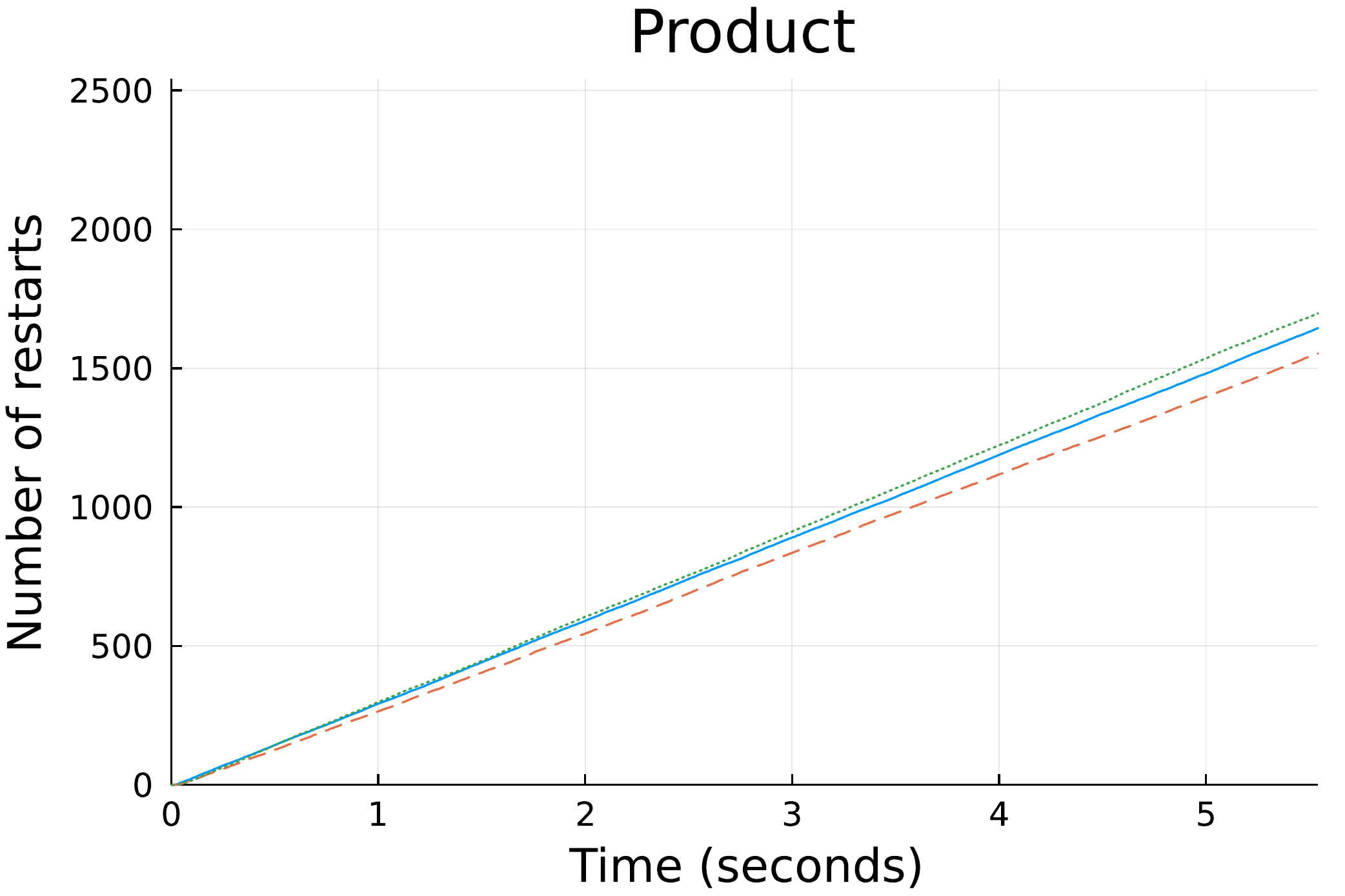}}
    \end{subfigure}
    \begin{subfigure}{0.325\textwidth}
      \centering
      \stackinset{l}{0.135\textwidth}{t}{0.06125\textwidth}
      {\includegraphics[width=0.4\textwidth]{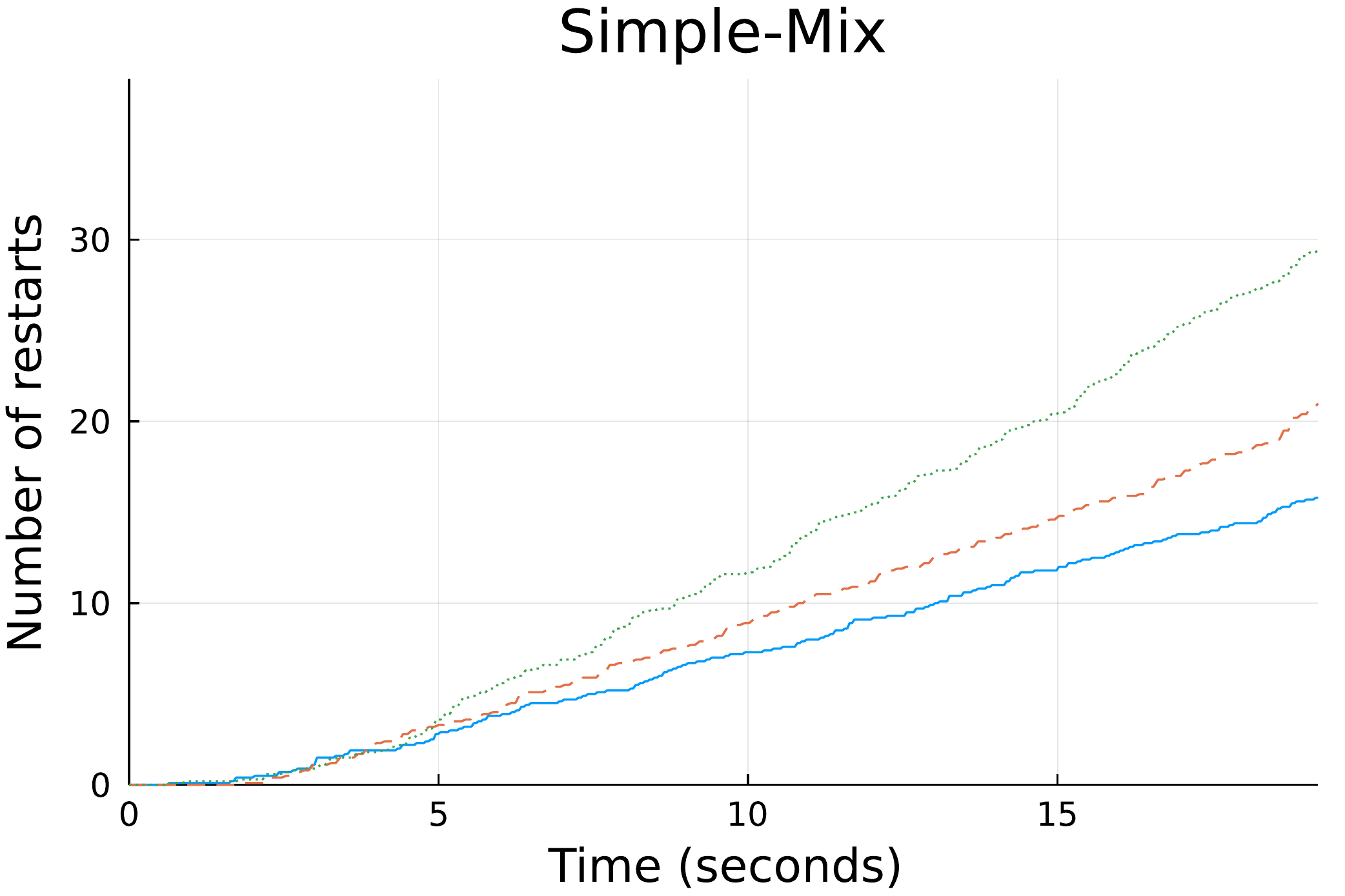}}
      {\includegraphics[width=\textwidth]{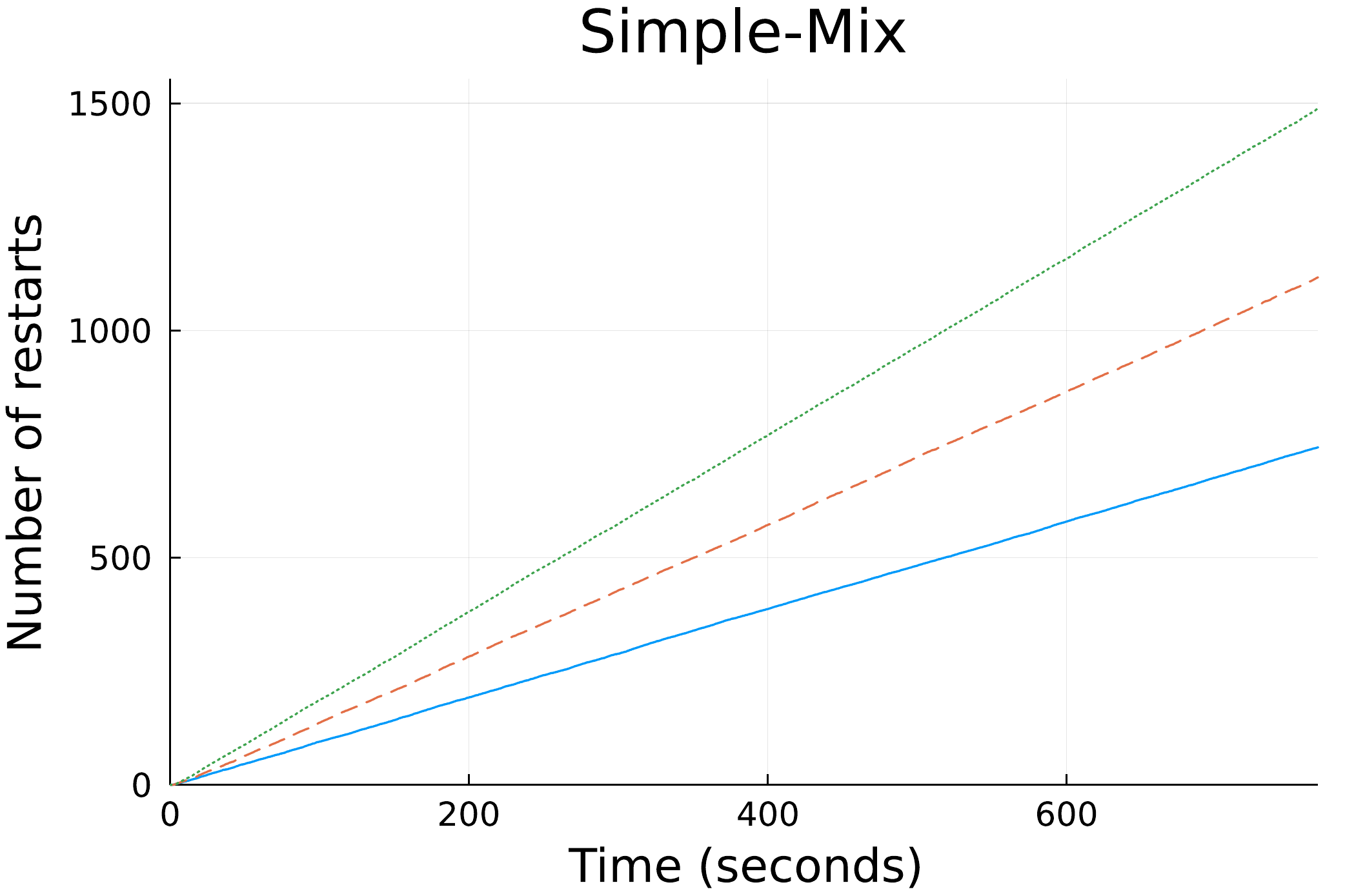}}
    \end{subfigure}
    \begin{subfigure}{0.325\textwidth}
      \centering
      \stackinset{l}{0.135\textwidth}{t}{0.06125\textwidth}
      {\includegraphics[width=0.4\textwidth]{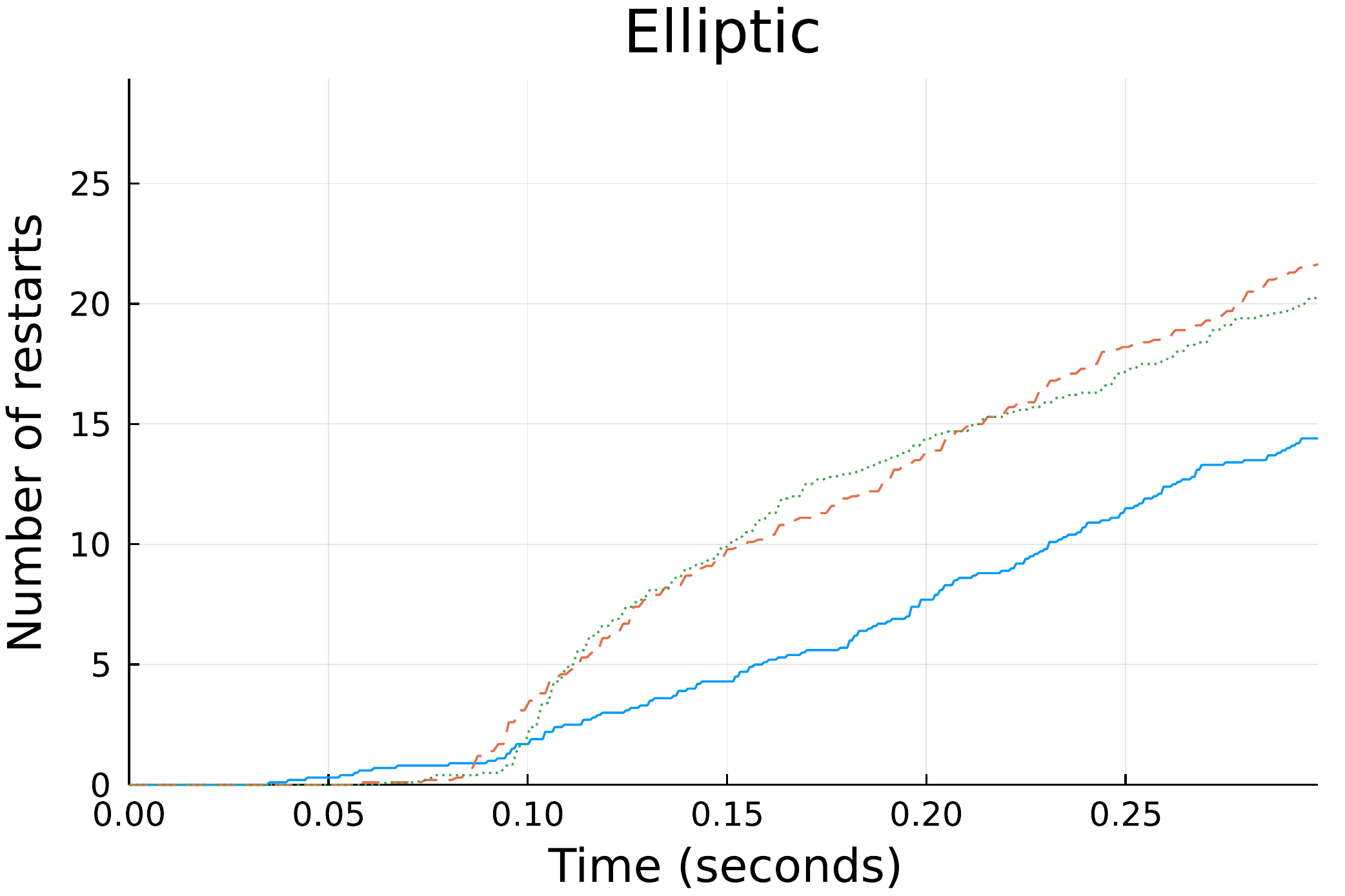}}
      {\includegraphics[width=\textwidth]{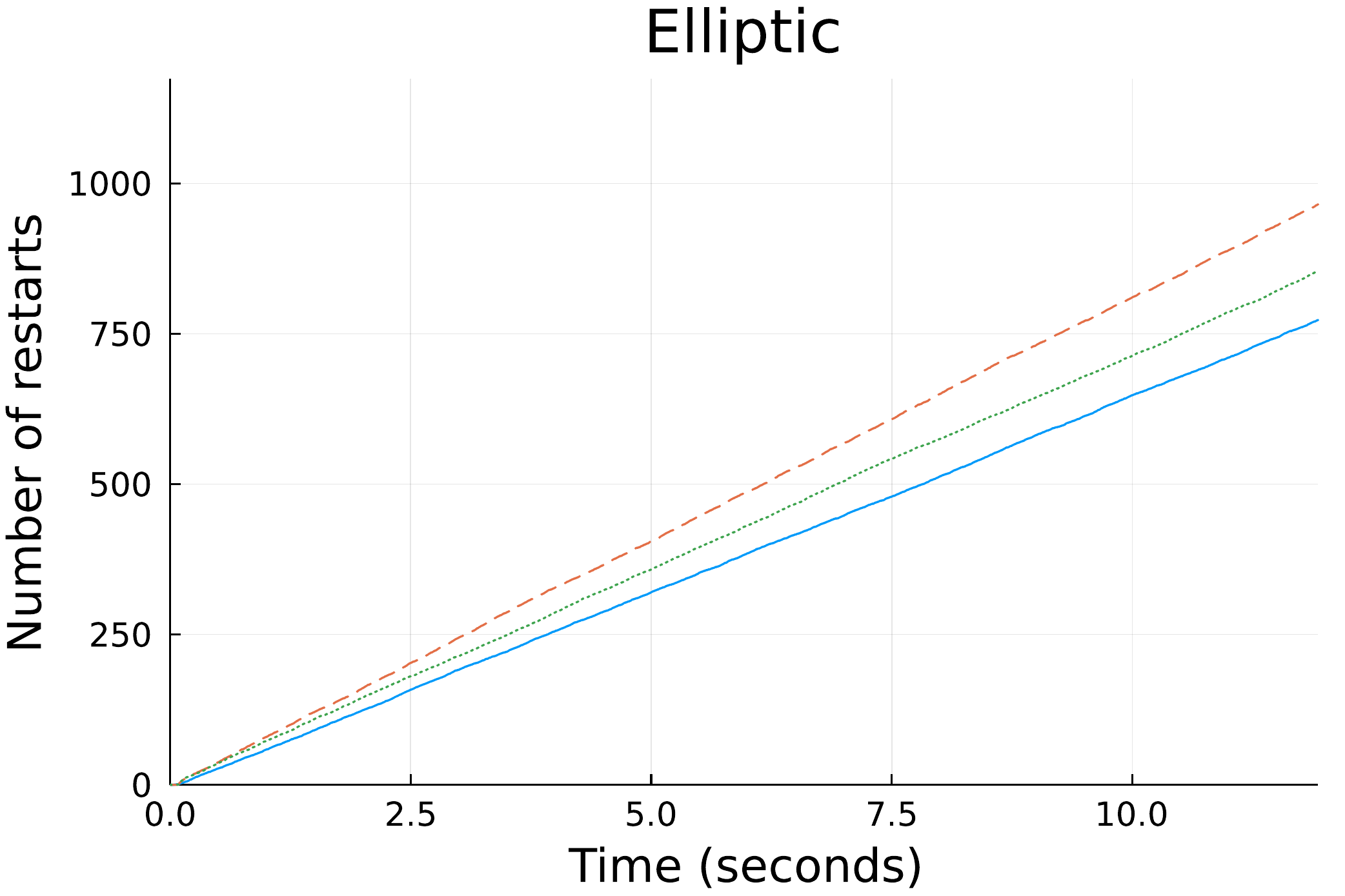}}
    \end{subfigure}
    \begin{subfigure}{0.325\textwidth}
      \centering
      \stackinset{l}{0.125\textwidth}{t}{0.06125\textwidth}
      {\includegraphics[width=0.4\textwidth]{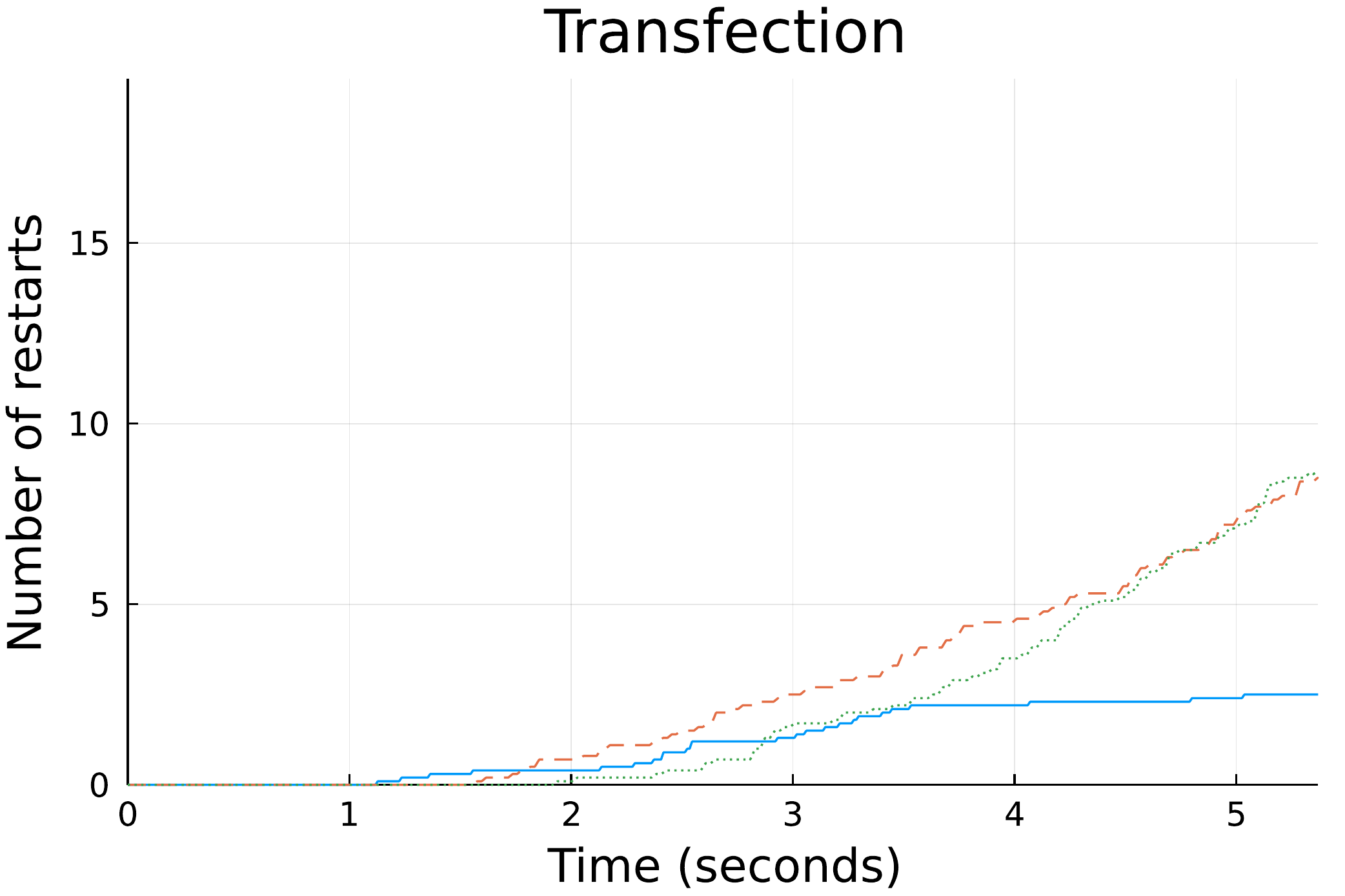}}
      {\includegraphics[width=\textwidth]{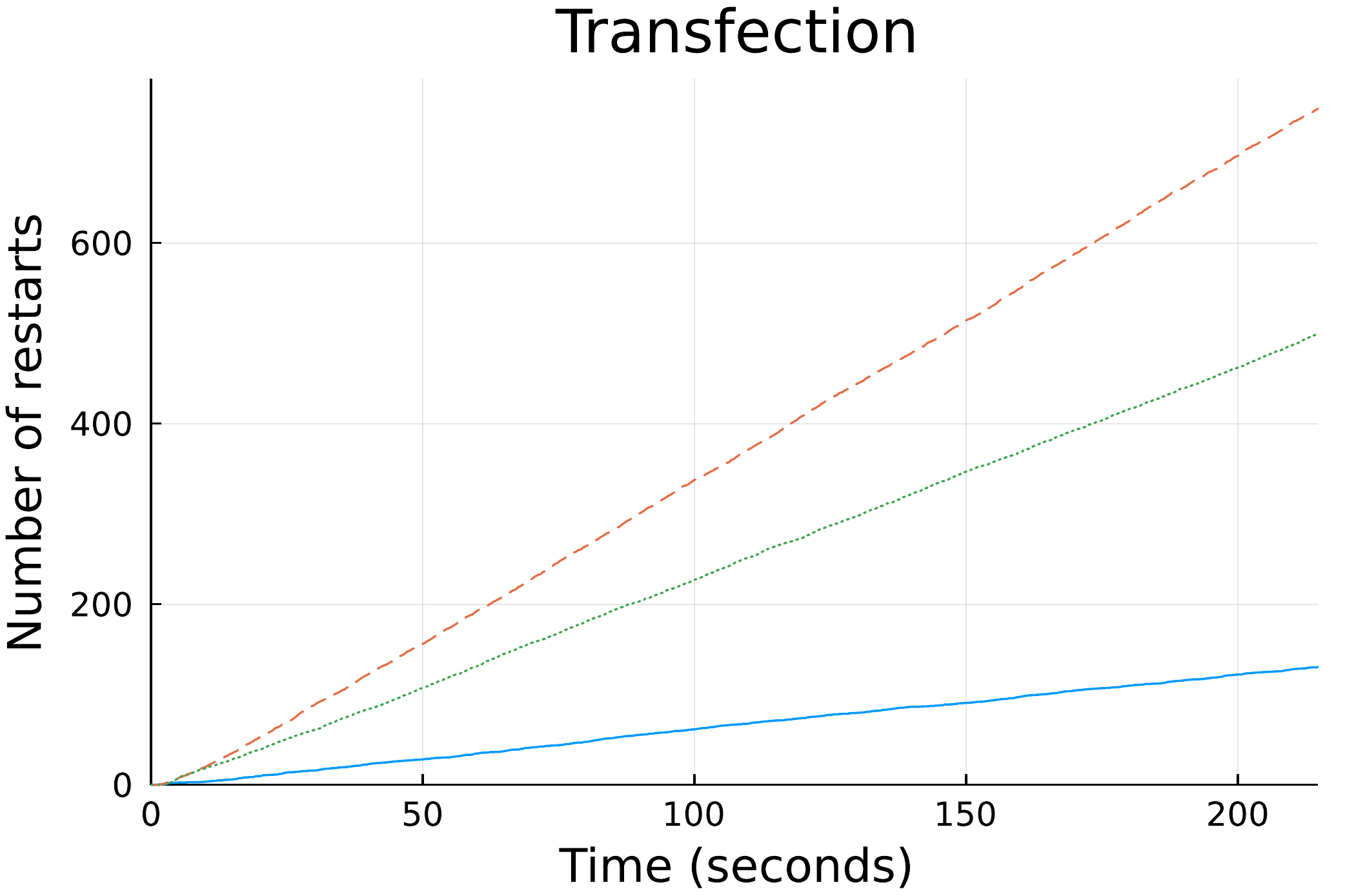}}
    \end{subfigure}
    \begin{subfigure}{0.325\textwidth}
      \centering
      \stackinset{l}{0.135\textwidth}{t}{0.06125\textwidth}
      {\includegraphics[width=0.4\textwidth]{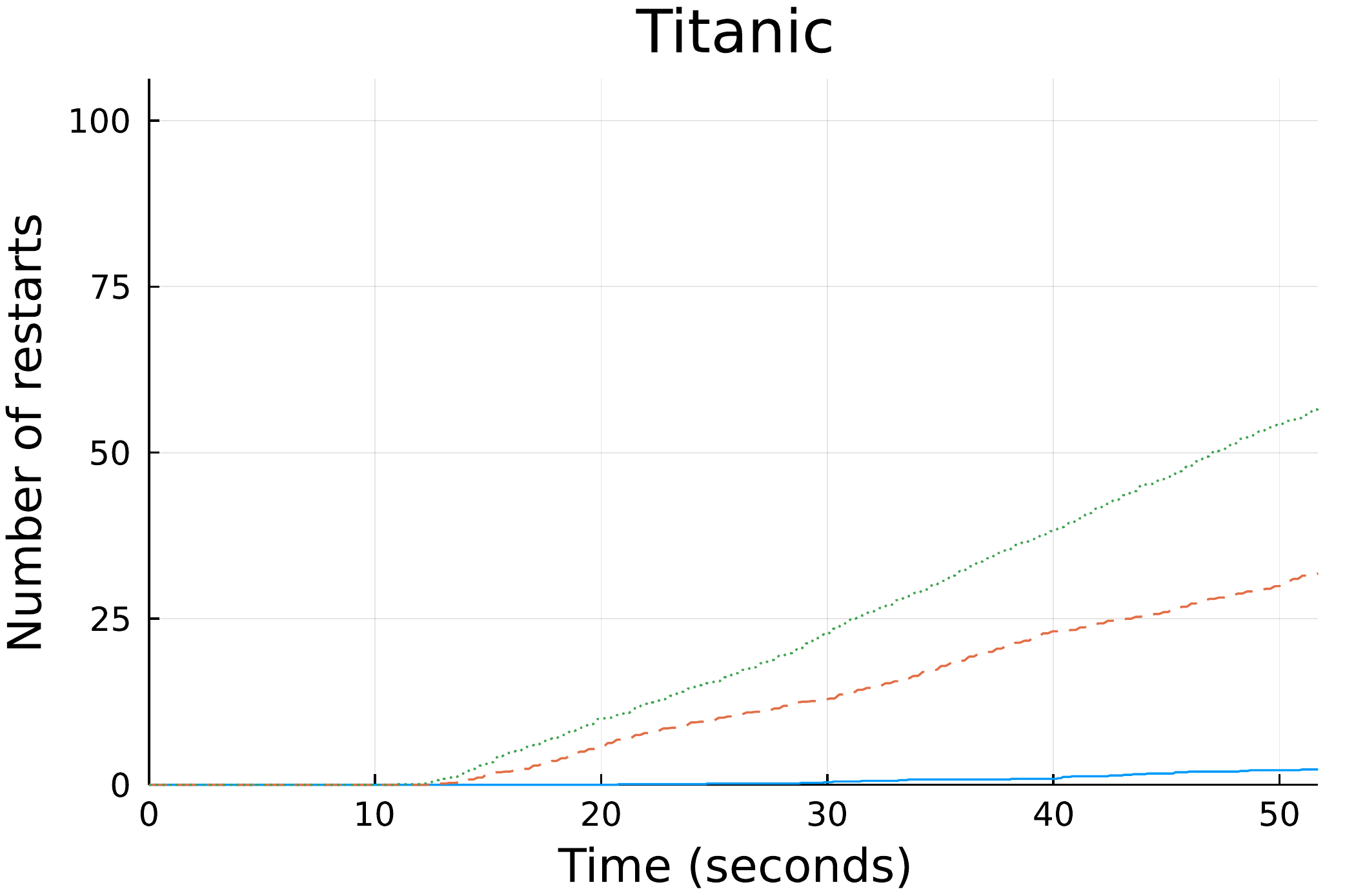}}
      {\includegraphics[width=\textwidth]{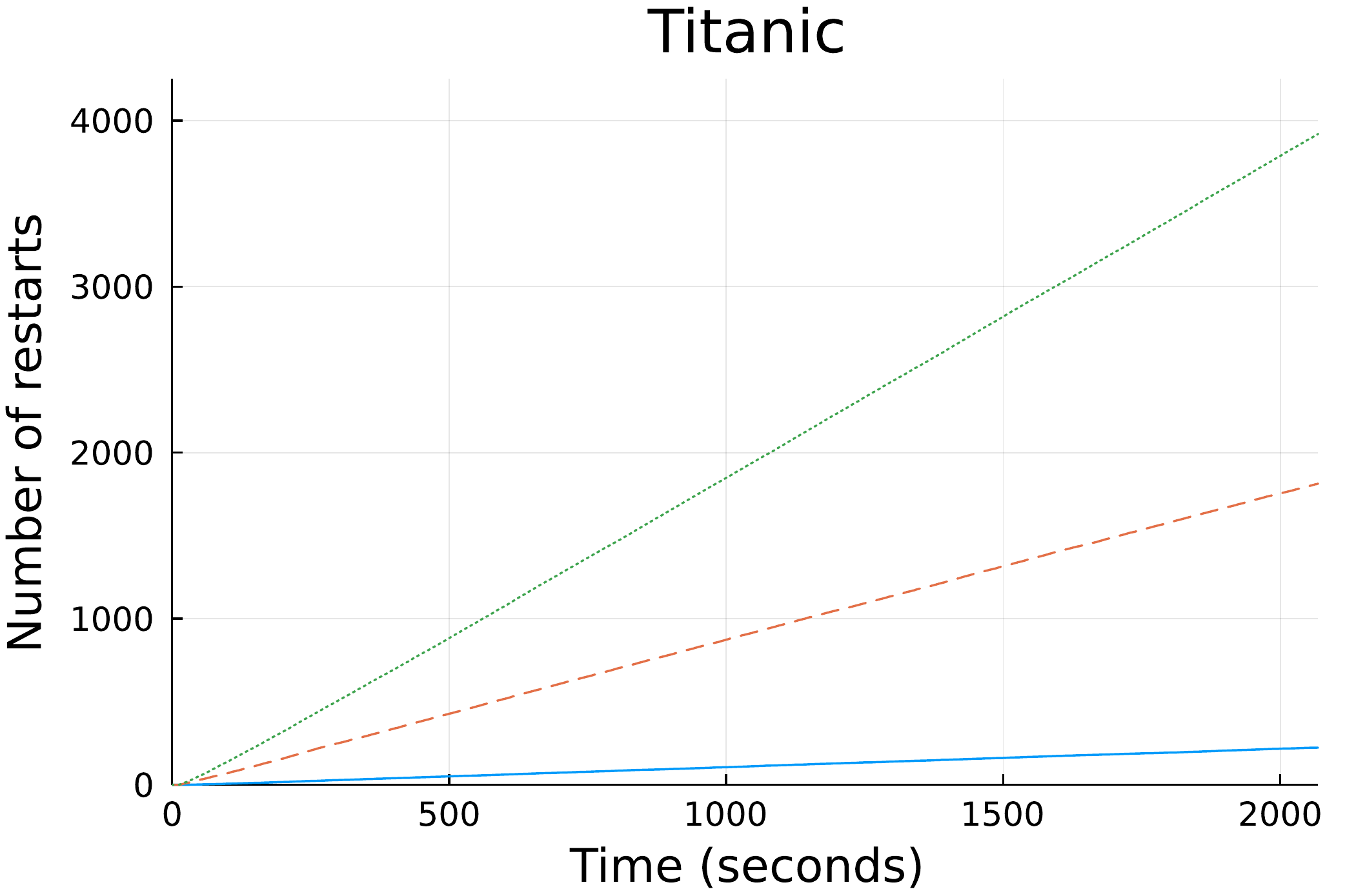}}
    \end{subfigure}
    \caption{{\footnotesize Number of restarts (higher is better) versus computation time in seconds for several models,
    with {\color{LimeGreen} \VPTF} in green, {\color{Bittersweet} \VPTD} in red, and the {\color{Cerulean} \NRPT} baseline in blue. The two variational PT methods generally provide a comparable or better
    rate of restarts per second. Insets highlight the initial 2.5\% of computation time, demonstrating that tuning of the variational references stabilizes quickly.}}
    \label{fig:sumroundtrips}
\end{figure}

We begin by comparing the communication efficiency of both
\VPTD and \VPTF to a
state-of-the-art existing PT method, \NRPT \cite{syed2019nrpt},
which uses a single, fixed reference.
Note that there is also a computational trade-off between the two variational families (but we remind the reader that both still yield convergence of the target chain to the posterior). In particular,
\texttt{VPT\_full} offers more flexibility---and hence a potentially 
lower GCB---at the cost of a higher computational cost per iteration and 
more variational parameters to fit, while \VPTD has
the same asymptotic computational cost per iteration as standard PT methods.
We explore this tradeoff in \cref{fig:sumroundtrips}.

We observe that both \VPTD and \VPTF often substantially improve the restart rate
compared to the \NRPT baseline, and at worst achieve similar performance.
The choice between \VPTF and \VPTD depends on the problem: for example, 
in low-dimensional problems such as \texttt{Challenger} and \texttt{Simple-mix}, we
observe that the full covariance in \VPTF is worth its additional cost per iteration. 
The situation is reversed in the \texttt{Transfection} problem. 
If one is pressed to select one PT variational family, we recommend \VPTD as a safe default in light of \cref{thm:restart_rate} and of its computational cost per iteration asymptotically equivalent to \NRPT.

\begin{figure}[t]
	\centering
	\begin{subfigure}[b]{0.49\textwidth}
		\centering
		\caption{Convergence of GCB estimates}
		\includegraphics[width=0.49\textwidth]{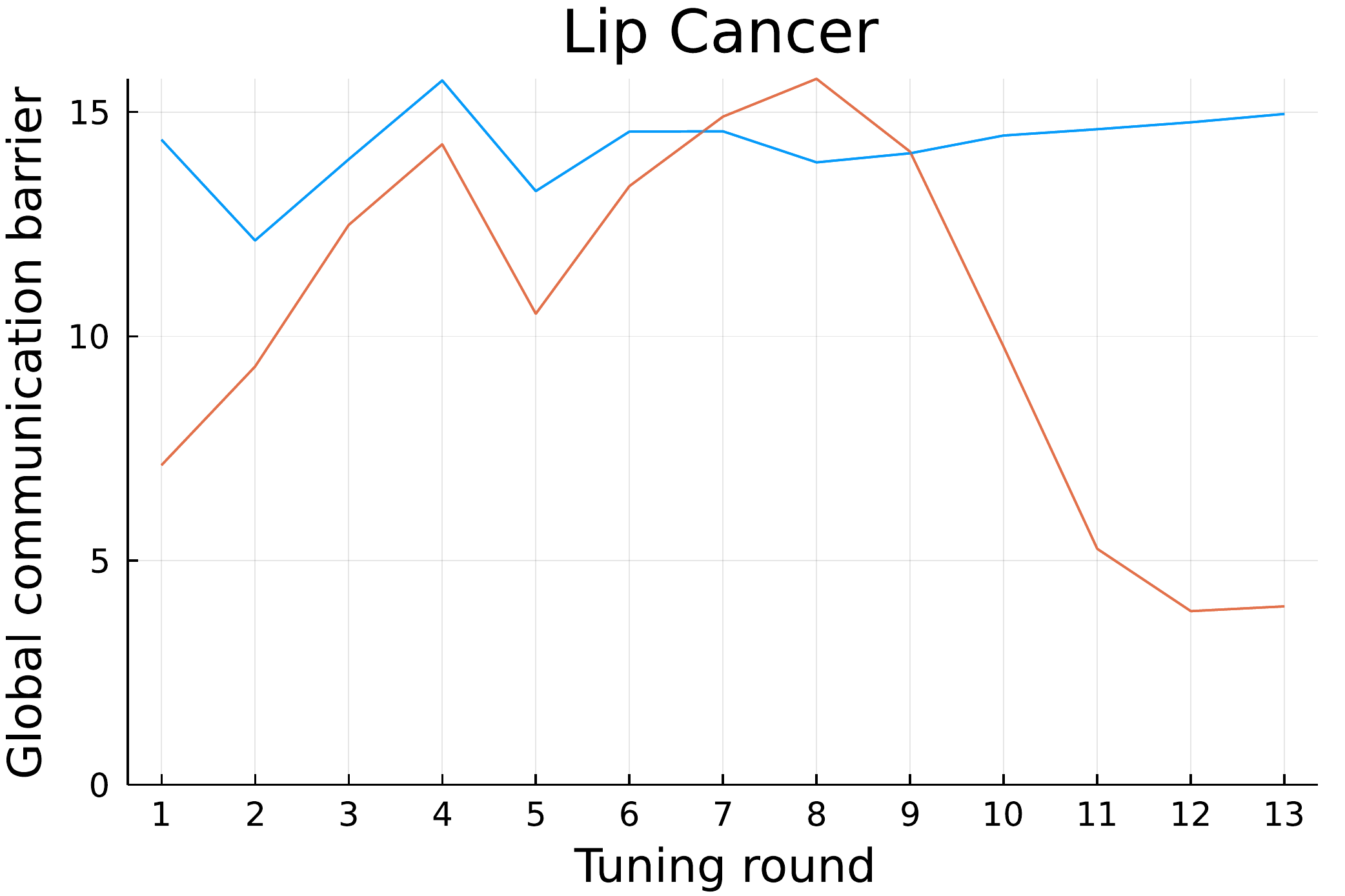}
		\includegraphics[width=0.49\textwidth]{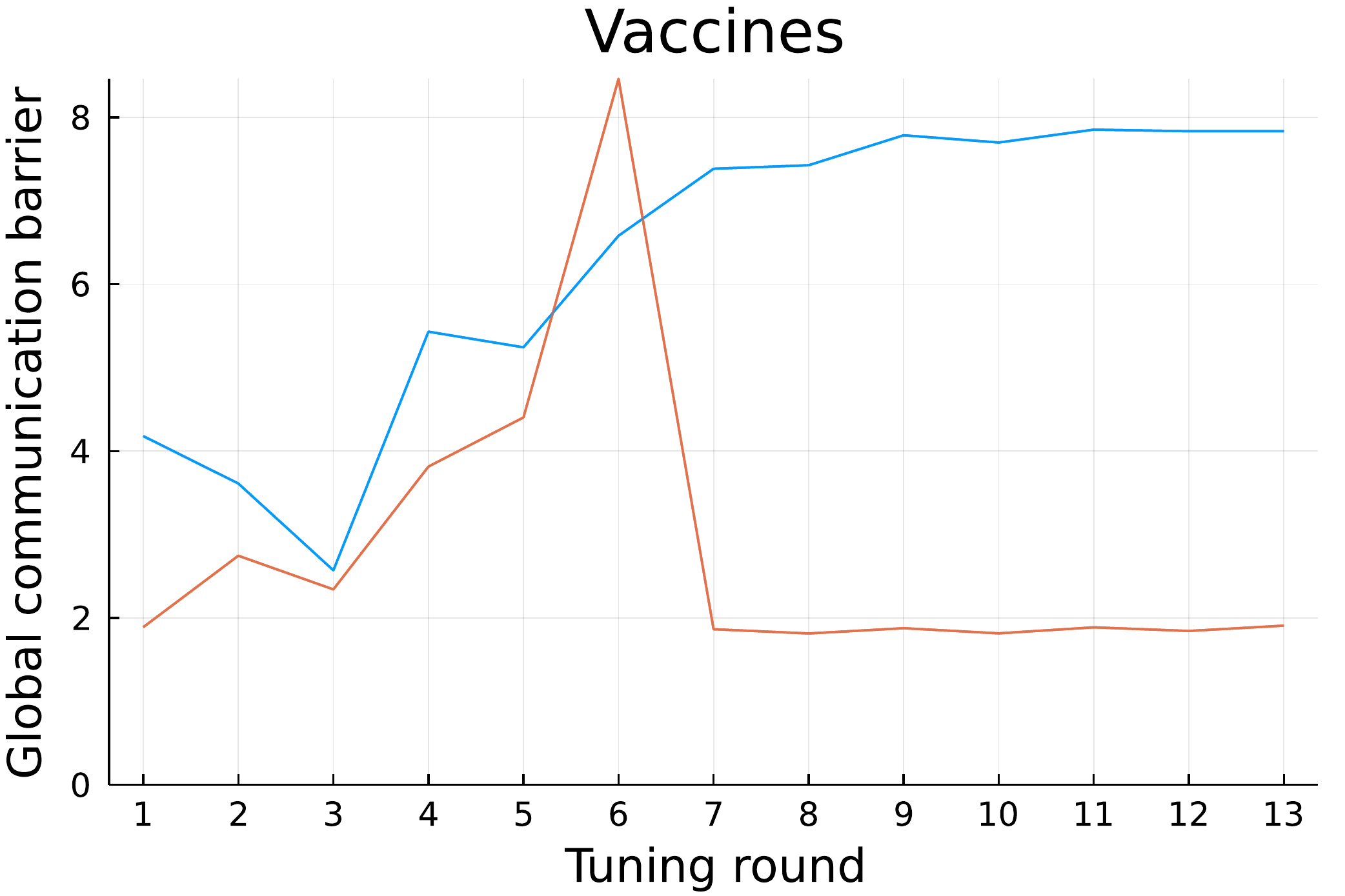}
	\end{subfigure}
	\begin{subfigure}[b]{0.49\textwidth}
		\centering
		\caption{Markovian score climbing results}
		\includegraphics[width=0.49\textwidth]{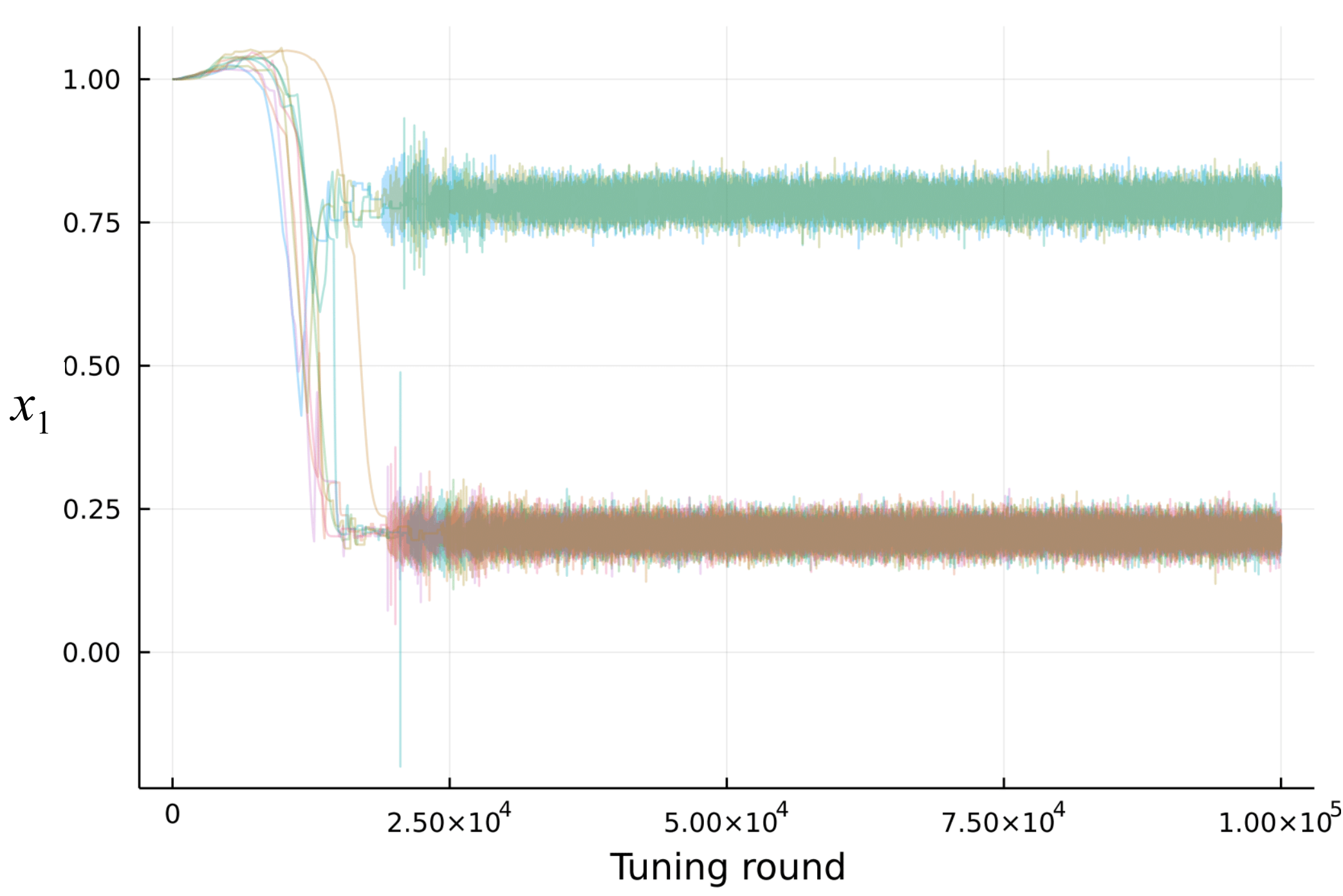}
		\includegraphics[width=0.49\textwidth]{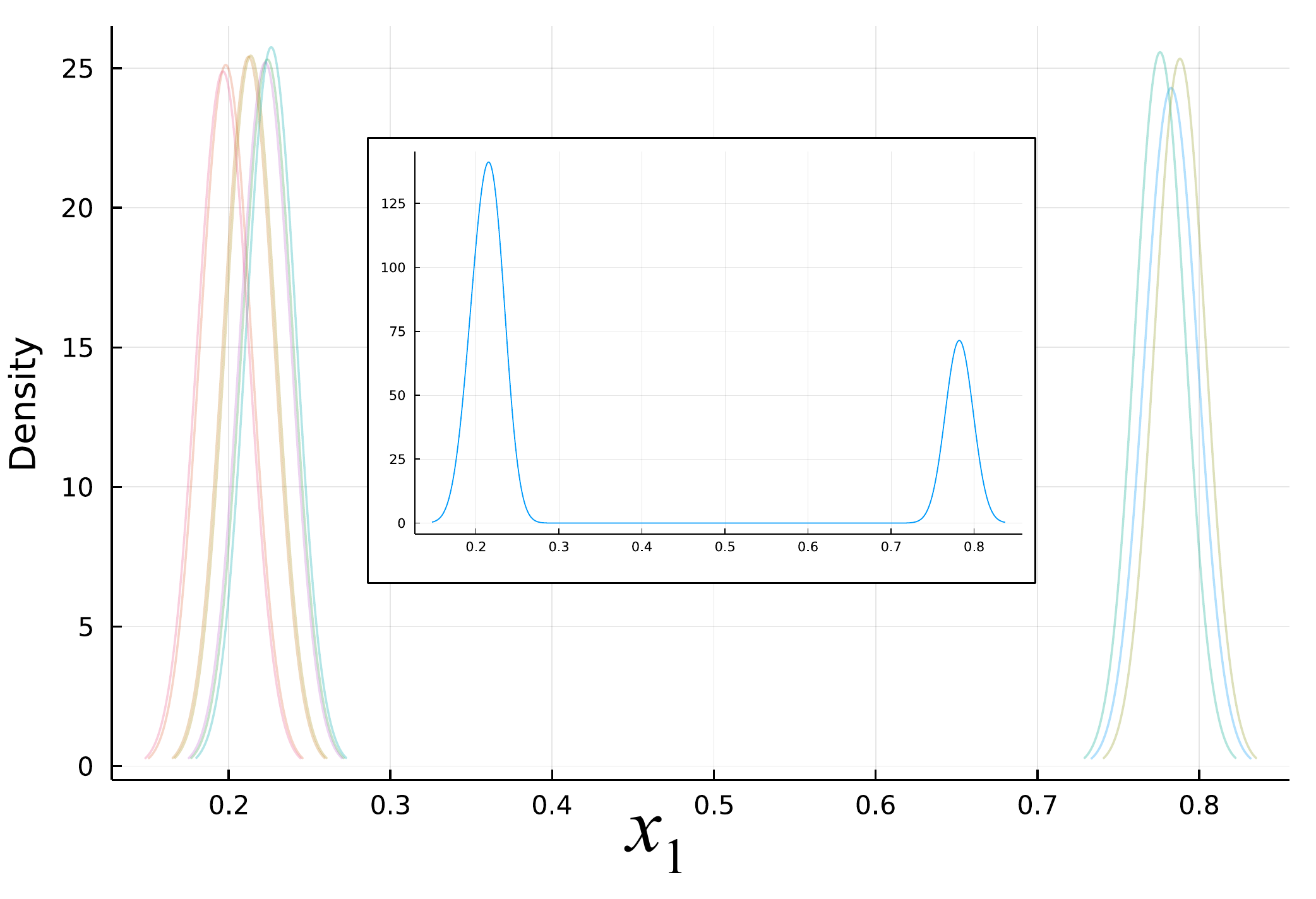}
	\end{subfigure}
	\caption{{\footnotesize \textbf{(a)} GCB for {\color{Bittersweet} \VPTD} (red)
	        and {\color{Cerulean} \NRPT} (blue) versus tuning round. \textbf{(b, left)} Tuning 
			variational parameters using \MSC with 10 replications (colours). 
			The mean of the variational distribution for the model parameter $x_1$  
			in the \texttt{Transfection} model is presented. The true marginal posterior 
			distribution of $x_1$ is bimodal (\cref{fig:demo}). We see here 
			that \MSC chooses one of the two modes for the estimation of 
			the mean parameter. \textbf{(b, right)} Variational Gaussian 
			approximation of
			the same parameter in the \texttt{Transfection} model produced by 10 
			\MSC runs with different seeds (colours). The aggregate of the 
			different runs is shown as an inset, cf.\ \cref{fig:demo}.}}
	\label{fig:GCB}
\end{figure}

We also note that the tuning procedure converges relatively quickly. We show the
number of restarts for the first 2.5\% of computation time as insets in \cref{fig:sumroundtrips},
and find that the number of restarts for the three methods can be
distinguished early on. 
We also show in \cref{fig:GCB}(a) that the GCB estimates converge in a small number
of rounds for two additional representative problems (\texttt{Lip Cancer} 
and \texttt{Vaccines}). Note also that the GCB
is substantially lower when a variational reference is introduced.

\subsection{Moment matching outperforms stochastic optimization}
\label{sec:MM_SGD}

\begin{figure}[t]
    \centering
    \includegraphics[width=\textwidth]{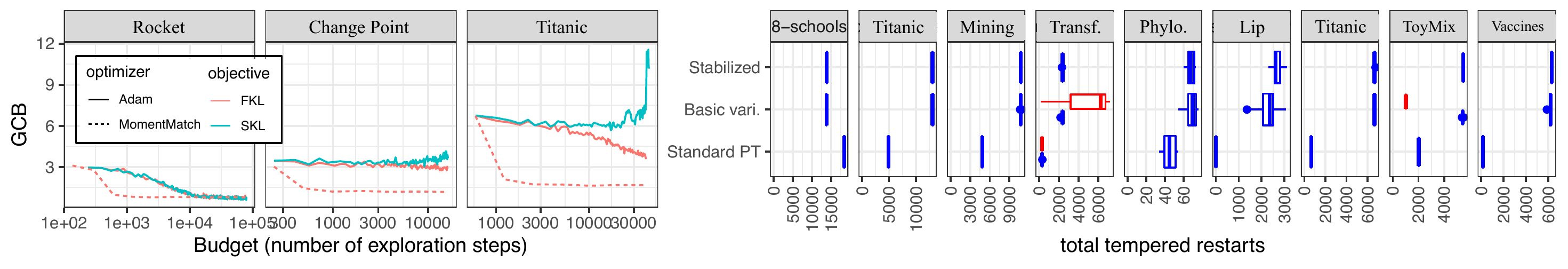}
    \caption{{\footnotesize \textbf{Left:} comparison of stochastic 
    		optimization and moment matching. An optimizer parameter setting
    		that works well for the \texttt{Rocket} problem 
    		(Adam+FKL/SKL, step size scale 0.1) does not generalize well to other 
    		problems (\texttt{Change-Point}, \texttt{Titanic}).
    		In contrast, moment matching reliably finds a well-tuned reference in 
    		all 14 problems considered in this paper. \textbf{Right:}
    		the same plot as in \cref{fig:demo}, but on a larger selection of models.}}
    \label{fig:topos-box-condensed}
\end{figure}

Next, we compare the proposed moment matching procedure described in \cref{sec:variational_PT}
to several other stochastic optimization schemes to tune the variational reference. In
contrast to moment matching---which is free of tuning hyper-parameters---we
show in \cref{sec:additional-stoch-gradient-vs-moment} that 
stochastic optimization schemes require extensive hyper-parameter tuning, 
including step size schedule, choice of optimizer, and surrogate objective function. 
Moreover, we show in \cref{fig:topos-box-condensed} (left) that it is difficult
to specify a ``default'' hyper-parameter setting for stochastic optimization methods;
a setting that works well on one problem (\texttt{Rocket}) generally will not work
well on other problems (e.g., \texttt{Change Point}, \texttt{Titanic}).
In contrast, moment matching performed well on all 14 problems we considered 
without requiring tuning. 
Consequently, our moment matching procedure is a better candidate for 
integration in probabilistic programming languages (PPLs), in which users do not 
expect to be required to frequently change algorithmic tuning hyper-parameters.
To illustrate this point, we have extended an existing open source PPL, Blang \cite{Bouchard2021Blang},
to include our method tuned via moment matching (code available at 
\url{https://github.com/UBC-Stat-ML/bl-vpt}).

\subsection{Comparison to an externally tuned reference}
\label{sec:msc}

We also compare our moment-matching variational reference tuning procedure to
a reference tuned by an existing procedure outside of the PT context.
In particular, we tested Markovian score 
climbing (\MSC) \cite{naesseth_markovian_2020}---which also optimizes the forward
KL---for tuning a Gaussian reference with a diagonal covariance. 
Details of \MSC tuning, including sensitivity to stochastic optimization settings, 
can be found in \cref{sec:experiments_supplement}.
The results for the \texttt{Transfection} model are presented in \cref{fig:GCB}.
In contrast to our stabilized moment matching approach (\cref{fig:demo}), tuning
the reference using \MSC in this example results in systematic catastrophic
forgetting of one of the modes (\cref{fig:GCB} (b)) in all 10 replicates.
We additionally refer readers to \cite{kim2022msc} for recent developments in score-based 
methods to minimize the forward KL divergence.

\subsection{Comparison of variational PT topologies}
\label{sec:comparison-variational-pt-strategies}

Finally, we compare the three algorithms introduced in
\cref{fig:reference_posterior} (\texttt{Basic Variational PT}, \texttt{Stabilized Variational PT} and \NRPT) on nine Bayesian inference problems (see \cref{fig:topos-box-condensed}, right). 
We set up the
algorithms so that the runtime per PT iteration has the same asymptotic complexity: this
is done by selecting a diagonal covariance matrix for the Gaussian variational
family and using the same total number of chains for all methods.

The results confirm the initial findings of \cref{fig:demo}: only 
the stabilized method always avoids catastrophic forgetting of modes.  
Moreover, in all but one example considered, we find that \texttt{Stabilized Variational PT} 
exhibits improved performance in terms of the number of restarts
compared to the \NRPT baseline (\cref{fig:topos-box-condensed}). The exception is the
\texttt{8-schools} problem, in which the posterior is not well approximated by a diagonal Gaussian. Even in this case, 
as suggested by \cref{thm:restart_rate}, the performance does not degrade
by more than a factor two. At the other end of the spectrum, for the \texttt{Vaccines} 
hierarchical model, the number of restarts increases $>$40-fold compared to
the \NRPT baseline, and for the spatial sparse conditional auto-regressive (CAR) model applied to the \texttt{Lip Cancer} problem, 
the performance jumps from zero restarts to $>$2300 restarts.

More results and details can be found in \cref{sec:additional-topo-comparisons}, including
alternative topological arrangements of variational PT algorithms, effective
sample size per second results, as well as global communication
barriers for the problems considered in this section.

\section{Conclusion}
This paper addressed sampling from a complex target distribution within the parallel tempering framework
by constructing a generalized annealing path connecting the posterior to an adaptively tuned variational
reference. Experiments in a wide range of realistic Bayesian inference scenarios demonstrate the large empirical
gains achieved by our method. Potential future work includes extending the gradient-free tuning methodology
to larger classes of variational families for the reference distribution.
Further, heavy-tailed distributions can violate \cref{def:regular_reference} and it is 
not clear without further examination what the implications are on the convergence
of the proposed algorithms.
Therefore, another possible direction for future work is to develop methods suitable for
heavy-tailed target distributions.

\section*{Acknowledgements}
NS acknowledges the support of a Vanier Canada 
Graduate Scholarship. 
ABC and TC acknowledge the support of an NSERC Discovery Grant.
We also acknowledge use of the ARC Sockeye computing platform from the
University of British Columbia.

\bibliographystyle{unsrtnat}
\bibliography{main.bib}

\section*{Checklist}


\begin{enumerate}

\item For all authors...
\begin{enumerate}
  \item Do the main claims made in the abstract and introduction accurately reflect the paper's contributions and scope?
    \answerYes{See the theoretical results (\cref{sec:variational_PT}) and the experiments (\cref{sec:experiments}).}
  \item Did you describe the limitations of your work?
    \answerYes{Limitations of the work are discussed in \cref{sec:stabilization}, where we assess the worst-case performance of the proposed method. We also note that chosen simulation examples are representative of various scenarios, including ones in which our method is comparable to a competitor.}
  \item Did you discuss any potential negative societal impacts of your work?
    \answerNA{}
  \item Have you read the ethics review guidelines and ensured that your paper conforms to them?
    \answerYes{}
\end{enumerate}

\item If you are including theoretical results...
\begin{enumerate}
  \item Did you state the full set of assumptions of all theoretical results?
    \answerYes{Theoretical assumptions are described and referenced in the theorem statements of \cref{sec:variational_PT}. They are laid out in full in \cref{sec:theoretical_conditions} and \cref{sec:largedataasymp}.}
        \item Did you include complete proofs of all theoretical results?
    \answerYes{Complete proofs are provided in Appendices A-E.}
\end{enumerate}

\item If you ran experiments...
\begin{enumerate}
  \item Did you include the code, data, and instructions needed to reproduce the main experimental results (either in the supplemental material or as a URL)?
    \answerYes{We include our code in public GitHub repositories. These repositories also contain the data or 
    instructions for obtaining the data.}
  \item Did you specify all the training details (e.g., data splits, hyperparameters, how they were chosen)?
    \answerYes{Experiments are described in detail in \cref{sec:experiments_details}.}
        \item Did you report error bars (e.g., with respect to the random seed after running experiments multiple times)?
    \answerYes{Instead of error bars, results of experiments that were repeated multiple times were displayed graphically to allow for approximate visual estimates of uncertainty. See, for example, \cref{fig:sumroundtrips}.}
        \item Did you include the total amount of compute and the type of resources used (e.g., type of GPUs, internal cluster, or cloud provider)?
    \answerYes{The description of the resources used is provided in \cref{sec:experiments_details}.}
\end{enumerate}

\item If you are using existing assets (e.g., code, data, models) or curating/releasing new assets...
\begin{enumerate}
  \item If your work uses existing assets, did you cite the creators?
    \answerYes{See \cref{sec:experiments_details} for the list of models used and citations for the data sets.}
  \item Did you mention the license of the assets?
    \answerYes{Licenses are mentioned in \cref{sec:experiments_details}, which summarizes the models and data sets used.}
  \item Did you include any new assets either in the supplemental material or as a URL?
    \answerYes{We include our code in public GitHub repositories. These repositories also contain the data or 
    instructions for obtaining the data.}
  \item Did you discuss whether and how consent was obtained from people whose data you're using/curating?
    \answerYes{Licenses are described in \cref{sec:experiments_details}.}
  \item Did you discuss whether the data you are using/curating contains personally identifiable information or offensive content?
    \answerNA{}
\end{enumerate}

\item If you used crowdsourcing or conducted research with human subjects...
\begin{enumerate}
  \item Did you include the full text of instructions given to participants and screenshots, if applicable?
    \answerNA{}
  \item Did you describe any potential participant risks, with links to Institutional Review Board (IRB) approvals, if applicable?
    \answerNA{}
  \item Did you include the estimated hourly wage paid to participants and the total amount spent on participant compensation?
    \answerNA{}
\end{enumerate}

\end{enumerate}

\newpage


\appendix

\section{PT-suitable reference family}
\label{sec:theoretical_conditions} 

\begin{assumption}[PT-suitable reference]
	\label{def:regular_reference}
	We say $\pi_0$ is PT-suitable for the target $\pi_1$ if:
	\begin{enumerate}
		\item (Full support): $\supp(\pi_0)=\supp(\pi_1)$.
		\item (Regularity): The log-likelihood ratio between $\pi_1$ and $\pi_0$, $\ell(x)=\log \frac{\pi_1(x)}{\pi_0(x)}$ satisfies,
		\[
		\label{eq:annealing_path}
		\max\{\E_0[|\ell|^3], \, \E_{1}[|\ell|^3]\} < \infty.
		\]
		where we denote $\E_0$, and $\E_1$ as the expectation with respect to $\pi_0$ and $\pi_1$ respectively.
	\end{enumerate}
We say that a family $\calQ = \{ q_\phi : \phi \in \Phi\}$ is PT-suitable for the target $\pi_1$ if for all $\phi$ the conditions above hold with $q_\phi$ in place of $\pi_0$.
\end{assumption}

\begin{assumption}[Efficient local exploration]
\label{assump:ELE}
Suppose $\pi_0$ is a PT-suitable reference for $\pi_1$, with log-likelihood $\ell(x)=\log \pi_1(x)-\log \pi_0(x)$ and schedule $\calB_N$.
Let $\bfX_t=(X^0_t,\dots,X^N_t)$ be the PT chain stationary with respect to \[\pis(\bfx)=\prod_{n=0}^N\pi_{\beta_n}(x^n)\] 
and $K_{\beta_n}$ be the $\pi_{\beta_n}$-stationary Markov kernel for the local exploration step. We will say $\bfX_t$ efficiently explores locally if \cite[Section 3.3]{syed2019nrpt},
\begin{enumerate}
    \item Stationarity: $\bfX_0 \sim \pis$.
    \item Efficient local exploration (ELE) : For all $n$ and $t$, if $\bar{X}^n_t\sim K_{\beta_n}(X^n_t,\dee\bar{x})$, then $\ell(X^n_t)$ is independent of $\ell(\bar{X}^n_t)$.
\end{enumerate}
\end{assumption}

\section{Large-data Asymptotics}\label{sec:largedataasymp}
\subsection{Conditional convergence in distribution}
Suppose $(\calX,d_\calX)$ is a metric space and let $X, X_1, X_2, \ldots$ be random variables taking values in $\calX$. Define a sequence of $\sigma$-algebras $(\calF_m)_{m=1}^\infty$ such that $\calF_m\subset\calF_{m+1}$. As $m\to\infty$, we say $X_m|\calF_m$ converges in distribution to $X$, denoted
$X_m | \calF_m \xrightarrow{d} X$, if for all bounded and continuous $f:\calX\to\reals$,
\[ 
\E[f(X_m) | \calF_m] \xrightarrow[m\to\infty]{a.s.} \EE[f(X)].
\]

Similarly, we define conditional convergence in probability, denoted $X_m | \calF_m \xrightarrow{p} X$, if for all $\epsilon>0$,
\[
\Pr(d_\calX(X_m,X)>\epsilon|\calF_m)\xrightarrow[m\to\infty]{a.s.}0.
\]

\begin{lemma}[Conditional portmanteau lemma]
\label{lem:portmanteau}
The following are equivalent:
\begin{enumerate}
\item $\displaystyle X_m|\calF_m\xrightarrow{d} X$ as $m\to\infty$.
\label{lem:portmanteau_continuous}
\item $\displaystyle \E[f(X_m)|\calF_m] \xrightarrow{a.s.} \E[f(X)]$  as $m\to\infty$, 
for all bounded Lipschitz functions $f:\calX\to \reals$.
\label{lem:portmanteau_lipschitz}
\item $\displaystyle \Pr[X_m\in A|\calF_m]\xrightarrow{a.s} \Pr[X\in A]$ as
$m\to\infty$ for all $A\subset\calX$ such that $\Pr[X\in\partial A]=0$.
\label{lem:portmanteau_distribution}
\end{enumerate}
\end{lemma}
The proof of this Lemma is identical to the portmanteau lemma for weak 
convergence by replacing probabilities/expectations with conditional probabilities/expectations (for example, see \cite[Section 2.1]{van2000asymptotic}).

\begin{lemma}
\label{lem:conditional_convergence}
Suppose $X,X_1,X_2,\dots$ and $X', X_1, X_2, \ldots$ are $\calX$-valued random variables. Then, the following hold:
\begin{enumerate}
    \item If $X_m|\calF_m\xrightarrow{d}X$ as $m\to\infty$ then $X_m\xrightarrow{d}X$.
    \item If $X_m|\calF_m\xrightarrow{p}X$ as $m\to\infty$ then $X_m\xrightarrow{p}X$.
    \item If $X_m|\calF_m\xrightarrow{p}X$ as $m\to\infty$ then $X_m|\calF_m\xrightarrow{d}X$.
    \label{lem:conditional_convergence_distribution}
    \item If $X_m|\calF_m\xrightarrow{d}X$ as $m\to\infty$, and $X$ is a constant a.s., then $X_m|\calF_m \xrightarrow{p}X$.
    \label{lem:conditional_convergence_constant}
    \item Fatou's lemma: If $X_m|\calF_m\xrightarrow{d} X$ as $m\to\infty$, then for all $f:\calX\to [0,\infty)$,
    \[
    \liminf_{m\to\infty} \E[f(X_m)|\calF_m] \geq \E[f(X)], \quad a.s.
    \]
    \label{lem:conditional_convergence_fatou}
    
    \item Continuous mapping theorem: If $\calX'$ is a metric space and
    $g:\calX \to \calX'$ is a continuous function, then 
    \[
     X_m|\calF_m\xrightarrow[m\to\infty]{d}X &\quad\Longrightarrow \quad  g(X_m) |\calF_m \xrightarrow[m\to\infty]{d} g(X).
    \]
    \label{lem:conditional_convergence_continuous}
    
    \item Slutsky's theorem: If $X_m|\calF_m\xrightarrow{d}X$ and $d_\calX(X_m',X_m)|\calF_m \xrightarrow{p}0$, then $X'_m|\calF_m\xrightarrow{d} X$. 
    \label{lem:conditional_convergence_slutsky}

    \item Suppose $\calX=\reals^d$, with $X_m|\calF_m\xrightarrow{d}X$. Suppose $A,A_1,\dots\in \reals^{d \times d}$ such that $A, A_m\in\calF_m$ and as $m\to\infty$ $A_m\xrightarrow{a.s.} A$, where $A$ is a constant. Then, 
    \[ 
    A_m X_m| \calF_m &\xrightarrow[m\to\infty]{d} A X.
    \]
    \label{lem:conditional_convergence_slutsky2}
\end{enumerate}
\end{lemma}

\bprfof{\cref{lem:conditional_convergence}}
\begin{enumerate}
    \item For any bounded and continuous $f$,
    \[ 
    \lim_{m \to \infty} \E[f(X_m)] 
    = \E\left[\lim_{m \to \infty} \E[f(X_m)|\calF_m]\right] 
    = \E[f(X)],
    \]
    We can exchange the expectation and limit by the dominated convergence theorem.
    
    \item For $\epsilon > 0$,
    \[
      \Pr(d_{\calX}(X_m,X) > \epsilon)
      &= \E[\Pr(d_{\calX}(X_m,X)  > \epsilon | \calF_m)] 
    \]
    Since $\Pr(d_{\calX}(X_m,X)  > \epsilon | \calF_m)\xrightarrow{a.s.} 0$ as $m\to\infty$, the result follows from the dominated convergence theorem.
    
    \item Let $f$ be a $\kappa$-Lipschitz function bounded by $M$. Let $\epsilon>0$,
    \[
    |\E[f(X_m)-f(X)|\calF_m]|
    &\leq \E[|f(X_m)-f(X)| 1(d_{\calX}(X_m,X)\leq \epsilon)|\calF_m]\\
    &  \quad+\E[|f(X_m)-f(X)| 1(d_{\calX}(X_m,X)>\epsilon)|\calF_m]\\
    &=\kappa\epsilon+2M\Pr(d_{\calX}(X_m,X)>\epsilon|\calF_m).
    \]
    Since $X_m\xrightarrow{p}X$ as $m\to\infty$, we have $\Pr(d_{\calX}(X_m,X)>\epsilon|\calF_m)\xrightarrow{a.s.}0$, therefore
    \[
    \lim_{m\to\infty}|\E[f(X_m)-f(X)|\calF_m]|\leq \kappa\epsilon,\quad a.s.
    \]
    The result follows by taking $\epsilon\to 0$.
    \item Since $X$ is a.s. constant there exits $x_0$ such that $\Pr(X=x_0)=1$. Then for all $\epsilon>0$, if $A_\epsilon=\{x:d_\calX(x,x_0)> \epsilon\}$, we have $\Pr(X\in A_\epsilon)=0$. Since $X_m|\calF\xrightarrow{d}X$, we have 
    \[
    \Pr(d_\calX(X_m,X)>\epsilon|\calF_m)
    =\Pr(X_m\in A_\epsilon|\calF_m)
    \xrightarrow[m\to\infty]{a.s.} \Pr(X\in A_\epsilon)=0.
    \]
    
    \item We adapt the proof of Fatou's lemma that holds for random variables that 
    converge in distribution instead of almost surely adapted from \cite[Lemma 5.11]{kallenberg2021foundations}.
    
    For any $K>0$, we have $x\to x\wedge K$ is a bounded and continuous function. 
    Since $X_m|Y_m\xrightarrow{d} X$, this implies that, almost surely,
    \[
    \liminf_{m\to\infty}\E[X_m|\calF_m]
    \geq \lim_{m\to\infty}\E[X_m\wedge K|\calF_m]=\E[X\wedge K].
    \]
    Since this is true for any $K$, and $X\wedge K\xrightarrow{a.s.} X$ as $K\to\infty$, by 
    the monotone convergence theorem, 
    \[
    \liminf_{m\to\infty}\E[X_m|\calF_m]\geq \lim_{K\to\infty}\E[X\wedge K]=\E[X].
    \]
    
    \item Fix any bounded and continuous $f:\calX'\to\reals$. Because $f \circ g:\calX\to\real$ is a bounded and continuous function, and $X_m|\calF_m\xrightarrow{d}X$, we have,
    \[ 
    \E[f(g(X_m)) | \calF_m] \xrightarrow[m\to\infty]{a.s.} \E[f(g(X))].
    \]  
    
    \item 
    
    Let $f$ be a $\kappa$-Lipschitz function bounded by $M$. 
    By triangle inequality,
    \[
    |\E[f(X_m')-f(X)|\calF_m]|
    &\leq \E[|f(X_m')-f(X_m)||\calF_m]|+|\E[f(X_m)-f(X)|\calF_m]|.
    \]
    Since $X_m\xrightarrow{d}X$, we have $\E[f(X_m)-f(X)|\calF_m]\xrightarrow{a.s.}0$. Also for all $\epsilon>0$, let $A_{\epsilon,m}=\{d_\calX(X_m,X_m')\leq \epsilon\}$. Note that
    \[
    \E[|f(X_m)-f(X_m')||\calF_m]
    &= \kappa\E[|f(X_m)-f(X_m')|1(A_{\epsilon,m})|\calF_m]\\
    &\quad+\E[|f(X_m)-f(X_m')|1(A_{\epsilon,m}^c)|\calF_m]\\
    &\leq \kappa\epsilon + 2M\Pr(A^c_{\epsilon,m}|\calF_m)
    \]
    Since $d_\calX(X_m,X_m')|\calF_m \xrightarrow{p}0$, we have $\Pr(A^c_{\epsilon,m}|\calF_m)\xrightarrow{a.s}0$. Also since $X_m|\calF_m\xrightarrow{d}X$, we have $\E[f(X_m)-f(X)|\calF_m]\xrightarrow{a.s.}0$. 
    
    \item Fix $\epsilon > 0$. Note that
    \[
      P(\|A_m - A\| > \epsilon | \calF_m)
      &= \EE[\mathbbm{1}(\|A_m - A\| > \epsilon) | \calF_m] \\
      &= \mathbbm{1}(\|A_m - A\| > \epsilon) \\
      &\to 0 \quad a.s.
    \]
    This implies $A_m|\calF_m\xrightarrow{p}A$ as $m\to\infty$.
    
    Note that $(X_m, A) | \calF_m \xrightarrow{d} (X, A)$ by the continuous mapping theorem with $x\to (x,A)$. Next, note that $\|(X_m,A_m) - (X_m,A)\| = \|A_m - A\|$, where use the matrix element-wise Euclidean norm. We are given that $A_m| \calF_m \xrightarrow{p} A$. We now show that $(X_m, A_m) | \calF_m \xrightarrow{d} (X,A)$. To this end, note that
    \[
      (X_m, A) | \calF_m \xrightarrow{d} (X, A), \qquad 
      (0, A_m) | \calF_m \xrightarrow{p} (0, A).
    \]
    Because $\|(X_m, A_m) - (X_m, A)\| = \|(0, A_m - A)\| = \|A_m - A\|$ and 
    $A_m |\calF_m \xrightarrow{p} A$, we have by Slutsky's theorem
    $(X_m, A_m) | \calF_m \xrightarrow{d} (X, A)$. The result follows by an application of the continuous mapping theorem with the function $(x,A)\to Ax$.
\end{enumerate}
\eprfof

\subsection{Model assumptions}
\label{sec:proof_prop}
The following sets of assumptions are only used to prove the large-data limit
results of  
\cref{prop:GCB_infty_posterior}, \cref{prop:GCB_zero_posterior}, and
\cref{prop:GCB_zero_posterior_2}. We suppose the data is 
$\data=\{Y_i\}_{i=1}^m$ drawn i.i.d.\ from distribution $L(y;x_0)\dee y$, where 
$L(y;x)$ defines a statistical model parametrized by $x\in\cal X$ where $\calX$ 
is an open subset of $\reals^d$. We denote the log-likelihood function $\ell$ 
and $\ell_m$ for the model and data respectively as,
\[
\ell(y;x)= \log L(y;x),\quad 
\ell_m(x) &= \sum_{i=1}^m \ell(Y_i;x) 
\]
We denote $\MLE$ to be the maximum likelihood estimator
\[
\MLE&\in\argmax_x \ell_m(x).
\]
We will use $\ell'$ and $\ell''$ to denote the gradient and Hessian of $\ell$ 
with respect to $x$, and use $I(x)$ to denote the Fisher information 
matrix,
\[ 
I(x) = -\EE[\ell''(Y;x)]= -\int \ell''(y;x)L(y;x)\dee y,
\]
and $I_m(x)$ for the observed information, 
\[
I_m(x) = -\ell_m''( x) = -\sum_{i=1}^m \ell''(Y_i;x).
\]
We will always use a subscript $m$ to indicate that the quantity is dependent on the data.

Given a prior $\pi_0$ distribution over $\calX$, we define the posterior, 
$\pi_{1,m}$ with density conditional on $\data$,
\[
\pi_{1,m}(x)\propto \pi_0(x)\prod_{i=1}^m L(Y_i;x)= \pi_0(x)\exp(\ell_m(x)).
\]

We will use $\pi_{\beta,m}$ to denote the power posterior
\[
\pi_{\beta,m}\propto \pi_0(x)\exp(\beta \ell_m(x)).
\]

For the remainder of this section we will assume the following regularity 
conditions.

\vspace{0.1in}
\begin{assumption}\ 
	
\label{assump:regularity}
\begin{enumerate}
    \item Euclidean state space:  $\calX\subset\reals^d$ has an open set
    containing
    $x_0$.  \label{assump:Euclidean}
    \item Continuity of prior density: The prior density $\pi_0$ is continuous 
    and positive in 
    a neighbourhood of $x_0$. \label{assump:continuity_prior}
    \item Regularity of log-likelihood: There is $K>0$, such that for 
    $\|x-x_0\|\leq K$, $\ell(y;x)$ continuously 3 times differentiable, and 
    there is a $M(y)$ such that,
    \[
    |\ell'''(y;x)|\leq M(y), &\quad \int M(y)L(y;x_0)\dee y<\infty.
    \]
    \label{assump:likelihood_regularity}
    \item Score at the MLE: For all $m$, $\MLE$ exists, is unique, and 
    $\ell_m'(\MLE) = 0$ almost surely. \label{assump:score_MLE}
    \item Strong consistency of MLE: $\MLE \xrightarrow{a.s.} x_0$ as
    $m\to\infty$. 
    \label{assump:MLE_consistency}
    \item Fisher information: $I(x)$ is positive definite and continuous on a
    neighbourhood of $x_0$.
    \label{assump:Fisher}
    \item PT-Suitable: For all $m$, 
    both $\pi_0$ and $\calQ$ are almost surely PT-suitable for the posterior $\pi_{1,m}$ (see
    \cref{def:regular_reference}).
    \item Efficient local exploration: For all $m$, the PT chain with target
    $\pi_{1,m}$  
    and references in $\{\pi_0\}\cup\calQ$, efficiently explore locally almost surely (see
    \cref{assump:ELE}).
    \item Bernstein-von Mises: For $0 < \beta \leq 1$ and $X_{\beta,m} \sim \pi_{\beta,m}$ and $\data \stackrel{iid}{\sim} L(\cdot;x_0)$, 
    \[ m^{1/2} (X_{\beta,m} - \MLE)|\data \xrightarrow{d} Z,\] where 
$Z=N(0,\beta^{-1} I(x_0)^{-1})$.
\label{lemma:Bernstein_beta_Xbeta}
\end{enumerate}
\end{assumption}
Note that \cref{assump:regularity}.\ref{lemma:Bernstein_beta_Xbeta} at $\beta = 1$ can be satisfied by 
introducing appropriate regularity conditions. See the paper ``Asymptotic Normality, Concentration, and Coverage of Generalized Posteriors'' by Miller (2021) for some possible conditions. For $0 < \beta < 1$, the result for the power posterior holds by noting that
the tempered log-likelihood is $\beta \cdot \ell$ and by invoking Bernstein-von Mises results on the tempered log-likelihood under model misspecification (where the true data generating mechanism is based on the non-tempered likelihood). Such results for model
misspecification are also available in the mentioned paper.

\subsection{Preliminary results}
We start off with an expansion of the log-likelihood, $\ell$, about the MLE, 
$\MLE$. Define
\[  Q_m(x) &= -\frac{1}{2} \cdot (x - \MLE)^\top I_m(\MLE)(x - \MLE). 
\label{eq:Qn} 
\]
We bound the difference between the log-likelihood at the MLE and the 
second-order term in the expansion of the log-likelihood.

\vspace{0.1in}

\begin{lemma} 
\label{lem:likelihood_diff}
\begin{enumerate}
Suppose \cref{assump:regularity} holds. Then,
\item $\displaystyle m^{-1}I_m(\MLE) \xrightarrow{a.s.} I(x_0)$ as $m\to\infty$.
\label{lem:likelihood_diff_fisher}
\item For all $\|x-x_0\|<K/2$, we have a.s. there is an $\bar{m}$ large enough 
such that for $m\geq \bar{m}$ 
\[
  \ell_m(x) =\ell_m(\MLE)+Q_m(x) + \epsilon_m(x) \label{eq:likelihood_diff}
\]
for some $\epsilon_m(x)$ satisfying 
\[
|\epsilon_m(x)| &\leq \frac{M m}{3} \cdot \|x - \MLE\|^3,
\]
and $M=\int M(y)L(y;x_0)\dee y<\infty$. 
\label{lem:likelihood_diff_taylor}
\item For any sequence of random variables, $X_m$, such that 
$m^{1/2}(X_m - \MLE)|\data\xrightarrow{d} X$ for some random variable $X$, 
we have 
\[
\epsilon_m(X_m)|\data &\xrightarrow{p} 0,\\
\pi_0(X_m)|\data &\xrightarrow{p} \pi_0(x_0).
\]
\label{lem:likelihood_diff_error}
\end{enumerate}
\end{lemma}
\bprfof{\cref{lem:likelihood_diff}}
\begin{enumerate}
\item 
By the triangle inequality,
\[
\|m^{-1}I_m(\MLE)-I(x_0)\|
&\leq \|m^{-1}I_m(\MLE)-m^{-1}I_m(x_0)\|\\
&\quad+\|m^{-1}I_m(x_0)-I(x_0)\|.
\]
We will now show that each term converges to 0 a.s.

Since $\MLE\xrightarrow{a.s.}x_0$, we have a.s., for $m$ large enough  $\|x_0-\MLE\|<K$. Therefore by the mean-value theorem,
\[
\|m^{-1}I_m(\MLE)-m^{-1}I_m(x_0)\|
&=\frac{1}{m}\sum_{i=1}^m\|\ell''(Y_i;\MLE)-\ell''(Y_i;x_0)\|\\
&\leq \frac{1}{m}\sum_{i=1}^m M(Y_i)\|\MLE-x_0\|.
\]
Since $\MLE\xrightarrow{a.s.}x_0$, and $\frac{1}{m}\sum_{i=1}^m M(Y_i)\xrightarrow{a.s.}\int M(Y)L(y;x_0)\dee y<\infty$, we have 
\[
\|m^{-1}I_m(\MLE)-m^{-1}I_m(x_0)\|\xrightarrow[m\to\infty]{a.s.}0.
\]

By the strong law of large numbers we have 
\[
m^{-1}I_m(x_0)=\frac{1}{m}\sum_{i=1}^m\ell''(Y_i;x_0) \xrightarrow[m\to\infty]{a.s.} \int \ell''(y;x_0)L(y;x_0)\dee y=I(x_0).
\]
This implies
\[
\|m^{-1}I_m(x_0)-I(x_0)\|\xrightarrow[m\to\infty]{a.s.}0.
\]

\item We use a second-order expansion around $\MLE$ and
\cref{assump:regularity}.\ref{assump:score_MLE} to get,
\[ 
  \ell_m(x) 
  &= \ell_m(\MLE) + (x - \MLE)^\top \ell'_m(\MLE) + Q_m(x) +\epsilon_m(x)\\
  &= \ell_m(\MLE) + Q_m(x) +\epsilon_m(x),
\]
for some $\epsilon_m(x)$ satisfying,
\[|\epsilon_m(x)|\leq \frac{M_m}{6}\|x-\MLE\|^3.\] 
where $M_m=\sup\{|\ell_m'''(\xi)|: \|\xi-\MLE\|<\|x-\MLE\|\}$. We are done if 
we can show that $M_m\leq 2Mm$. 

Suppose $\|\xi-\MLE\|<\|x-\MLE\|$. Then, by triangle inequality,
\[
\|\xi-x_0\|
&\leq \|\xi-\MLE\|+\|\MLE-x_0\|\\
&\leq \|x-\MLE\|+\|\MLE- x_0\|\\
&\leq \|x-x_0\|+\|x_0-\MLE\|+\|\MLE-x_0\|\\
& < \frac{K}{2}+2\|x_0-\MLE\|
\]
By the strong consistency of the MLE 
\cref{assump:regularity}.\ref{assump:MLE_consistency}, we have a.s. there is an 
$m_0$ large enough such that for $m\geq m_0$, $\|\MLE-x_0\|<K/4$ and thus 
$\|\xi-x_0\|<K$. This means that 
\cref{assump:regularity}.\ref{assump:likelihood_regularity} implies,
\[
|\ell_m'''(\xi)|\leq \sum_{i=1}^M |\ell'''(Y_i;\xi)|\leq \sum_{i=1}^m M(Y_i).
\]
Since $Y_i\sim L(y;x_0)\dee y$, the law of large numbers implies 
$m^{-1}\sum_{i=1}^m M(Y_i)$ converges a.s. to $M$ as $m\to\infty$. Almost 
surely, there is an $\bar{m}\geq m_0$ such that for $m\geq \bar{m}$,
\[
\frac{1}{m}\sum_{i=1}^m M(Y_i)<2M<\infty.
\]
Therefore, $M_m\leq 2mM$, which completes the proof.

\item Note that $\epsilon_m(X_m)$ satisfies almost surely for $m$ large enough,
\[ 
|\epsilon_m(X_m)|\leq \frac{M}{3 m^{1/2}} \|m^{1/2} (X_m -\MLE) \|^3.
\]
By the conditional continuous mapping theorem,
\cref{lem:conditional_convergence}.\ref{lem:conditional_convergence_continuous},
\[
 \|m^{1/2} (X_m - \MLE) \|^3|\data \xrightarrow{d} \|X\|^3,
 \] 
and $\epsilon_m(X_m)|\data\xrightarrow{d} 0$ and hence also 
$\epsilon_m(X_m)|\data \xrightarrow{p} 0 $ and $\epsilon_m(X_m) \xrightarrow{p} 0$ by
\cref{lem:conditional_convergence}.\ref{lem:conditional_convergence_constant}.

Also note that $m^{1/2} (X_m - \MLE)|\data \xrightarrow{d} X$ implies
$X_m-\MLE|\data \xrightarrow{p}0$, and hence by the continuous mapping theorem,
$\pi_0(X_m)|\data \xrightarrow{p}\pi_0(x_0).$
\end{enumerate}

\eprfof

Next, we claim that the second-order term in the expansion of the 
log-likelihood around the MLE evaluated at an appropriate random variable 
converges to a transformed chi-squared random variable. This result will be 
used repeatedly in the proofs of \cref{prop:GCB_infty_posterior}, 
\cref{prop:GCB_zero_posterior}, and \cref{prop:GCB_zero_posterior_2}.
\begin{lemma}
\label{lemma:Bernstein_beta}
Suppose \cref{assump:regularity} holds. If $X_{\beta,m}, X_{\beta,m}' 
\sim \pi_{\beta,m}$, are independent conditioned on $\data$, then for all 
$0 < \beta \leq 1$ we have as $m \to \infty$:
\begin{enumerate}

\item $\eps_m(X_{\beta,m})|\data \xrightarrow{p} 0$. 
\label{lemma:Bernstein_beta_error}
\item $\displaystyle		
Q_m(X_{\beta,m})|\data \xrightarrow{d} -Q/(2\beta)$, where 
$Q\sim \chi^2_d $.
\label{lemma:Bernstein_beta_Q}
\item $\displaystyle		
Q_m(X_{\beta,m}) - Q_m(X_{\beta,m}')|\data \xrightarrow{d} 
\frac{1}{2\beta}(Q - Q'),$ 
where $Q, Q' \iidsim \chi^2_d$.
\label{lemma:Bernstein_beta_Q_diff}
\item $\displaystyle	 \ell_m(X_{\beta,m}) - \ell_m(X_{\beta,m}')|\data 
\xrightarrow{d} \frac{1}{2\beta}(Q - Q'),$, where $Q, Q' \iidsim \chi^2_d$.
\label{lemma:Bernstein_beta_ell_diff}
\end{enumerate}
\end{lemma}

\bprfof{\cref{lemma:Bernstein_beta}}
\begin{enumerate}

\item This follows immediately \cref{assump:regularity}.\ref{lemma:Bernstein_beta_Xbeta}.
and \cref{lem:likelihood_diff}.\ref{lem:likelihood_diff_error}.

\item Note that we can decompose $Q_m$ as
\[ 
Q_m(X_{\beta,m})  
&= -\frac{1}{2} [m^{1/2} (X_{\beta,m} - \MLE)]^\top 
\left[ \frac{1}{m}I_m(\MLE)\right] 
[m^{1/2} (X_{\beta,m} -\MLE)]. 
\]
By \cref{assump:regularity}.\ref{lemma:Bernstein_beta_Xbeta} we have,
\[
m^{1/2} (X_{\beta,m} - \MLE)|\data\xrightarrow[m\to\infty]{d} N(0,\beta^{-1} I^{-1}(x_0)),
\]
and $m^{-1} I_m(\MLE)\xrightarrow{a.s.} I(x_0)$ by
\cref{assump:regularity}.\ref{assump:Fisher}. By \cref{lem:conditional_convergence}.\ref{lem:conditional_convergence_slutsky} 
and 
\ref{lem:conditional_convergence}.\ref{lem:conditional_convergence_continuous}
we get,
\[
Q_m(X_{\beta,m}) |\data &\xrightarrow[m\to\infty]{d} -Q/(2\beta),
\]
where $Q \sim \chi^2_d$. 
\item Note that for any $0 < \beta \leq 1$, by using the arguments as above,
\[
    Q_m(X_{\beta,m})|\data &\xrightarrow{d} -Q/(2\beta) \\
    Q_m(X_{\beta,m}')|\data &\xrightarrow{d} -Q/(2\beta),
\]
where $Q \sim \chi^2_d$. Each of the $X_{\beta,m}, X_{\beta,m}'$ are assumed to 
be conditionally independent given the data $\data$, and therefore
\[
   Q_m(X_{\beta,m}) - Q_m(X_{\beta,m}')|\mathbf{Y}_m 
   \xrightarrow[m\to\infty]{d} (Q - Q')/(2\beta), 
\]
where $Q, Q' \sim \chi^2_d$ are independent.

\item Finally, we employ
\cref{lem:likelihood_diff}.\ref{lem:likelihood_diff_taylor} 
and triangle inequality to get, 
\[
\left| 
[\ell_m(X_{\beta,m})-\ell_m(X_{\beta,m}')] -  
[Q_m(X_{\beta,m}) -Q_m(X_{\beta,m}')]
\right|
\leq |\epsilon_m(X_{\beta,m})|+|\epsilon_m(X_{\beta,m}')|
\]
By part \cref{lemma:Bernstein_beta}.\ref{lemma:Bernstein_beta_error}, we have
$\epsilon_m(X_{\beta,m})|\data\xrightarrow{p} 0$ and $\epsilon_m(X_{\beta,m}')|\data \xrightarrow{p} 0$ so
\[
  \left[\ell_m(X_{\beta,m}) - \ell_m(X_{\beta,m}')\right] - 
  \left[Q_m(X_{\beta,m}) - Q_m(X_{\beta,m}')\right] |\data \xrightarrow{p} 0.
\]
By \cref{lemma:Bernstein_beta}.\ref{lemma:Bernstein_beta_Q_diff} and the
conditional Slutsky's theorem 
\cref{lem:conditional_convergence}.\ref{lem:conditional_convergence_slutsky},
\[
\ell_m(X_{\beta,m}) - \ell_m(X_{\beta,m}')|\data
\xrightarrow[m\to\infty]{d} (Q - Q')/(2\beta).
\]

\end{enumerate}
\eprfof

\subsection{Proof of \cref{prop:GCB_infty_posterior}}
\bprfof{\cref{prop:GCB_infty_posterior}}
The asymptotic restart rate, $\tau_m$, is related to the GCB by
\[
\tau_m = \frac{1}{2+2\Lambda(\pi_0,\pi_{1,m})},
\]
where $\Lambda(\pi_0,\pi_{1,m})$ is the GCB between the prior $\pi_0$ and the 
posterior $\pi_{1,m}$. This implies that 
\[
0 \leq \limsup_{m\to\infty}\tau_m
\leq \frac{1}{2+2\liminf_{m\to\infty}\Lambda(\pi_0,\pi_{1,m})}.
\]
Therefore, we are are done if we can show that 
$\liminf_{m\to\infty}\Lambda(\pi_0,\pi_{1,m})=\infty$ almost surely.

Suppose $X_{\beta,m},X_{\beta,m}'\sim\pi_{\beta,m}$ are independent conditioned 
on $\data$. Then, 
\[
\Lambda(\pi_0,\pi_{1,m})
=\frac{1}{2}\int_0^1\E[|\ell_m(X_{\beta,m})-\ell(X_{\beta,m}')||\data]\dee\beta.
\]
Since the integrand is positive, for any $\delta > 0$, 
\[
\Lambda(\pi_0,\pi_{1,m})
\geq \frac{1}{2}\int_\delta^1
\E[|\ell_m(X_{\beta,m})-\ell(X_{\beta,m}')||\data]\dee\beta
\]
By taking the limit infimum of both sides, and using Fatou's lemma,
\[
\liminf_{m\to\infty}\Lambda(\pi_0,\pi_{1,m})
&\geq \frac{1}{2}\int_\delta^1\liminf_{m\to\infty}
\E[|\ell_m(X_{\beta,m})-\ell(X_{\beta,m}')||\data]\dee\beta.
\]
From \cref{lemma:Bernstein_beta}.\ref{lemma:Bernstein_beta_ell_diff} and the
conditional continuous mapping theorem
\cref{lem:conditional_convergence}.\ref{lem:conditional_convergence_continuous} 
applied to $x\to|x|$, we have 
\[
|\ell_m(X_{\beta,m})-\ell(X_{\beta,m}')| |\data
\xrightarrow[m\to\infty]{d} \frac{|Q-Q'|}{2\beta},\quad Q,Q'\sim\chi^2_d. 
\]
By \cref{lem:conditional_convergence}.\ref{lem:conditional_convergence_fatou}, 
almost surely we have
\[
\liminf_{m\to\infty}\Lambda(\pi_0,\pi_{1,m})
\geq \frac{1}{2} \int_{\delta}^1 
\frac{1}{2\beta} \EE\left[|Q - Q'|\right] \, d\beta
=-\frac{1}{4}\EE\left[|Q - Q'|\right]\log(\delta).
\]
Since this is true for all $\delta>0$, and since the right hand side increases 
to infinity at $\delta\to 0$, we have almost surely,
\[
\liminf_{m\to\infty}\Lambda(\pi_0,\pi_{1,m})
\geq \lim_{\delta\to 0}-\frac{1}{4}\EE\left[|Q - Q'|\right]\log(\delta)=\infty.
\]
\eprfof

\subsection{Proof of \cref{prop:GCB_zero_posterior}}

\begin{lemma}
\label{lemma:rejection_bound}
 Assume $\pi_0$ is a PT-suitable reference for $\pi_1$, with log-likelihood
 $\ell$.
\begin{enumerate}
    \item If $L_\beta = \ell(X_\beta)$, for $X_\beta \sim \pi_\beta$, then
    $L_\beta$ is stochastically non-decreasing in $\beta$, i.e.\ for all $y\in\reals$,
    $\Pr(L_\beta> y)$ is a non-decreasing function of $\beta$.
    
    \item Let $r(\beta,\beta')$ be a mean rejection rate for a swap between 
    $\pi_\beta,\pi_{\beta'}$, 
 \[
 r(\beta,\beta')
 =1-\E\left[1\wedge
 \frac{\pi_\beta(X_{\beta'})\pi_{\beta'}(X_\beta)}
 {\pi_\beta(X_{\beta})\pi_{\beta'}(X_{\beta'})}\right],\quad
 (X_\beta,X_{\beta'})\sim\pi_{\beta}\times\pi_{\beta'}.
 \]
 If $[a,b]\subset[a',b']\subset [0,1]$, then $r(a,b)\leq r(a',b')$.
\end{enumerate}
\end{lemma}

\bprfof{\cref{lemma:rejection_bound}}
\begin{enumerate}
\item 	Fix $y\in\reals,$ and $0\leq \beta<\beta'\leq 1$. We want to show that 
\[0\leq \Pr[L_{\beta'}> y]-\Pr[L_\beta>y]
=\E[f(L_\beta,L_{\beta'})],\]
where $f(l,l')=1(l'>y)-1(l>y)$. We have
\[
\E[f(L_\beta,L_{\beta'})]
&=\int f(\ell(x),\ell(x'))\frac{1}{Z(\beta)}\exp(\beta \ell(x))
\frac{1}{Z(\beta')}\exp(\beta' \ell(x'))\dee x \, \dee x'\\
&=\frac{Z(\beta)}{Z(\beta')}\E[f(L,L')\exp(\delta L')],
\]
where $L,L'\deq \ell(X_\beta)$ are independent and $\delta=\beta'-\beta>0$.
Now, notice that
\[
f(l,l')
&=1(l'>y)-1(l>y)\\
&=1(l'>y)1(l\leq y)-1(l>y)1(l'\leq y).
\]
Therefore, we have
\[
\E[f(L,L')\exp(\delta L')]
&=\E[1(L'>y)1(L\leq y)\exp(\delta L')]\\
&\quad-\E[1(L>y)1(L'\leq y)\exp(\delta L')]\\
&=\E[1(L'>y)1(L\leq y)(\exp(\delta L')-\exp(\delta L))]\\
&\geq 0,
\]
where the second to last line used the fact that $L,L'$ are i.i.d. 
\item 
To simplify notation, suppose first $a = a'$. Denote the cumulative distribution
function of $L_\beta$ by $F_\beta$. From (a), we have that $F_b \ge F_{b'}$. It 
follows that we can construct a random variable $L'$ which is equal in 
distribution to $L_{b'}$ and such that $L_b(\omega) \le L'(\omega)$ for all 
outcomes $\omega$ in the probability space. This is achieved by setting $L' = 
F^{-1}_{L_{b'}} \circ F_{L_b} \circ L_b$, where $F^{-1}$ denotes the generalized 
inverse cumulative function. To see why, note that $F_{L_b} \circ L_b$ is 
uniformly distributed, being a probability integral transform. Hence 
$F^{-1}_{L_{b'}}$ applied to that uniform yields a $L_{b'}$-distributed random 
variable. The inequality $L_b(\omega) \le L'(\omega)$ follows from $F_b \ge 
F_{b'}$. 

Next, define $f(\delta)=1-1\wedge \exp(-\delta)$, which is an increasing function
in $\delta$. Hence, $f((b-a)(L_b - L_a) \le f((b-a)(L' - L_a)$ for all outcomes,
and so 
\[
r(a,b) = \E[f((b-a)(L_b - L_a))] \le \E[f((b-a)(L' - L_a))] = r(a, b'),
\]
where in the last equality we also used that $L'$ is independent of $L_a$, being 
a deterministic transformation of the random variable $L_b$. Finally, if $a < a'$,
use $r(a, b) \le r(a, b') \le r(a', b')$ where the last inequality is obtained
using a very similar argument as above.
\end{enumerate}
\eprfof

\begin{lemma}\label{lemma:perfect_restart}
 Suppose $\calQ$ is almost surely a PT-suitable reference family for all targets $\pi_{1,m}$. Also assume that for all $m$, the PT chain with target $\pi_{1,m}$ and references $\calQ$, efficiently explore locally almost surely. Given a (random) sequence
 $q_m\in\calQ$, if
$\alpha_m$ is the average acceptance probability between $q_m$ and $\pi_{1,m}$,
then if $\alpha_m\xrightarrow{p}1$, then for any $\calB_N$, we have
$\tau_m(\calB_N)\xrightarrow{p}\frac{1}{2}$ as $m\to\infty$.

\end{lemma}

\bprfof{\cref{lemma:perfect_restart}}

For any schedule $\calB_{n}=(\beta_{n})_{n=0}^N$, let $\{r_{n,m}\}_{n=0}^{N-1}$
be the average rejection rates between components $n$ and $n+1$. For any $n$, 
we have by \cref{lemma:rejection_bound},
\[
0\leq r_{n,m}\leq 1-\alpha_m.
\]
Since $\alpha_m\xrightarrow{p}1$, we have $r_{n,m}\xrightarrow{p} 0$ as
$m\to\infty$ and
\[
\tau_m(\calB_N)=\frac{1}{2+2\sum_{n=0}^{N-1}\frac{r_{n,m}}{1-r_{n,m}}}
\xrightarrow[m\to\infty]{p} \nicefrac{1}{2}.
\]
\eprfof

\begin{lemma}\label{lemma:normality_ref_MLE}
Suppose \cref{assump:regularity} holds and $X_{0,m}\sim N(\MLE,I_m(\MLE)^{-1})$.
Then, as $m\to\infty$,
\[
m^{1/2}(X_{0,m}-\MLE)|\data\xrightarrow{d} \distNorm(0,I^{-1}(x_0)).
\]
\end{lemma}
\bprfof{\cref{lemma:normality_ref_MLE}}
Since $X_{0,m}\sim N(\MLE,I_m(\MLE)^{-1})$,
\[
  m^{1/2} \cdot \frac{I_m^{1/2}(\MLE)}{m^{1/2}} \cdot 
  I^{-1/2}(x_0) (X_{0,m} - \MLE) | \mathbf{Y}_m 
  \sim \distNorm(0, I^{-1}(x_0)).
\]
By \cref{lem:likelihood_diff}.\ref{lem:likelihood_diff_fisher} and
\cref{assump:regularity}.\ref{assump:MLE_consistency}, we have 
\[ 
  \frac{I_m^{1/2}(\MLE)}{m^{1/2}} \cdot I^{-1/2}(x_0)
  \xrightarrow[m\to\infty]{a.s.} \mathbb{I}_d.
\]
By \cref{lem:conditional_convergence}.\ref{lem:conditional_convergence_slutsky} 
it follows,
\[\label{eq:asymptotic_normality_0}
  m^{1/2} (X_{0,m} - \MLE) | \data \xrightarrow{d} \distNorm(0, I^{-1}(x_0)).
\]
\eprfof

\bprfof{\cref{prop:GCB_zero_posterior}}

Let $q_m = N(\MLE, I^{-1}_m(\MLE))\in\calQ$. The acceptance probability $\alpha_m$ is 
\[
  \alpha_m &= \EE\left[1\wedge A_m(X_{0,m},X_{1,m})|\data \right],
\]
where $A_m(X_{0,m},X_{1,m})$ is the acceptance ratio,
\[ 
 A_m(X_{0,m},X_{1,m}) 
  &=\frac{q_m(X_{1,m}) \cdot \pi_{1,m}(X_{0,m})}{q_m(X_{0,m}) 
  \cdot \pi_{1,m}(X_{1,m})} \\
  &= \frac{q_m(X_{1,m})}{q_m(X_{0,m})} \cdot \frac{\pi_0(X_{0,m})}{\pi_0(X_{1,m})}
  \cdot \frac{\exp(\ell_m(X_{0,m}))}{\exp(\ell_m(X_{1,m}))}. 
\]
Note that up to an additive constant
$\log q_m(x)= Q_m(x)$, and so
\[
  \log q_m(X_{1,m}) - \log q_m(X_{0,m})
  &= Q_m(X_{1,m}) - Q_m(X_{0,m}).
\]
Therefore the log-acceptance ratio satisfies
\[
  \log A_m(X_{0,m},X_{1,m})
  &=  \log \pi_0(X_{0,m}) - \log \pi_0(X_{1,m}) 
  + \epsilon_m(X_{0,m}) -\epsilon_m(X_{1,m}),
\]
where $\epsilon_m$ was defined in \cref{lem:likelihood_diff}.

Since we have the asymptotic normality of $X_{0,m}$ and
$X_{1,m}$ conditioned on $\data$ by \cref{lemma:Bernstein_beta}.\ref{lemma:Bernstein_beta_Xbeta}, and \cref{lemma:normality_ref_MLE}, we can invoke \cref{lem:likelihood_diff}.\ref{lem:likelihood_diff_error},
\[\label{eq:consitency_epsilon_prior_MLE}
\epsilon_m(X_{0,m}),\epsilon_m(X_{1,m})|\data&\xrightarrow[m\to\infty]{p} 0,\\
\log\pi_0(X_{0,m}),\log\pi_0(X_{1,m})|\data&\xrightarrow[m\to\infty]{p} \log\pi_0(x_0).
\]
Combining \eqref{eq:consitency_epsilon_prior_MLE} we have the acceptance ratio satisfies,
\[
A_m(X_{0,m},X_{1,m})|\data\xrightarrow{p}1.
\] 
To conclude, note that $1\wedge A_m(X_{0,m},X_{1,m}) \leq 1$, 
so by the dominated convergence theorem,
\[
\alpha_m
=\E[1\wedge A_m(X_{0,m},X_{1,m})|\data]
\xrightarrow[m\to\infty]{a.s.}1,
\]
and hence $\alpha_m\xrightarrow{a.s.}1$. The result follows from
\cref{lemma:perfect_restart}.
\eprfof

\subsection{Proof of \cref{prop:GCB_zero_posterior_2}}

Suppose $\mu_m$ and $\Sigma_m$ are posterior mean and variance conditional 
to $\data$,
\[
\mu_m=\E[X_{1,m}|\data],\quad
\Sigma_m=\var[X_{1,m}|\data],
\]
where $X_{1,m}\sim \pi_{1,m}$. We introduce a final set of assumptions that are
required for the proof of 
\cref{prop:GCB_zero_posterior_2}. 
\begin{assumption}
\label{assump:posterior_moments}
\begin{enumerate}
    \item Posterior mean and MLE: As $m\to\infty$, 
    $m^{1/2} (\mu_m - \MLE) \xrightarrow{a.s.} 0$.
    \label{assump:posterior_moments_mean}
    \item Posterior variance and Fisher information: For all $m$, $\Sigma_m$ is
    almost surely positive definite and $m \Sigma_m \xrightarrow{a.s.} I^{-1}(x_0)$
    as $m\to\infty$.
    \label{assump:posterior_moments_variance}
\end{enumerate}
\end{assumption}
For such results in the univariate case with convergence in probability, see \cite{severini1991comparison} and
\cite{johnson1970asymptotic}.

Given $q_m'=N(\mu_m,\Sigma_m)$, define,
\[
Q'_m(x)&=-\frac{1}{2}(x-\mu_m)^\top\Sigma^{-1}_m(x-\mu_m).
\]
\begin{lemma}
\label{lem:Qn_limit_posterior_mean}
Suppose \cref{assump:regularity} and \cref{assump:posterior_moments} hold. If
$X'_{0,m}\sim q'_m$ and $X_{1,m}\sim\pi_{1,m}$, then as $m\to\infty$,
\begin{enumerate}
\item $m^{1/2}(X'_{0,m}-\MLE)|\data\xrightarrow{d} \distNorm(0,I^{-1}(x_0))$,
\label{lem:Qn_limit_posterior_mean_X_0}
\item $\displaystyle Q_m'(X_{1,m})- Q_m(X_{1,m}) \xrightarrow{p} 0$,
\label{lem:Qn_limit_posterior_mean_Q_1}
\item $\displaystyle  Q_m'(X'_{0,m})- Q_m(X'_{0,m})\xrightarrow{p} 0.$
\label{lem:Qn_limit_posterior_mean_Q_0}
\end{enumerate}
\end{lemma}
\bprfof{\cref{lem:Qn_limit_posterior_mean}}
\begin{enumerate}

\item Note that since $X_{0,m}'\sim \distNorm(\mu_m, \Sigma_m)$,
\[
  \frac{\Sigma_m^{-1/2}}{m^{1/2}} 
  \cdot I^{-1/2}(x_0)\cdot  m^{1/2}(X_{0,m} - \mu_m) | \mathbf{Y}_m 
  \sim \distNorm(0, I^{-1}(x_0)).
\]
By \cref{assump:posterior_moments},
\[
  \frac{\Sigma_m^{-1/2}}{m^{1/2}} \cdot I^{-1/2}(x_0) 
  \xrightarrow[m\to\infty]{a.s.} \mathbb{I}_d.
\]
Therefore using Slutsky's theorem,
\cref{lem:conditional_convergence}.\ref{lem:conditional_convergence_slutsky}, 
it follows that
\[
  m^{1/2} (X_{0,m}' - \mu_m) | \data 
  \xrightarrow[m\to\infty]{d} \distNorm(0, I^{-1}(x_0)).
\]
Finally, the result follows from Slutsky's theorem using 
\cref{assump:posterior_moments}.\ref{assump:posterior_moments_mean}.

\item Using the definition of $Q(x)$ and $Q'(x)$ we obtain the following
decomposition,
\[
Q'_m(x)-Q_m(x)= \epsilon^0_{m}+\epsilon^1_m(x)+\epsilon^2_m(x)
\]
where,
\[
\epsilon^0_m&=-(\mu_m - \MLE)^\top \Sigma_m^{-1} (\mu_m - \MLE) \\
\epsilon^1_m(x)&=-2 (x - \MLE)^\top \Sigma_m^{-1} (\MLE - \mu_m) \\
\epsilon^2_m(x)&=  -[m^{1/2} (x - \MLE)]^\top 
  \cdot [m^{-1} (\Sigma_m^{-1} - I_m(\MLE))] 
  \cdot [m^{1/2} (x - \MLE)]
\]

Now, using \cref{assump:posterior_moments}, it follows that
$\epsilon_m^0\xrightarrow{p}0$ as $m\to\infty$,
\[
  \epsilon^0_{m} 
  &= -[m^{1/2} (\mu_m - \MLE)]^\top [m \Sigma_m]^{-1} [m^{1/2} (\mu_m - \MLE)]\\
  &= o_p(1) \cdot O_p(1) \cdot o_p(1) \\
  &= o_p(1).
\]

Therefore we are done if we can show (1) $\epsilon^1_m(X_{1,m})\xrightarrow{p}0$
and (2) $\epsilon^2_m(X_{1,m})\xrightarrow{p}0$ as $m\to\infty$. 

(1) follows from \cref{assump:posterior_moments} and
\cref{lemma:Bernstein_beta}.\ref{lemma:Bernstein_beta_Xbeta} since as
$m\to\infty$,
\[
  \epsilon^1_m(X_{1,m}) 
  &= -2 [m^{1/2} (X_{1,m} - \MLE)]^\top 
  [m \Sigma_m]^{-1} [m^{1/2} 
  (\MLE - \mu_m)]\\
  &= O_p(1) \cdot O_p(1) \cdot o_p(1)  \\
  &= o_p(1).
\]
Finally for (2) notice that by \cref{assump:posterior_moments}, and  
$m^{-1}(\Sigma_m^{-1}-I_m(\MLE))\xrightarrow{p}0$. Using
\cref{lemma:Bernstein_beta}.\ref{lemma:Bernstein_beta_Xbeta} as $m\to\infty$,
\[
  \epsilon^2_{m}(X_{1,m})\\
  &= -[m^{1/2} (X_{1,m} - \MLE)]^\top 
  [m^{-1} (\Sigma_m^{-1} - I_m(\MLE))] 
  [m^{1/2} (X_{1,m} - \MLE)]\\
  &= O_p(1)\cdot o_p(1)\cdot O_p(1)\\
  &= o_p(1)
\]

\item Similar to the proof of
\cref{lem:Qn_limit_posterior_mean}.\ref{lem:Qn_limit_posterior_mean_Q_1}, 
we are done if we can show (3) $\epsilon^1_m(X'_{0,m})\xrightarrow{p}0$ and 
(4) $\epsilon^2_m(X'_{0,m})\xrightarrow{p}0$ as $m\to\infty$.

To show (3) we use 
\cref{lem:Qn_limit_posterior_mean}.\ref{lem:Qn_limit_posterior_mean_X_0} and
\cref{assump:posterior_moments},
\[
  \epsilon_m^1(X'_{0,m})
  &= -2 [m^{1/2} (X'_{0,m} - \MLE)]^\top 
  [m \Sigma_m]^{-1} 
  [m^{1/2} (\MLE - \mu_m)]\\
  &= O_p(1) \cdot O_p(1) \cdot o_p(1) \\
  &= o_p(1).
\]
Finally, to obtain (4) follows from
\cref{lem:Qn_limit_posterior_mean}.\ref{lem:Qn_limit_posterior_mean_X_0} and
$m^{-1}(\Sigma_m^{-1}-I_m(\MLE))\xrightarrow{p}0$,
\[
  \epsilon^2_m(X'_{0,m})
  &= [m^{1/2} (X'_{0,m} - \MLE)]^\top 
  [m^{-1} (\Sigma_m^{-1} - I_m)] 
  [m^{1/2} (X'_{0,m} - \MLE)]
  \\
  &= O_p(1) \cdot o_p(1) \cdot O_p(1) \\
  &= o_p(1).
\]
\end{enumerate}
\eprfof

\bprfof{\cref{prop:GCB_zero_posterior_2}}
Suppose $X_{1,m}\sim\pi_{1,m}$ and 
$X_{0,m}'\sim q'_m = N(\mu_m, \Sigma_m)\in\calQ$ are independent conditioned on
$\data$. Let $\alpha'_m=\E[1\wedge A_m'(X'_{0,m},X_{1,m})]$ be the average
acceptance probability, where $A'_m(X_{0,m}',X_{1,m})$ is the acceptance ratio,
\[
 A'_m(X'_{0,m},X_{1,m}) 
  &=\frac{q'_m(X_{1,m}) \cdot \pi_{1,m}(X'_{0,m})}
  {q'_m(X'_{0,m}) \cdot \pi_{1,m}(X_{1,m})} \\
  &= \frac{q'_m(X_{1,m})}{q'_m(X'_{0,m})} 
  \cdot \frac{\pi_0(X'_{0,m})}{\pi_0(X_{1,m})}   
  \cdot \frac{\exp(\ell_m(X'_{0,m}))}{\exp(\ell_m(X_{1,m}))}.
\]

Note that up to an additive constant $\log q'_m(x)= Q'_m(x)$, and by
\cref{lem:Qn_limit_posterior_mean}, we have as $m\to\infty$,
\[
  \log q'_m(X_{1,m}) - \log q'_m(X'_{0,m})
  &= Q'_m(X_{1,m}) - Q'_m(X_{0,m}),\\
  &= Q_m(X_{1,m})- Q_m(X_{0,m})+o_p(1).
\]
Therefore we have the log-acceptance ratio satisfies,
\[
  \log A'_m(X'_{0,m},X_{1,m})
  &=  \log \pi_0(X'_{0,m}) - \log \pi_0(X_{1,m}) 
  + \epsilon_m(X'_{0,m}) -\epsilon_m(X_{1,m})+o_p(1).
\]
where $\epsilon_m$ was defined in \cref{lem:likelihood_diff}. By
\cref{lemma:Bernstein_beta}.\ref{lemma:Bernstein_beta_Xbeta}, and
\cref{lem:Qn_limit_posterior_mean}.\ref{lem:Qn_limit_posterior_mean_X_0} we have
$m^{1/2}(X_{1,m}-\MLE)$ and $m^{1/2}(X'_{0,m}-\MLE)$ are asymptotically normal
conditioned on $\data$. Therefore by
\cref{lem:likelihood_diff}.\ref{lem:likelihood_diff_error},
\[
\epsilon_m(X'_{0,m}),\epsilon_m(X_{1,m})&\xrightarrow[m\to\infty]{p} 0\\
\log\pi_0(X'_{0,m}),\log\pi_0(X_{1,m})&\xrightarrow[m\to\infty]{p} \log\pi_0(x_0),
\]
and the acceptance ratio
$A'_m(X'_{0,m},X_{1,m})\xrightarrow{p}1$.

To conclude, note that $1\wedge A'_m(X'_{0,m},X_{1,m}) \leq 1$, so by dominated
convergence theorem,
\[
\lim_{m\to\infty}\E[\alpha_m]
=\lim_{m\to\infty}\E[1\wedge A_m(X_{0,m},X_{1,m})]
=1,
\]
and hence $\alpha'_m\xrightarrow{p}1$.
\eprfof

\newpage

\section{Proof of \cref{thm:adapt_tune}}

\bprfof{\cref{thm:adapt_tune}}

Since $\eta$ is bounded there is a $K>0$ such that $\eta(x)=(\eta_1(x),\dots,\eta_d(x))$ where $\eta_i:\calX\to [-K,K]$. Fix $0 < \epsilon < \frac{1}{2}$. Suppose $(\bfX_{t,r})_{t=1}^{T_r}$ are 
the draws from the chain parameterized by $\hphi_{r}$ at round $r$. Define for each $i=1,\dots,d$,
\[
A_{r,i}
=\left\{\left|\frac{1}{T_r}\sum_{t=1}^{T_r}\eta_i(X^N_{t,r})-\E_1[\eta_i]\right|
>\delta_r\right\},
\]
where $\delta_r$ is a sequence to be determined. By assumption, we have
$|\eta_i(x)|\leq K$ for some $K>0$. We can therefore apply Theorem 1 of \cite{fan2021hoeffding}
(Hoeffding's inequality for Markov chains) to get that
\[
    P\left(A_{r,i}|\hat\phi_{r} \right)
    &\leq 2 \exp\left( 
    -\frac{\Gap(\hat{\phi}_{r})}{2-\Gap(\hat\phi_{r})}
    \cdot\frac{\delta_r^2T_r}{2K^2}\right)\\
    &\leq 2 \exp\left( -\frac{\kappa}{2-\kappa} \cdot 
    \frac{\delta_r^2 T_r}{2K^2} \right).
\]
The last inequality used the assumption that $\Gap(\phi)$ is bounded below
by $\kappa$ on $\Phi$. By taking expectations over $\hat\phi_{r}$ and by setting
$\delta_r=d^{-1}T_r^{-1/2+\epsilon}$, we obtain
\[ 
P(A_{r,i}) \leq 2 \exp\left( -\frac{\kappa}{2-\kappa} 
\cdot \frac{T_r^\epsilon}{2 d^2K^2} \right).
\]
Since $T_r=\Omega(2^r)$, we have by the ratio test, 
\[ \sum_{r=1}^\infty P(A_{r,i}) < \infty.\]
From the Borel-Cantelli lemma it follows that for each $i$, $P(A_{r,i} \text{ i.o.}) =0$, and 
thus a.s. there exists $R_i(\epsilon)$ and such that for 
all $r \geq R_i(\eps)$,
\[
\left\|\frac{1}{T_r}\sum_{t=1}^{T_r}\eta_i(X^N_{t,r}) - \E_1[\eta_i]\right\| 
\leq \frac{1}{d}T_r^{-\frac{1}{2}+\epsilon}.
\]
Since $\hphi_{r+1}$ is chosen to satisfy,
\[
\E_{\hat\phi_{r+1}}[\eta]=\frac{1}{T_r}\sum_{t=1}^{T_r}\eta(X^N_{t,r}),
\]
by the triangle inequality, we have for all $r>\max\{R_1(\eps),\dots,R_d(\eps)\}=R(\eps)$,
\[
\|\E_{\hat\phi_{r+1}}[\eta] - \E_1[\eta]\| \leq  T_r^{-\frac{1}{2}+\epsilon}
\]
In particular, since $\E_1[\eta]=\E_{\phi_\KL}[\eta]$, we have a.s.
\[
\lim_{r\to\infty}\E_{\hat\phi_{r}}[\eta] =\E_{\phi_\KL}[\eta].
\]
Since $\calQ$ is an exponential family of full-rank, convergence in the mean of the sufficient
statistic is equivalent to the convergence of the the natural parameters. That is,
\[
\hat\phi_{r}\xrightarrow[r\to\infty]{a.s.}\phi_\KL.
\]
\eprfof

\section{Upper bounds on the GCB}
\label{sec:upper_bounds_GCB}

\subsection{Proof of \cref{thm:GCB_KL}}

Suppose $X_{\phi,\beta},X'_{\phi,\beta}\sim \pi_{\phi,\beta}$ are indepdendent. 
By Jensen's inequality and Result 4 in \cite{dabak2002relations},
\[
    \Lambda(q_\phi, \pi_1)
    &= \frac{1}{2} \int_0^1 \EE[|\ell_\phi(X_{\phi,\beta}) - 
    \ell_\phi(X_{\phi,\beta})|] \, d\beta \\
    &\leq \frac{1}{2} \left(\int_0^1 
    \EE[(\ell_\phi(X_{\phi,\beta}) - \ell_\phi(X_{\phi,\beta}'))^2] \, 
    d\beta\right)^{1/2} \\
    &= \sqrt{\frac{1}{2}\text{SKL}(q_\phi, \pi_1)} \label{eq:GCB_upper_bound_SKL}.
\]
Also, because $\phi_\KL$ minimizes the forward KL divergence, it follows that
\[ \EE_{\phi_\KL}[\eta] = \EE_{\pi_1}[\eta].\]
In particular, by taking a dot product with $\phi$ for any $\phi \in \Phi$
it holds that
\[ 
  \EE_{\phi_\KL}[\log q_\phi] - \EE_{\pi_1}[\log
  q_\phi] = \EE_{\phi_\KL}[\log h] - \EE_{\pi_1}[\log h]. \label{eq:EE_q_equals_EE_pi}
\]
From \cref{eq:EE_q_equals_EE_pi},
\[
    \left|\text{SKL}(q_{\phi_\KL}, \pi_1)\right|
    &= \left|\EE_{\phi_\KL}\left[\log 
    \frac{\pi_1}{q_{\phi_\KL}}\right] - \EE_{\pi_1}\left[\log 
    \frac{\pi_1}{q_{\phi_\KL}}\right]\right| \\
    &= \abs[2]{ \EE_{\phi_\KL}\left[\log \pi_1\right] - 
    \EE_{\pi_1}\left[\log \pi_1\right] + \EE_{\pi_1}[\log h] - 
    \EE_{\phi_\KL}[\log h] }\\
    &= \abs[2]{ \EE_{\phi_\KL}\left[\log \pi_1 - \log 
    q_{\phi_0}\right] - \EE_{\pi_1}\left[\log 
    \pi_1 - \log q_{\phi_0}\right] } \\
    &\leq \EE_{\phi_\KL}\left[\left|\log \pi_1 - \log 
    q_{\phi_0}\right|\right] + \EE_{\pi_1}\left[\left|\log 
    \pi_1 - \log q_{\phi_0}\right|\right] \\
    &\leq \EE_{\phi_\KL}\left[g\right] + 
    \EE_{\pi_1}\left[g\right] \\
    &\leq M_1 + M_2. \label{eq:SKL_upper_bound_M}
\]
Therefore, combining \cref{eq:GCB_upper_bound_SKL} and \ref{eq:SKL_upper_bound_M},
\[
  \Lambda(q_\phi, \pi_1) \leq \sqrt{\frac{1}{2}(M_1 + M_2)}.
\]

\subsection{Example with multivariate normal distributions}

We consider here two simple examples to verify that the upper bound 
given by \cref{thm:GCB_KL} is small enough for practical purposes. 

To put into perspective the numerical values obtained in the examples below, note
that the GCBs measured by \cite{syed2019nrpt} in 17 problems were in the range
0.4--88.

\begin{example}
	Suppose that $\pi_1 \sim N(0, \Sigma_1)$ and $q_\phi \sim N(0, \Sigma_0(\phi))$
	where
	\[
	\Sigma_1 &= \begin{bmatrix} 1 & \rho \\ \rho & 1 \end{bmatrix} 
	\qquad \text{and} \qquad
	\Sigma_0(\phi) = \begin{bmatrix} \phi_1 & 0 \\ 0 & \phi_2 \end{bmatrix}
	\]
	for some $0 \leq \rho < 1$. By applying \cref{thm:GCB_KL} we have
	\[ \Lambda(q_{\phi_\text{KL}}, \pi_1) \leq \sqrt{-\frac{1}{2}\log(1-\rho^2) 
		+ \frac{\rho}{1-\rho}}.\]
	Substituting $\rho = 0.9, 0.95, 0.99$ provides GCB upper bounds of
	approximately
	$3.14, 4.49, 10.05$, respectively, while we obtained values of  
	$\Lambda(q_{\phi_\text{KL}}, \pi_1)$ of $\approx 0.8$, $1.0$ and $1.5$ using
	our stabilized moment matching algorithm.
\end{example}

\begin{proof}
Based on moment-matching, $\phi_\KL = (1, 
1)'$ and therefore $\Sigma_0(\phi_\KL)=\mathbb{I}_2$ is the $2\times 2$ identity 
matrix. 

We have that
\[
\left| \log \pi_1(x) - \log q_{\phi_\text{KL}} (x) \right|
&= \left| -\frac{1}{2} \log(1 - \rho^2) - \frac{1}{2} x^\top \Sigma_1^{-1} x
+ \frac{1}{2} x^\top \mathbb{I}_2 x \right| \\
&= \left| -\frac{1}{2} \log(1 - \rho^2) - \frac{1}{2} x^\top 
\left(\Sigma_1^{-1} - \mathbb{I}_2\right) x \right| \\
&\leq -\frac{1}{2} \log(1 - \rho^2) + \frac{1}{2} \|x\|^2 \max 
\{|\lambda_\text{max}|, |\lambda_\text{min}|\} \\
&= -\frac{1}{2} \log(1 - \rho^2) + \frac{1}{2} \|x\|^2 \max 
\left\{\frac{\rho}{1-\rho}, \frac{\rho}{1+\rho}\right\} \\
&= -\frac{1}{2} \log(1 - \rho^2) + \frac{1}{2} \|x\|^2 \frac{\rho}{1-\rho} 
\\
&=: g(x)
\]
by the min-max eigenvalue theorem. It can then be verified that 
\[M_1 = M_2 = -\frac{1}{2}\log(1-\rho^2) + \frac{\rho}{1-\rho}.\]
\end{proof}

\begin{example}
	Suppose that the target is a mixture of normal distributions so that 
	\[
	\pi_1 \sim 0.5 \cdot N(-\mu, 1) + 0.5 \cdot N(\mu, 1) 
	\qquad \text{and}\qquad q_\phi \sim N(0, \phi),
	\]
	for some $\mu$. We estimate the expectations in the upper bound of the GCB
	using Monte Carlo draws from 
	$q_{\phi_\text{KL}}$ and $\pi_1$. For $\mu = 5, 10, 100$ we find that 
	$\Lambda(q_{\phi_\text{KL}}, \pi_1)$ is upper bounded by approximately 1.7, 
	3.2, and 32, respectively (up to Monte Carlo estimation error based on
	1,000,000 Monte Carlo simulation draws for $\mu = 5, 10$, and 100,000,000 
	Monte Carlo simulations for $\mu = 100$), while we obtained values of  
	$\Lambda(q_{\phi_\text{KL}}, \pi_1)$ of $\approx 2.3$, $2.8$ and $4.2$ using
	our stabilized moment matching algorithm.
\end{example}

\begin{proof}
Based on moment-matching, we obtain that $\phi_\text{KL} = \mu^2 + 1$. 
Denote the pdf of the standard normal density by $\phi(\cdot)$. Then,
\[
    \left| \log \pi_1(x) - \log q_{\text{KL}}(x)\right|
    &= \left| \log\left(  \frac{1}{2} \phi(x + \mu) + \frac{1}{2} \phi(x - 
    \mu)\right) - \frac{1}{\sqrt{\mu^2+1}} \cdot 
    \phi\left(\frac{x}{\sqrt{\mu^2+1}}\right)\right| \\
    &=: g(x).
\]
\end{proof}

\newpage

\section{Proof of \cref{thm:restart_rate}}
The proof of this theorem is almost identical to the proof of Theorem 1 in 
\cite{syed2019nrpt}. The main difference is that we study two delayed renewal processes
simultaneously instead of one. 

First, define the index process for the $j$-th machine for $j=0, 1, \ldots, \barN$ 
as $(n_t(j), \epsilon_t(j))$ where $n_t(j) \in \{0, 1, \ldots, \barN\}$, $\epsilon_t(j) 
\in \{-1, 1\}$, and $\barN = N_\phi + N$. Here, $n_t(j)$ denotes the annealing parameter 
index for the $j$-th machine at iteration $t$ and $\epsilon_t(j) = 1$ if after iteration 
$t$ the annealing parameter on machine $j$ will be proposed to increase to index
$n_t(j) + 1$. Otherwise, $\epsilon_t^j = -1$. In particular, machine $j$ is storing annealing parameter $\bar{\beta}_{n_t(j)}$, and therefore  $n_t(j)=0,N_\phi,\bar{N}$ means the machine $j$ is at the annealing parameter corresponding to $q_\phi,\pi_1,\pi_0$, respectively.

Informally, a restart occurs when a sample from \textit{either} one of 
the two references reaches the target distribution chain. Because the 
target distribution is placed between the two references, we see that we can 
count the number of restarts by defining two delayed renewal processes and 
summing the number of restarts for each renewal process. We ensure that 
we are not double-counting any restarts by introducing two processes instead 
of one.

Machine $j$ undergoes a restart from $q_\phi$ when $n_t(j)$ goes from $0$ to $N_\phi$. 
Similarly, we will say machine $j$ undergoes a restart from $\pi_0$ when $n_t(j)$ goes 
from $\barN$ to $N_\phi$. We define $\calT_{\phi,t}(j)$ and $\calT_{t}(j)$ to be the 
total number of restarts on machine $j$ from $q_\phi$ and $\pi_0$ respectively by time 
iteration $t$. We will denote the total number of restarts by time $t$ from $q_\phi$ and 
$\pi_0$ as $\calT_{\phi,t}$ and $\calT_t$ respectively, so that
\[
\calT_{\phi,t}=\sum_{j=0}^{\bar{N}}\calT_{\phi,t}(j),\quad
\calT_{t}=\sum_{j=0}^{\bar{N}}\calT_{t}(j)
\]

Formally, define
\[ T_{\phi,0}^{-}(j) = \inf \{t : (n_t(j), \epsilon_t(j)) = (0,-1) \}\]
and then recursively define for $k \geq 1$
\[
    T_{\phi,k}^{+}(j) 
    &= \inf \{t > T_{\phi,k-1}^{-}(j) : (n_t(j),\epsilon_t(j)) = (N_\phi,+1) \} \\
    T_{\phi,k}^{-}(j) 
    &= \inf \{n >  T_{\phi,k}^{+}(j) : (n_t(j), \epsilon_t(j)) = (0,-1) \}. 
\]
We note that $T_{\phi,k}^{+}(j)$ corresponds to the time of the $k$-th restart from
$q_\phi$ on machine $j$ and 
\[
\calT_{\phi,t}(j) &= \max \{k : T_{\phi,k}^+(j) \leq t \}.
\]
We have $\calT_{\phi,t}(j)$ is a delayed renewal process counting the number of times 
a sample travels from chain 0 (targeting $q_\phi$) to chain $N_1$ (targeting $\pi_1$)
with inter-arrival times $T_{\phi,k}(j)=T_{\phi,k}^{+}(j)-T_{\phi,k-1}^{+}(j)$

Similarly, define
\[ T_{0}^{+}(j) = \inf \{t : (n_t(j), \epsilon_t(j)) = (\barN,+1) \}\]
and then recursively define for $k \geq 1$
\[
    T_{k}^{-}(j) 
    &= \inf \{t > T_{k-1}^{+}(j) : (n_t(j),\epsilon_t(j)) = (N_\phi,-1) \} \\
    T_{k}^{+}(j) 
    &= \inf \{t >  T_{k}^{-}(j) : (n_t(j), \epsilon_t(j)) = (\bar{N},+1) \}. 
\]
We note that $T_{k}^{-}(j)$ corresponds to the time of the $k$-th restart from $\pi_0$ 
on machine $j$ and 
\[
\calT_{t}(j) &= \max \{k : T_{k}^-(j) \leq t \}.
\]
We have $\calT_{t}(j)$ is a delayed renewal process counting the number of times a 
sample travels from chain $\bar{N}$ 
(targeting $\pi_0$) to chain $N_1$ (targeting $\pi_1$) with inter-arrival times
$T_k(j)=T_{k}^{-}(j)-T_{k-1}^{-}(j)$.

Although it is possible for a sample to travel from $q_\phi$ to $\pi_1$ and 
then to $\pi_0$ before returning to $q_\phi$, note that we are not 
double-counting or missing any restarts by including two renewal processes. More
importantly, by introducing two renewal processes (instead of one), we ensure that the
times between successive restarts from a given reference on each machine are 
independent and identically distributed. 

In particular, under \cref{assump:ELE}, the inter-arrival times $\{T_{\phi,k}(j)\}_{k}$,
and $\{T_{k}(j)\}_{k}$ for $\calT_{\phi,t}(j)$ and $\calT_t(j)$ respectively are i.i.d. 
for each machine $j$ with distributions $T_\phi$ and $T$ respectively. The round trip 
rate is thus,
\[
    \bar\tau_\phi(\bar\calB_{\phi,\barN}) 
    &= \lim_{t \to \infty} \frac{1}{t} 
    \EE[\calT_{\phi,t} + \calT_{t}] \nonumber \\
    &=  \sum_{j=0}^{\barN} \lim_{t \to \infty} \frac{1}{t}
    \EE[\calT_{\phi,t}(j)] 
    + \sum_{j=0}^{\barN} \lim_{t \to \infty} \frac{1}{t}
    \EE[\calT_{t}(j)] \nonumber \\
    &= \frac{\barN+1}{\EE[T_\phi]} + \frac{\barN + 
    1}{\EE[T]}, \label{eq:tau}
\]
where the last equality follows from the renewal theorem. We find an expression for 
$\EE[T_\phi]$ and argue the form for $\EE[T]$ by symmetry. As in 
\cite{syed2019nrpt}, we omit the $j$ index for the machine number and define $T^{\pi_1}$ and 
$T^{q_\phi}$ as the first time on a machine where the index reaches $n_t=N_\phi$ and 
$n_t=0$ respectively: 
\[
    T^{\pi_1} &= \min\{t : (n_t, \epsilon_t) = (N_\phi, 1)\} \\
    T^{q_\phi} &= \min\{t : (n_t, \epsilon_t) = (0, -1)\}. 
\]
We define $A^{\pi_1}_{n,\epsilon}$ and $A^{q_\phi}_{n,\epsilon}$ to be the expected 
time for a machine with index process initialized at $(n_0,\epsilon_0)=(n,\epsilon)$, 
to reach $\pi_1$ and $q_\phi$ respectively,
\[
A_{n,\epsilon}^{\pi_1} &= \EE[T^{\pi_1} | n_0 = n, \epsilon_0 = \epsilon]\\
A_{n,\epsilon}^{q_\phi} &= \EE[T^{q_\phi} | n_0 = n, \epsilon_0 = \epsilon].
\]
We can decompose the expected inter-arrival time $\E[T_\phi]$ as at the expected time 
for a machine to travel from $q_\phi$ to $\pi_1$ plus the expected time to travel from 
$\pi_1$ to $q_\phi$,
\[
\label{eq:expectation_T}
    \EE[T_\phi] = A^{\pi_1}_{0,-1} + A^{q_\phi}_{N_\phi,+1}.
\]

Also, for notation convenience, we redefine $r_n$, as the probability of a swap is accepted and respectively rejected 
between machines with annealing parameter $\bar\beta_{n-1}$ and $\bar\beta_n$.

\begin{lemma}
\label{lemma:reference_to_target}
\[
\label{eq:a_uparrow}
    A_{0,-1}^{\pi_1} = N_\phi + 1 + 2\cdot\sum_{n=1}^{N_\phi} n\cdot\frac{r_n}{1-r_n}.
\]
\end{lemma}
\bprfof{\cref{lemma:reference_to_target}}
Note that by definition, $A_{0,-1}^{\pi_1} = \EE[T^{\pi_1} | (n_0,\epsilon_0) =(0,-)]$. 
Because it is impossible for the index process to reach chains
$N_\phi+1, N_\phi+2,\dots, \barN$ before time $T^\pi$, these chains do not 
enter the calculations for $A_{0,-1}^{\pi_1}$. Therefore \eqref{eq:a_uparrow} is the
same expression for ``$a_\uparrow^{0,-}$'' as in \cite{syed2019nrpt} 
but with $N$ replaced by $N_\phi$.
\eprfof

\begin{lemma}
\label{lemma:target_to_reference}
\[
\label{eq:a_downarrow}
    A^{q_\phi}_{N_\phi,+1} = 2(\bar{N}+1) - (N_\phi+1) + 2\cdot\sum_{n=1}^{N_\phi} (\bar{N}+1- n)\cdot\frac{r_n}{1-r_n}.
\]
\end{lemma}
\bprfof{\cref{lemma:target_to_reference}}
It follows from the proof of Theorem 1 in \cite{syed2019nrpt} that for 
$p\in \{\pi_1,q_\phi\}$, and $1\leq n\leq \bar{N}$, $A^p_{n,\epsilon}$ satisfies the 
following recursive relation,
\[
A^{p}_{n,+1}-A^p_{n-1,+1}&=r_n(A_{n,+1}^p-A_{n-1,-1}^p)-1 \label{eq:recursion_A+}\\
A^{p}_{n,-1}-A^p_{n-1,-1}&=r_{n-1}(A_{n,+1}^p-A_{n-1,-1}^p)+1.\label{eq:recursion_A-}
\]
If we define $C_n^p$ and $D_n^p$,
\[
C_n^p&=A_{n,+1}^p+A_{n-1,-1}^p\label{eq:def_C}\\
D_n^p&=A_{n,+1}^p-A_{n-1,-1}^p.\label{eq:def_D}
\]
By adding and subtracting \eqref{eq:recursion_A+} and \eqref{eq:recursion_A-}, we get 
the joint recursion in $C^p_n$ and $D^p_n$, for $n=1,\dots, \bar{N}$:
\[
C^p_{n+1}-C_n^p=r_{n+1}D^p_{n+1}+r_nD^p_n,\label{eq:recursion_C}\\
(1-r_{n})D_n^p=(1-r_{n+1})D_{n+1}^p+2\label{eq:recursion_D}.
\]

If the machine's index process is initialized at $(n_0,\epsilon_0)=(0,-1)$, then 
$A^{q_\phi}_{0,-1}=0$. We can then substitute in $n=1$ into \eqref{eq:def_C} and 
\eqref{eq:def_D} to get,
\[
C_1^{q_\phi}=D_1^{q_\phi}.\label{eq:recursion_C_1}
\]

Similarly if the machine's index process is initialized at $(n_0,\epsilon_0)=(\barN,+1)$,
then  
$A_{\bar{N},+1}^{q_\phi}=1+A_{\bar{N},-1}^{q_\phi}$. We can substitute this into 
\eqref{eq:recursion_A-} for $n=\barN$ to get,
\[
(1-r_{\barN})D^{q_\phi}_{\barN}=2.\label{eq:recursion_D_N}
\]
Using recursion \eqref{eq:recursion_D}, we find for $n=1,\dots, \barN$,
\[
(1-r_n)D^{q_\phi}_n=2(\bar{N}+1-n).
\]

By adding \eqref{eq:def_C} and \eqref{eq:def_D}, that for $1\leq n\leq \bar{N}$,
\[\label{eq:recursion_D_N_solution}
2A^p_{n,+1}=C_n^p+ D_n^p.
\]
We can decompose $C_n^{q_\phi}$ as a telescoping sum, and use \eqref{eq:recursion_C} 
within initial condition \eqref{eq:recursion_C_1} to get the following expression for 
$2A^{q_\phi}_{n,+1}$ in terms of $D_n^{q_\phi}$
\[
    2 A^{q_\phi}_{N_{\phi},+1}
    &= C^{q_\phi}_n + D^{q_\phi}_n \\
    &= C^{q_\phi}_1 + D^{q_\phi}_n + \sum_{n=1}^{N_{\phi}-1} (C^{q_\phi}_{n+1} - 
    C^{q_\phi}_n) \\
    &= D^{q_\phi}_1 + D^{q_\phi}_{N_{\phi}} + \sum_{n=1}^{N_{\phi}-1} (r_{n+1} 
    D^{q_\phi}_{n+1} + r_{N_{\phi}} D^{q_\phi}_{N_{\phi}}) \\
    &= (1-r_1) D^{q_\phi}_1 + (1-r_{N_{\phi}}) D^{q_\phi}_{N_{\phi}} 
    + 2 \sum_{n=1}^{N_{\phi}} r_n 
    D^{q_\phi}_n\\
    &= 2\bar{N} + 2(\barN+1-{N_{\phi}})
    +2 \sum_{n=1}^{N_{\phi}} 2 (\bar{N}+1-n) \frac{r_n}{1-r_n}.
\]
We arrived at the last line by using \eqref{eq:recursion_D_N_solution}. Therefore, 
by dividing by $2$ we arrive at our result.
\eprfof

By combining \cref{lemma:reference_to_target} and \cref{lemma:target_to_reference}, 
and using \cref{eq:expectation_T},
\[
\EE[T_\phi] 
&= A^{\pi_1}_{0,-1} + A^{q_\phi}_{N_\phi,+1}.\\
&= 2(\bar{N}+1)+ 2(\barN+1) \sum_{n=1}^{N_\phi} \frac{r_n}{1-r_n}.
\label{eq:expectation_T_closed}
\]
By symmetry, we can repeat the same calculation to compute $\E[T]$, the expected
restart time for $\pi_0$ to get,
\[
\label{eq:expectation_U_closed}
    \EE[T] = 2(\bar{N}+1)+ 2(\barN+1) \sum_{n=N_{\phi}+1}^{\bar{N}} \frac{r_n}{1-r_n}.
\]
Therefore, \cref{eq:expectation_T_closed} and \cref{eq:expectation_U_closed} 
in \cref{eq:tau} imply that
\[
\bar\tau_\phi(\bar\calB_{\phi,\barN})
&=\frac{\bar{N}+1}{\E[T_\phi]}+\frac{\bar{N}+1}{\E[T]}\\
&=\frac{1}{2 + 2 \sum_{n=1}^{N_\phi} \frac{r_n}{1-r_n}} +
\frac{1}{2 + 2 \sum_{n=N_\phi+1}^{\barN} \frac{r_n}{1-r_n}}\\
&=\tau_\phi(\calB_{\phi,N_\phi})+\tau(\calB_{N}).
\]

Finally, it follows from Theorem 3 in \cite{syed2019nrpt}, that as
$\|\calB_N\|,\|\calB_{\phi,N_\Phi}\|\to 0$ we have 
\[
\lim_{N_\phi\to\infty}\tau_\phi(\calB_{\phi,N_\phi})
&=\frac{1}{2+2\Lambda(q_\phi,\pi_1)}\\
\lim_{N\to\infty}\tau(\calB_N)
&=\frac{1}{2+2\Lambda(\pi_0,\pi_1)}.
\]  
Therefore, as $\|\barB_{\phi,\barN}\|\to 0$ chain-asymptotic restart rate satisfies,
\[
\lim_{\bar{N}\to\infty}\bar\tau_\phi(\bar\calB_{\phi,\barN}) = \frac{1}{2 + 2 
\Lambda(q_\phi, \pi_1)} + \frac{1}{2 + 2 \Lambda(\pi_0, \pi_1)}.
\]

\newpage

\section{Details of experiments}
\label{sec:experiments_details}

\begin{table}[t]
    \centering
    \begin{tabular}{rrccc}
    	\toprule
        &Inference problem & $n$ & $d$ &  $\hat\Lambda(\pi_0, \pi_1)$ \\ 
      \midrule 
	      Synthetic &Product & 100,000 & 2 & 3.7 \\
	      &Simple-Mix (collapsed) & 300 & 5 & 4.3 \\
	      &Elliptic & 100,000 & 2 & 4.4 \\
	      &Toy-Mix & NA & 1 & 9.3 \\
	    \midrule
        Real data&Transfection & 52 & 5 & 6.4 \\ 
        &Titanic & 887 & 9 & 7.7  \\
        &Rockets (collapsed) & 5,667 & 2 & 3.4 \\       
        &Challenger & 23 & 2 & 4.2  \\
        &Change-point & 109 & 7 & 3.3\\  
        &Vaccines & 77,828 & 12 & 7.8 \\
        &Lip Cancer & 536 & 60 & 15  \\
        &Pollution & 22,548 & 275 & 56.9  \\
        &8 schools & 8 & 10 & 0.8  \\
        &Phylogenetic inference & 249 & 10,395 & 7.0  \\
        &Spring failure (improper prior) & 10 & 1 & Not defined  \\
       \bottomrule
       \vspace{0.1in}
    \end{tabular} 
    \caption{Summary of the models considered in this paper. The sample size $n$, number of model parameters $d$, and fixed reference GCB $\hat\Lambda(\pi_0, \pi_1)$ (defined for models with a proper prior distribution). The label ``collapsed'' is used for the models that are based on \cite{syed2019nrpt} but with some latent variables analytically marginalized. }
    \label{tab:models}
\end{table}

\subsection{ODE parameters (mRNA transfection data)}\label{sec:ode}
The first data set that we consider (``Transfection'') is a time series data set based on
mRNA transfection data and an ordinary differential equation (ODE) model \cite{ballnus2017comprehensive}. The data are 
observations $O_t$ at times $t=1,2,\ldots,52$ modelled as $O_t | k_{m_0}, 
\delta, \beta, t_0, \sigma \sim N(\mu, \sigma^2)$ where $k_{m_0}, \delta, \beta,
t_0, \sigma$ are parameters and 
\[ \mu = \frac{k_{m_0}}{\delta - \beta} \left( 1 - \exp(-(\delta - \beta) \cdot 
(t - t_0)\right) \cdot \exp(-\beta \cdot (t-t_0)).\]
The priors placed on the parameters are all log-uniform (i.e., distributed 
according to $10^{U}$ where $U$ denotes a random variable with a uniform distribution on 
$[a,b]$). Specifically, $k_{m_0}, \delta, \beta \sim \text{LogUniform}(-5, 5)$, 
$t_0 \sim \text{LogUniform}(-2, 1)$, and $\sigma \sim \text{LogUniform}(-2,2)$, 
with all parameters a priori independent. We use 50 PT chains, unless stated otherwise. Data is included in the supplement under \texttt{bl-vpt/data/m\_rna\_transfection/processed.csv}.

\subsection{GLM (Titanic data)}
\label{sec:titanic} 
Next, we consider a binary generalized linear model (``Titanic'') with binary 
response data, $Y_i$, indicators of survival for the $i$-th Titanic passenger, 
along with several covariates, $\bfx_i$. We assume that $Y_i | \beta_0, \beta_1,
\bfx_i \sim \text{Bernoulli}(1/(1+\exp(-(\beta_0+\beta_1^\top \bfx_i))))$. Note 
that this example is similar to the one used by \cite{syed2019nrpt} although our
prior differs in that we assume that $\beta_0, \beta_{1,1}, \beta_{1,2}, \ldots,
\beta_{1,7} \sim \text{Cauchy}(\sigma)$ with scale parameter $\sigma$. 
Further, 
$\sigma \sim \text{Exp}(1)$. All parameters are assumed to be independent in the
prior. We use 30 PT chains, unless stated otherwise. 
Data is included in the supplement under \texttt{bl-vpt/data/titanic}.

\subsection{Unidentifiable product}
The unidentifiable product model (``Product'') is an artificial data set of size
$n=100,000$. The number of failures of an experiment, $n_f$, given the number of
trials $n_t$ and parameters $x$ and $y$ is modelled as $n_f | n_t, x, y \sim 
\text{Bin}(n_t, x \cdot y)$. We place priors $X,Y \sim U(0,1)$ and set $n_t = 
100,000$. We observe $n_f = 50,000$ failures. For this model, $d=2$. Due to 
identifiability, the posterior concentrates on a thin curve in the square 
$[0,1]\times [0,1]$. We use 15 PT chains, unless stated otherwise.

\subsection{Mixture model}
We consider a multi-modal posterior that arises from a normal mixture model 
(``Simple-Mix''). We model data $X_1, X_2, \ldots, X_m | \mu_1, \mu_2, \sigma_1, 
\sigma_2, \pi \sim \pi \cdot N(\mu_1, \sigma_1^2) + (1-\pi) \cdot N(\mu_2, 
\sigma_2^2)$ with priors $\mu_1, \mu_2 \sim N(150, 100^2)$, $\sigma_1, \sigma_2 
\sim U(0, 100)$, and $\pi \sim U(0,1)$. 
Due to the 
label-switching problem, the resulting posterior is multi-modal. 
In all the experiments we marginalize over the mixture model indicator variables
in contrast to \cite{syed2019nrpt}. 
We use 10 PT chains, unless stated otherwise. 
Data is included in the supplement under \texttt{VariationalPT/data/simple-mix.csv}.

\subsection{Bayesian hierarchical model}
We also consider a Bayesian hierarchical model based on 
rocket launch failure data (``Rockets''). The number of rocket launch failures, $f_r$, along 
with the total number of rocket launches, $l_r$, for $R=367$ types of rockets is
obtained. Given the probability of rocket launch failure for rocket $r$, we 
model $f_r | \pi_r, l_r \sim \text{Bin}(l_r, \pi_r)$. Given parameters $m$ and 
$s$, we model $\pi_r | m, s \sim \text{Beta}(ms, (1-m)s)$ and use the 
hyper-prior $m \sim U(0,1)$ and $s \sim \text{Exp}(0.1)$ (rate parameter). 
In this example we perform inference over each of the $\pi_r$ and $m, s$ so that $d = 
369$ and $n=5,667$. 
In all the experiments we marginalize over the random effects 
in contrast to \cite{syed2019nrpt}. 
We use 10 PT chains, unless stated otherwise. 
Data is included in the supplement under \texttt{bl-vpt/data/failure\_counts.csv}.

\subsection{Weakly identifiable elliptic curve}
A weakly identifiable model is also considered (``Elliptic''). This data set is an artificial data set with the 
number of failures of an experiment, $n_f$, given the number of trials $n_t$ and
parameters $x, y$ modelled as $n_f | n_t, x, y \sim \text{Bin}(n_t, p(x,y))$ 
where
\[
    p'(x,y) &= y^2 - x^3 + 2x - 0.5 \\
    p(x,y) &= \begin{cases} 
      0 & p'(x,y) < 0 \\
      1 & p'(x,y) > 1 \\
      p'(x,y) & \text{ otherwise}.
    \end{cases}
\]
We observe $n_t = 100,000$ trials and $n_f = 50,000$ failures. We use the prior 
$X, Y \sim U(-3, 3)$. We use 30 PT 
chains, unless stated otherwise.

\subsection{GLM (Challenger data)}\label{sec:challenger}
Another GLM but applied to Challenger shuttle O-ring data (``Challenger''). 
The responses consist of binary indicator variables for incidents, $Y_i$, for 23
shuttle launches at temperatures $x_i$. We model the responses as $Y_i | x_i, 
\beta_0, \beta_1 \sim \text{Bern}(1/(1+\exp(-(\beta_0 + \beta_1 x_i))))$ and use 
the prior $\beta_0, \beta_1 \sim N(0, 100)$. 
We use 15 PT 
chains, unless stated otherwise.
Data is included in the supplement under \texttt{bl-vpt/data/challenger}.

\subsection{Additional Blang models}\label{sec:additional-blang} \label{sec:change-point}

We have implemented the models in Sections \ref{sec:ode}--\ref{sec:challenger} both 
procedurally in Julia and declaratively in the Blang probabilistic programming language (PPL) 
\cite{Bouchard2021Blang} to ensure agreement of the results. 
We then implemented additional models in Blang, briefly described in this section, to illustrate that the proposed method 
can be seamlessly incorporated into a PPL. 
To see more details on these additional models (as well as those described in Sections \ref{sec:ode}--\ref{sec:challenger}), see their PPL source code in the 
sub-directory 
\texttt{bl-vpt/src/main/java/ptbm/models/} of the supplement. 

\begin{figure}[t]
  \centering
  \includegraphics[width=0.5\textwidth]{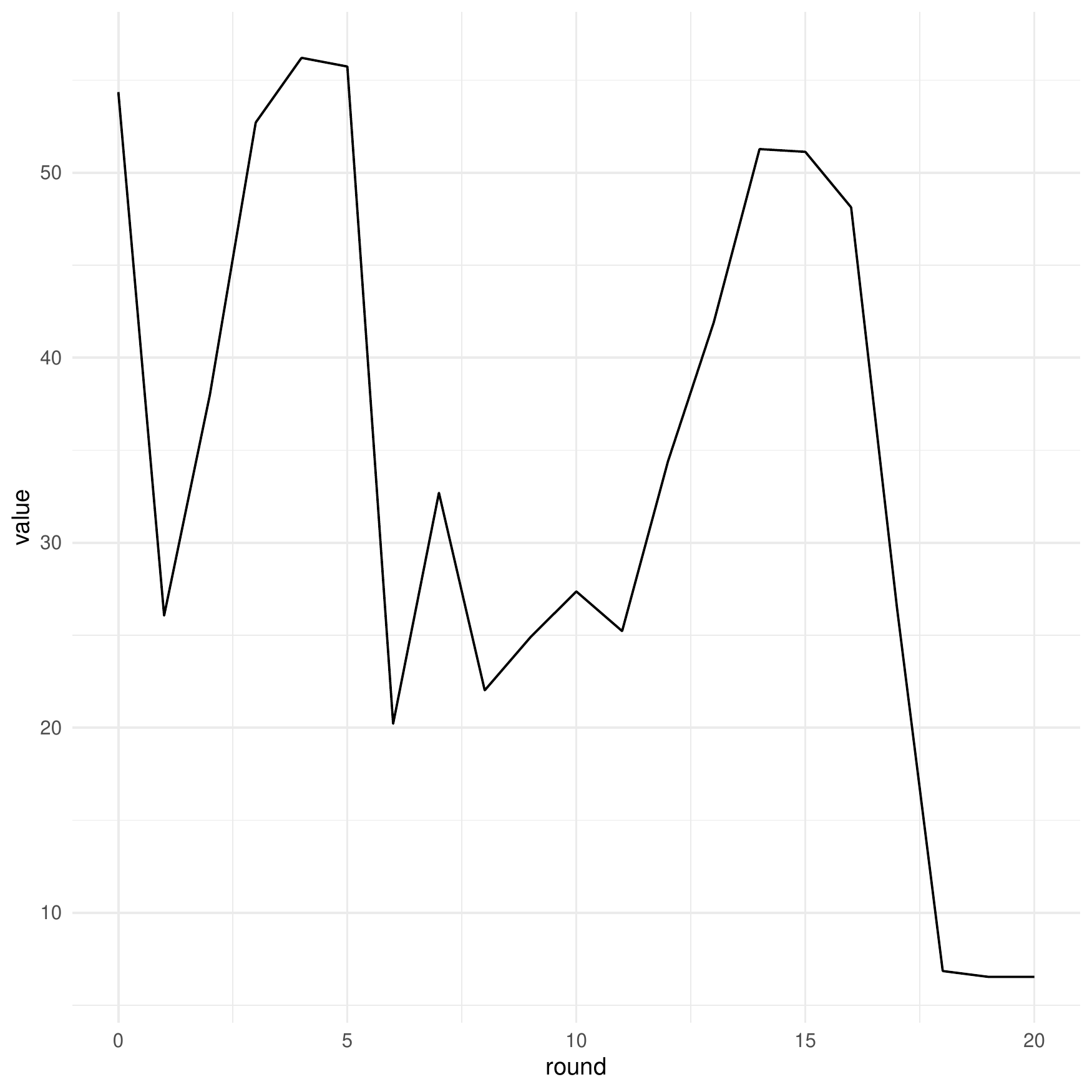}
  \caption{Global communication barrier for the Pollution model in the variational reference chain during each of the tuning rounds. The GCB for the fixed
    reference chain is estimated to be 57.}
  \label{fig:GCB_Pollution}
\end{figure}

\begin{figure}[!htbp]
	\centering
	\begin{subfigure}{0.6\textwidth}
		\centering
		\includegraphics[width=\textwidth]{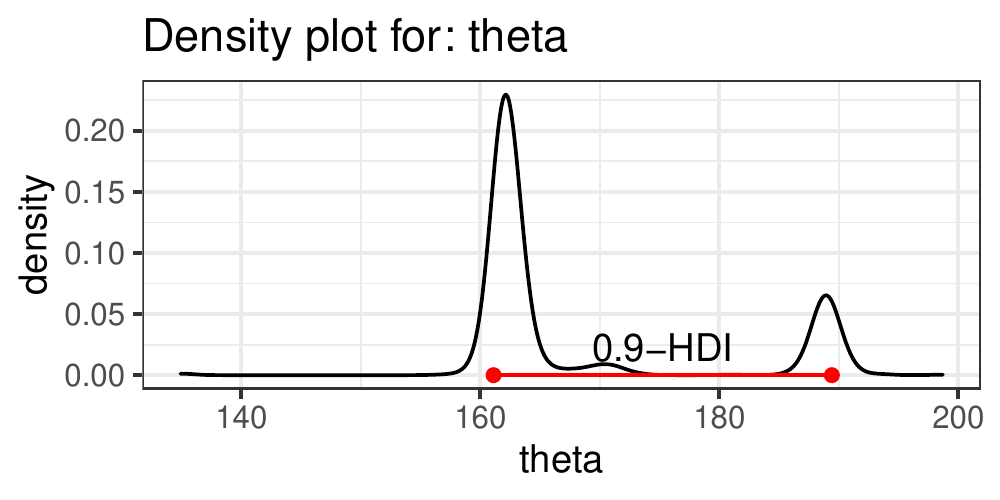}
	\end{subfigure}
	\begin{subfigure}{0.49\textwidth}
		\centering
		\includegraphics[width=\textwidth]{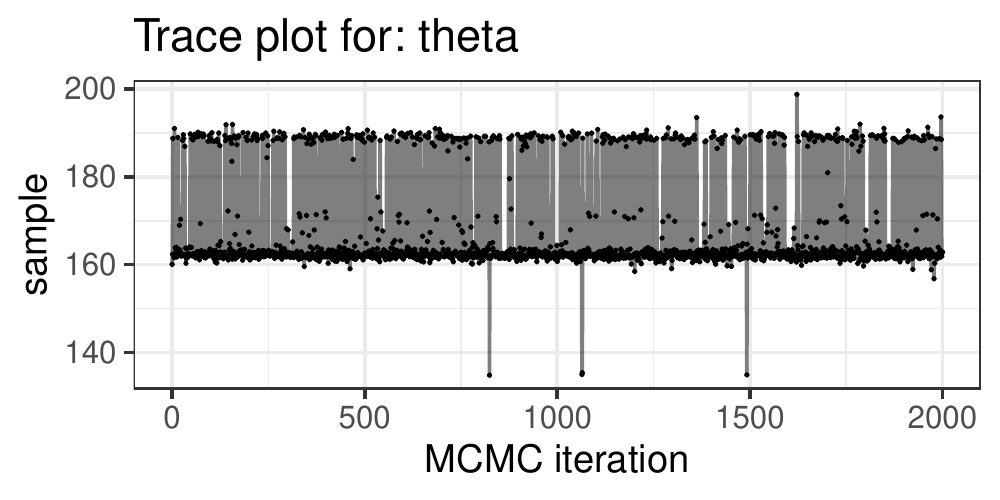}
	\end{subfigure}
	\begin{subfigure}{0.49\textwidth}
		\centering
		\includegraphics[width=\textwidth]{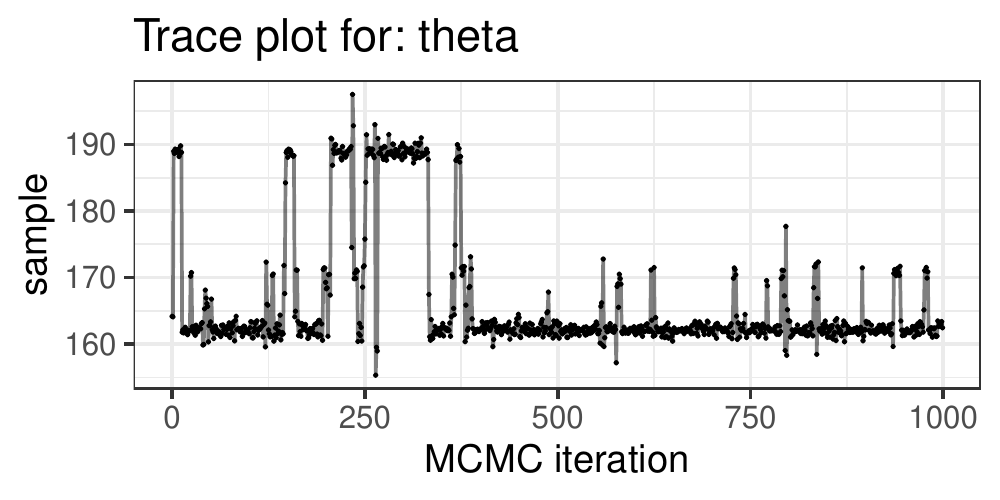}
	\end{subfigure}
		\begin{subfigure}{0.3\textwidth}
			\centering
			\includegraphics[width=\textwidth]{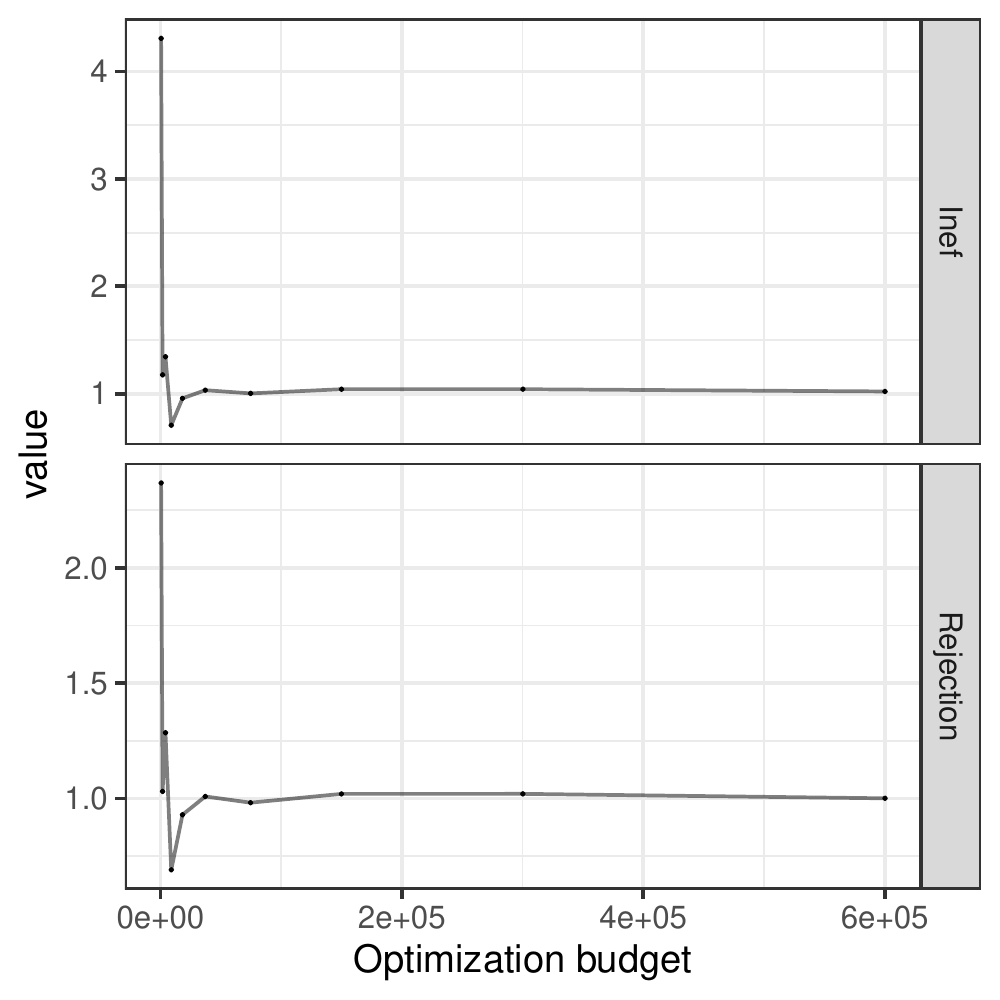}
		\end{subfigure}
		\begin{subfigure}{0.3\textwidth}
			\centering
			\includegraphics[width=\textwidth]{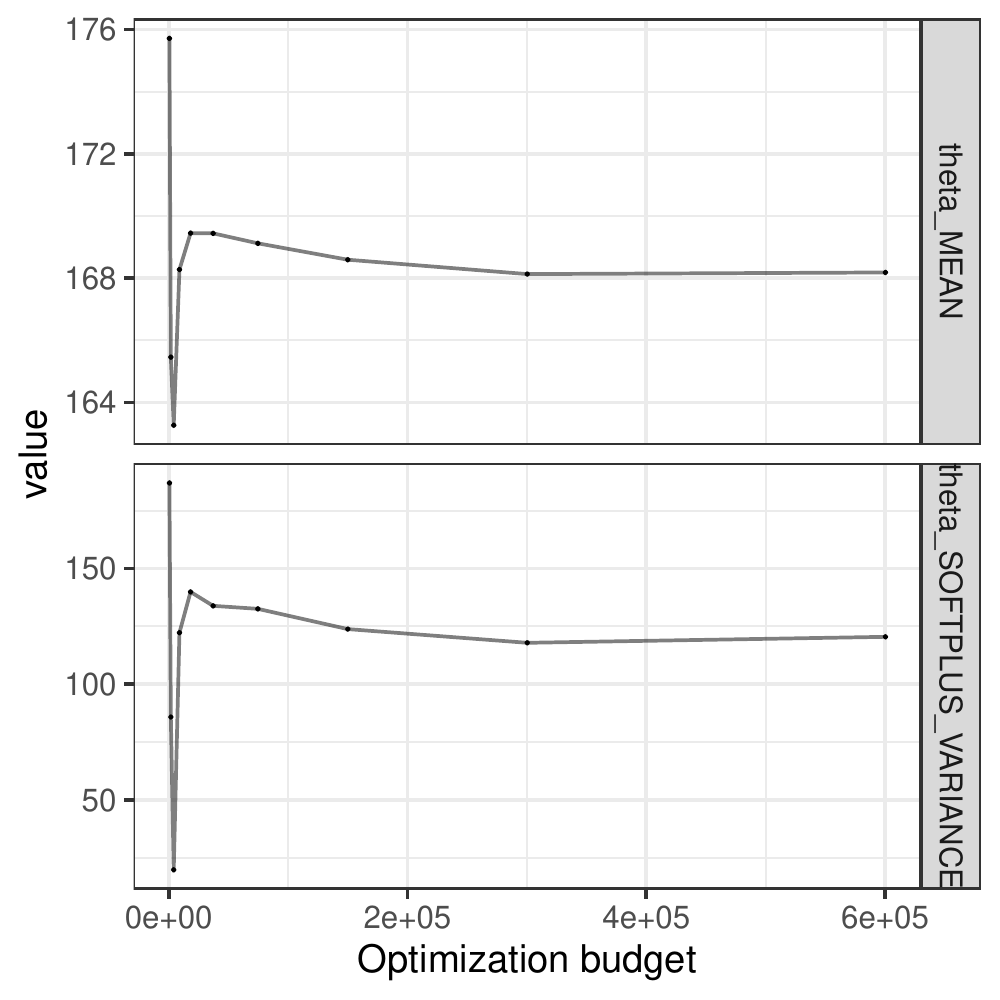}
		\end{subfigure}
		\begin{subfigure}{0.3\textwidth}
			\centering
			\includegraphics[width=\textwidth]{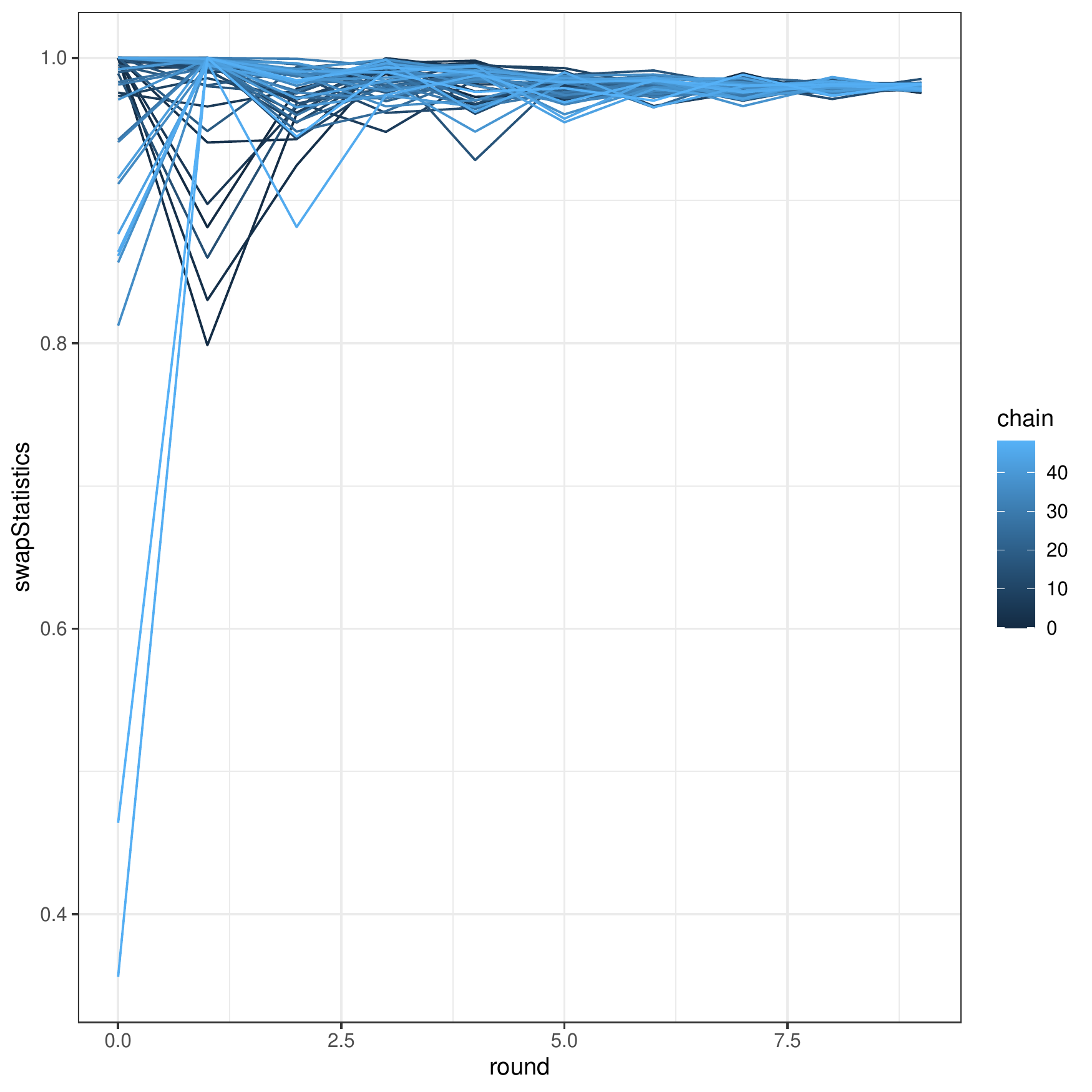}
		\end{subfigure}
	\begin{subfigure}{0.3\textwidth}
		\centering
		\includegraphics[width=\textwidth]{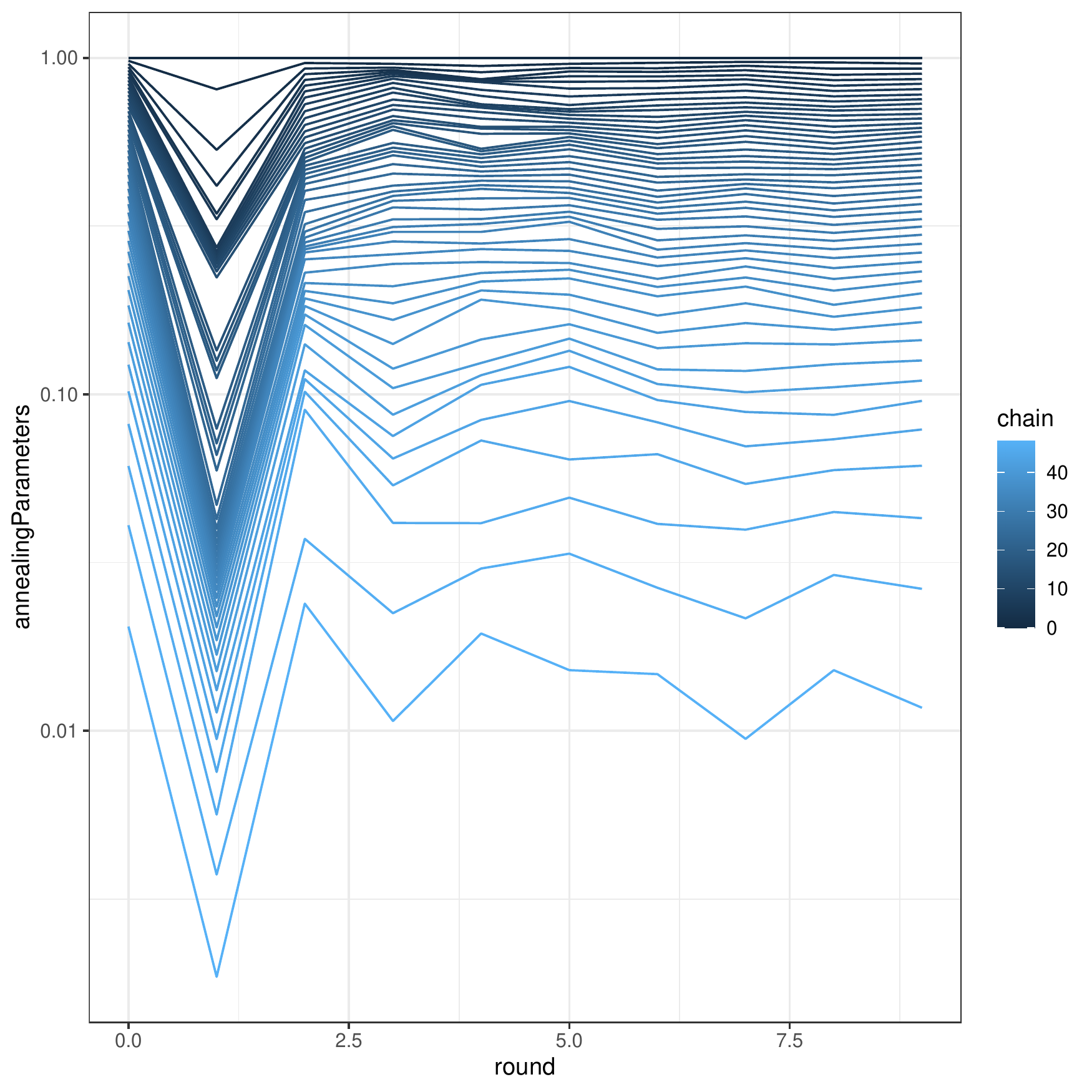}
	\end{subfigure}
	\begin{subfigure}{0.3\textwidth}
		\centering
		\includegraphics[width=\textwidth]{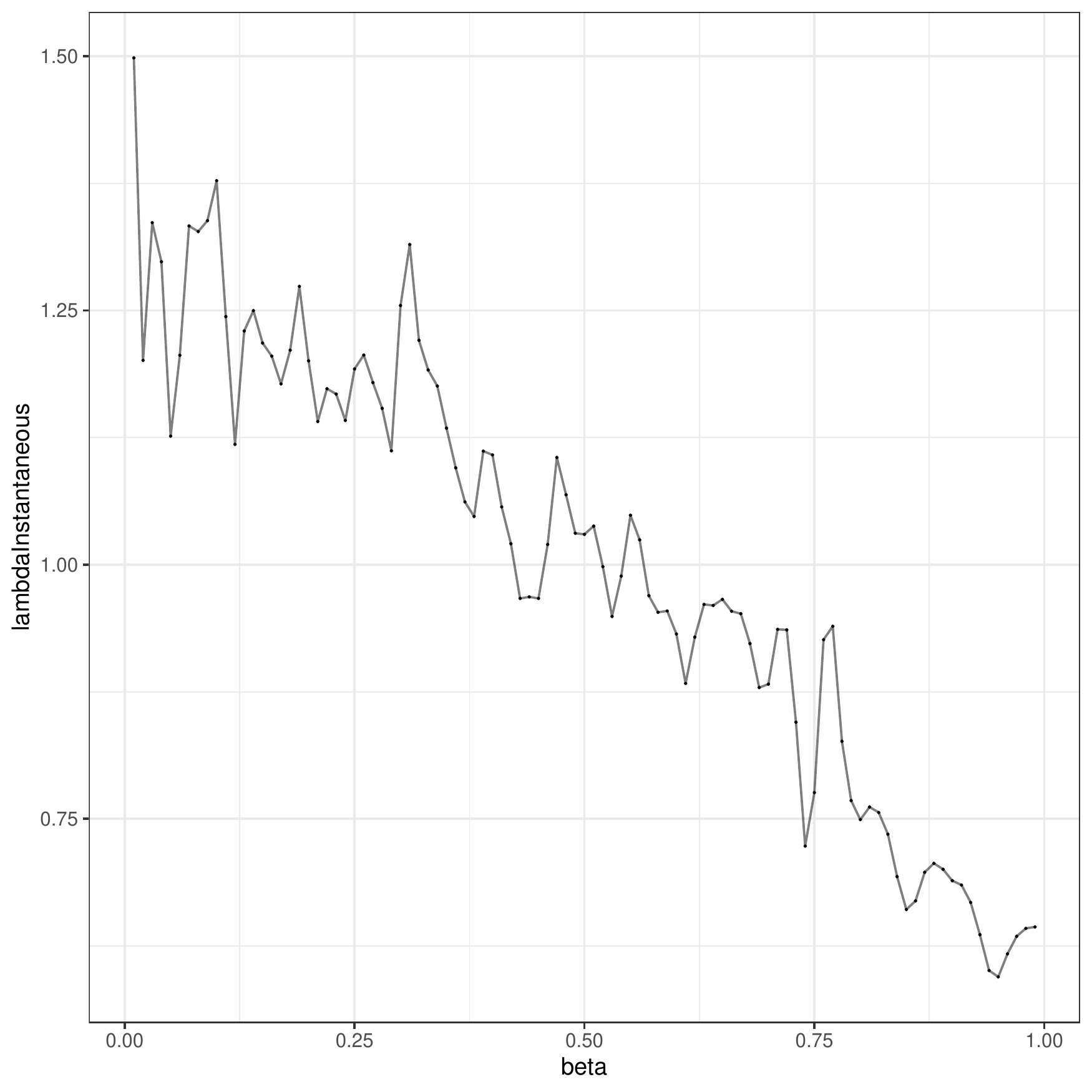}
	\end{subfigure}
	\begin{subfigure}{0.3\textwidth}
		\centering
		\includegraphics[width=\textwidth]{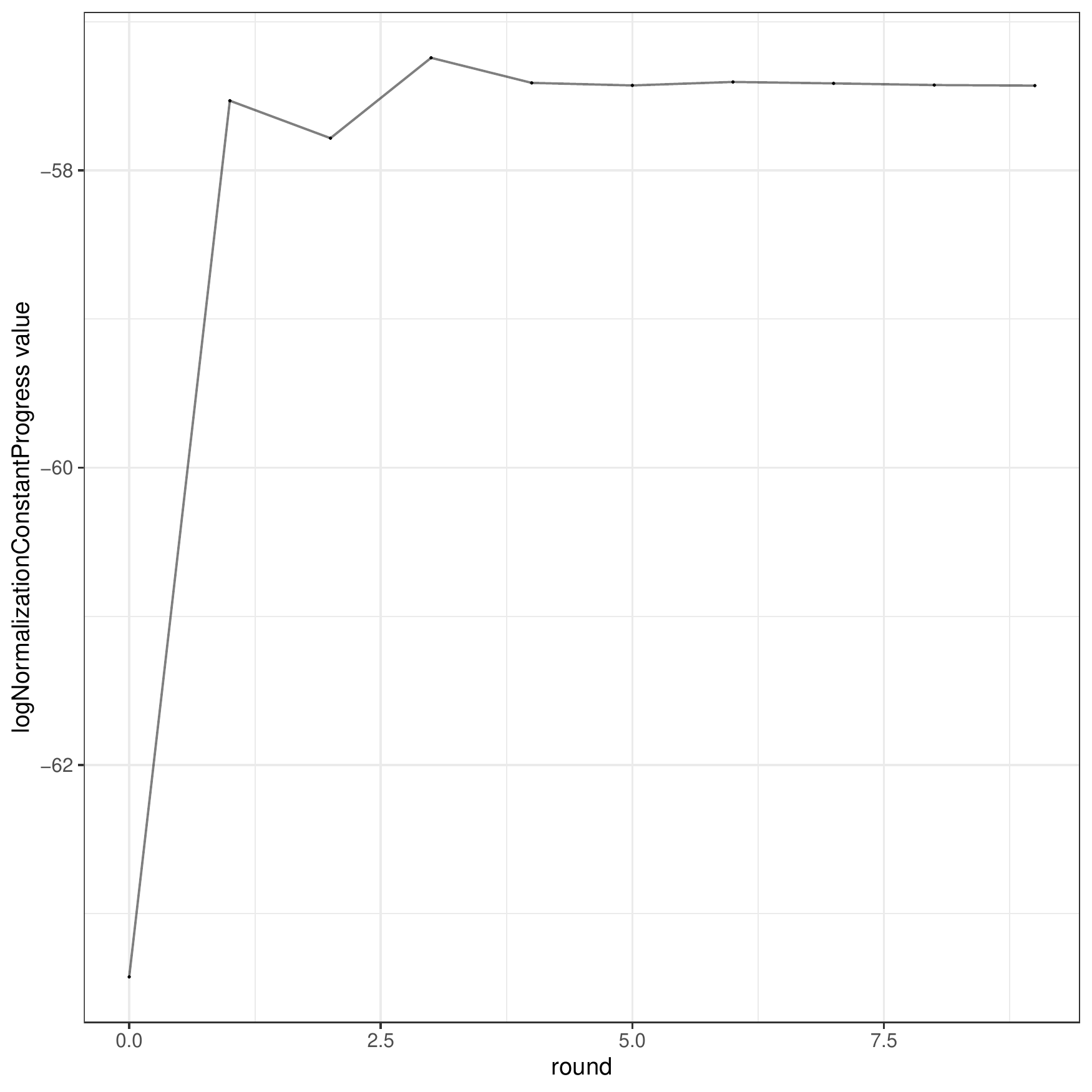}
	\end{subfigure}
	\caption{{\footnotesize Stabilized variational PT applied to a density estimation 
	problem with a Cauchy likelihood and an improper prior (Section~\ref{sec:additional-blang}).
	  {\bf First row:} post-burn-in approximation of the posterior distribution over the location
	  parameter $\theta$. Multi-modality is clearly visible. {\bf Second row, left:} 
	  trace plot of the same parameter obtained from stabilized variational PT. 
	  The stabilized variational PT algorithm is able to frequently switch modes. 
	  {\bf Second row, right:} same trace plot but obtained from a single chain MCMC 
	  algorithm with the same exploration kernel. {\bf Third row, left:} estimates of 
	  two objective functions as a function of the adaptation rounds (``Inef'' refers to 
	  the sum of the rejection odds, ``Rejection,'' to the sum of rejection rates). 
	  {\bf Third row, center:} value of the variational parameters as a function of 
	  the adaptation round. The facet ``theta\_MEAN'' refers to the mean of the 
	  variational normal distribution approximating the posterior on $\theta$, and 
	  ``theta\_SOFTPLUS\_VARIANCE'', to the (reparameterized) variance parameter of the 
	  same variational distribution. {\bf Third row, right:} average swap acceptance 
	  probabilities for the $N=50$ chains as a function of the adaptation round. 
	  {\bf Bottom row, left:} each chain's $\beta_i$ as a function of the adaptation round. 
	  {\bf Bottom row, center:} estimated local communication barrier. Even with the 
	  relatively large number of chains used in this example, it appears intrinsically less
	  smooth than for the other examples considered in this paper. This may be due to the 
	  heavier likelihood tails used in this example. 
	  {\bf Bottom row, right:} estimated normalization constant based on the stepping 
	  stone method as a function of the adaptation round.}}
	\label{fig:cauchy}
\end{figure}

\begin{description}
	\item[Lip Cancer and Pollution:] a sparse Conditional Auto Regressive (CAR) Poisson regression spatial model, 
		constructed as in \url{https://mc-stan.org/users/documentation/case-studies/mbjoseph-CARStan.html} (full model available in \texttt{bl-vpt/src/main/java/ptbm/models/SparseCAR.bl}). For the 
		Lip Cancer problem, we use the dataset documented in the R package \texttt{CARBayesdata}, see 
		\texttt{bl-vpt/data/scotland\_lip\_cancer} in the supplement for more information on preprocessing. 
    The Pollution problem is based on the same model but with the bigger dataset documented in \texttt{bl-vpt/data/pollution\_health}, also obtained from the \texttt{CARBayesdata} package on CRAN, which is under a GPL-2 / GPL-3 license. 
		This provides some examples of how the global communication barrier can be decreased in a realistic data analysis scenario: in the first dataset, from 
15 to 6, and in the second, from 57 to 7.  For the Pollution dataset, we found the GCB to decrease substantially after considering a larger number of tuning 
rounds with 4 million PT scans. The GCB with the variational reference during each of the tuning rounds for this model is presented in \cref{fig:GCB_Pollution}.

	\item[Vaccines:] the data (\texttt{bl-vpt/data/vaccines/data.csv}) consists in the following Phase III COVID-19 vaccines clinical trials: Pfizer-BioNTech, Moderna-NIH, and two AZ-Oxford trials (`South Africa B.1.351' and `Combined'). The number of cases in each arm of 
	each trial is modelled using a binomial. In each control arm, the probability parameter is set to a trial-specific incidence parameter. In each `treatment' arm, the probability parameter is set to the same incidence parameter multiplied by one minus a trial-specific efficacy parameter. While the incidence and efficacy parameters are trial specific, the parameters of these distributions are tied across the four trials (full model available in \texttt{bl-vpt/src/main/java/ptbm/models/Vaccines.bl}).
	\item[Change-point:] a time series model of discrete counts. We assume one change point uniformly distributed on 
	the time series. The observations at each time point is negative binomial distributed with one set of parameters for the observations before the change point and one set for after the change point (full model available in \texttt{bl-vpt/src/main/java/ptbm/models/Mining.bl}). This model is applied to the dataset of annual numbers of accidents due to disasters in British coal mines for years from 1850 to 1962, considered in Carlin et al. (1992)  (\texttt{bl-vpt/data/mining-disasters.csv}).
	\item[8 schools:] the classical problem from Rubin (1981), see \texttt{bl-vpt/src/main/java/ptbm/models/EightSchools.bl} and \texttt{bl-vpt/data/eight-schools.csv}.
	\item[Phylogenetics:] The same setup as the \emph{phylogenetic species tree problem} from \cite{syed2019nrpt} (\texttt{bl-vpt/src/main/java/ptbm/models/PhylogeneticTree.bl}). The data consists in aligned mtDNA sequences (\texttt{bl-vpt/data/FES\_8.g.fasta}). We only fit variational distributions to the real-valued evolutionary parameters---there is work on constructing variational families for phylogenetic trees \cite{baele_genealogical_2016, karcher_variational_2021}, but we leave the combination of these families with our inference method for future work. 
	\item[Spring failure:] The density estimation problem for spring failure data described in Davison (2003), example 4.2, with a Cauchy likelihood (\texttt{bl-vpt/src/main/java/ptbm/models/ImproperCauchy.bl}). We use the Lebesgue improper prior for the Cauchy likelihood's  location parameter. This illustrates another use case for our method: when standard PT is applied to a model with an improper prior, one has to select $\beta_0 > 0$ such that $\beta > \beta_0$ guarantees that $\pi_\beta$ is a (normalizable) probability distribution, moreover it is in general not possible to get i.i.d.\ samples from $\pi_{\beta_0}$. In contrast, our method works with $\beta_0 = 0$ and allows i.i.d.\ sampling from the variational chain. In the stabilized variant, for the non-adaptive ``leg'' one can use a fixed variational parameter instead of $\pi_0$. Results for this model are shown in \cref{fig:cauchy}.
	\item[ToyMix:] the density $f(x) = 0.5 \phi(x; -r, 0.01) + 0.5 \phi(x; r, 0.01)$ where $\phi(\cdot; \mu, \sigma^2)$ is the normal density (\texttt{bl-vpt/src/main/java/ptbm/models/ToyMix.bl}). We use $r = 10$. 
\end{description}

\subsection{Markovian score climbing}
For the \texttt{Transfection} model, we implemented the Markovian score climbing (MSC) algorithm of \cite{naesseth_markovian_2020}.
In the MSC algorithm, there are three main tuning parameters: the number of tuning rounds $K$, the number of samples used
in each tuning round $S$, and the step size (or step size sequence) $\epsilon$. The inputs to the algorithm
also include the starting values of the variational parameters $\lambda_0$, and the starting sample observation $z_0$.
For the mean-field normal variational references, we use the marginal means and log standard deviations as the variational
parameters so that all parameters can take on values in $\reals$. 
We initialize each of the log standard deviation parameters to $\log(100.0)$. For the mean parameters, we initialize them 
to $(1.0, 1.0, 1.0, 0.32, 1.0)$, which are also the initial values provided to the MCMC samplers in the PT algorithms.
We use $K=100,000$ tuning rounds with $S=100$ and a step size of $\epsilon = 0.0005$. The initial sample provided is the vector of 
initial mean parameters. The MSC experiments are run 10 times for each selection of simulation settings.

Initially, we tried larger values of the step size parameters, starting at $\epsilon = 0.1$. However, we observed that the estimated
variational parameters jumped around a large range of values. We therefore decreased the value of $\epsilon$ until there were no convergence issues
(i.e., no exceptionally large gradients and the variational parameter estimates seemed to converge based on plots of the parameter estimates against the 
tuning round number). We also ran the experiment with a smaller step size of $\epsilon = 0.00025$. However, we noticed that the variational parameter estimates
displayed a similar behaviour: the parameter estimate would center on only one of the two modes.

\subsection{Additional Details}\label{sec:additional-details}
We use the same number of chains in each leg of variational PT with 
two reference distributions. For ESS calculations in Julia, we use all samples from the target distribution, including those
obtained during annealing schedule and variational reference tuning rounds. For the implementation of 
variational PT with two references, we employ two separate PT chains: one with a fixed reference and one with a variational reference. During each tuning round, samples from
the target distribution in each chain are pooled together to estimate the variational parameters 
for the reference distribution. In Blang, we implement the topology in which the target 
distributions between the instances of PT are connected. More information about these different
topologies can be found in \cref{sec:additional-topo-comparisons}. 
In the Julia implementation of the variational PT algorithm, the covariance matrices 
for the Gaussian reference are estimated using the functions \texttt{std()} and \texttt{cov()}.
The estimates of the variance therefore differ by a factor of $n/(n-1)$ 
compared to the MLE estimates where $n$ is the sample size for parameter estimation.

All experiments use the same exploration kernel, namely a slice sampling algorithm with ``doubling and shrinking'' \cite[Sections 4.1--4.2]{neal_slice_2003}.
We performed preliminary experiments with HMC. However, we found that having to tune one HMC sampler for each annealed chain was onerous. 
We leave the problem of adapting several annealed exploration kernels to future work.
Slice sampling in contrast does not have sensitive tuning parameters that require tuning. 
For the initial state in each chain for the simulations of \cref{sec:VPT_versus_NRPT}, the same state is 
used for all PT initializations for the different seeds. The initial state was chosen to lie 
in a region with positive density with respect to the reference and target distributions. 

Unless mentioned otherwise, all PT algorithms use Deterministic Even Odd (DEO) swaps and the NRPT adaptive schedule algorithm. 
See \cite{syed2019nrpt} for details. 
Julia simulations were run on an Intel i9-10900K processor with 32 GB of RAM.
We also acknowledge use of the ARC Sockeye computing platform 
(Gold and Silver Xeon 2.1 GHz processors) from the University of British Columbia.

\subsection{Additional experimental results}
\label{sec:experiments_supplement}

\subsubsection{Additional plots}
\label{sec:additional_plots}
\begin{figure}[!htbp]
    \centering
    \begin{subfigure}{0.325\textwidth}
      \centering
      \includegraphics[width=\textwidth]{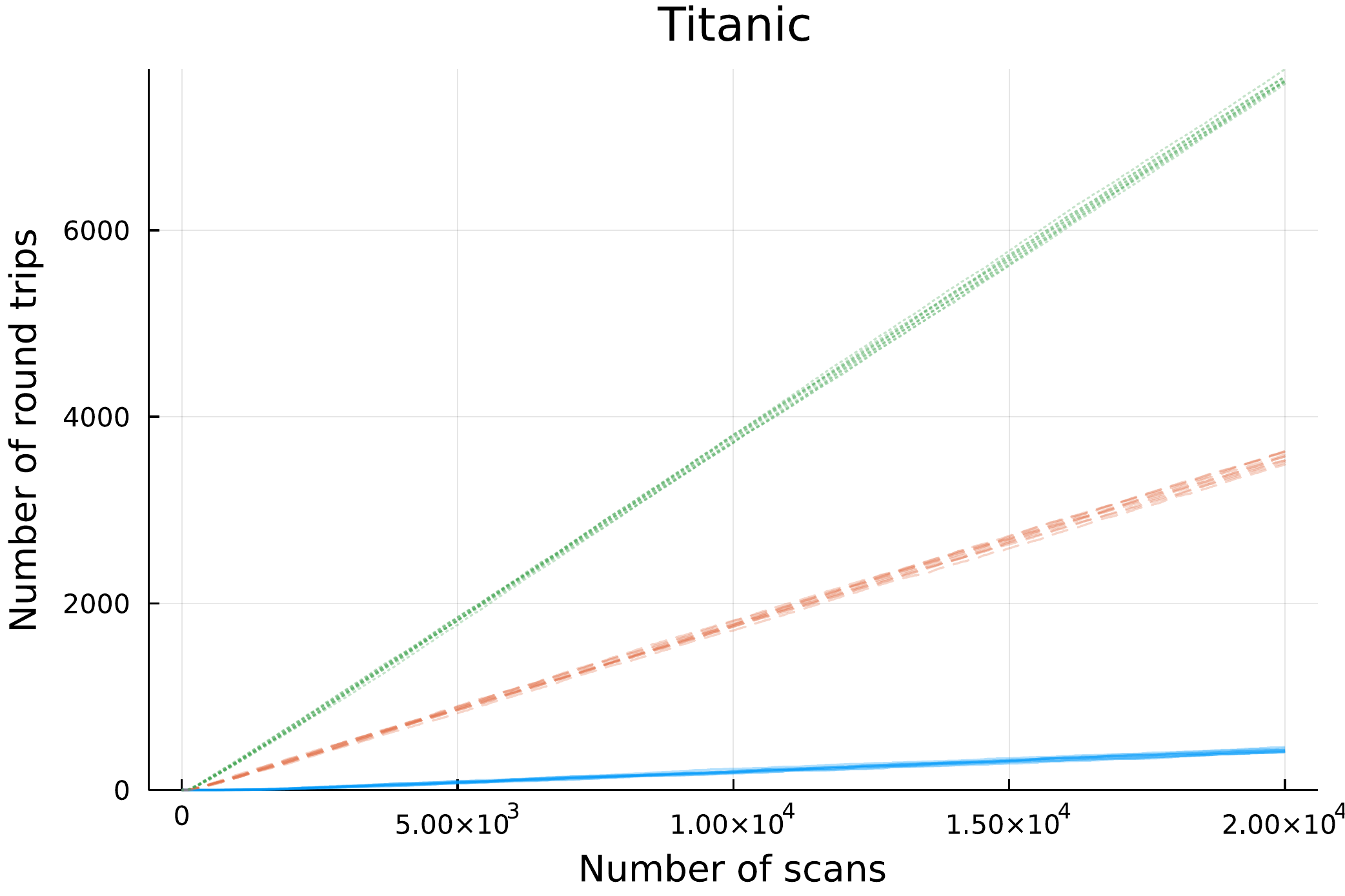}
    \end{subfigure}
    \begin{subfigure}{0.325\textwidth}
      \centering
      \includegraphics[width=\textwidth]{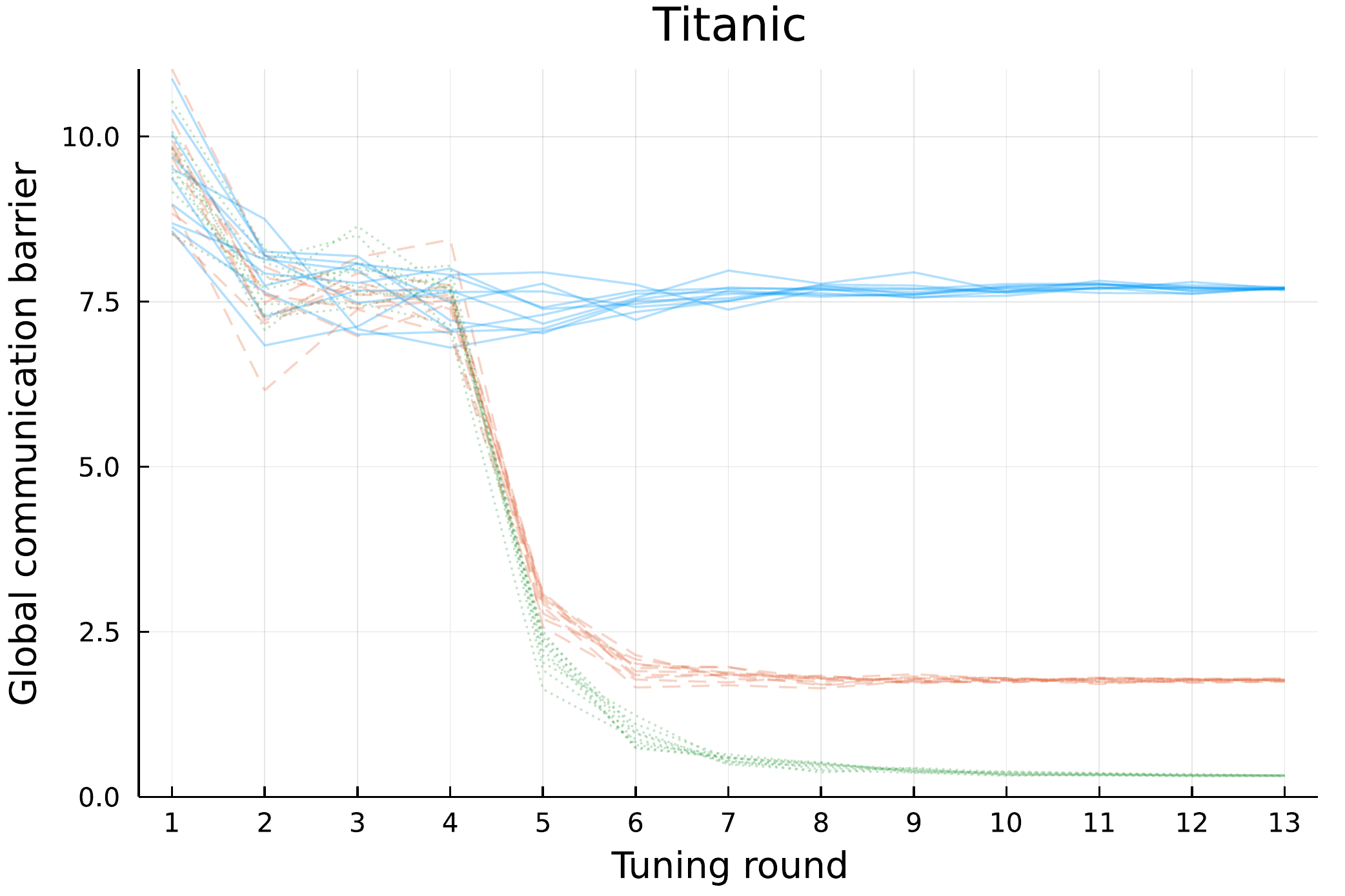}
    \end{subfigure}
    \begin{subfigure}{0.325\textwidth}
      \centering
      \includegraphics[width=\textwidth]{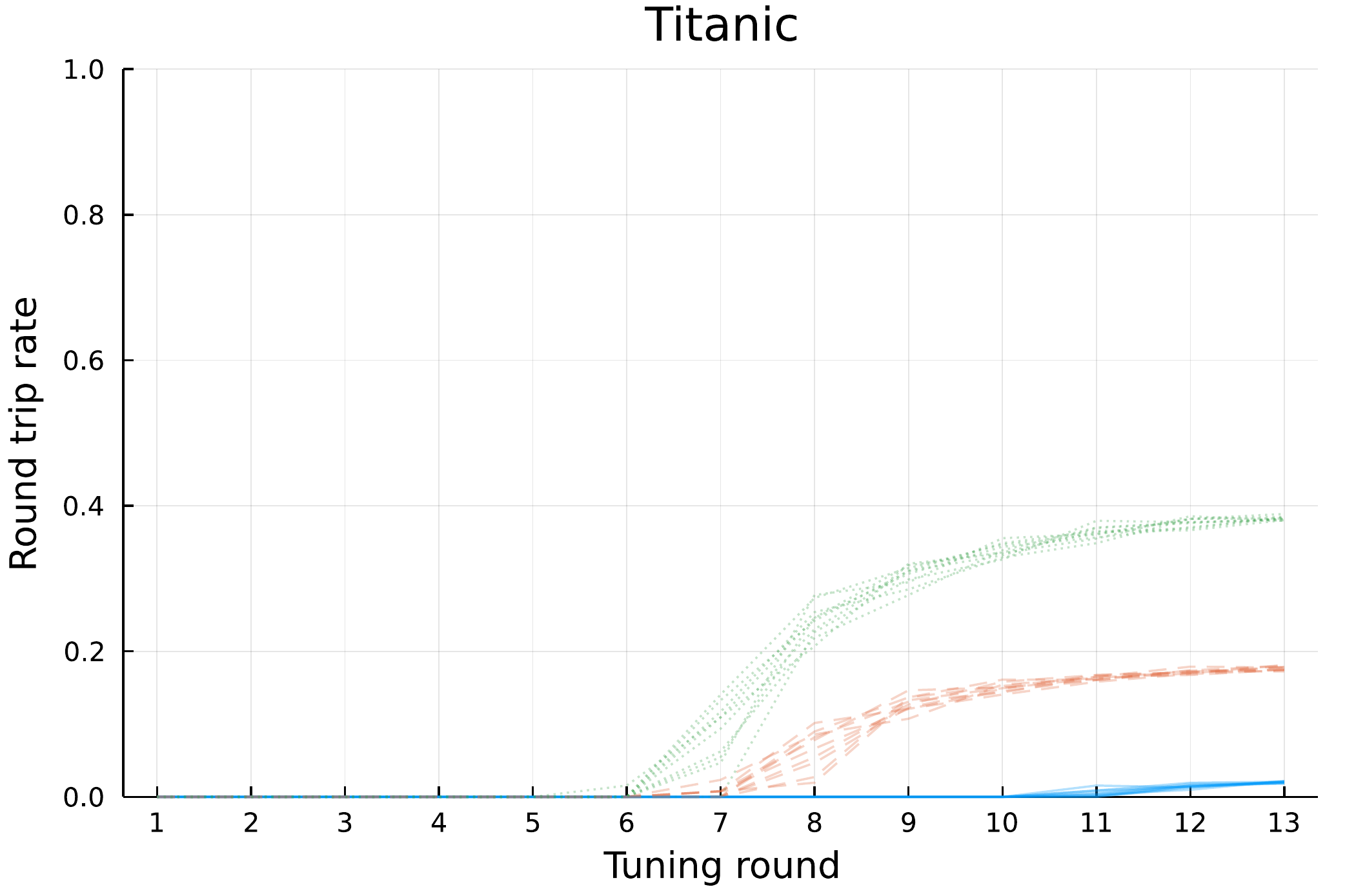}
    \end{subfigure}
    \begin{subfigure}{0.325\textwidth}
      \centering
      \includegraphics[width=\textwidth]{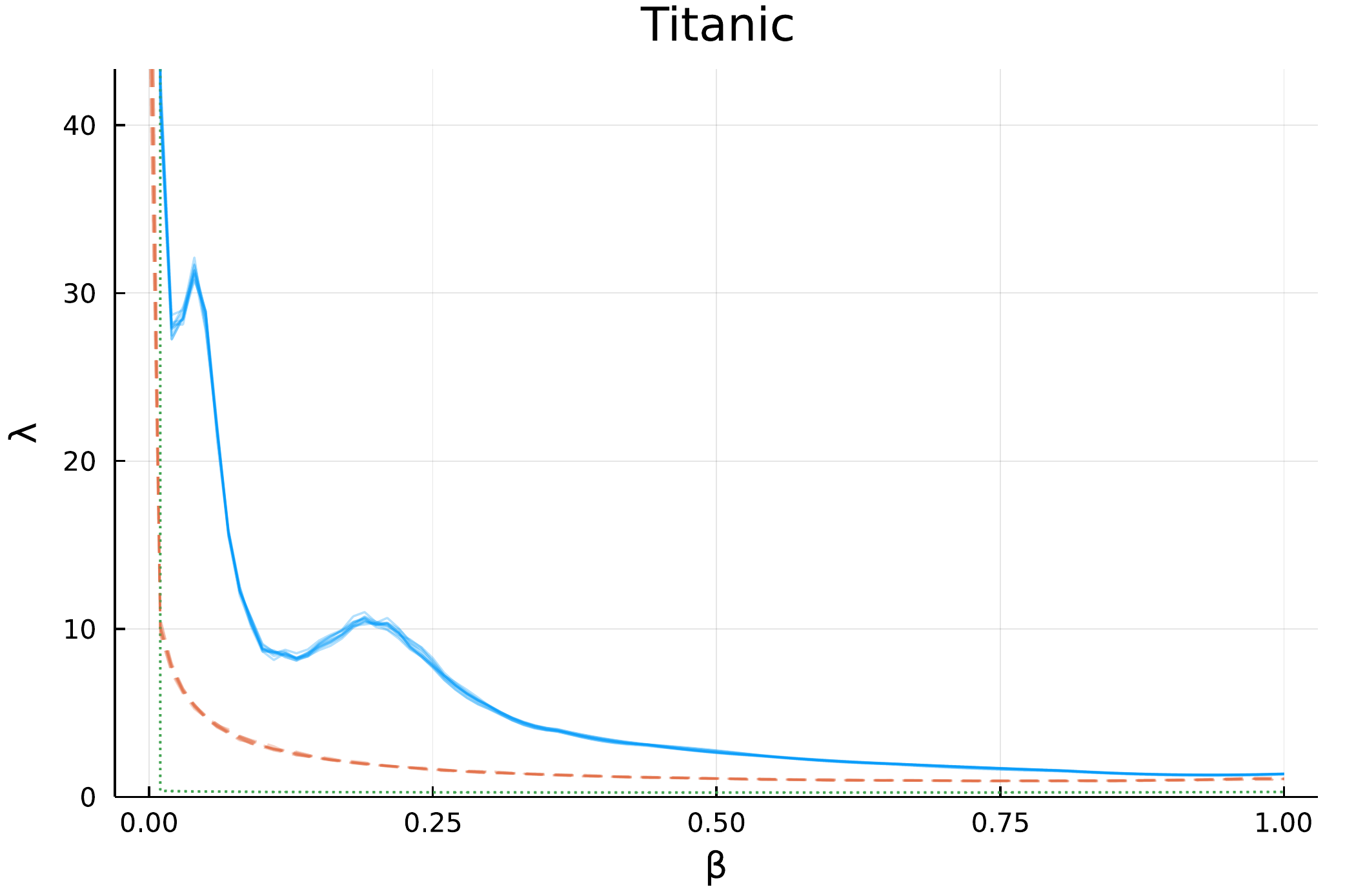}
    \end{subfigure}
    \begin{subfigure}{0.325\textwidth}
      \centering
      \includegraphics[width=\textwidth]{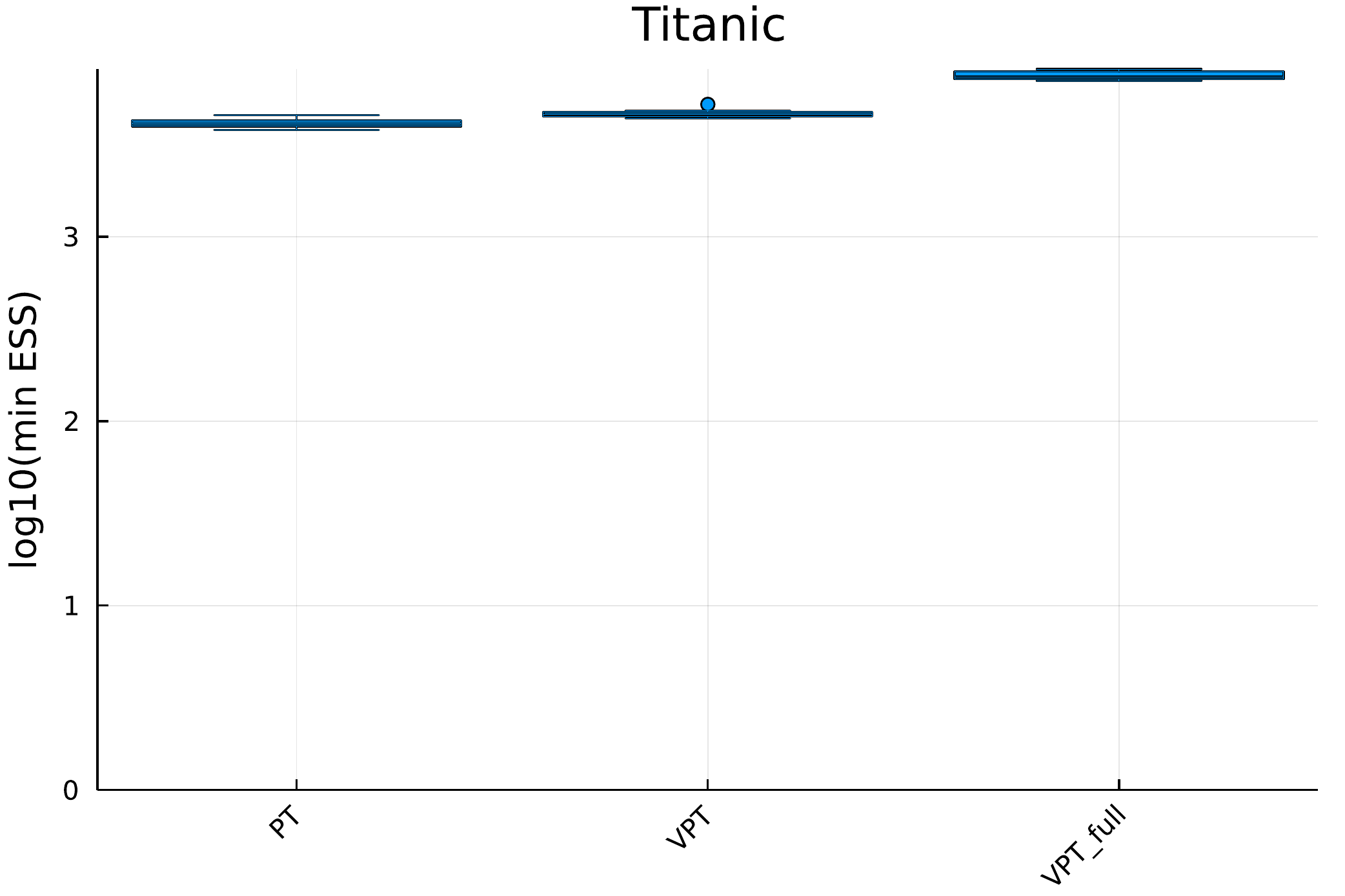}
    \end{subfigure}
    \begin{subfigure}{0.325\textwidth}
      \centering
      \includegraphics[width=\textwidth]{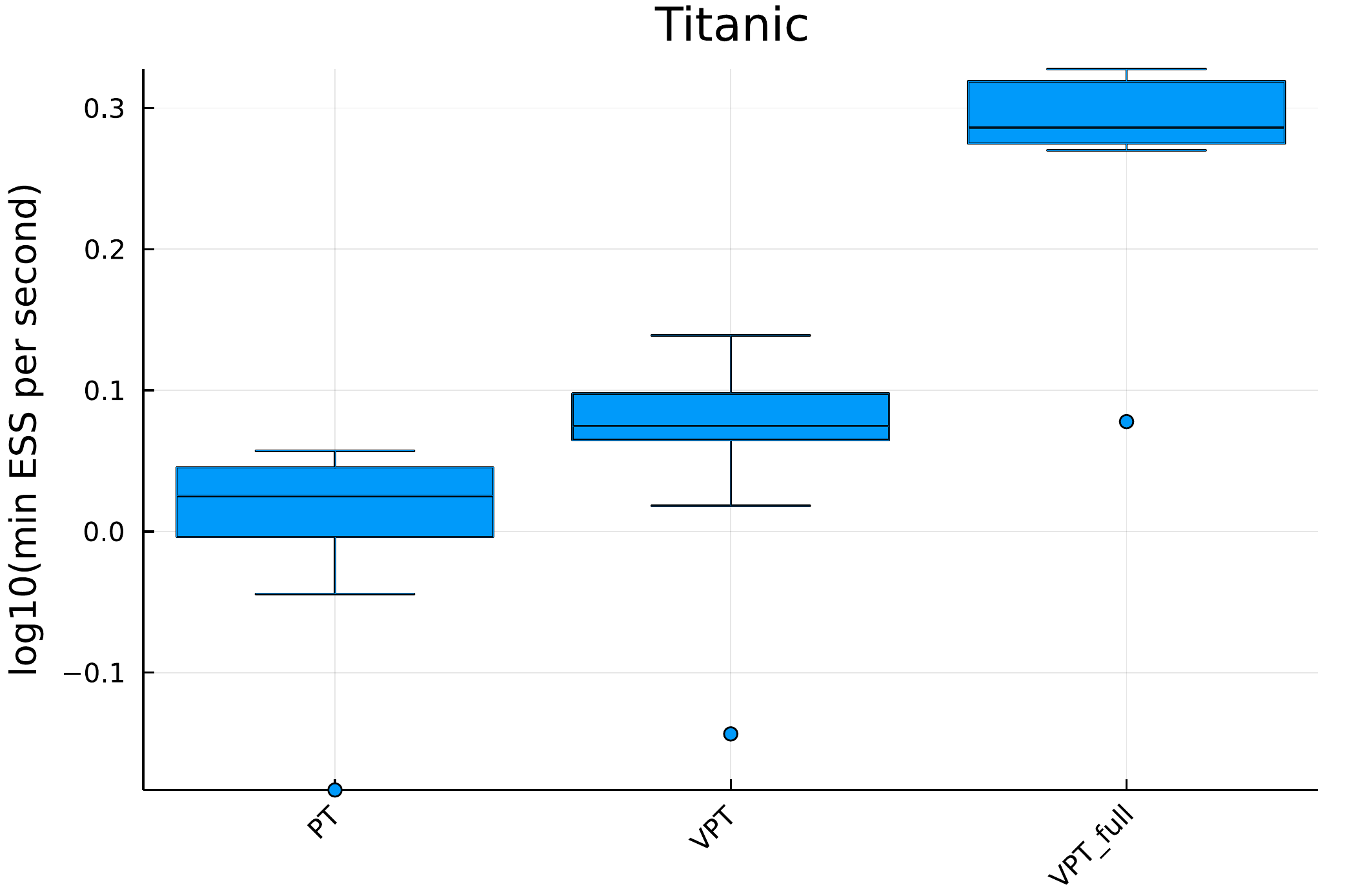}
    \end{subfigure}
    \begin{subfigure}{0.325\textwidth}
      \centering
      \includegraphics[width=\textwidth]{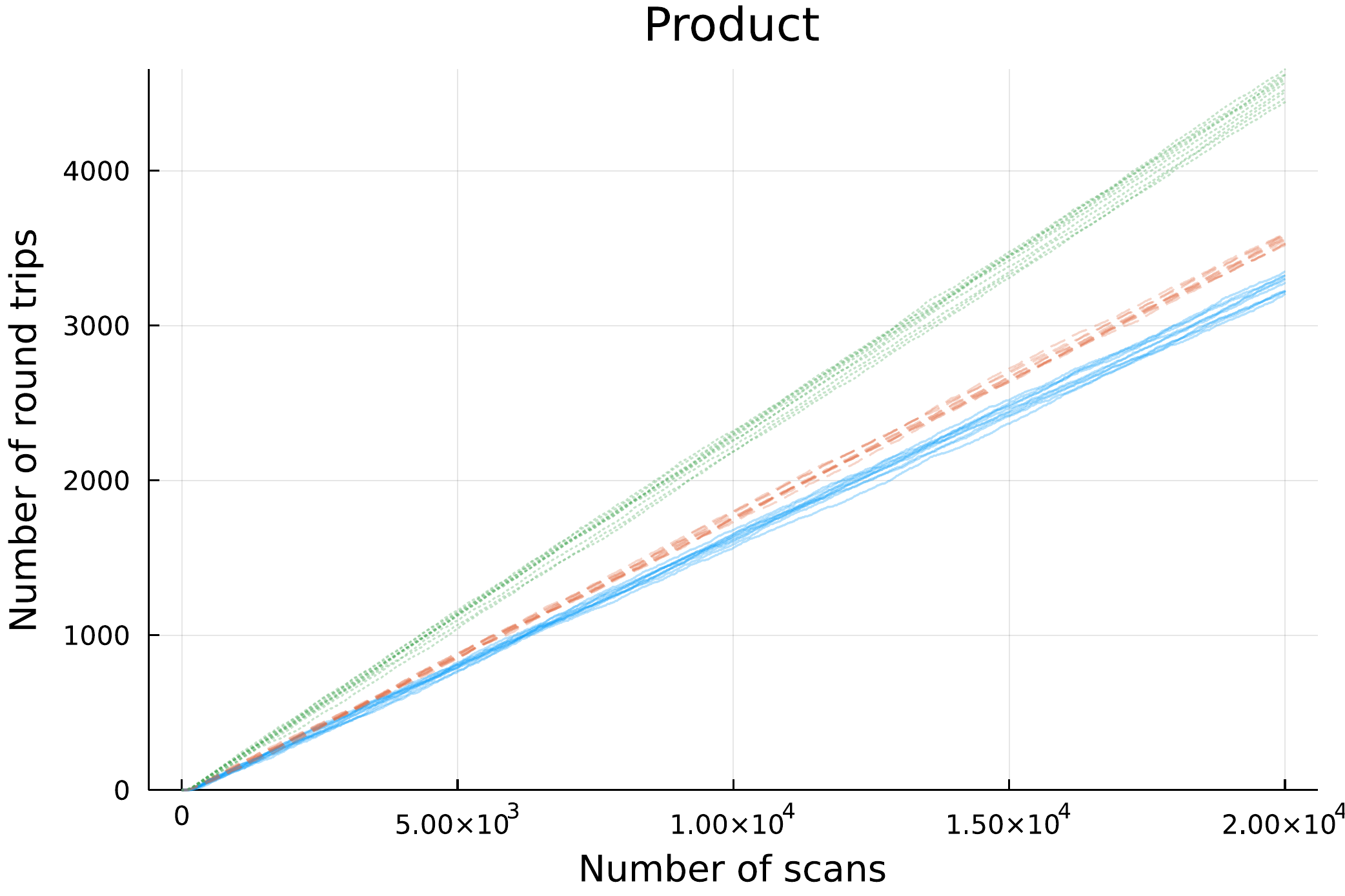}
    \end{subfigure}
    \begin{subfigure}{0.325\textwidth}
      \centering
      \includegraphics[width=\textwidth]{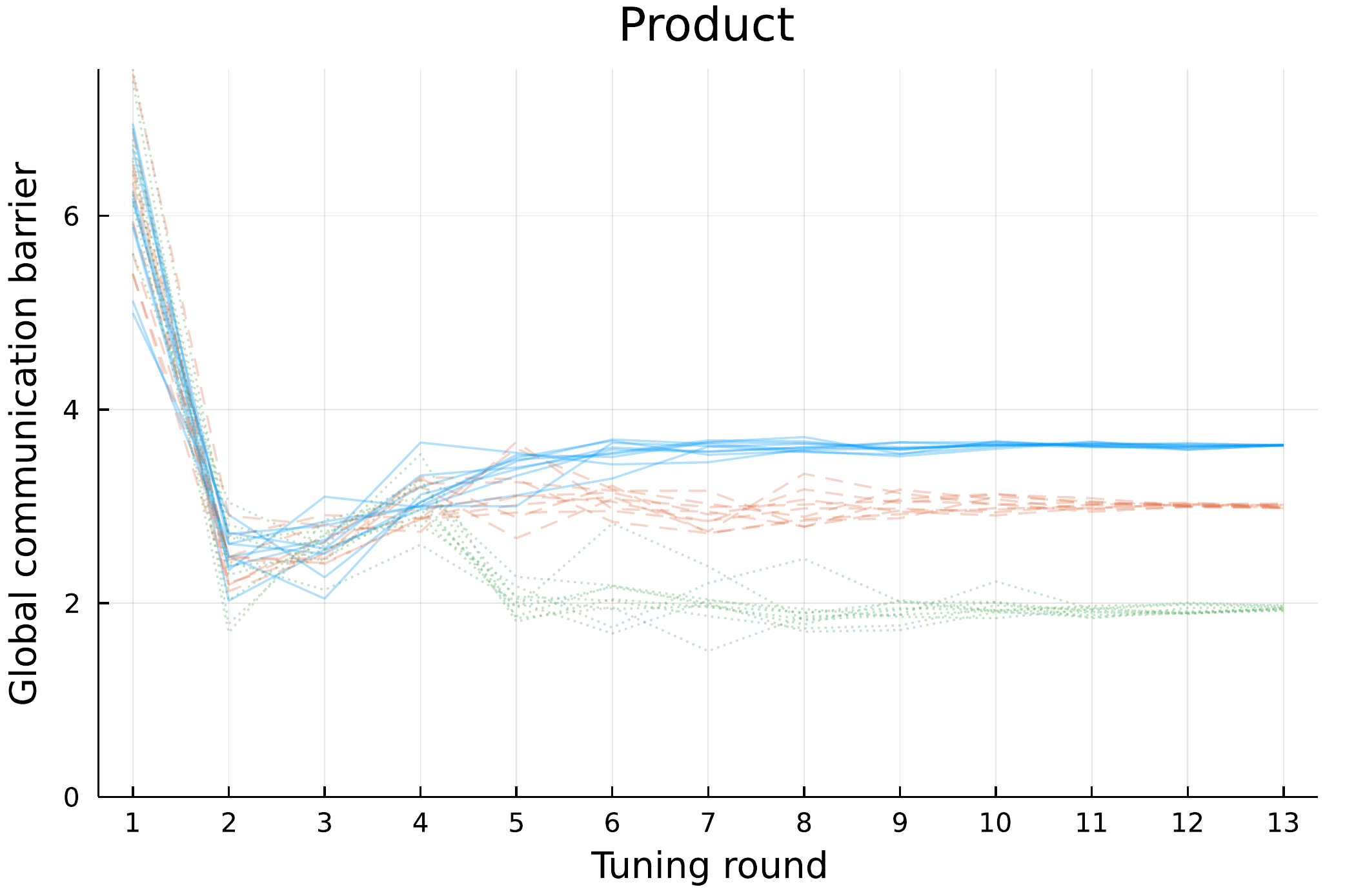}
    \end{subfigure}
    \begin{subfigure}{0.325\textwidth}
      \centering
      \includegraphics[width=\textwidth]{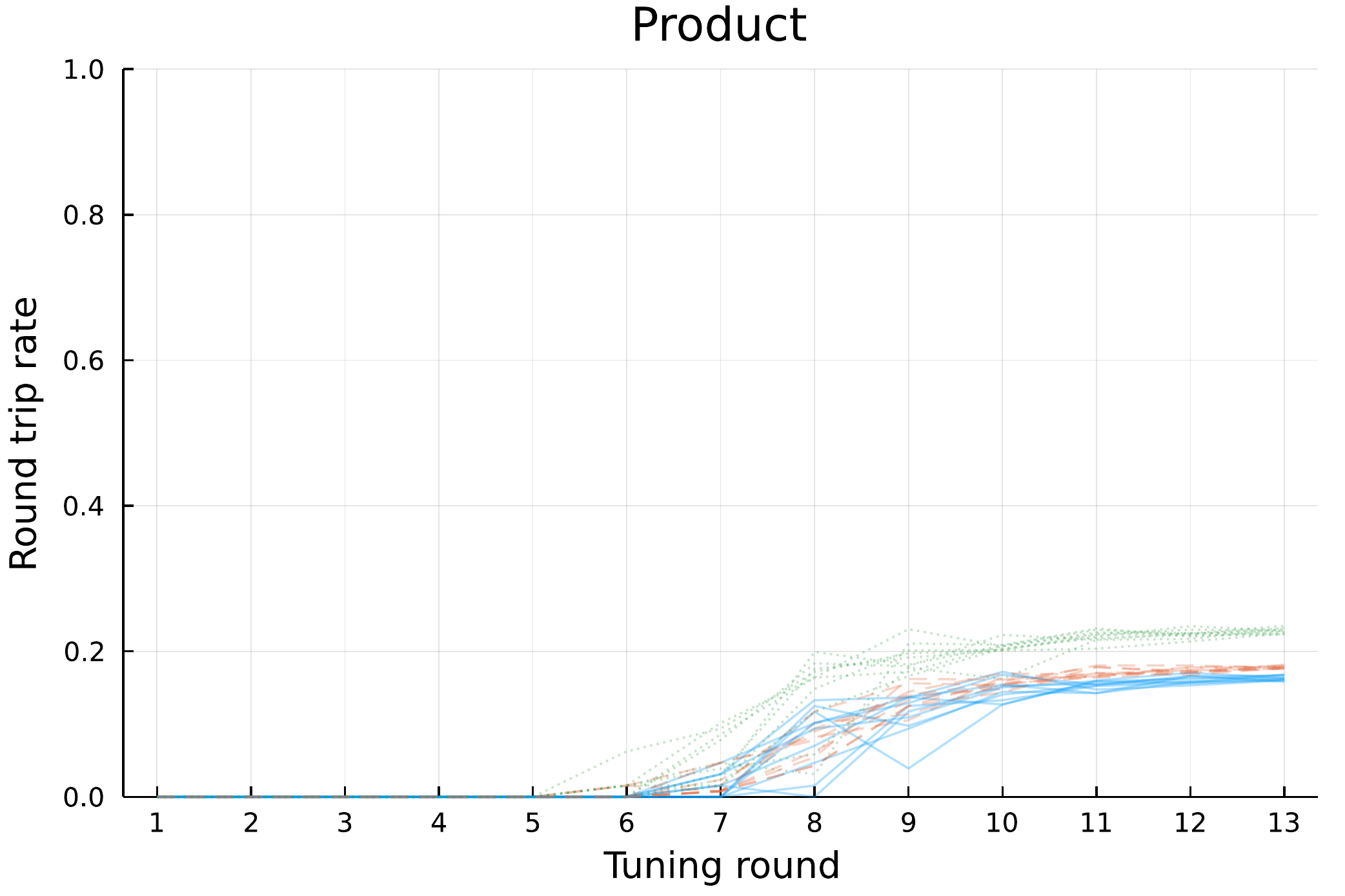}
    \end{subfigure}
    \begin{subfigure}{0.325\textwidth}
      \centering
      \includegraphics[width=\textwidth]{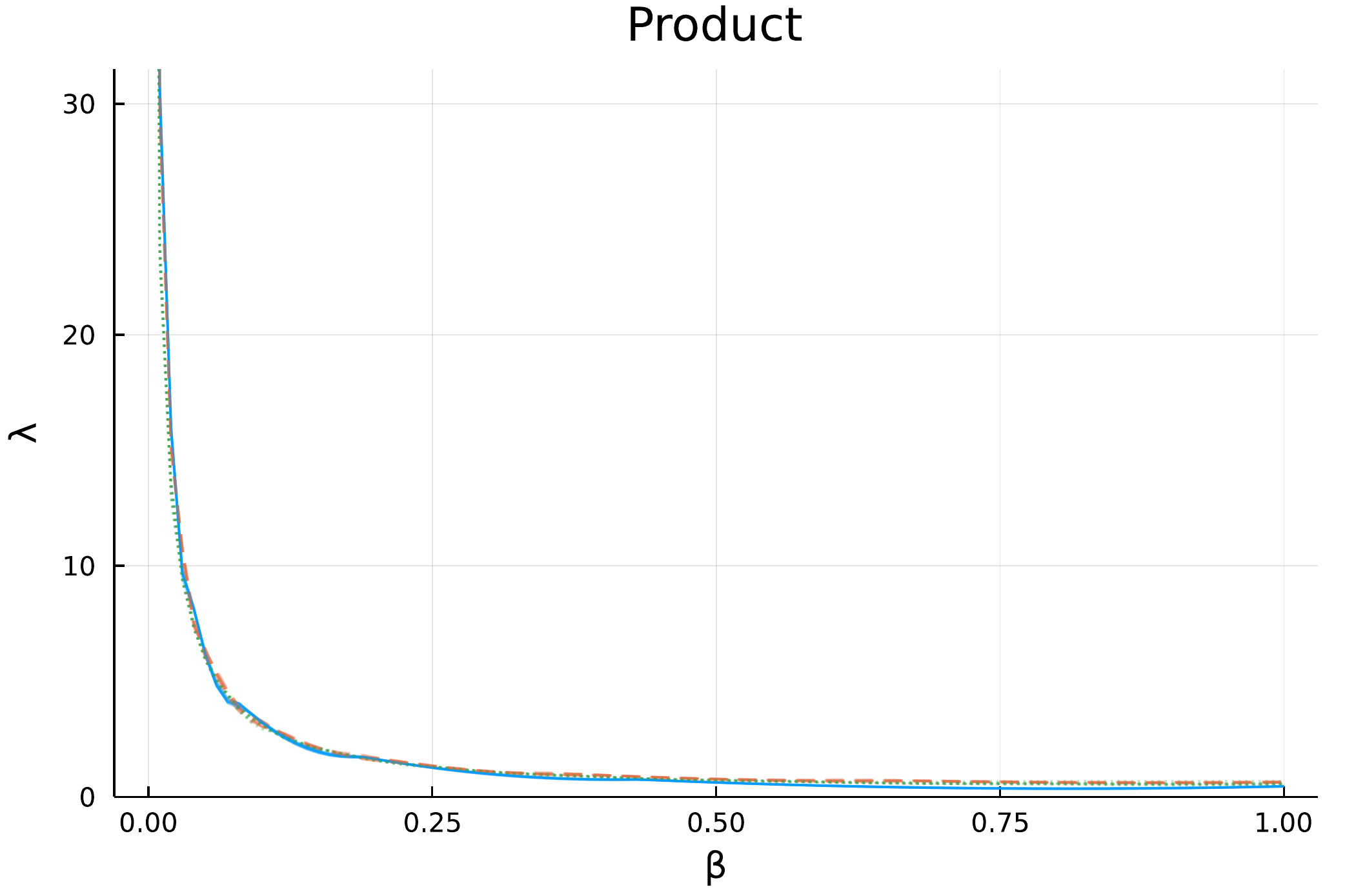}
    \end{subfigure}
    \begin{subfigure}{0.325\textwidth}
      \centering
      \includegraphics[width=\textwidth]{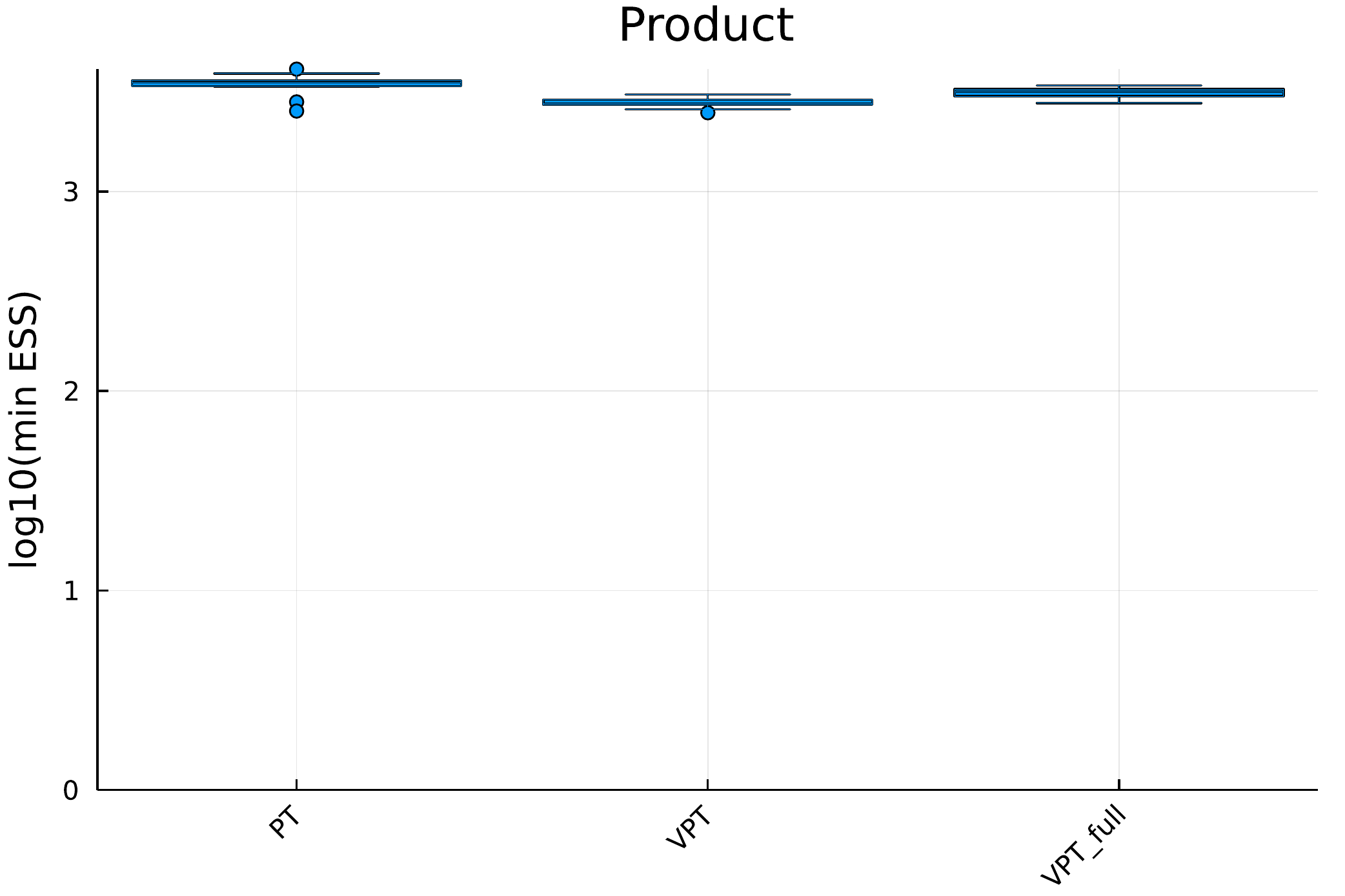}
    \end{subfigure}
    \begin{subfigure}{0.325\textwidth}
      \centering
      \includegraphics[width=\textwidth]{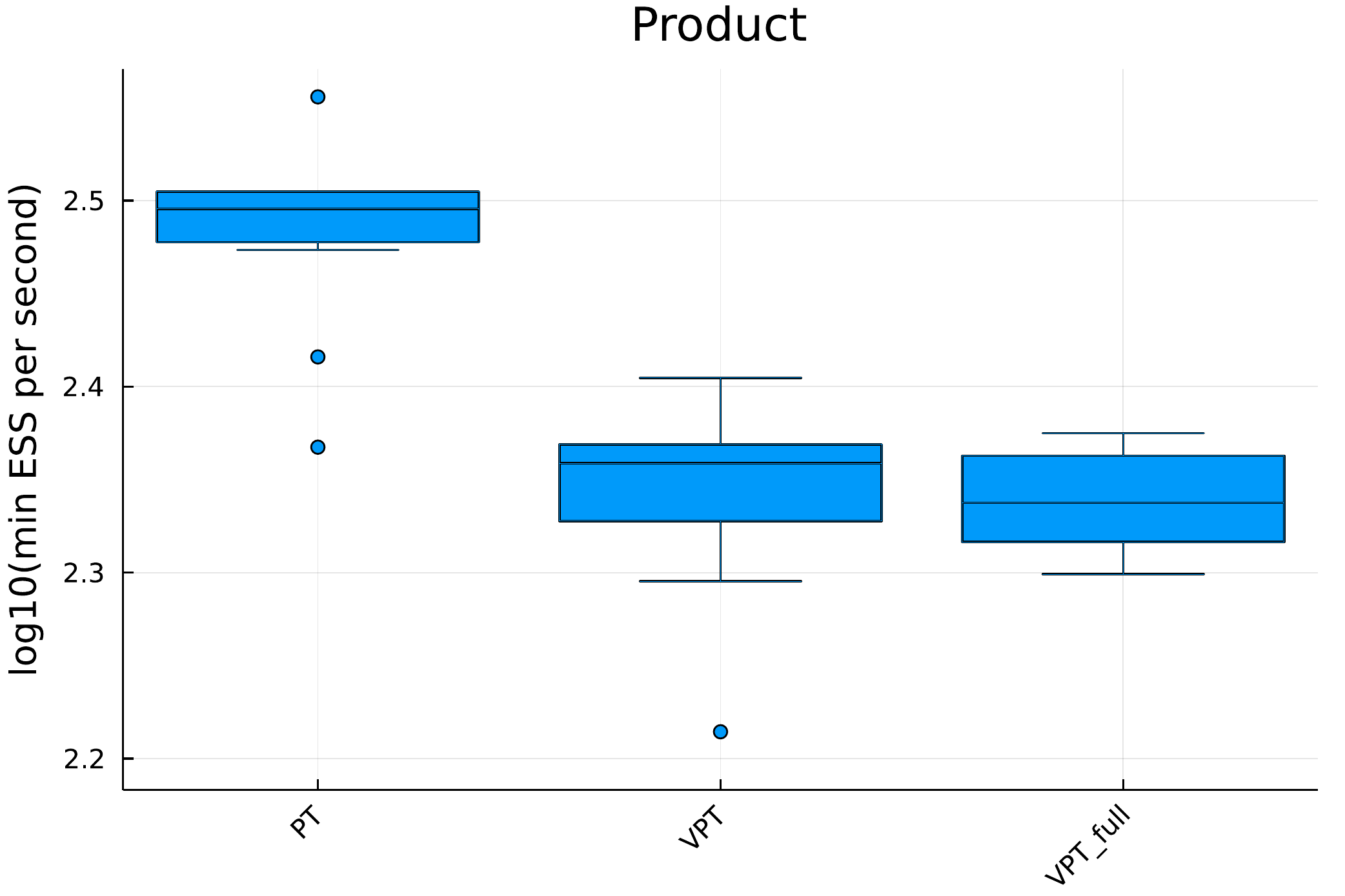}
    \end{subfigure}
    \begin{subfigure}{0.325\textwidth}
      \centering
      \includegraphics[width=\textwidth]{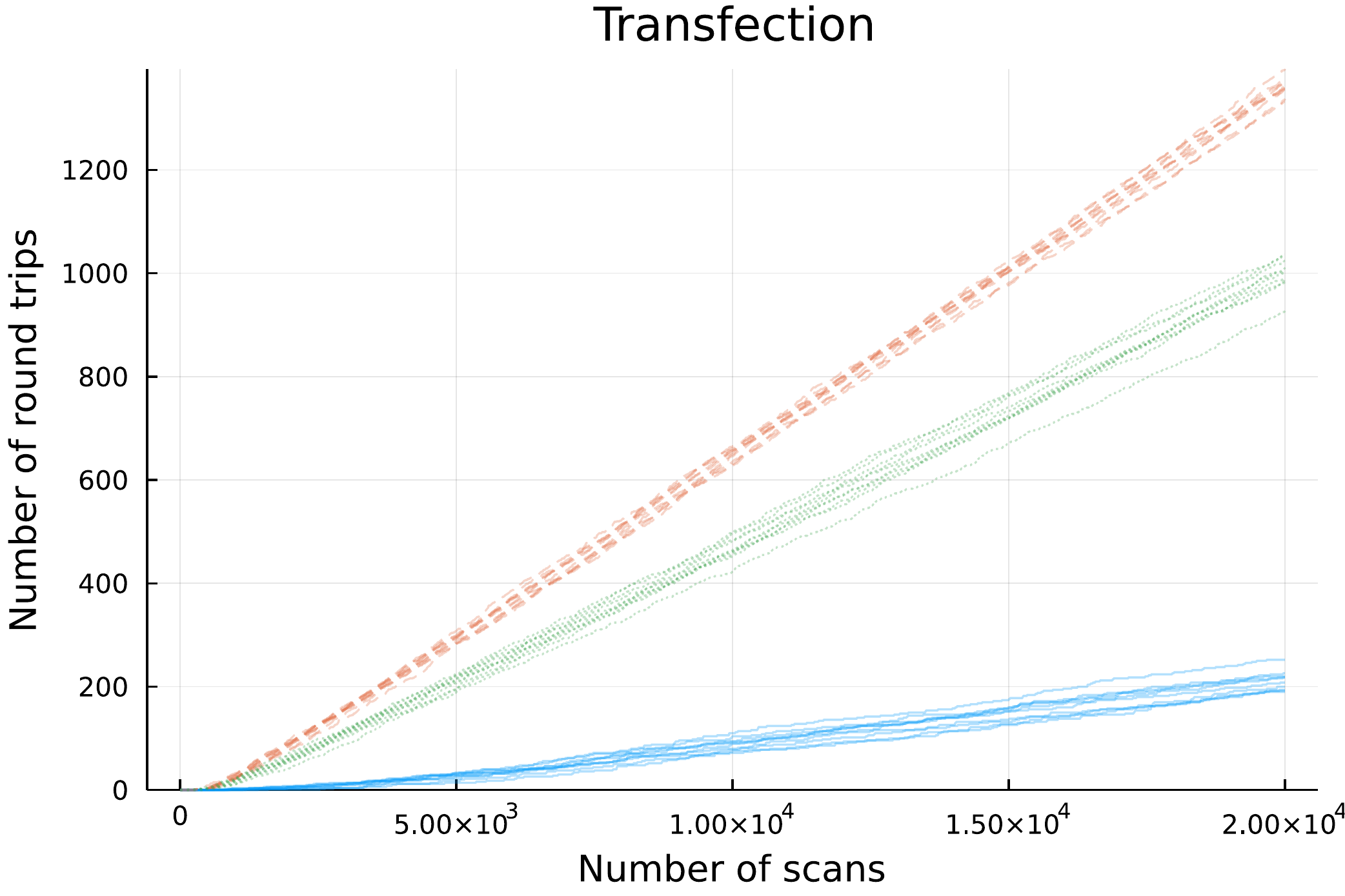}
    \end{subfigure}
    \begin{subfigure}{0.325\textwidth}
      \centering
      \includegraphics[width=\textwidth]{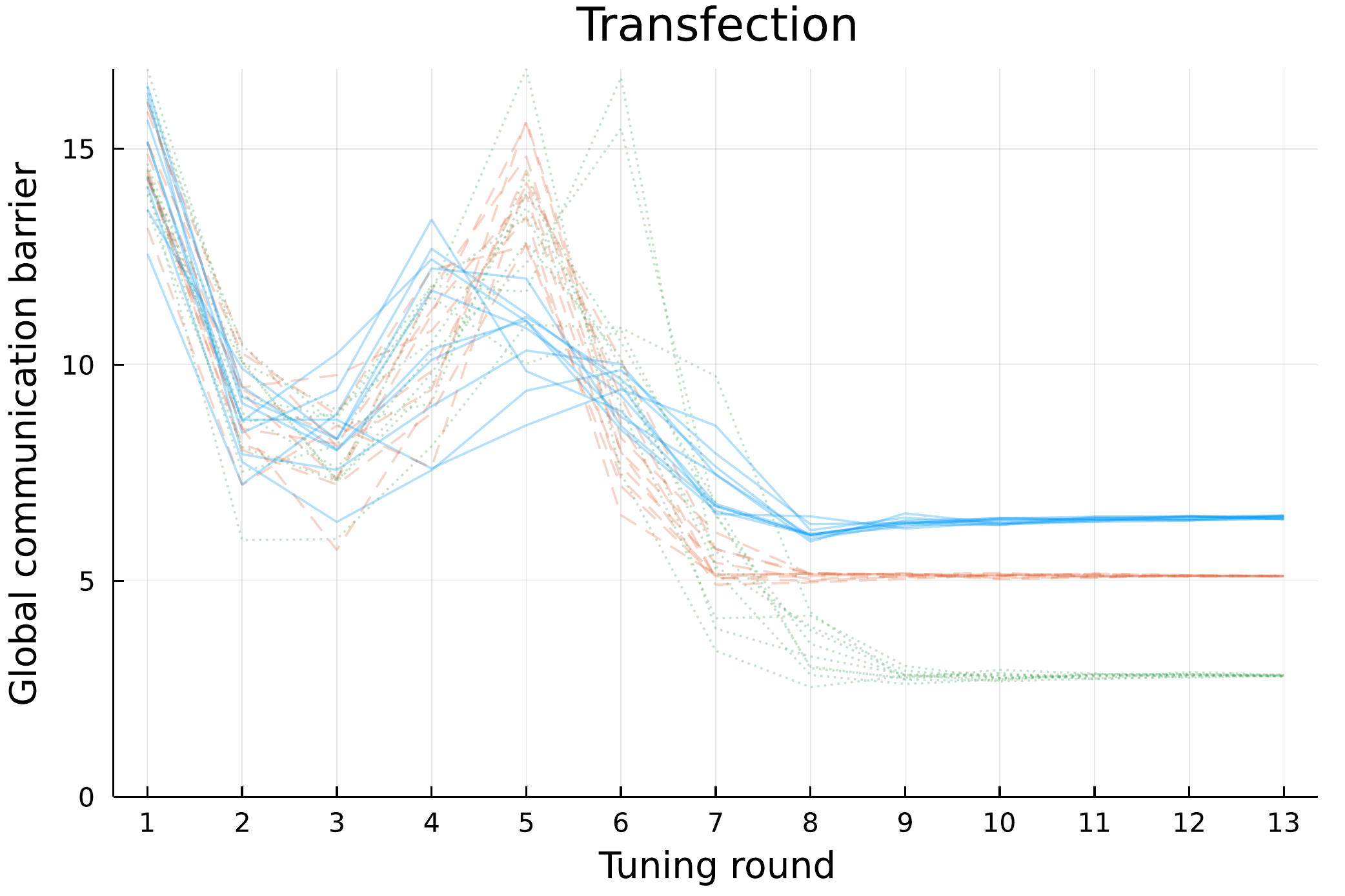}
    \end{subfigure}
    \begin{subfigure}{0.325\textwidth}
      \centering
      \includegraphics[width=\textwidth]{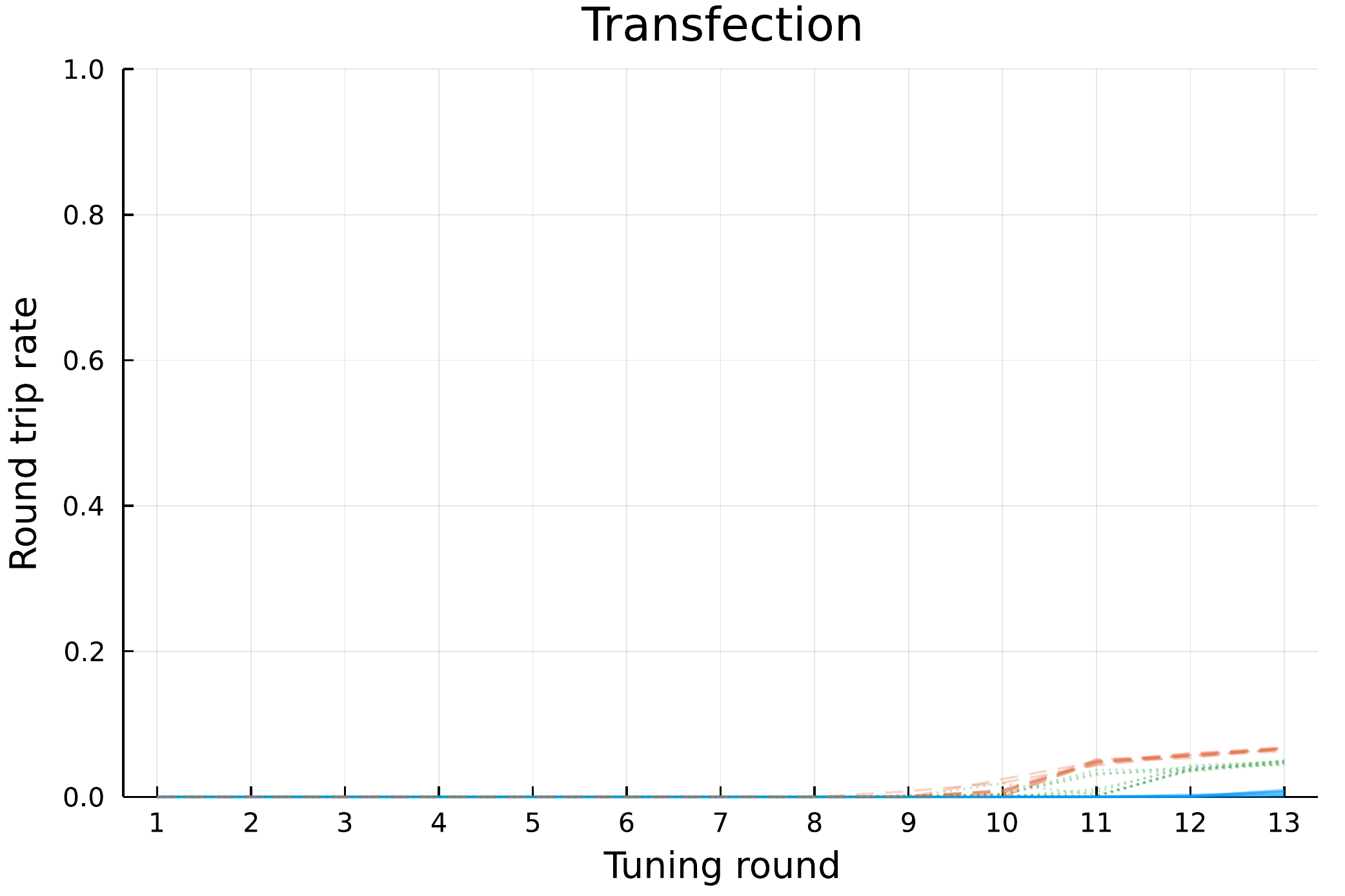}
    \end{subfigure}
    \begin{subfigure}{0.325\textwidth}
      \centering
      \includegraphics[width=\textwidth]{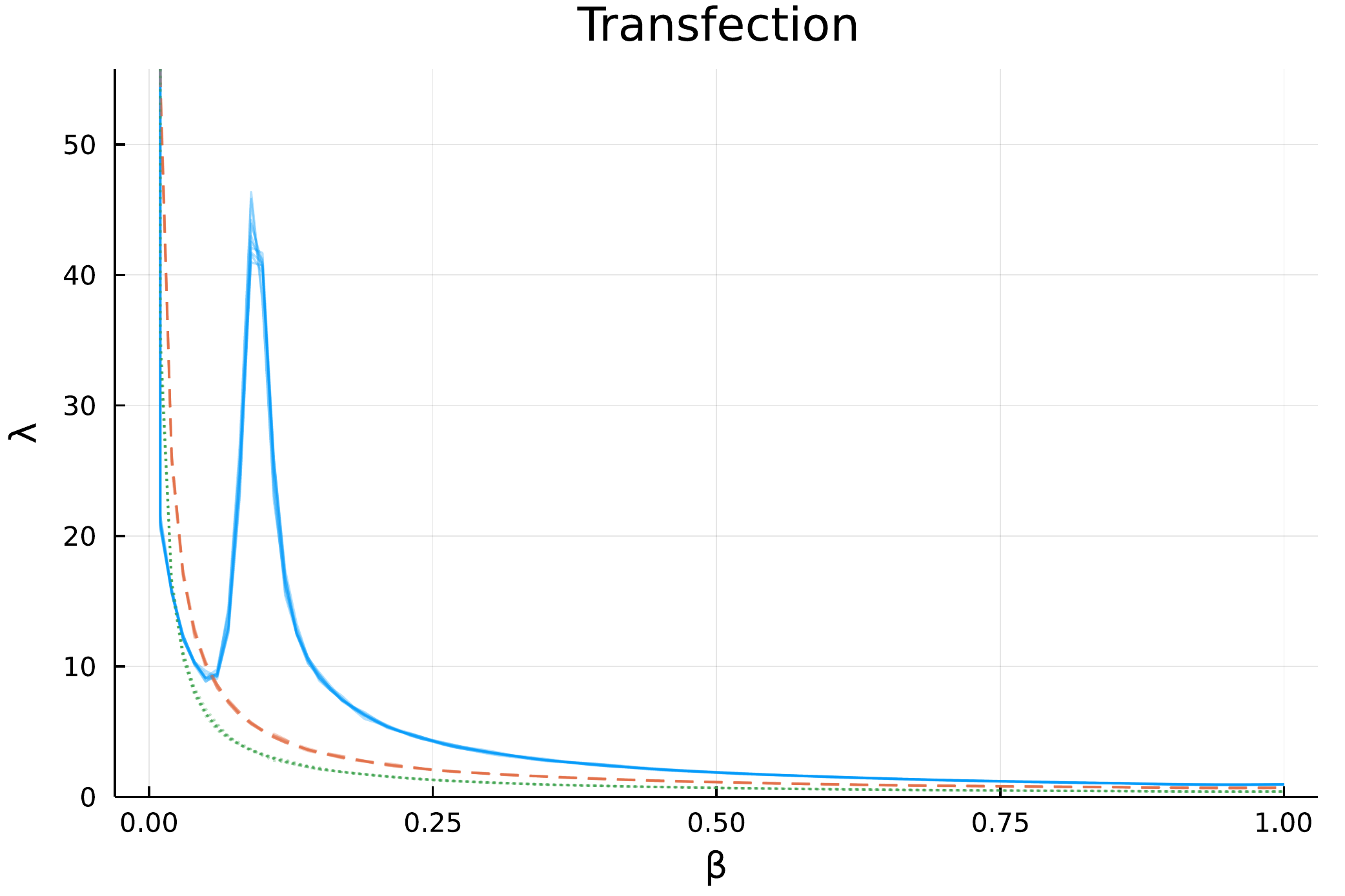}
    \end{subfigure}
    \begin{subfigure}{0.325\textwidth}
      \centering
      \includegraphics[width=\textwidth]{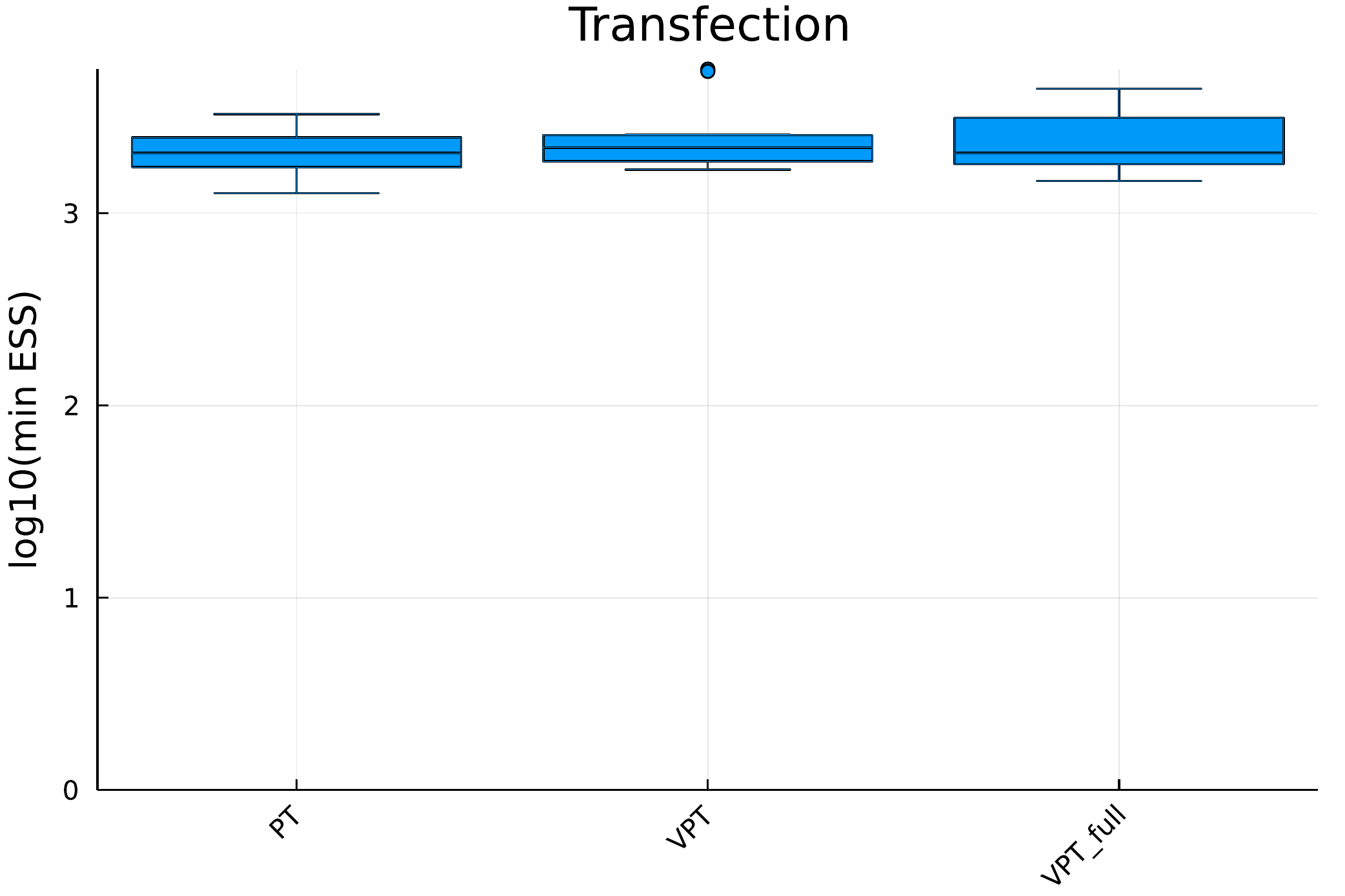}
    \end{subfigure}
    \begin{subfigure}{0.325\textwidth}
      \centering
      \includegraphics[width=\textwidth]{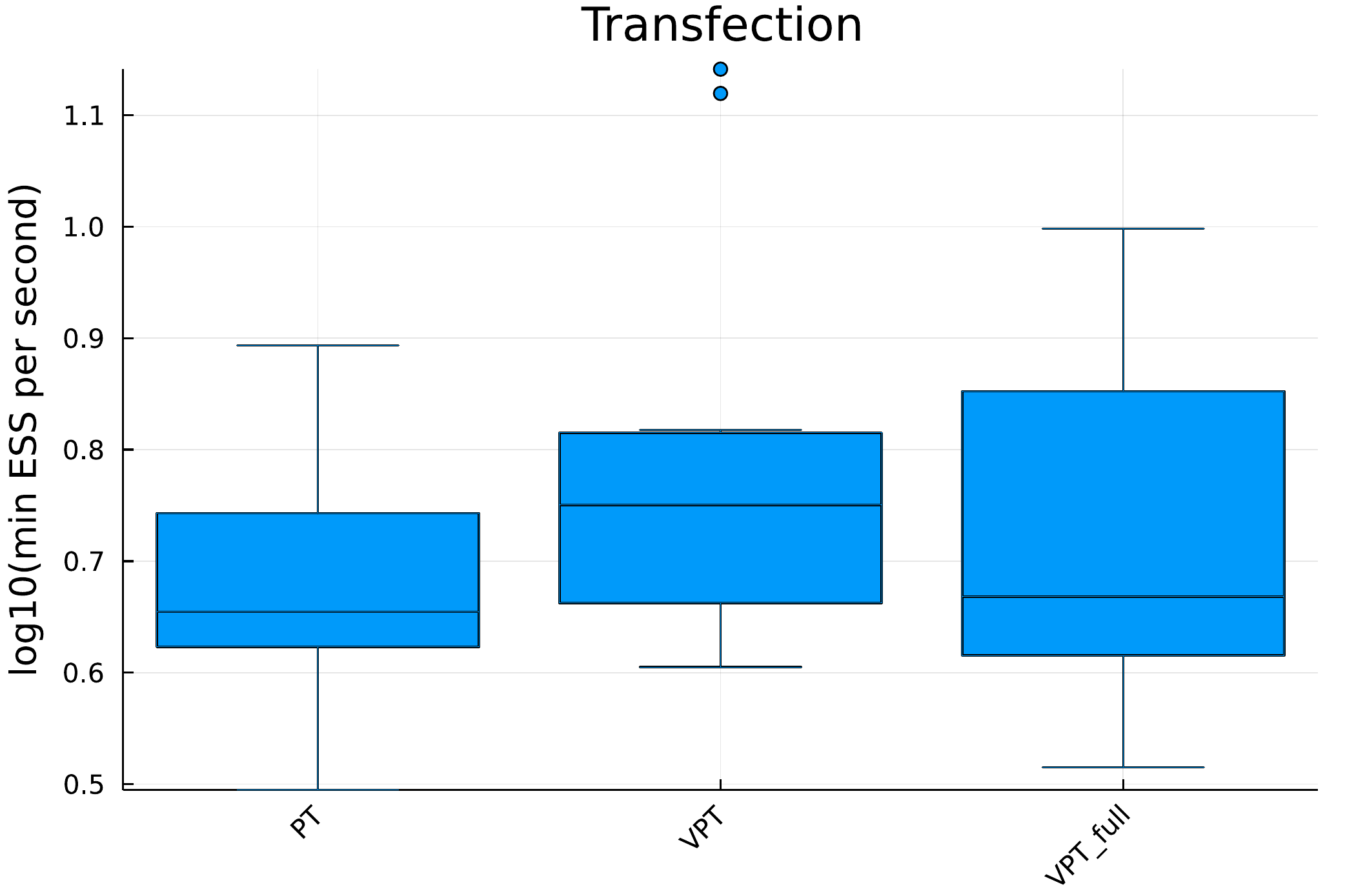}
    \end{subfigure}
    \caption{Number of round trips, GCB, round trip rate, local communication barrier (LCB), ESS, and ESS per second. Green/red: Full-covariance/mean-field variational PT. Blue: NRPT.}
    \label{fig:additional_plots}
\end{figure}

\begin{figure}[t]
    \centering
    \begin{subfigure}{0.325\textwidth}
      \centering
      \includegraphics[width=\textwidth]{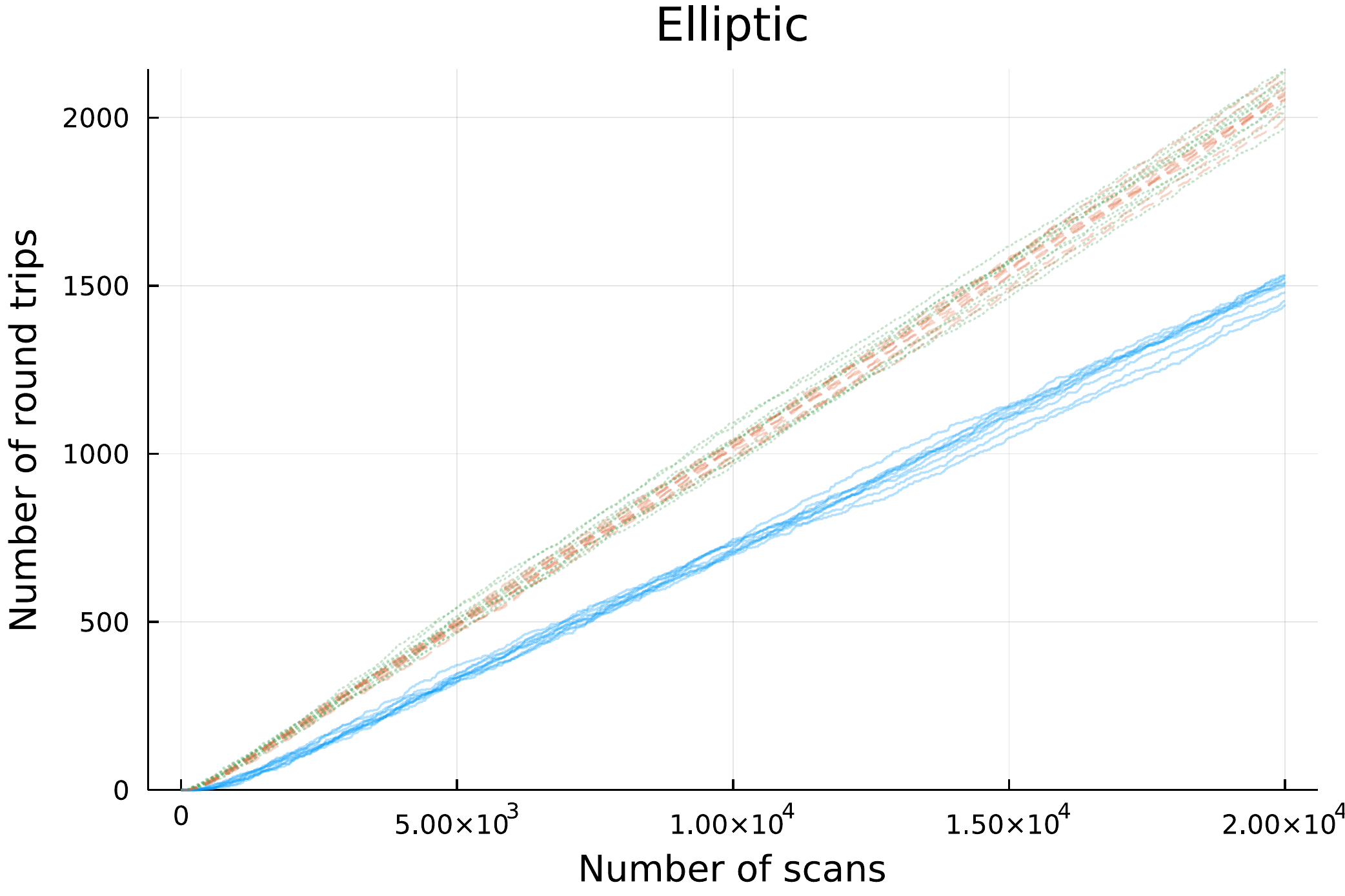}
    \end{subfigure}
    \begin{subfigure}{0.325\textwidth}
      \centering
      \includegraphics[width=\textwidth]{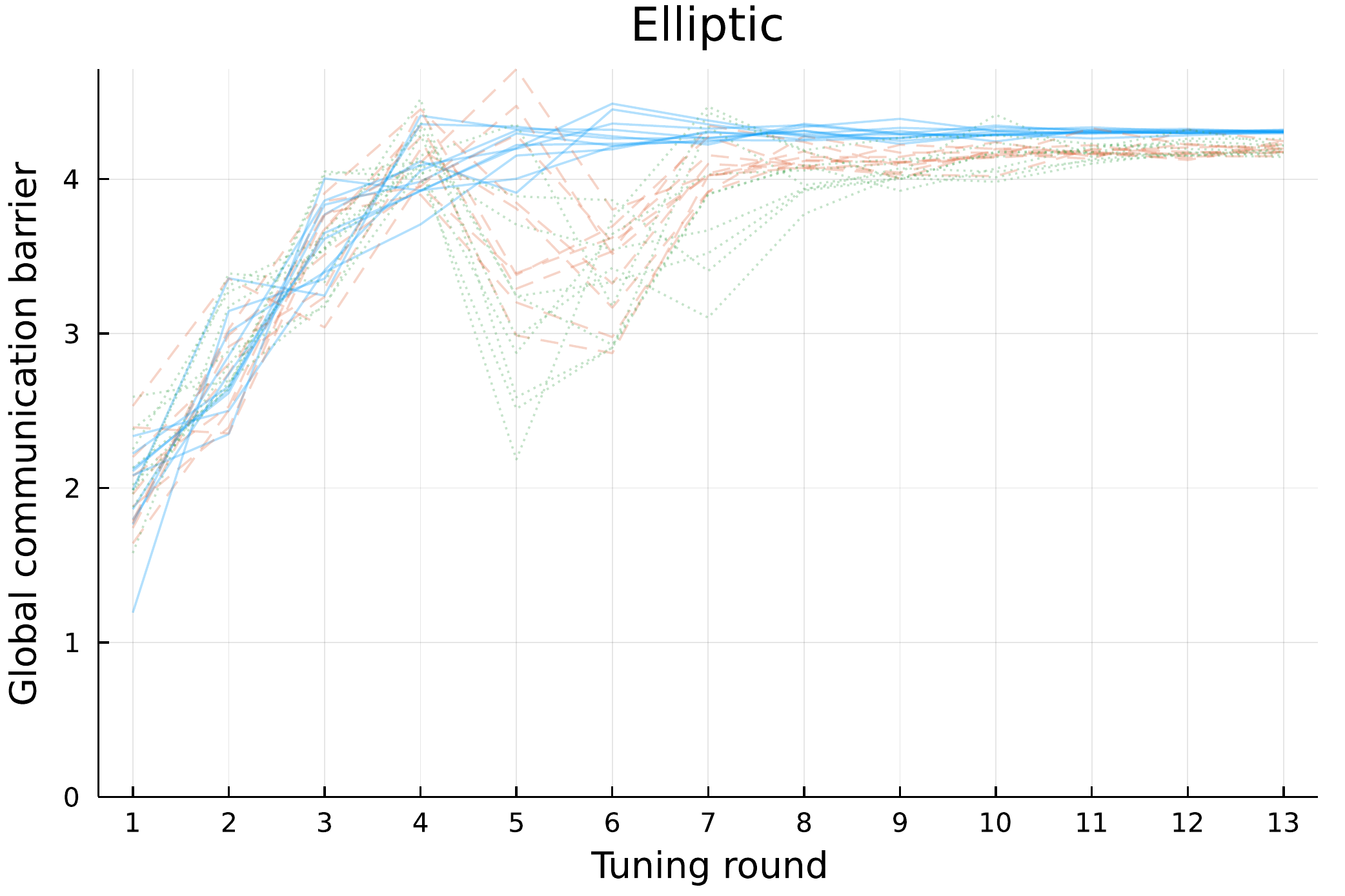}
    \end{subfigure}
    \begin{subfigure}{0.325\textwidth}
      \centering
      \includegraphics[width=\textwidth]{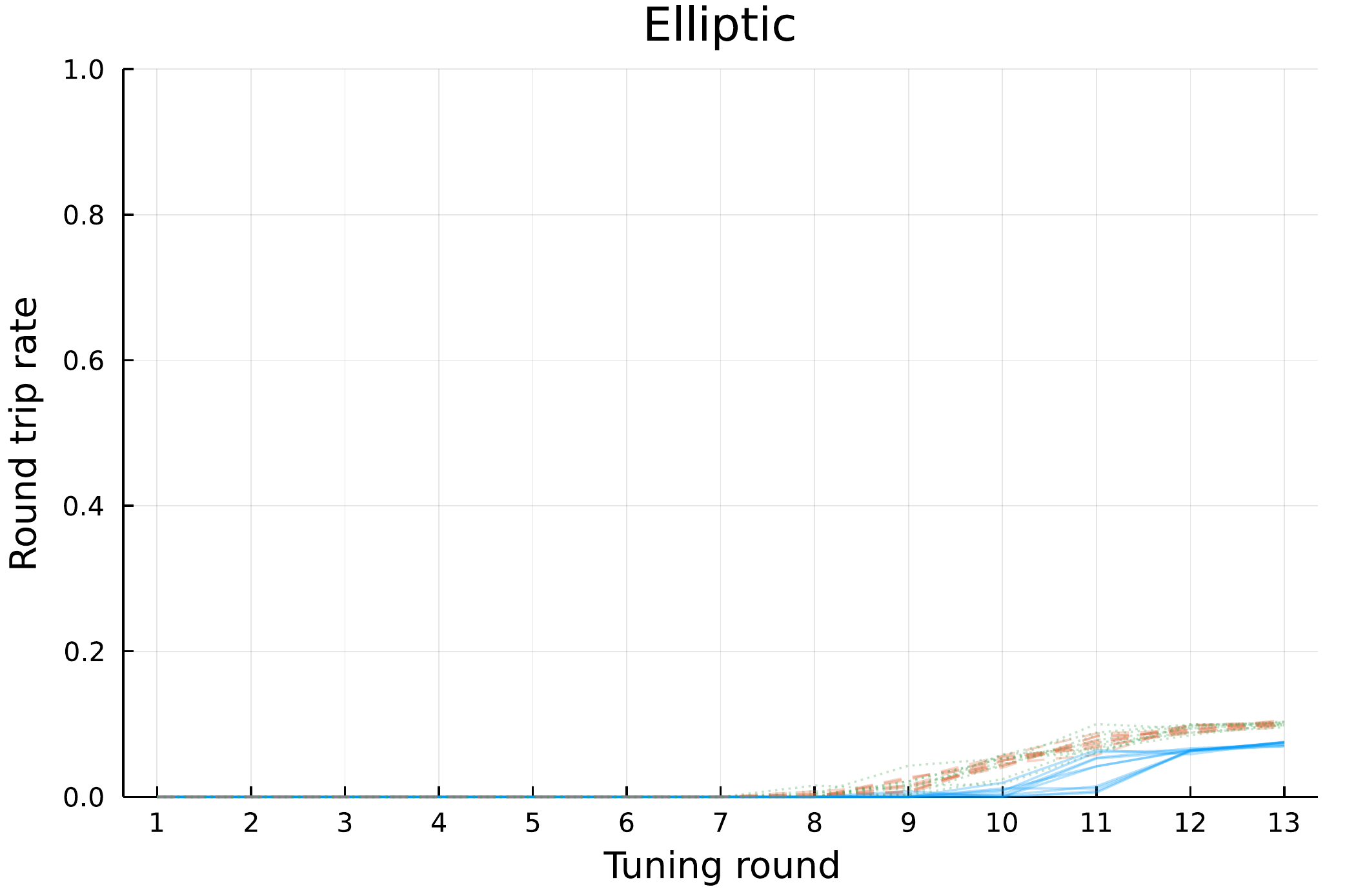}
    \end{subfigure}
    \begin{subfigure}{0.325\textwidth}
      \centering
      \includegraphics[width=\textwidth]{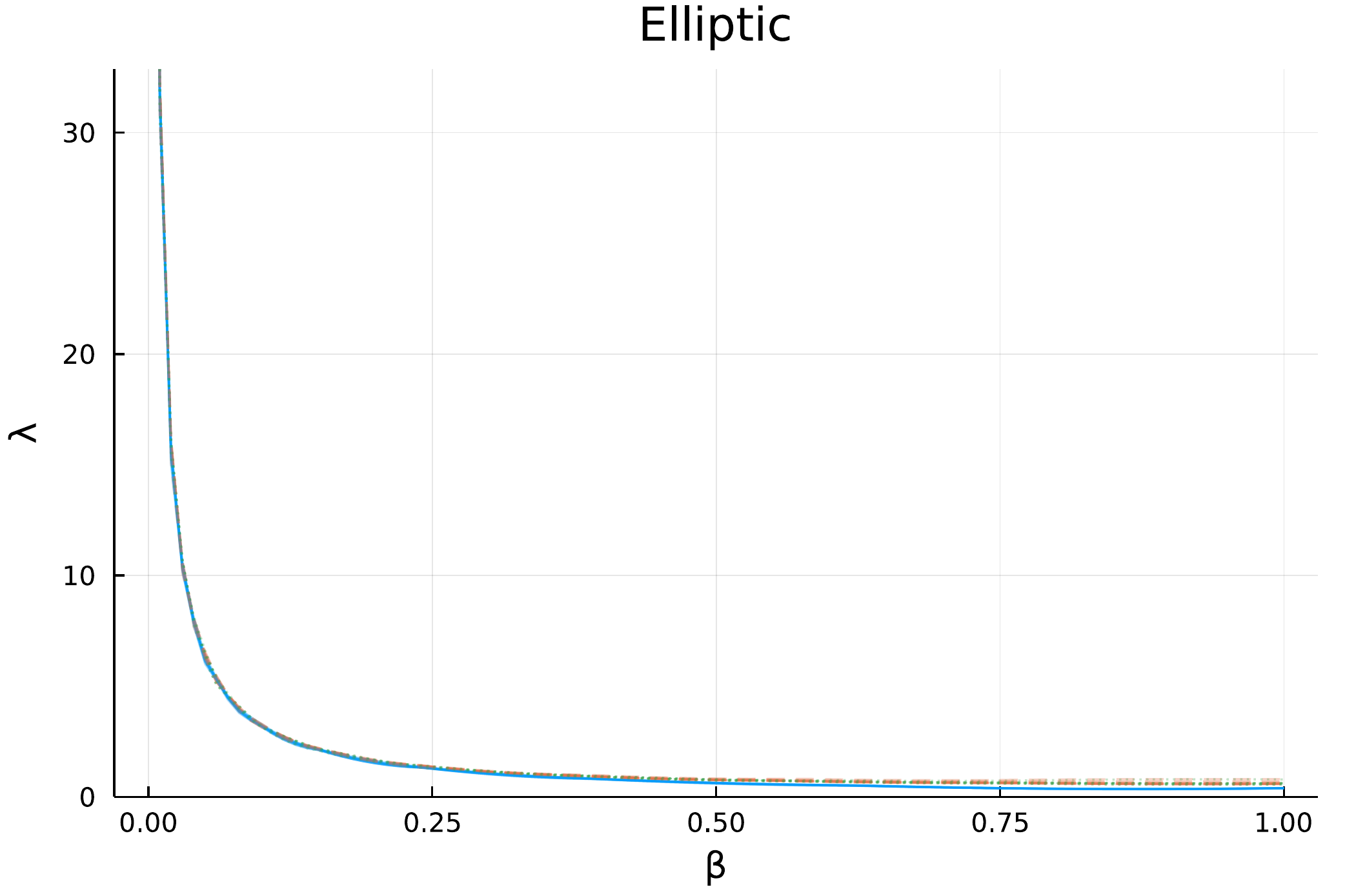}
    \end{subfigure}
    \begin{subfigure}{0.325\textwidth}
      \centering
      \includegraphics[width=\textwidth]{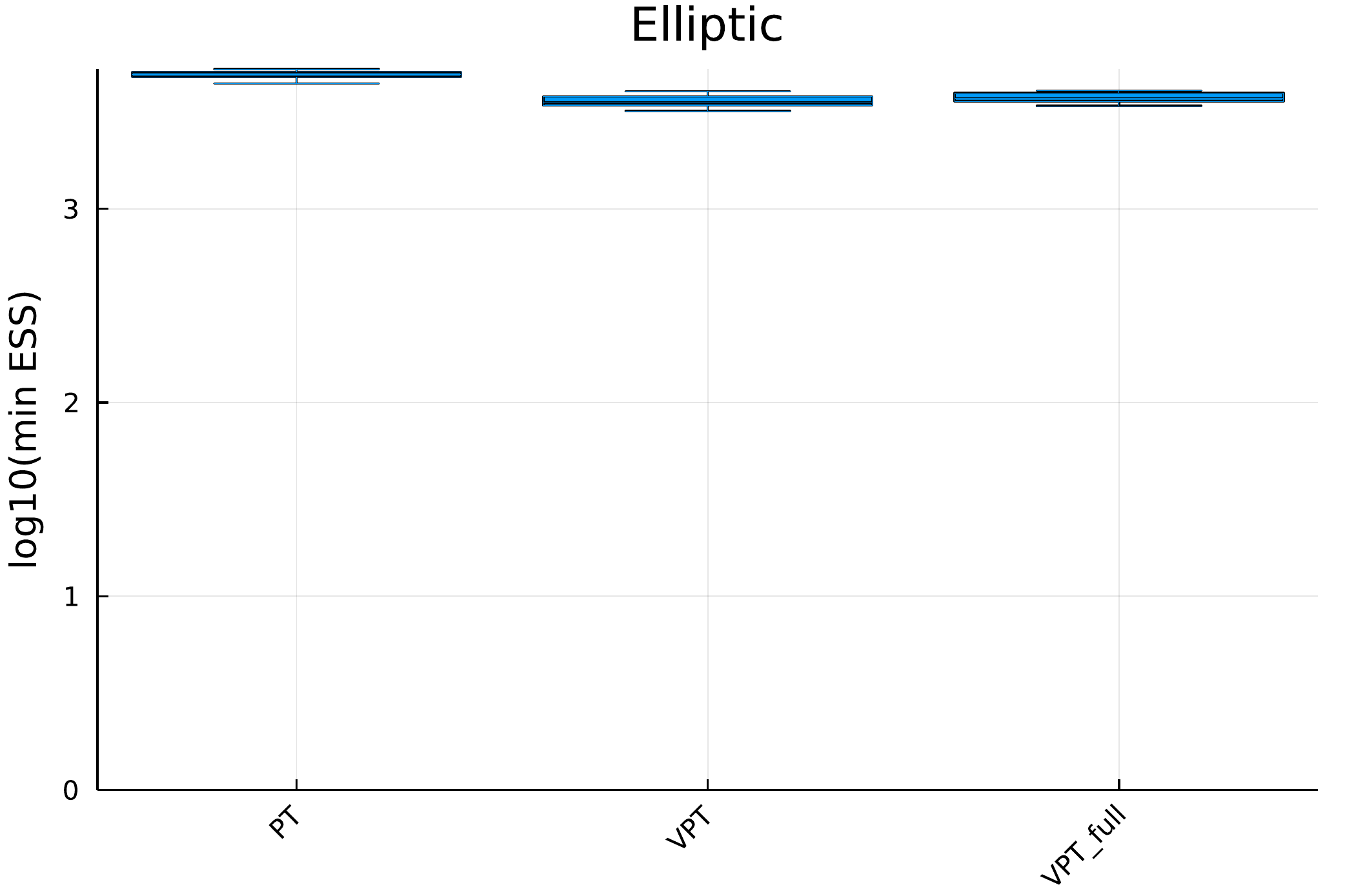}
    \end{subfigure}
    \begin{subfigure}{0.325\textwidth}
      \centering
      \includegraphics[width=\textwidth]{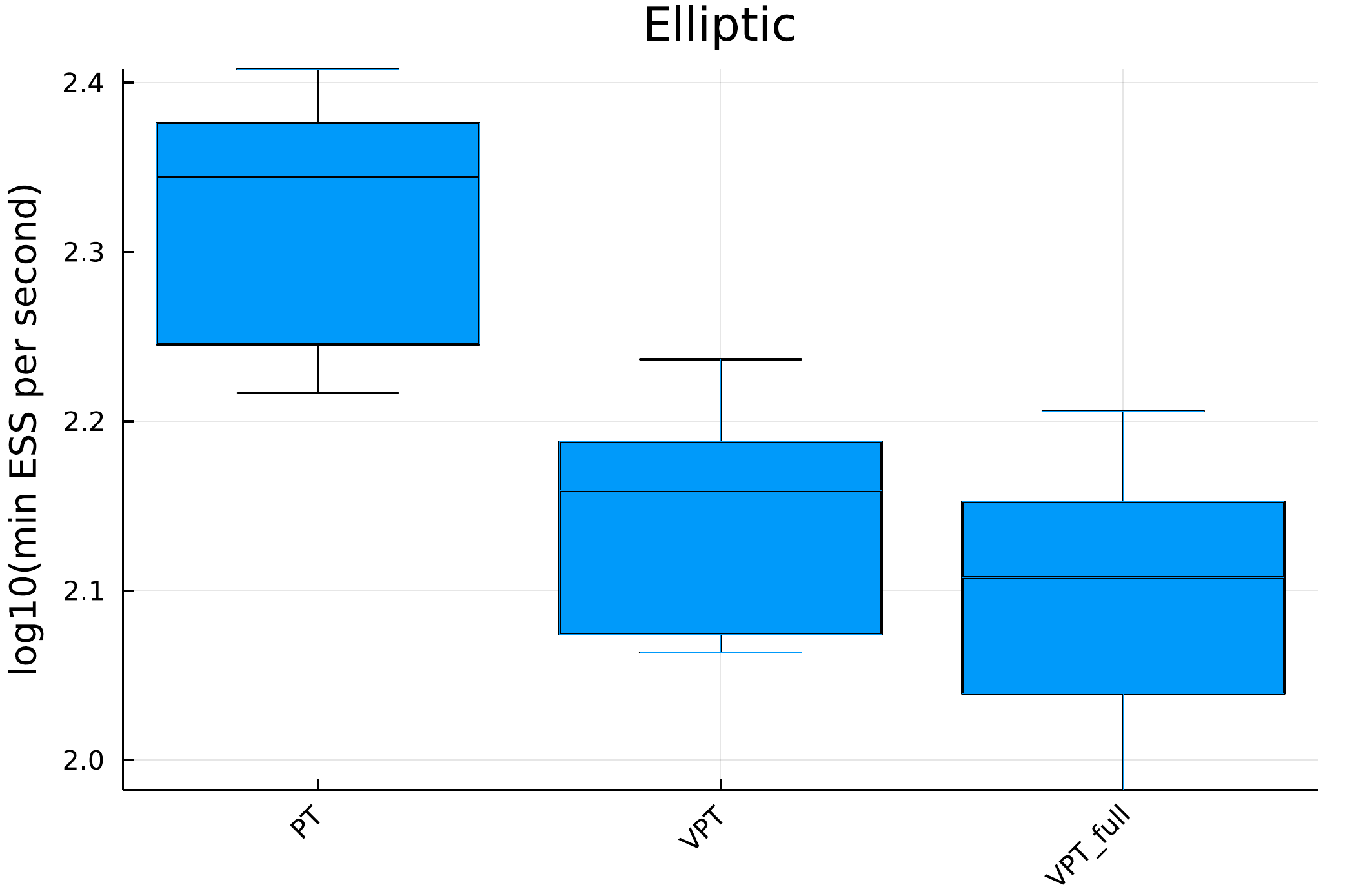}
    \end{subfigure}
    \begin{subfigure}{0.325\textwidth}
      \centering
      \includegraphics[width=\textwidth]{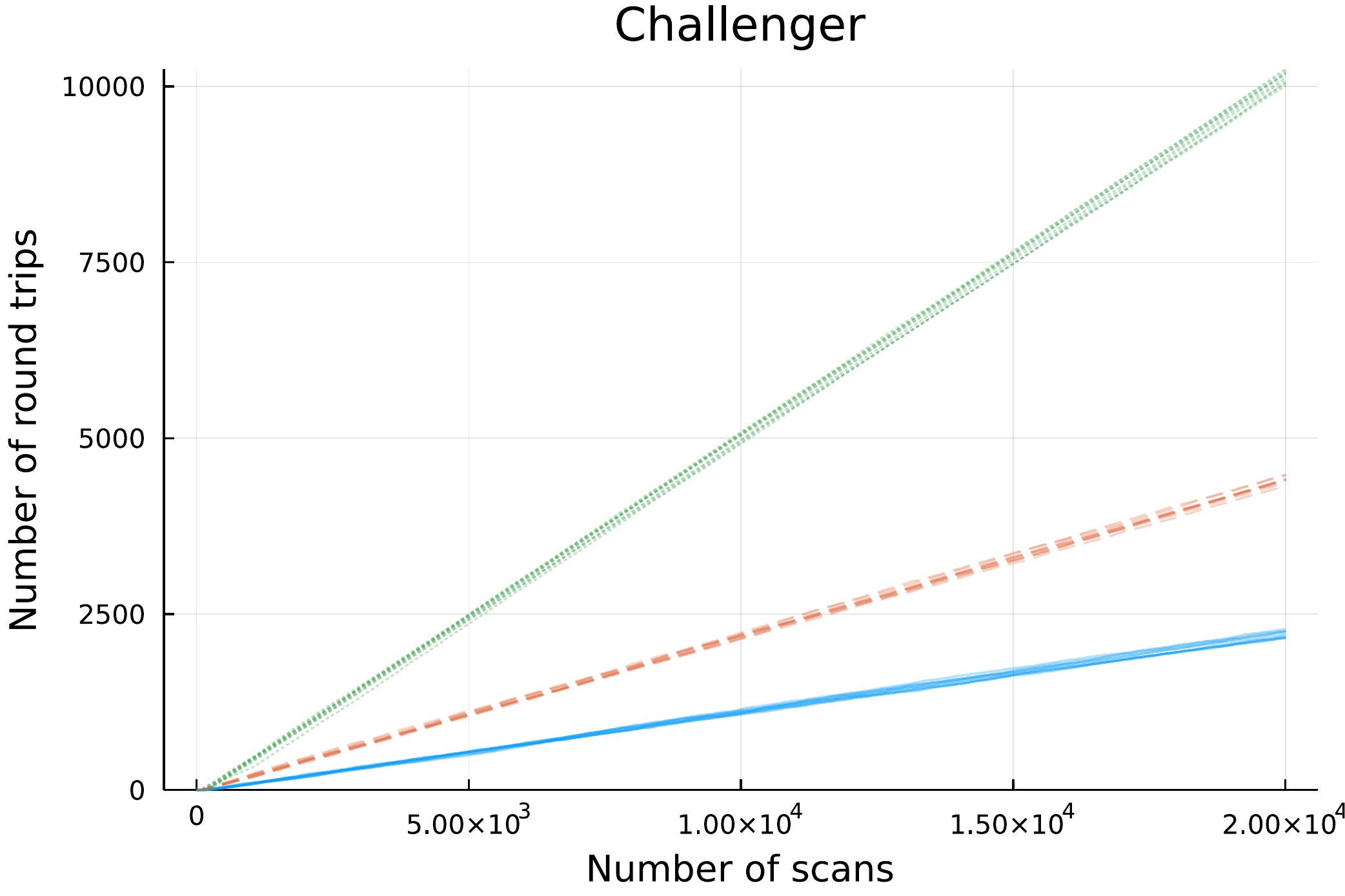}
    \end{subfigure}
    \begin{subfigure}{0.325\textwidth}
      \centering
      \includegraphics[width=\textwidth]{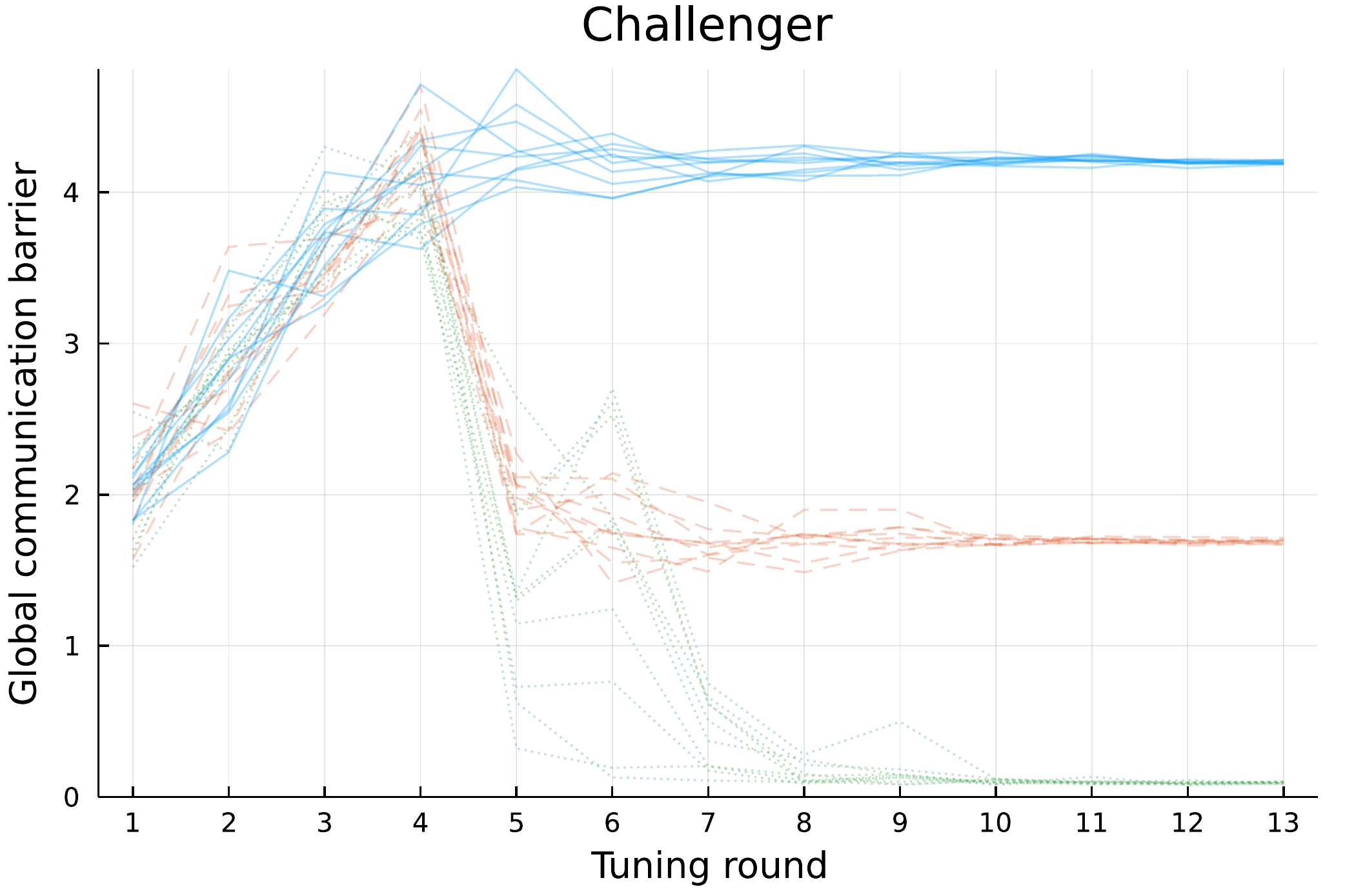}
    \end{subfigure}
    \begin{subfigure}{0.325\textwidth}
      \centering
      \includegraphics[width=\textwidth]{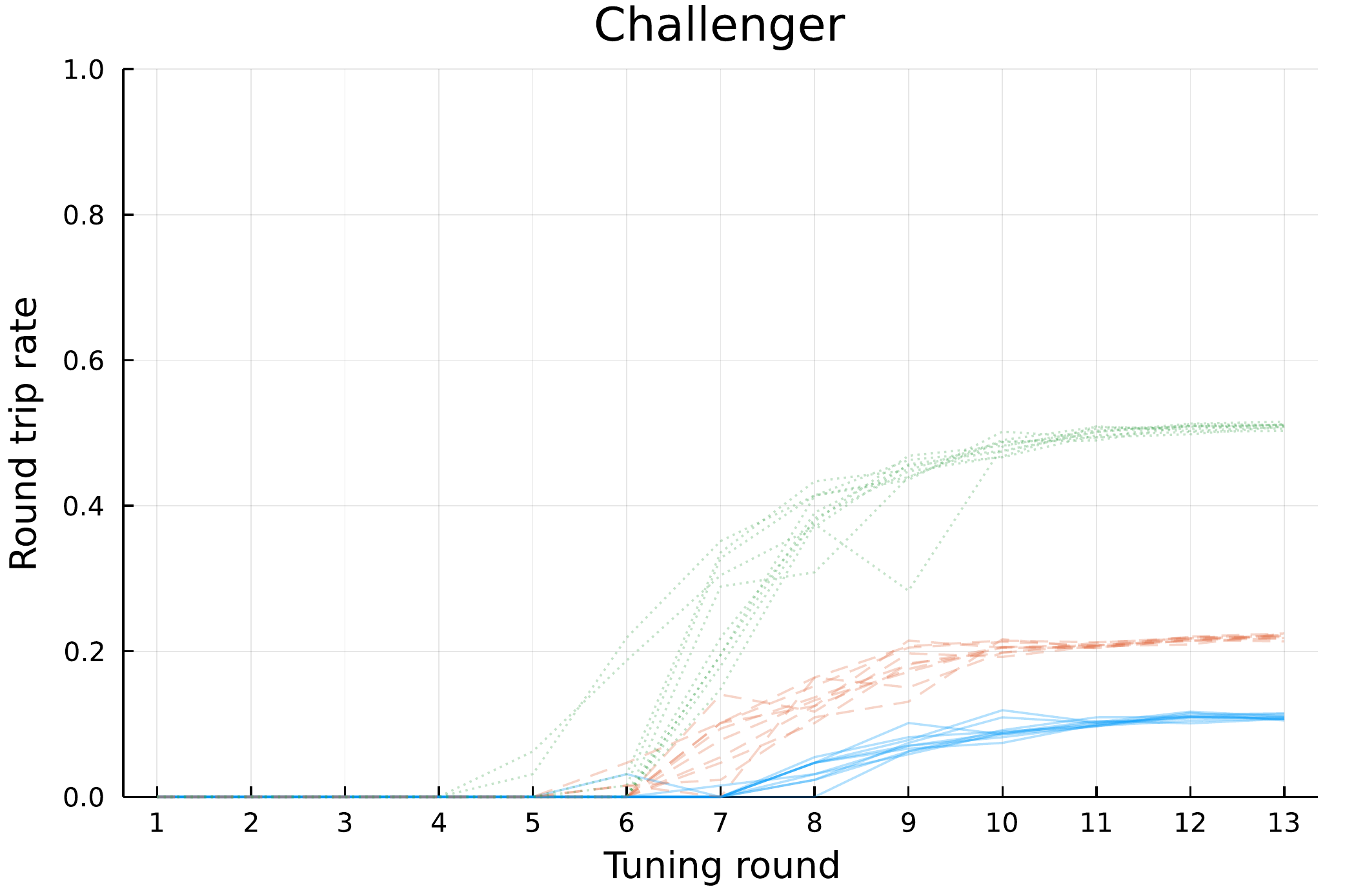}
    \end{subfigure}
    \begin{subfigure}{0.325\textwidth}
      \centering
      \includegraphics[width=\textwidth]{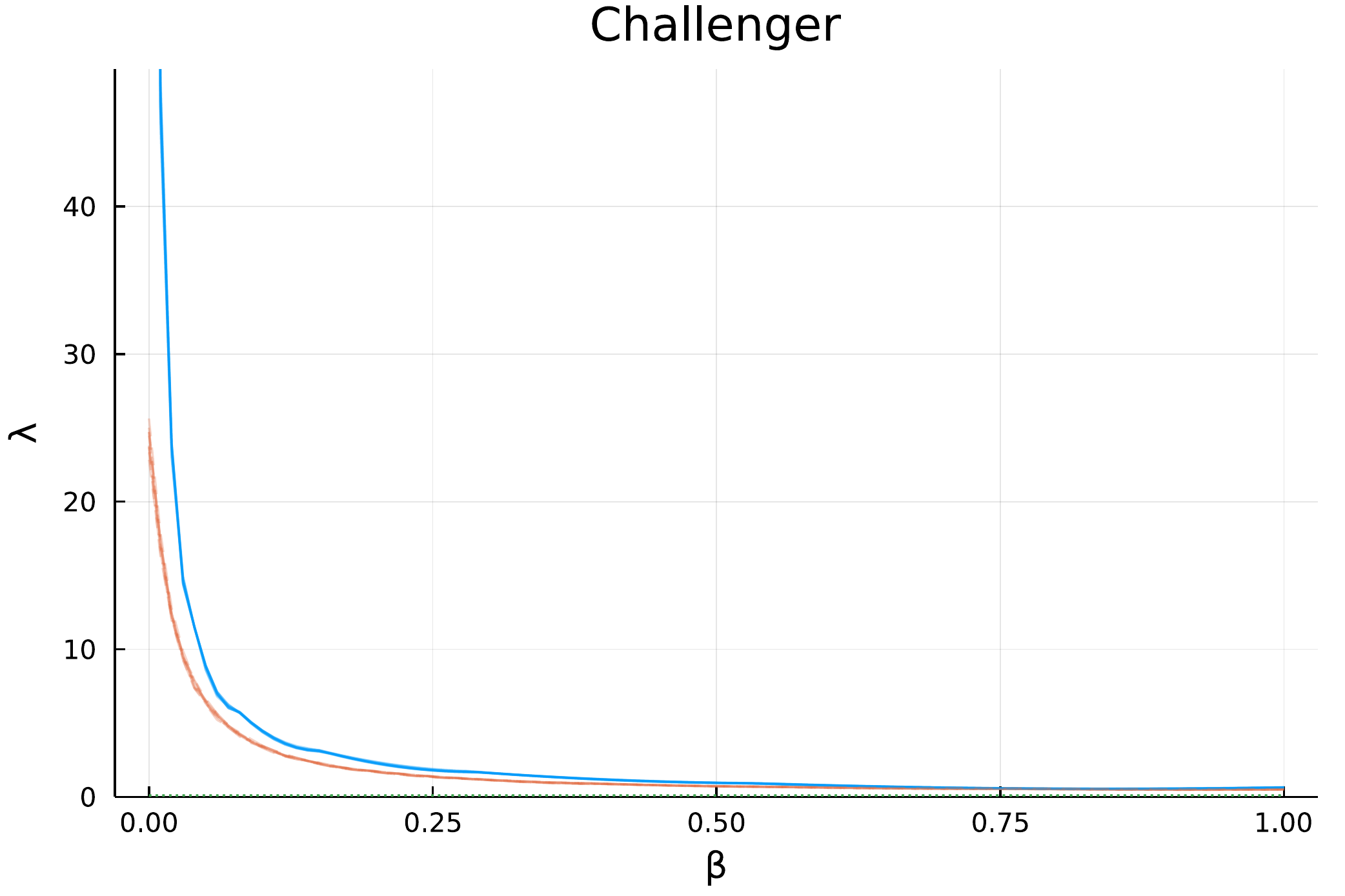}
    \end{subfigure}
    \begin{subfigure}{0.325\textwidth}
      \centering
      \includegraphics[width=\textwidth]{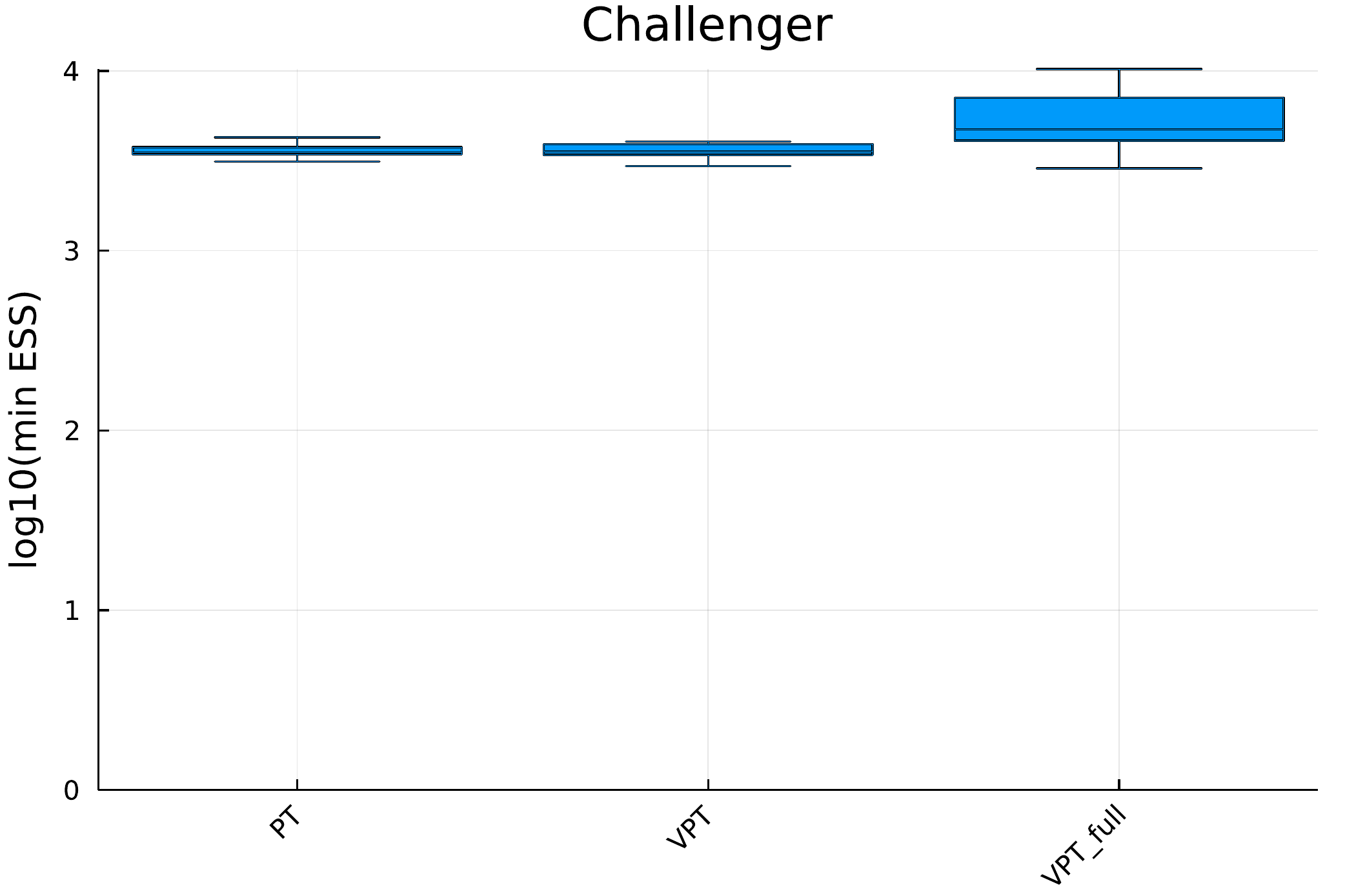}
    \end{subfigure}
    \begin{subfigure}{0.325\textwidth}
      \centering
      \includegraphics[width=\textwidth]{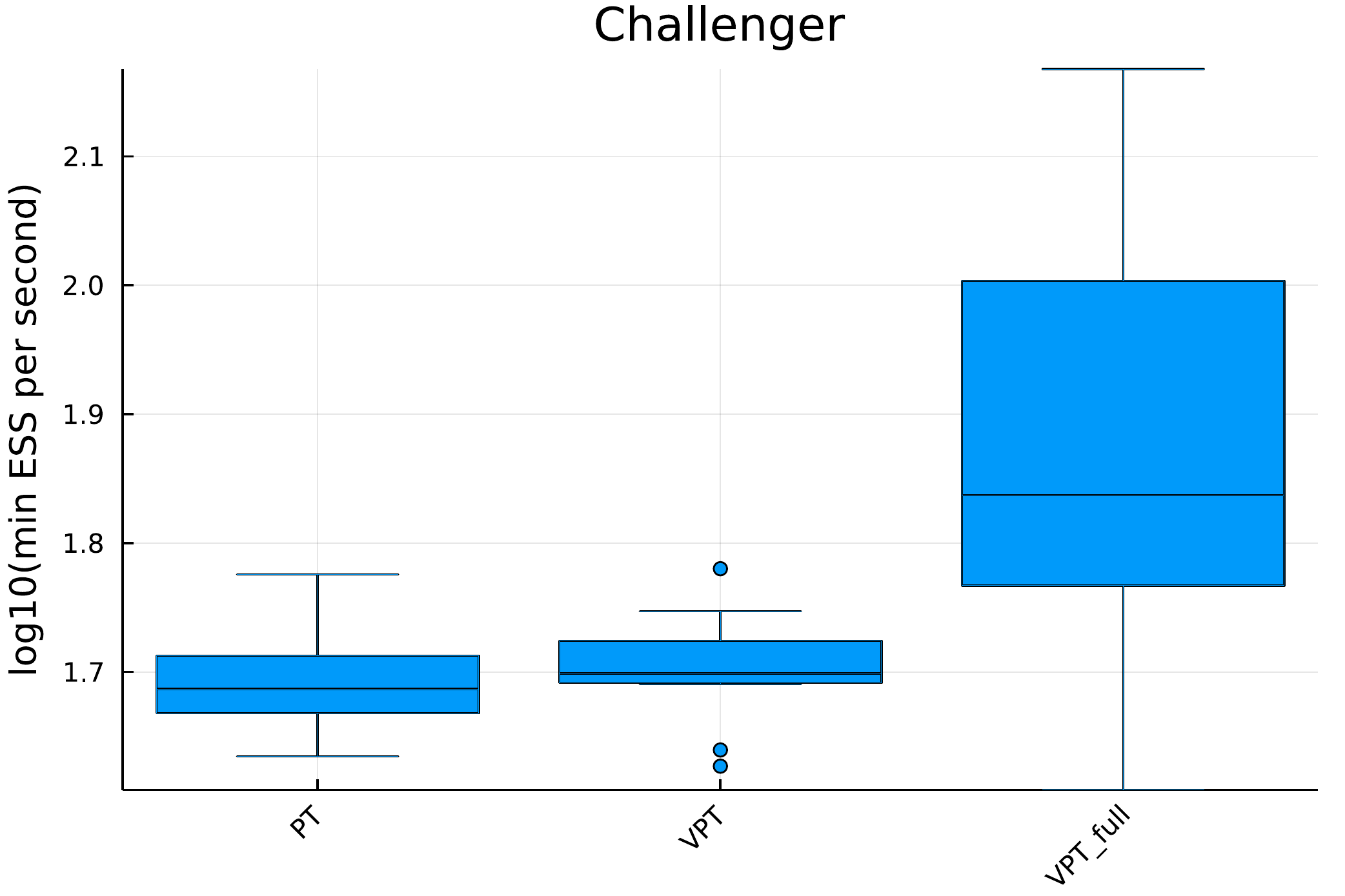}
    \end{subfigure}
    \begin{subfigure}{0.325\textwidth}
      \centering
      \includegraphics[width=\textwidth]{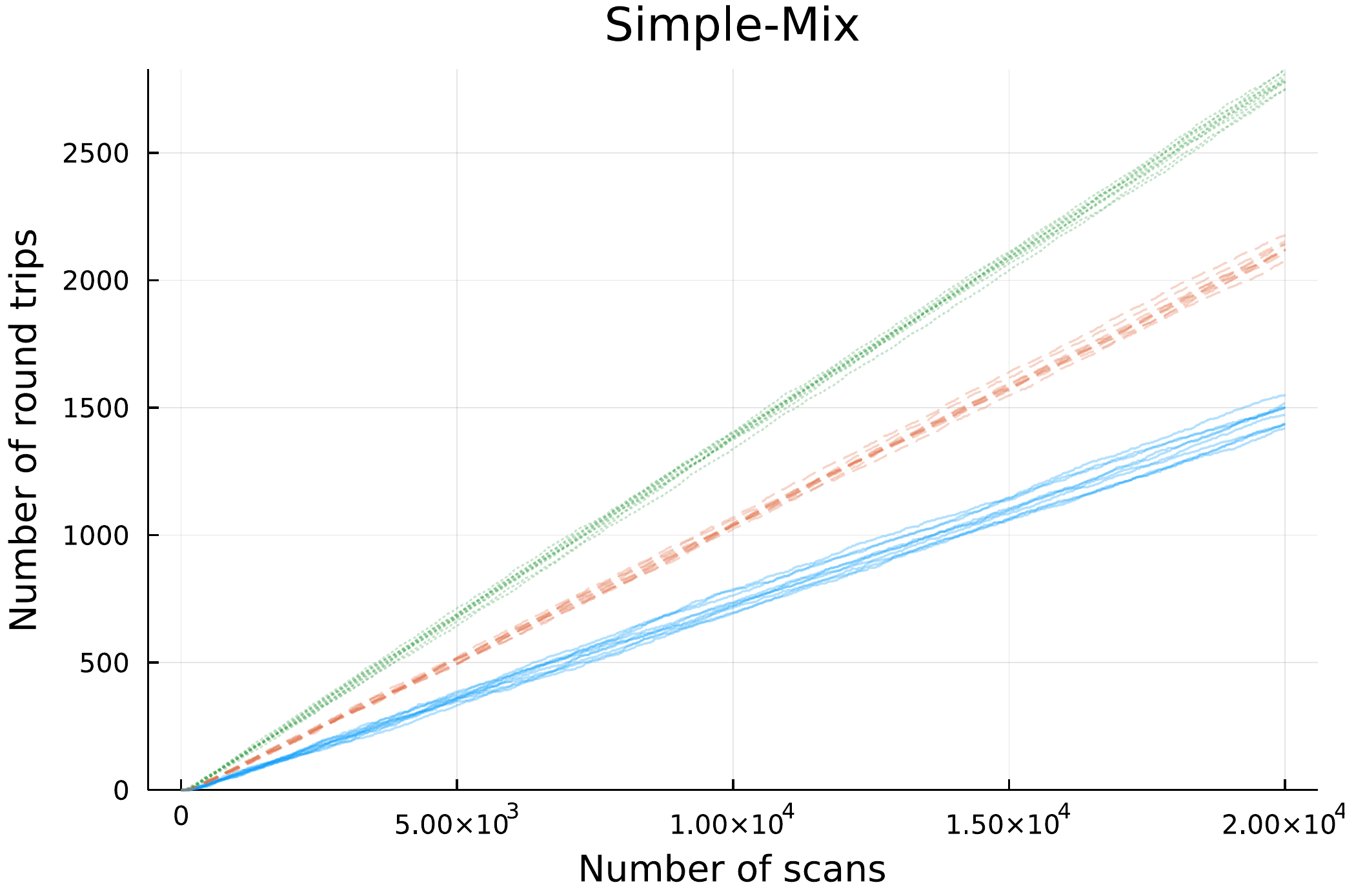}
    \end{subfigure}
    \begin{subfigure}{0.325\textwidth}
      \centering
      \includegraphics[width=\textwidth]{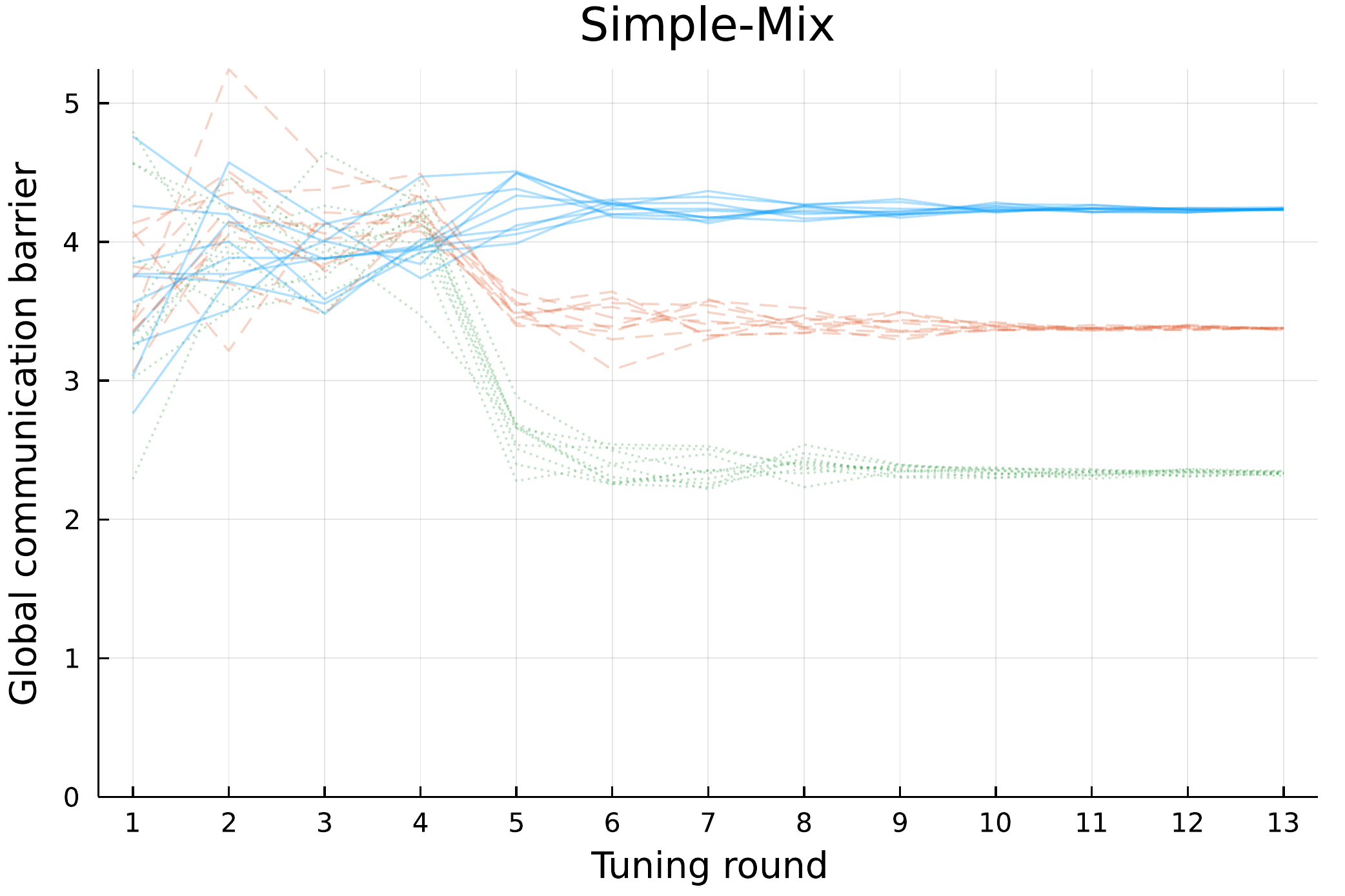}
    \end{subfigure}
    \begin{subfigure}{0.325\textwidth}
      \centering
      \includegraphics[width=\textwidth]{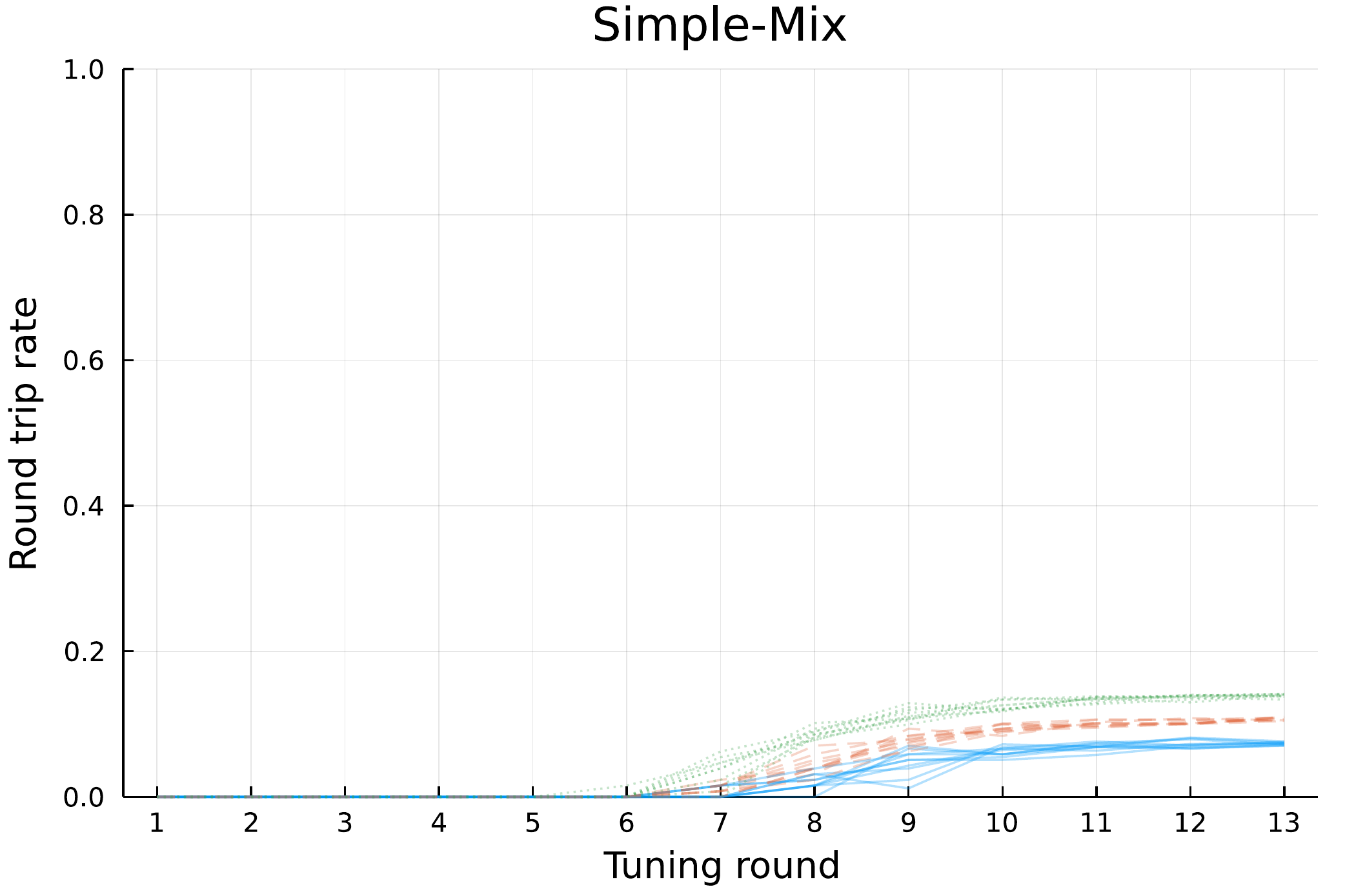}
    \end{subfigure}
    \begin{subfigure}{0.325\textwidth}
      \centering
      \includegraphics[width=\textwidth]{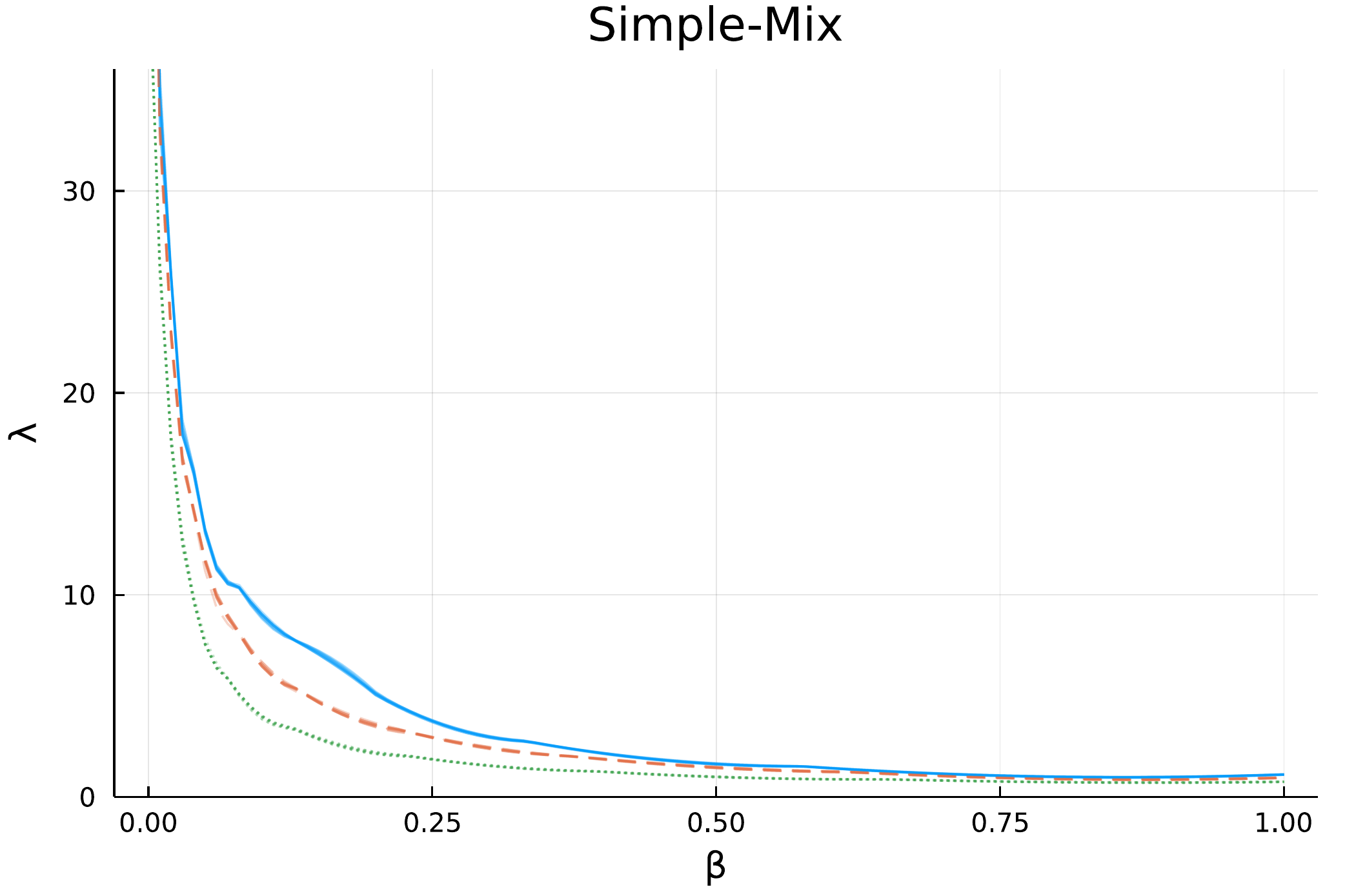}
    \end{subfigure}
    \begin{subfigure}{0.325\textwidth}
      \centering
      \includegraphics[width=\textwidth]{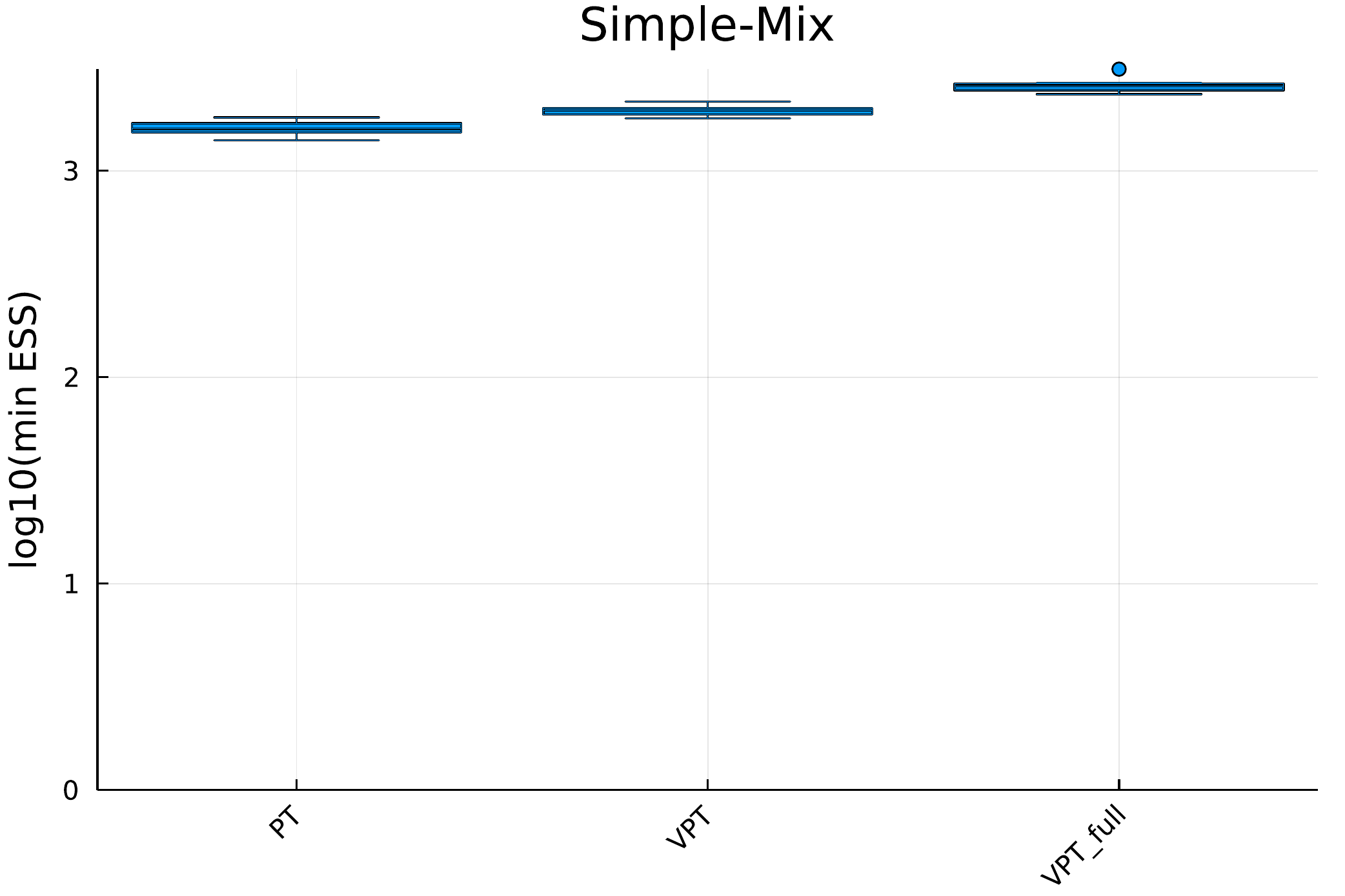}
    \end{subfigure}
    \begin{subfigure}{0.325\textwidth}
      \centering
      \includegraphics[width=\textwidth]{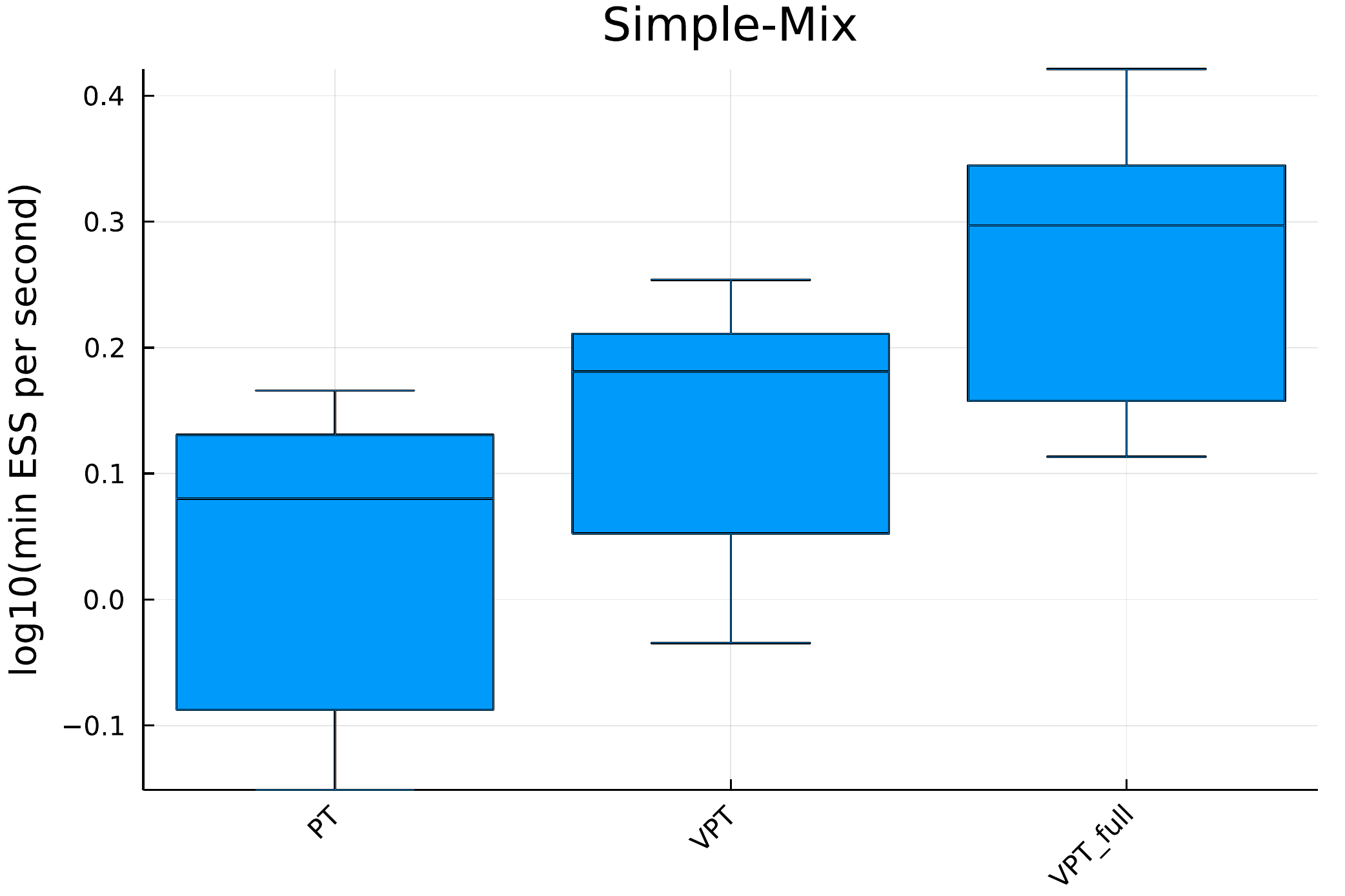}
    \end{subfigure}
    \caption{Number of round trips, GCB, round trip rate, LCB, ESS, and ESS per second. Green/red: Full-covariance/mean-field variational PT. Blue: NRPT.}
    \label{fig:additional_plots_2}
\end{figure}

Additional plots accompanying the results of \cref{sec:VPT_versus_NRPT} are presented 
in \cref{fig:additional_plots} and \cref{fig:additional_plots_2}. Additional plots for the 
Cauchy example can be found in \cref{fig:cauchy}. For these figures, the local 
communication barrier (LCB) between $\pi_0$ and $\pi_1$ at $\beta \in [0,1]$  is
\[
\lambda(\pi_0, \pi_1, \beta) = \frac{1}{2} \mathbb{E}[|\ell(X_\beta) -
  \ell(X_\beta')|], \qquad \ell(x) 
  = \log \frac{\pi_1(x)}{\pi_0(x)}, \qquad X_\beta, X'_\beta \distiid \pi_\beta.
\]
The LCB can be viewed as an instantaneous communication barrier for each point along 
the path between $\pi_0$ and $\pi_1$. The integral of the LCB yields the GCB under appropriate 
moment conditions.

\begin{figure}
    \centering
    \begin{subfigure}{0.325\textwidth}
      \centering
      \includegraphics[width=\textwidth]{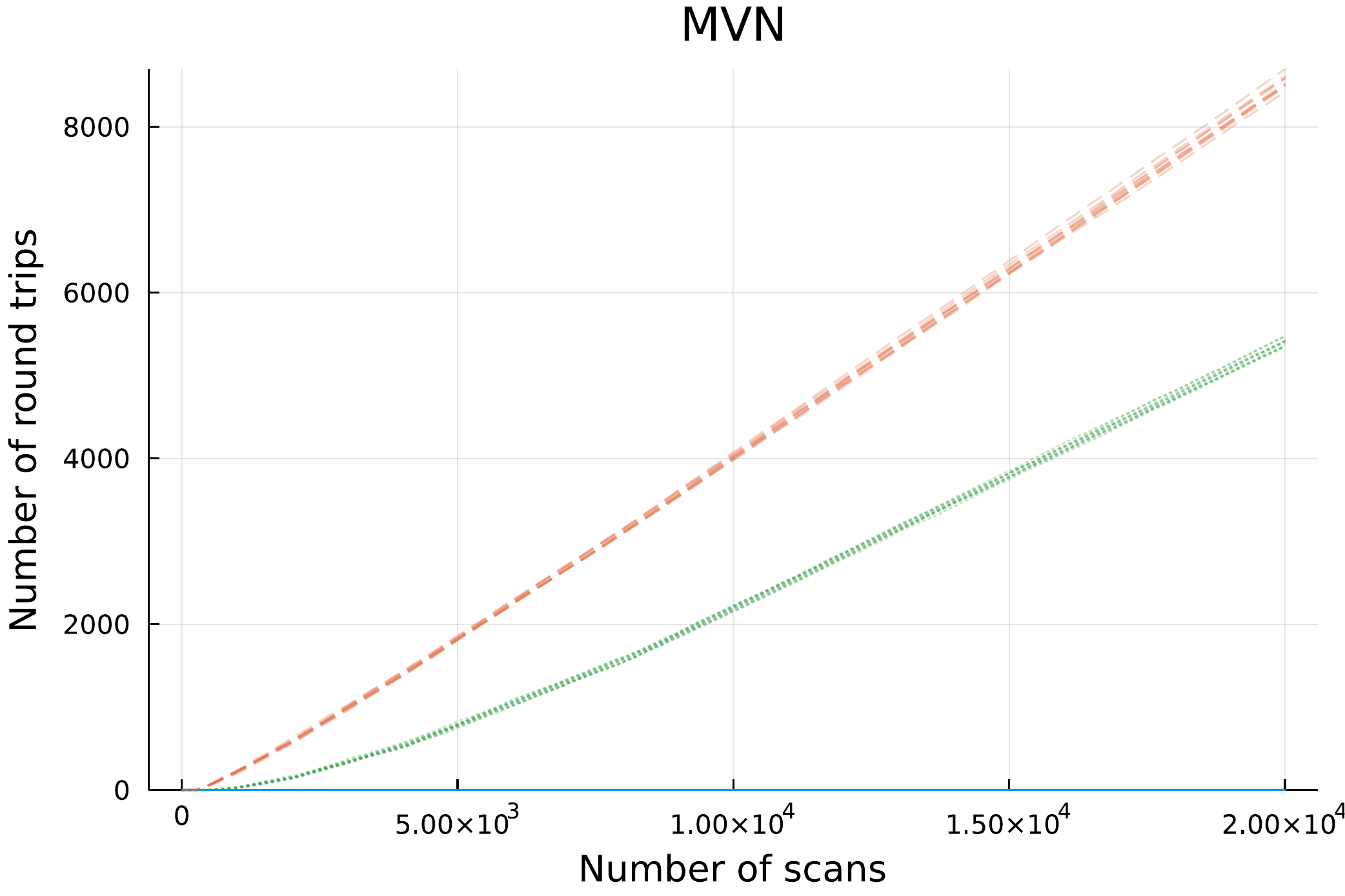}
    \end{subfigure}
    \begin{subfigure}{0.325\textwidth}
      \centering
      \includegraphics[width=\textwidth]{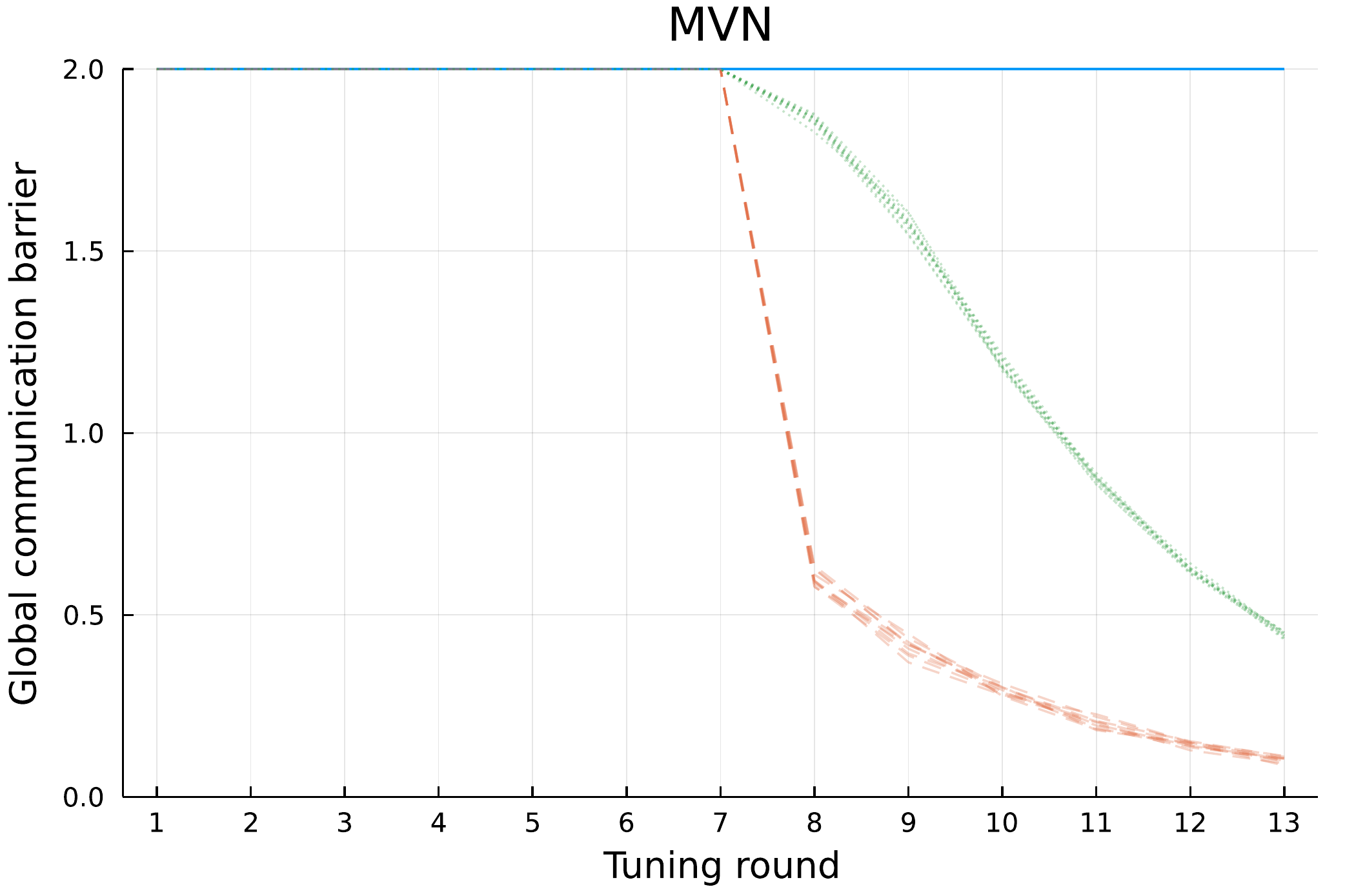}
    \end{subfigure}
    \begin{subfigure}{0.325\textwidth}
      \centering
      \includegraphics[width=\textwidth]{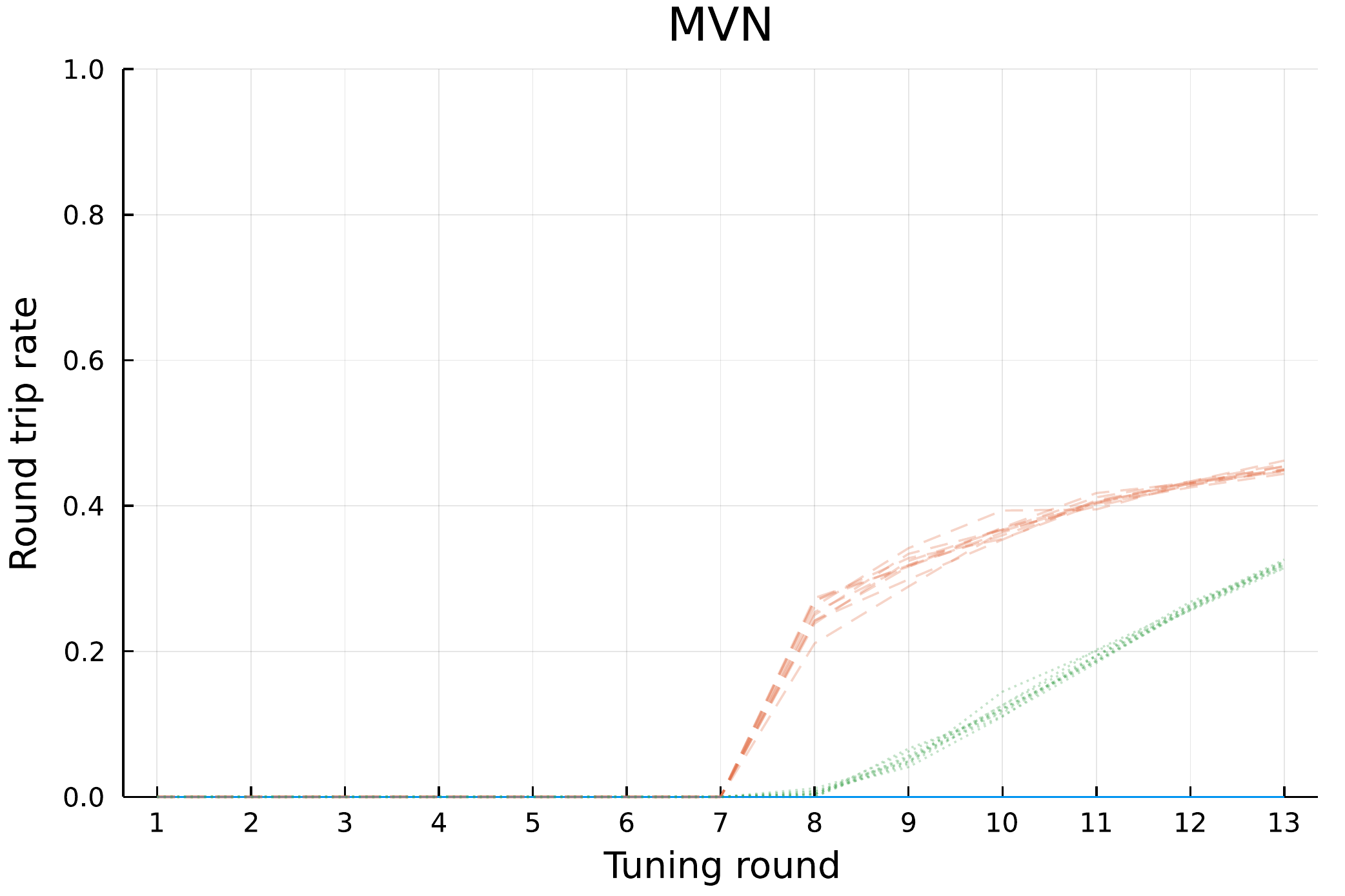}
    \end{subfigure}
    \begin{subfigure}{0.325\textwidth}
      \centering
      \includegraphics[width=\textwidth]{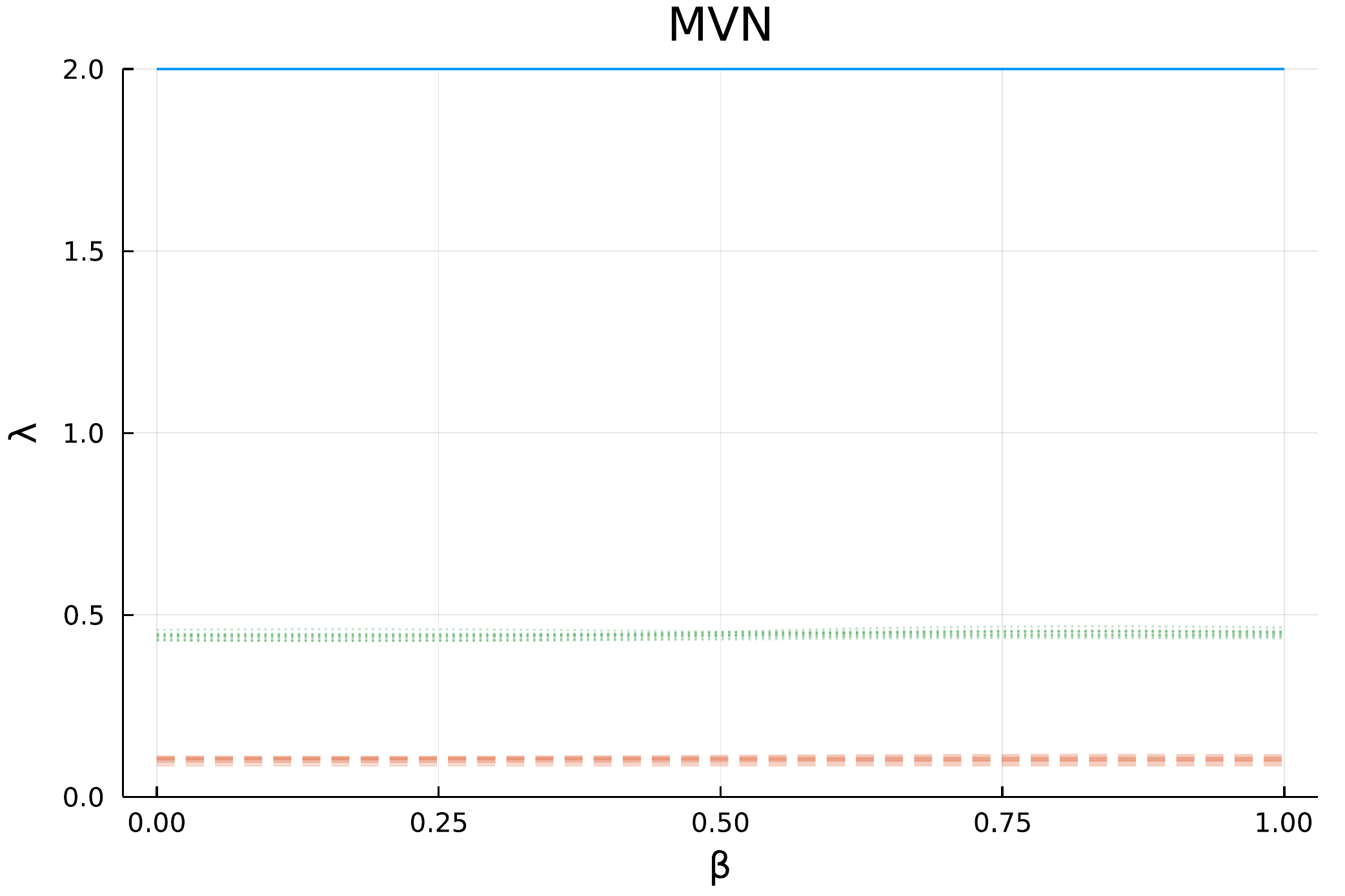}
    \end{subfigure}
    \begin{subfigure}{0.325\textwidth}
      \centering
      \includegraphics[width=\textwidth]{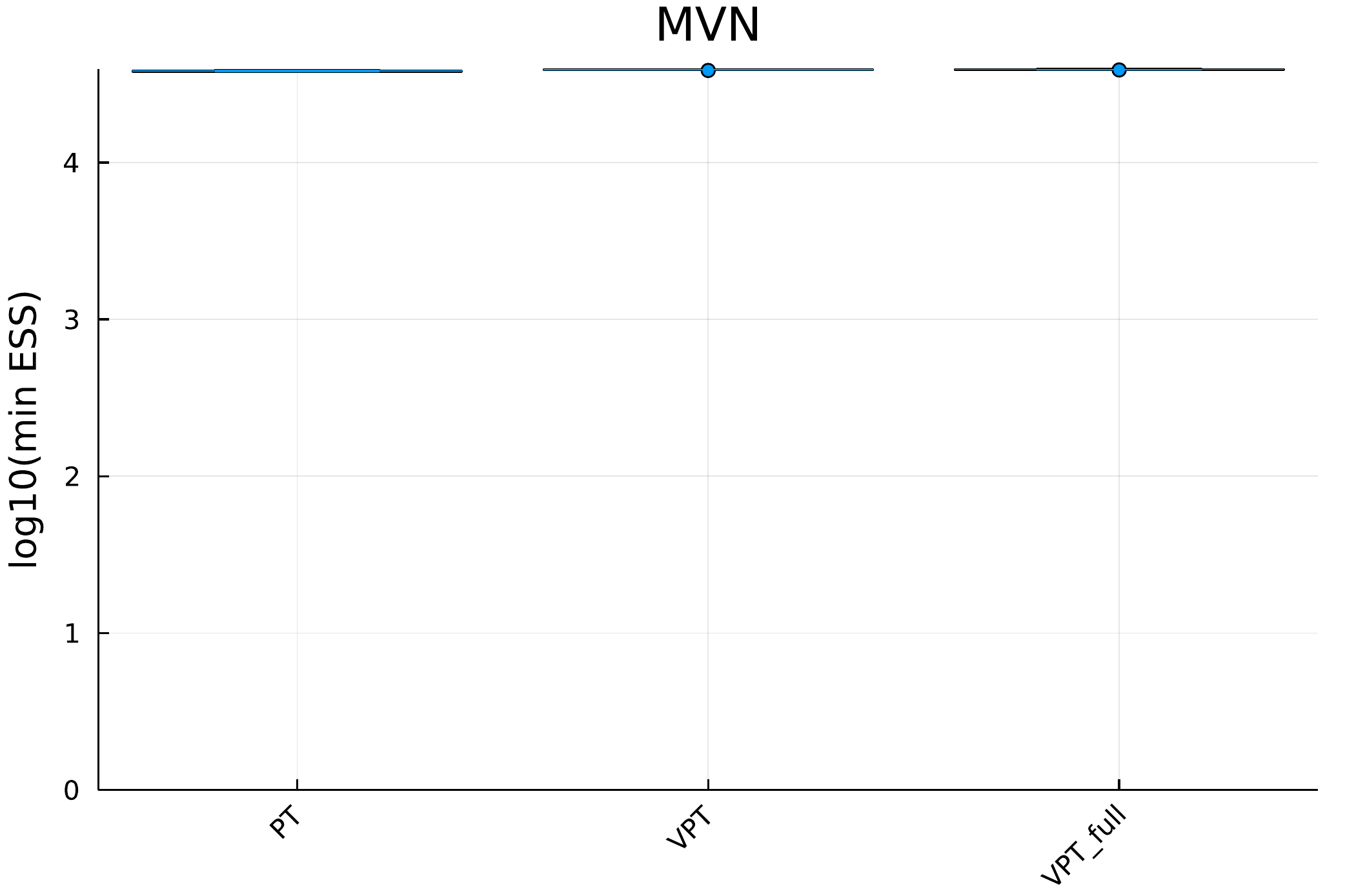}
    \end{subfigure}
    \begin{subfigure}{0.325\textwidth}
      \centering
      \includegraphics[width=\textwidth]{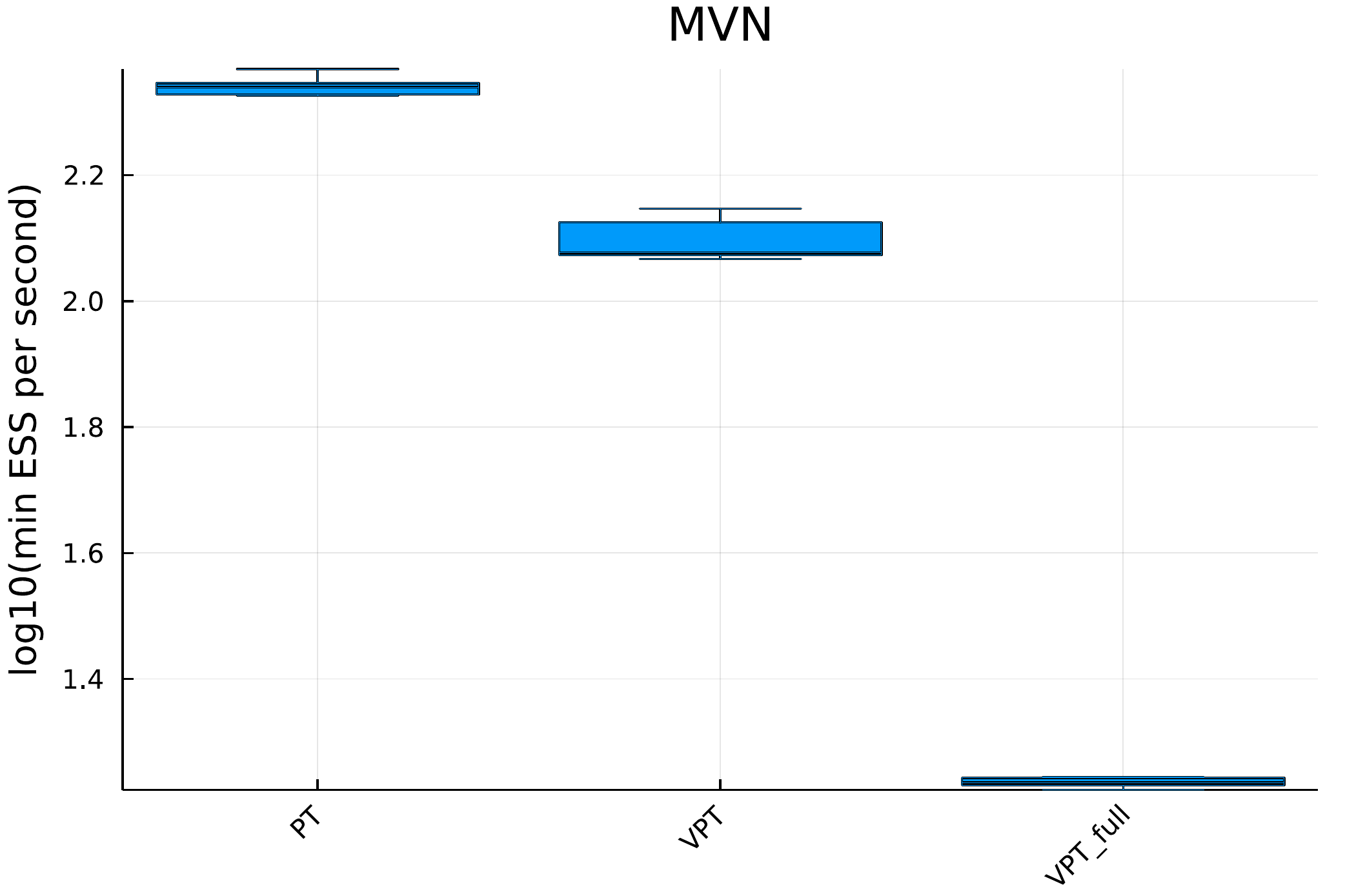}
    \end{subfigure}
    \begin{subfigure}{0.325\textwidth}
      \centering
      \includegraphics[width=\textwidth]{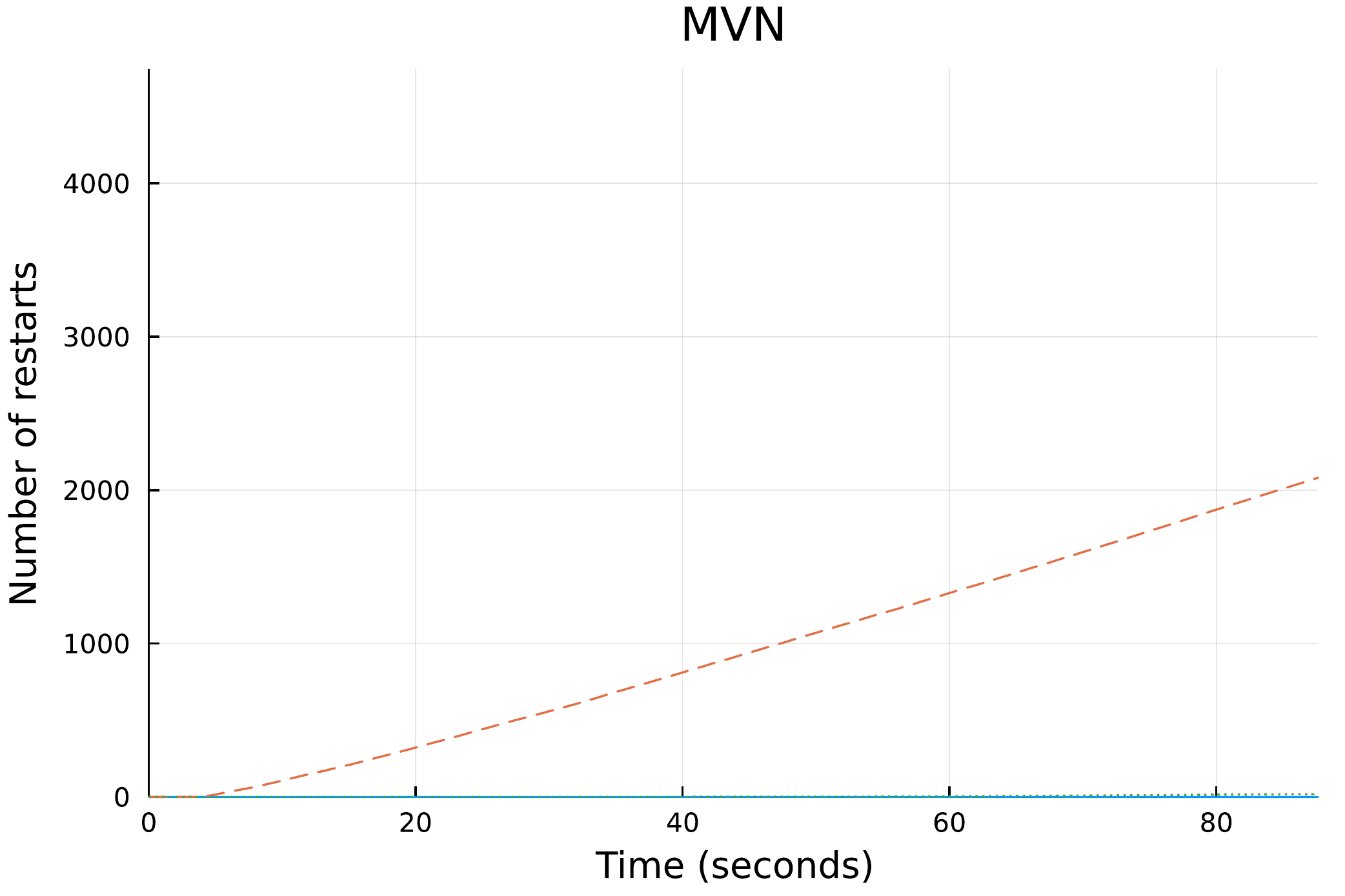}
    \end{subfigure}
    \caption{Results for the 100-dimensional multivariate normal target distribution.}
    \label{fig:MVN_restarts_time}
\end{figure}

\subsubsection{Additional results comparing stochastic gradient optimization and moment matching}
\label{sec:additional-stoch-gradient-vs-moment}

This section provides additional details for the results described in \cref{sec:MM_SGD}. 

For the stochastic gradient experiments, we considered optimizing several surrogate functions in addition to the global communication barrier (GCB). The first one is the SKL objective, which was used in \cite{syed2021optimized} as a GCB surrogate for optimizing the path between a fixed reference and the target. An estimator of the gradient of the SKL is derived in \cite{syed2021optimized}. We also considered a straightforward variant of this estimator for the forward KL objective. 
Even when considering a surrogate objective function, we measure performance using the quantity we are ultimately interested in, the GCB, estimated using the sum of the communication rejection rates.
Methods optimizing the GCB are labelled `Rejection' to emphasize that they are technically optimizing the sum of rejections. 

We compared the following optimization algorithms: the moment matching method described in \cref{sec:variational_PT}, basic stochastic gradient descent (SGD), where we update variational parameters using the formula $\phi^{(i+1)} \gets \phi^{(i)} - \alpha (i+1)^{-0.6} g^{(i)}$, following \cite[Section  4]{miasojedow_adaptive_2013}, where $g^{(i)}$ is the stochastic gradient estimate, and Adam \cite{kingma2015adam}. We found both SGD and Adam to be sensitive to the scale $\alpha$ of the updates (where $\alpha$ is defined above for SGD, and a similar parameter appears in Adam, also denoted $\alpha$  in  \cite{kingma2015adam}), which we varied in the grid $1/100, 1/10, 1, 10$ (this grid was iteratively increased starting at 1 until the optima laid inside the grid). 
 
We re-parameterized the variance parameters to lie in the real line using a differentiable transformation, which we set to the soft-plus transformation, $\text{log1p}(\exp(x))$. 
All methods use the same exploration kernel (Section~\ref{sec:additional-details}), which is the expensive inner loop of all optimization methods considered, so we use the number of exploration kernel applications as a model for computational time (abscissa for all plots in this subsection).
To set the number of chains, we used the heuristic in \cite[Section 5.3]{syed2019nrpt}, $N^* \approx 2 \Lambda$. We estimated $\Lambda \approx 3.5$ and used 8 chains for these experiments. 
For the stochastic optimization methods, we based the estimator of the gradient on 20 samples each interspersed with one scan, where a scan consists in one application of the local exploration kernel to each variable in each chains, and one set of odd or even swaps. 

Simultaneous optimization of the annealing schedule and the variational parameters is straightforward with our moment matching scheme: this is done using Algorithms 2 and 3 in \cite{syed2019nrpt} which, like our scheme,  proceed in rounds and hence can be performed in combination with round-based variational moment matching. 
Therefore, for all experiments involving moment matching, we learn an optimal schedule from scratch (i.e., initialized at an equally spaced schedule). 
For the stochastic optimization methods, simultaneous optimization of annealing schedules is more involved. 
The GCB is not a suitable objective, as when the number of chains gets larger, the GCB is asymptotically invariant to the schedule (provided the mesh size goes to zero, see \cite[Corollary 2]{syed2019nrpt}). 
It would be possible to use the scheme of \cite{miasojedow_adaptive_2013}, however this would add more tuning parameters. 
Instead, for these experiments, we provide all stochastic optimization methods with an optimally tuned schedule and exclude the cost of creating this optimal schedule from the computational budget calculation. 
As we shall see even with this head start, stochastic optimization does not perform as well as moment matching.

\begin{figure}[!htbp]
	\centering
	\includegraphics[width=\textwidth]{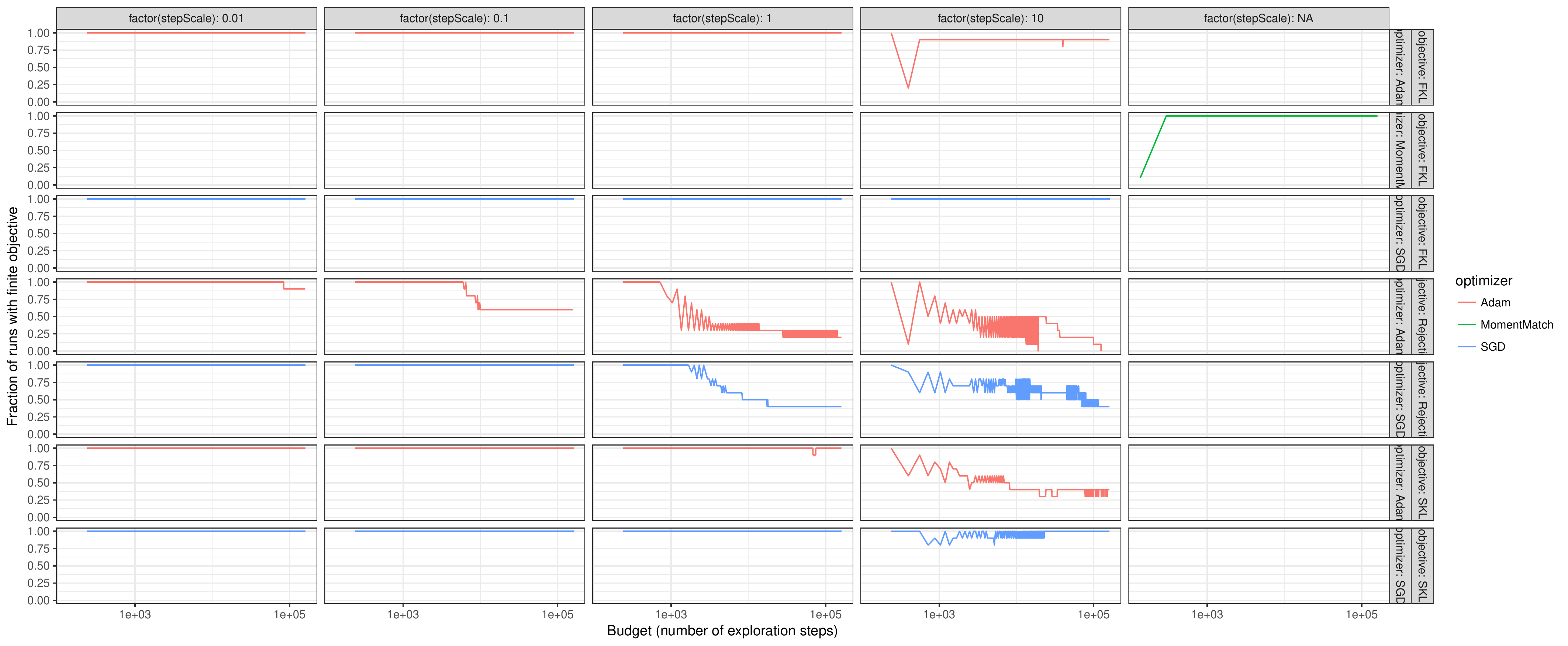}
	\caption{Fraction of the 10 replicates (each provided with a different random seed) that have a finite objective function at each optimization iteration. }
	\label{fig:opt-supp-isfinite}
\end{figure}

\begin{figure}[!htbp]
	\centering
	\includegraphics[width=\textwidth]{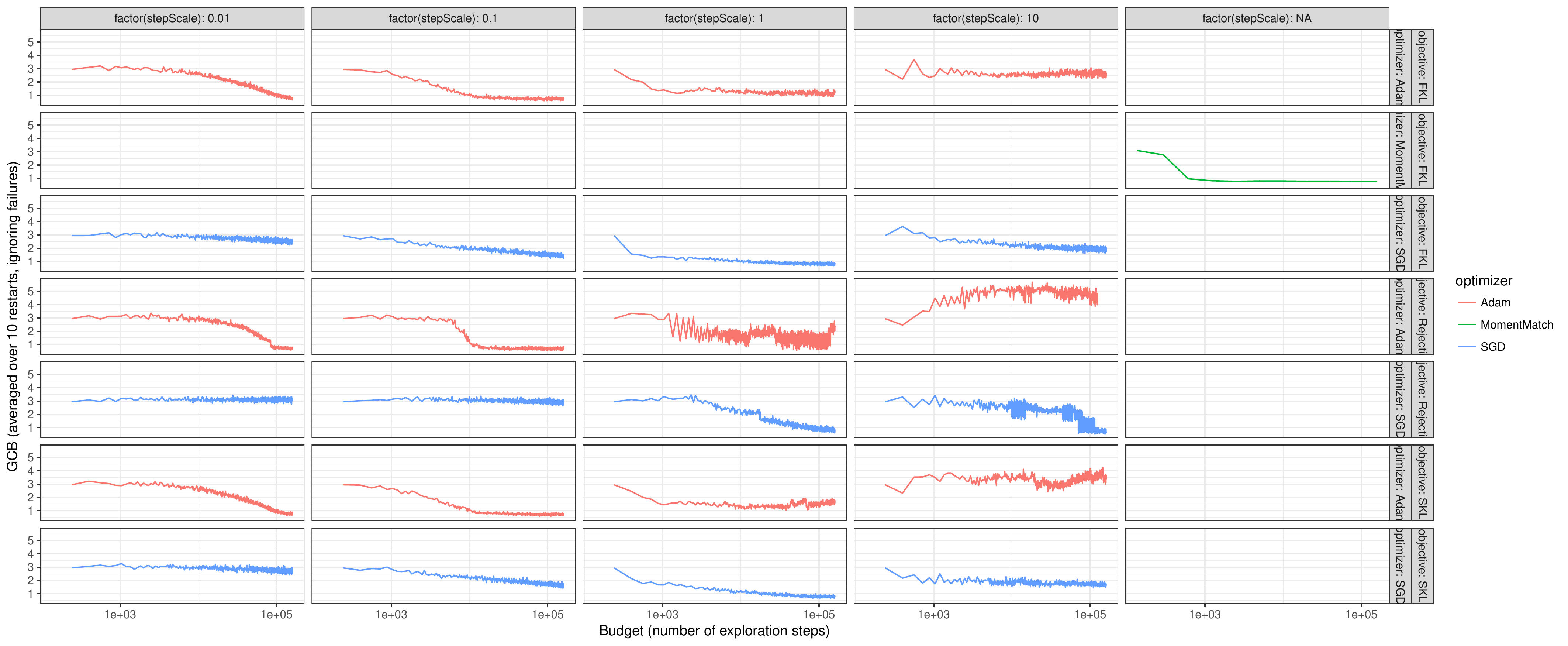}
	\caption{Aggregated performance of all optimization methods and surrogate functions considered in this work. }
	\label{fig:opt-supp-means}
\end{figure}

\begin{figure}[!htbp]
	\centering
	\includegraphics[width=\textwidth]{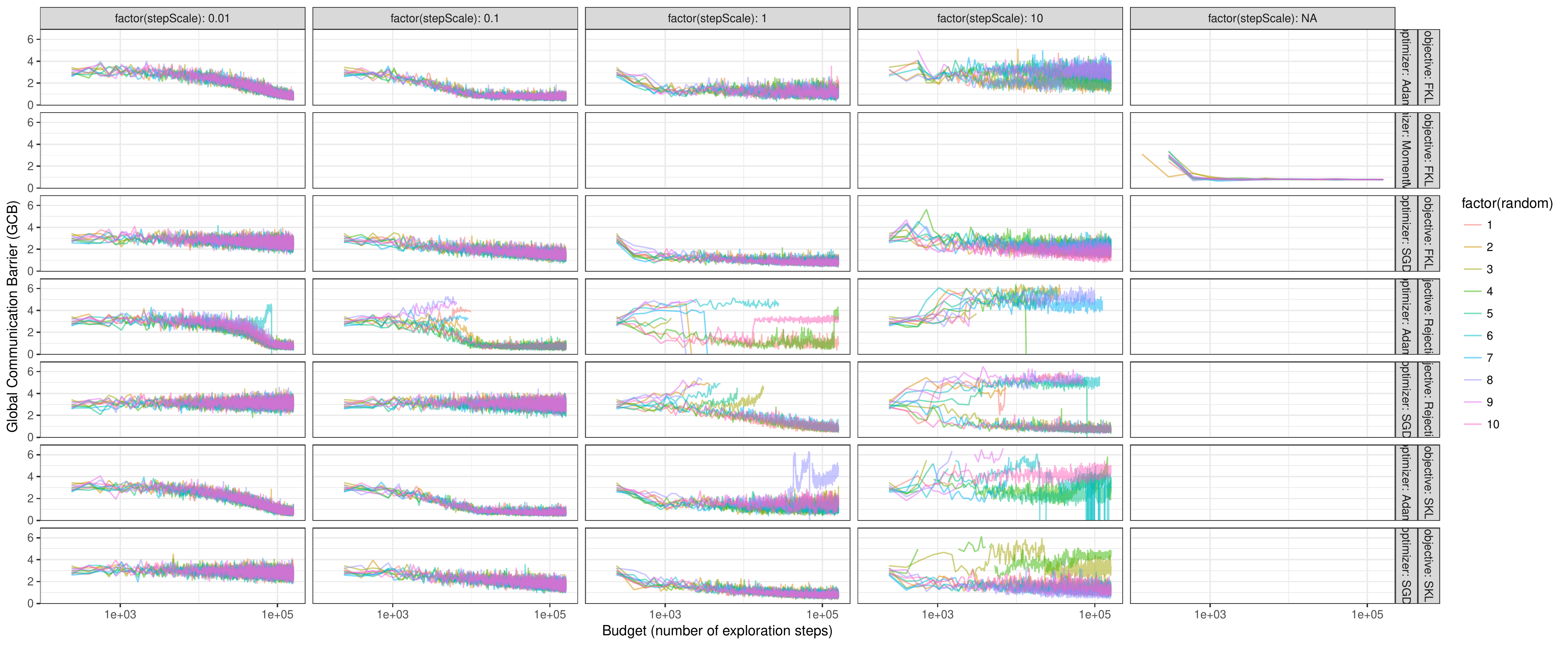}
	\caption{Performance traces of individual optimizers provided with different random seeds (colours). }
	\label{fig:opt-supp-monitoring}
\end{figure}

Our implementations of SGD and Adam are augmented with a form of error recovery: when a parameter update leads to a variational parameter where the GCB estimate is not finite (e.g. due to underflow or overflow), we roll back to the previous parameter value. 
Our implementation stores only one previous valid parameter setting to avoid keeping a history in memory. As a result, some of the stochastic optimization methods move to regions where with high probabilities the proposed next position leads to a rollback---this occurs especially when the step size scale is too large, as shown in \cref{fig:opt-supp-isfinite}. 
More precisely, this occurs with Adam+FKL for scales larger than 1, with Adam+GCB for all scales considered, with SGD+GCB for scales larger than 0.1, and for all SKL methods for scales larger than 1.
We show in \cref{fig:opt-supp-means} the mean GCB objective, averaged over the finite values (individual traces shown in \cref{fig:opt-supp-monitoring}). 

Among the configurations that did not lead to non-finite values, the following methods eventually achieved the same optimal GCB loss within the total optimization budget: Adam+FKL+0.01,  Adam+FKL+0.1, SGD+SKL+1, Adam+SKL+0.01, Adam+SKL+0.1, SGD+SKL+1. Among those, the fastest to reach that value were Adam+SKL+0.1 and Adam+FKL+0.1, with roughly equivalent performances.

We applied these best performing settings, Adam+SKL+0.1 and Adam+FKL+0.1, to two other problems to see how well they generalize. 
The two additional problems are: a Bayesian GLM 
model applied to the titanic dataset described in Section~\ref{sec:titanic} (with the only difference that we use normal priors here instead of Cauchy priors), and the change point problem described in Section~\ref{sec:change-point}.
We use the same experimental conditions as those described above for the Bayesian hierarchical model, with the difference that for the logistic regression problem we used a larger number of chains (20) to take into account the larger GCB ($\approx$8).
The results are shown in \cref{sec:MM_SGD} and demonstrate that stochastic optimization tuning does not generalize well from one problem to another, while moment matching performs well without requiring tuning. 
 
To replicate the results in this subsection, run, from the root of the repository \texttt{bl-vpt-nextflow}   
at \url{https://github.com/UBC-Stat-ML/bl-vpt-nextflow} 
the command \texttt{./nextflow run optimization.nf} and the results will be produced in the directory \texttt{bl-vpt-nextflow/deliverables/optimization}.

\subsubsection{Additional results comparing different PT topologies across several models}\label{sec:additional-topo-comparisons}
In this section, we perform an exhaustive empirical comparison of different PT algorithms incorporating a variational distribution.
We also include a baseline consisting of PT with a fixed reference.
Based on the results of the previous section, we perform optimization using moment matching. 

In all experiments in this subsection, we use a normal variational family with diagonal covariance.
This ensures that the running time is dominated by the local exploration kernels. 
We also ensure that the total number of chains is the same for all methods considered.
It follows that one scan has comparable running time for all methods considered in this subsection.

To describe the PT algorithms we use the following notation: we use {\bf T} to denote the target distribution; {\bf F}, to denote a fixed reference distribution; {\bf V}, to denote a variational distribution.
We use a star superscript to indicate which chain(s) are used to collect samples for moment matching when updating the variational distribution. 
Using this notation, the algorithms considered are:
\begin{description}
	\item[``F---T\;\;\;F---T'': ] two independent, non-interacting copies of standard PT. We use two copies in order to have a comparable running time per MCMC scan.
	\item[``V---T$^\star$\;\;F---T'': ] a basic variational approximation ({\bf V---T$^\star$}), with one independent, non-interacting standard PT, added to have comparable running time per MCMC scan. 
	\item[``V---T$^\star$\;\;F---T$^\star$'': ] a first type of stabilized variational approximation
	PT algorithm. 
	As in the algorithm above, it consists of two independent PT algorithms, one using a variational approximation and the other with a fixed reference. In contrast to the one above, the samples from the two PT algorithms are pooled at each round when performing moment-matching.
	\item[``V---T$^\star$---F'': ] the same algorithm as above, but where the states of the two target distributions are swapped at every second scan in order to maintain non-reversibility of the index process \cite[Section 3.4]{syed2019nrpt}.
	\item[``Reference'': ] the same algorithm as {\bf ``F---T\;\;\;F---T''}, but with ten times the number of MCMC scans as the other methods.
\end{description} 

In the main paper, for simplicity, {\bf ``F---T\;\;\;F---T''} is described as ``standard PT''; {\bf ``V---T$^\star$\;\;F---T''}, as ``basic variational PT''; and {\bf ``V---T$^\star$---F''}, as ``stabilized variational PT.'' As shown in this section {\bf ``V---T$^\star$---F''} and {\bf ``V---T$^\star$\;\;F---T$^\star$''} have similar performance in terms of restart rate, however the former is more natural and easier to analyze (\cref{thm:restart_rate}).

We consider the following models in this section, with the number of chains indicated in parentheses (number of chains selected to be approximately $2\Lambda$ as recommended in \cite[Section 5.3]{syed2019nrpt}):  the eight schools model ($N = 10$), a change point detection model applied on a dataset of mining incidents ($N = 10$), Bayesian estimation of ODE parameters for an mRNA transfection dataset ($N = 30$), Bayesian logistic regression applied to the titanic dataset ($N = 10$), two distinct Bayesian hierarchical models, one for prediction of rocket reliability ($N = 10$) and another tailored to estimating efficacy of multiple COVID-19 vaccines ($N = 20$), one sparse Conditional Auto-Regressive (CAR) spatial model applied to a lip cancer dataset ($N = 20$), one synthetic multi-modal example ($N = 20$), and one phylogenetic inference problem ($N = 20$). Refer to Sections~\ref{sec:ode}--\ref{sec:additional-blang} for more information.

 \begin{figure}[t]
 	\centering
 	\includegraphics[width=\textwidth]{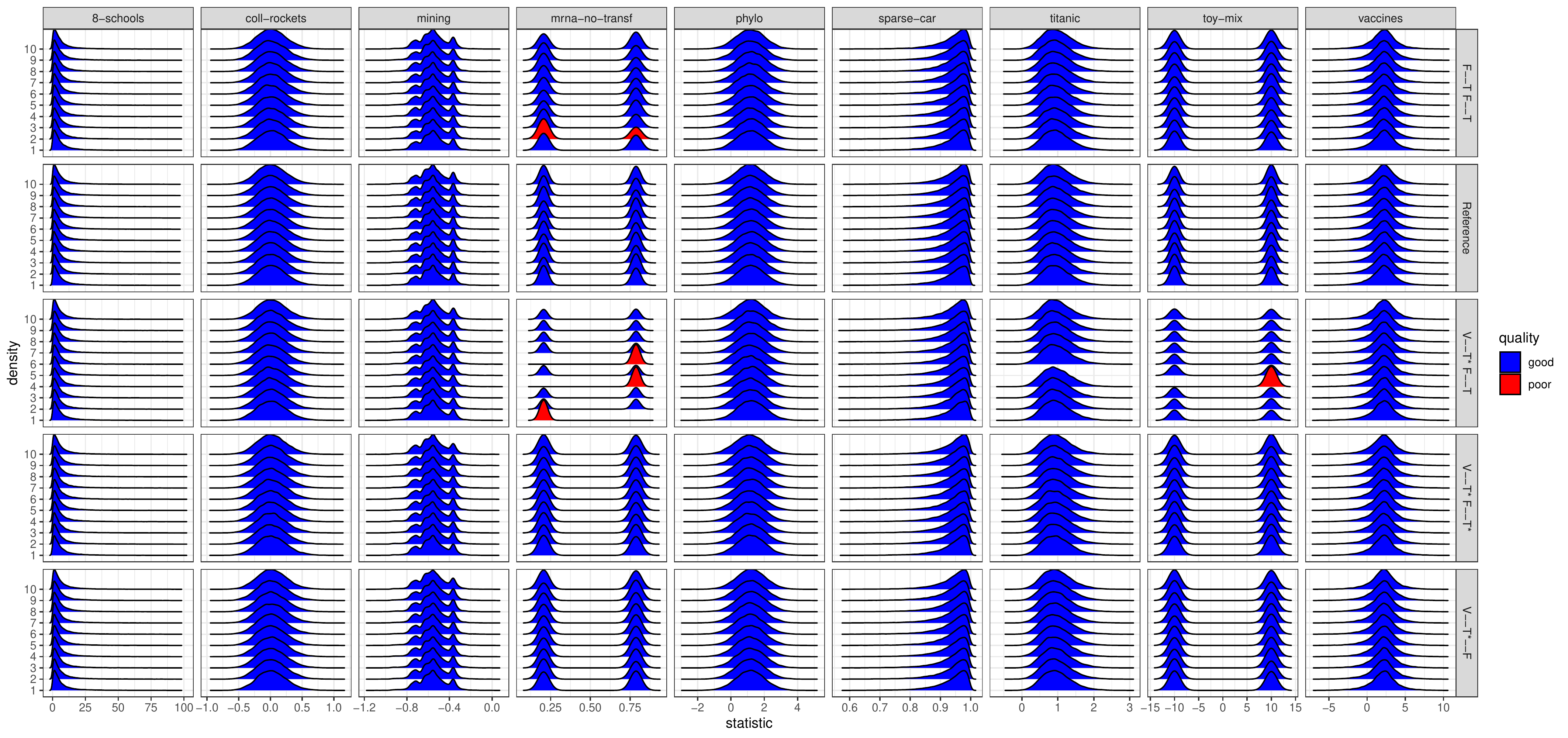}
 	\caption{For each model (column), we selected a key statistic or parameter. We show here the approximations of the marginal posterior distributions of these key statistics, for the different algorithms (rows) and 10 random seeds. Those in red have Kolmogorov-Smirnov (KS) distance greater than 0.1 compared to the {\bf ``Reference.''}}
 	\label{fig:many-posteriors}
 \end{figure}

For each model we first identified one key statistic: either the one of scientific interest or one 
such that its marginal posterior distribution is multi-modal. 
We ran each combination of algorithm and model 10 times with different random seeds. 
The approximations of the marginal posterior distributions are shown in 
\cref{fig:many-posteriors}.
Visual inspection shows that there are runs that clearly miss the multi-modal structure of the 
problem. 
To quantify this, we computed the Kolmogorov-Smirnov (KS) distance between the marginal posterior of this statistic obtained from the {\bf ``Reference''} run and that obtained from each algorithm. 
We mark in red the result from any random seed leading to a KS distance greater than 1/10. 

\begin{figure}[t]
 	\centering
 	\includegraphics[width=\textwidth]{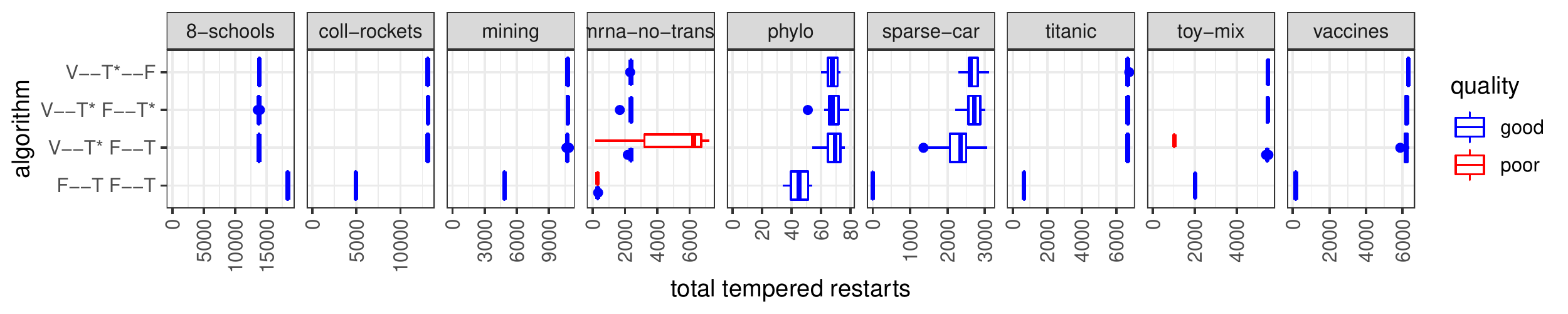}
 	\caption{Tempered restarts for different PT algorithms (rows) and models (facets). Each algorithm is executed 10 times with different random seeds. The results are also segregated based on the KS criterion described in \cref{fig:many-posteriors}.}
 	\label{fig:topo-restarts} 
 \end{figure}

Next, we looked at tempered restart count statistics for all combinations of models and algorithms, also splitting the results based on the above mentioned KS threshold. 
It is important to make this KS-based split, as runs that missed one of the mode may under-estimate the difficulty of the problem and hence over-estimate the number of tempered restarts. 
The results are shown in \cref{fig:topo-restarts} and \cref{table:restarts}.

\begin{table}[ht]
\centering
\scalebox{0.9}{
\begin{tabular}{llrrrr}
  \toprule
model & algorithm & \# KS$<$0.1 & Q1 & Q2 & Q3 \\ 
  \midrule
8-schools & F--T F--T &  10 & 18310 & 18382 & 18414 \\ 
  8-schools & V--T* F--T &  10 & 13728 & 13788 & 13809 \\ 
  8-schools & V--T* F--T* &  10 & 13718 & 13770 & 13797 \\ 
  8-schools & V--T*--F &  10 & 13797 & 13829 & 13852 \\ 
  coll-rockets & F--T F--T &  10 & 4876 & 4931 & 4953 \\ 
  coll-rockets & V--T* F--T &  10 & 13048 & 13105 & 13160 \\ 
  coll-rockets & V--T* F--T* &  10 & 13074 & 13128 & 13161 \\ 
  coll-rockets & V--T*--F &  10 & 13081 & 13108 & 13151 \\ 
  mining & F--T F--T &  10 & 4824 & 4848 & 4863 \\ 
  mining & V--T* F--T &  10 & 10704 & 10728 & 10745 \\ 
  mining & V--T* F--T* &  10 & 10700 & 10766 & 10823 \\ 
  mining & V--T*--F &  10 & 10716 & 10750 & 10798 \\ 
  mrna-no-transf & F--T F--T &   9 & 311 & 316 & 320 \\ 
  mrna-no-transf & V--T* F--T &   7 & 2366 & 2381 & 2384 \\ 
  mrna-no-transf & V--T* F--T* &  10 & 2361 & 2378 & 2414 \\ 
  mrna-no-transf & V--T*--F &  10 & 2366 & 2376 & 2385 \\ 
  phylo & F--T F--T &  10 &  40 &  45 &  51 \\ 
  phylo & V--T* F--T &  10 &  65 &  70 &  73 \\ 
  phylo & V--T* F--T* &  10 &  65 &  68 &  72 \\ 
  phylo & V--T*--F &  10 &  65 &  68 &  71 \\ 
  sparse-car & F--T F--T &  10 &   0 &   0 &   0 \\ 
  sparse-car & V--T* F--T &  10 & 2063 & 2348 & 2501 \\ 
  sparse-car & V--T* F--T* &  10 & 2556 & 2714 & 2878 \\ 
  sparse-car & V--T*--F &  10 & 2579 & 2607 & 2810 \\ 
  titanic & F--T F--T &  10 & 642 & 650 & 661 \\ 
  titanic & V--T* F--T &   9 & 6615 & 6662 & 6691 \\ 
  titanic & V--T* F--T* &  10 & 6649 & 6666 & 6715 \\ 
  titanic & V--T*--F &  10 & 6644 & 6661 & 6687 \\ 
  toy-mix & F--T F--T &  10 & 2010 & 2020 & 2038 \\ 
  toy-mix & V--T* F--T &   9 & 5443 & 5458 & 5469 \\ 
  toy-mix & V--T* F--T* &  10 & 5446 & 5491 & 5519 \\ 
  toy-mix & V--T*--F &  10 & 5474 & 5484 & 5529 \\ 
  vaccines & F--T F--T &  10 & 147 & 154 & 166 \\ 
  vaccines & V--T* F--T &  10 & 6118 & 6208 & 6264 \\ 
  vaccines & V--T* F--T* &  10 & 6168 & 6236 & 6262 \\ 
  vaccines & V--T*--F &  10 & 6288 & 6316 & 6336 \\ 
   \bottomrule
\end{tabular}
}
\caption{Quantiles (0.25, 0.5, 0.75) of the number of restarts for the subset of the runs with KS < 0.1.} 
\label{table:restarts}
\end{table}

In most of the models considered, we observed a large increase in the number of tempered restarts when going from standard PT to variational PT. We observed an increase of a factor $\sim$2.5 in the rocket model, $\sim$2.2 for the mining problem, $\sim$1.6 for the phylogenetic problem, $\sim$10.2 for the titanic problem, $\sim$40.3 for the vaccine problem (above speed-up estimates computed on the median column of \cref{table:restarts}). The gains are particularly impressive for the sparse CAR model applied to the lip cancer dataset, in which standard PT achieves 0 restarts while the median number of restarts for the variational methods are all higher than 2300 restarts. 

For the mRNA and toy mixture problems, only the stabilized algorithms ({\bf ``V---T$^\star$---F''} and {\bf ``V---T$^\star$\;\;F---T$^\star$''}) succeeded in avoiding catastrophic failures, and compared to standard PT led to an increase in median restarts of a factor $\sim$7.5 for the mRNA problem and of a factor $\sim$2.7 for the toy mixture problem. In the mRNA example, 3 out of 10 applications of the basic variational method, {\bf ``V---T$^\star$\;\;F---T''}, led to a catastrophic failure, and 1 out of 10 in the toy mixture example. Note that the number of restarts is overestimated in the failed runs, highlighting the importance of using a stabilized algorithm. 

The 8 schools problem provides an example where a variational approach based on a diagonal covariance matrix underperforms standard PT. In this example, the number of round trips decreased from a median of 18382 to a median of 13788, 13770, and 13829 for the three variational algorithms considered. But note that the drop in performance is less than 50\%, which agrees with the result in \cref{thm:restart_rate}. 

\begin{figure}[t]
 	\centering
 	\includegraphics[width=\textwidth]{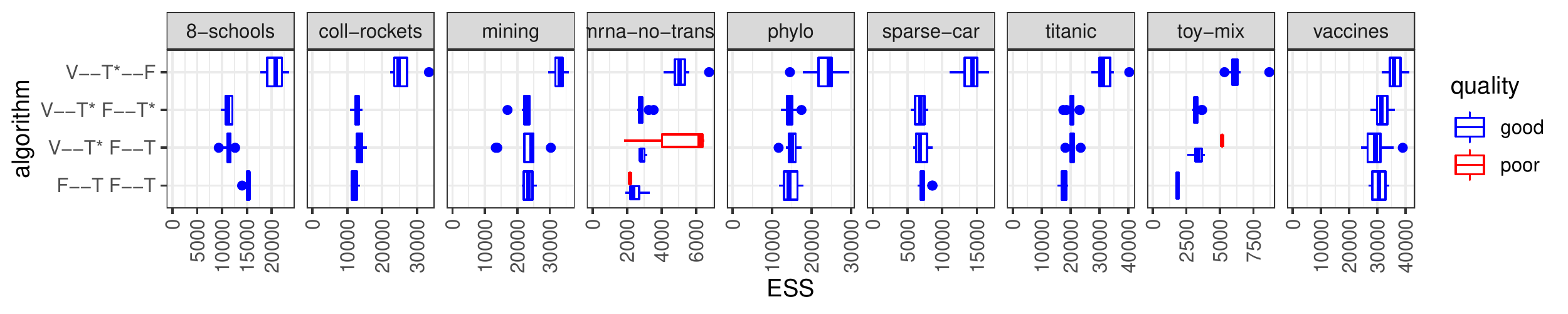}
 	\caption{Effective Sample Size (ESS) for each model's key statistic, for different PT algorithms (rows) and models (facets). Each algorithm is executed 10 times with different random seeds. The results are also segregated based on the KS criterion described in \cref{fig:many-posteriors}.}
 	\label{fig:topo-ess}
 \end{figure}

\cref{fig:topo-ess} and \cref{table:ESS} show performance in terms of the effective sample size (ESS) of each model's key statistic. 
The ESS is computed using the \texttt{mcmcse} R  package which implements a batch mean estimator. 
For the non-interacting variational variants, {\bf ``V---T$^\star$\;\;F---T''} and {\bf ``V---T$^\star$\;\;F---T$^\star$''}, we observe modest ESS gains compared to standard PT in the rocket, sparse CAR, mRNA and titanic problems, while the gains are larger for the toy mixture problem ($\sim$1.8 increase). 
For the interacting variational variant, {\bf ``V---T$^\star$---F''}, the gains are substantial in all problems considered. However, in contrast to the other algorithms where adding the ESS of the two copies is justified by independence, the ESS estimator may not be reliable in the case of {\bf ``V---T$^\star$---F''} given the interactions. 
The restart rate does not have this limitation so we recommend gauging the performance of the methods primarily based on their restart rate.

\begin{table}[ht]
\centering
\scalebox{0.9}{
\begin{tabular}{llrrrr}
  \toprule
model & algorithm & \# KS$<$0.1 & Q1 & Q2 & Q3 \\ 
  \midrule
8-schools & F--T F--T &  10 & 15011 & 15316 & 15506 \\ 
  8-schools & V--T* F--T &  10 & 11088 & 11290 & 11698 \\ 
  8-schools & V--T* F--T* &  10 & 10767 & 11222 & 12074 \\ 
  8-schools & V--T*--F &  10 & 19047 & 20811 & 22046 \\ 
  coll-rockets & F--T F--T &  10 & 11323 & 12013 & 12700 \\ 
  coll-rockets & V--T* F--T &  10 & 12799 & 13358 & 14301 \\ 
  coll-rockets & V--T* F--T* &  10 & 12315 & 12829 & 13350 \\ 
  coll-rockets & V--T*--F &  10 & 23473 & 24712 & 27128 \\ 
  mining & F--T F--T &  10 & 22004 & 23410 & 24588 \\ 
  mining & V--T* F--T &  10 & 22136 & 24382 & 24928 \\ 
  mining & V--T* F--T* &  10 & 22111 & 22799 & 23663 \\ 
  mining & V--T*--F &  10 & 31719 & 33186 & 33899 \\ 
  mrna-no-transf & F--T F--T &   9 & 2180 & 2360 & 2709 \\ 
  mrna-no-transf & V--T* F--T &   7 & 2755 & 2783 & 3013 \\ 
  mrna-no-transf & V--T* F--T* &  10 & 2688 & 2816 & 2881 \\ 
  mrna-no-transf & V--T*--F &  10 & 4787 & 5053 & 5350 \\ 
  phylo & F--T F--T &  10 & 13061 & 14249 & 16406 \\ 
  phylo & V--T* F--T &  10 & 14256 & 15036 & 15927 \\ 
  phylo & V--T* F--T* &  10 & 13798 & 14476 & 15071 \\ 
  phylo & V--T*--F &  10 & 21723 & 24421 & 25046 \\ 
  sparse-car & F--T F--T &  10 & 6945 & 7071 & 7372 \\ 
  sparse-car & V--T* F--T &  10 & 6234 & 6777 & 7817 \\ 
  sparse-car & V--T* F--T* &  10 & 6122 & 6853 & 7460 \\ 
  sparse-car & V--T*--F &  10 & 13244 & 14351 & 15011 \\ 
  titanic & F--T F--T &  10 & 16942 & 17567 & 18469 \\ 
  titanic & V--T* F--T &   9 & 20082 & 20601 & 21078 \\ 
  titanic & V--T* F--T* &  10 & 19982 & 20406 & 20871 \\ 
  titanic & V--T*--F &  10 & 29952 & 31216 & 33741 \\ 
  toy-mix & F--T F--T &  10 & 1766 & 1843 & 1936 \\ 
  toy-mix & V--T* F--T &   9 & 3186 & 3347 & 3631 \\ 
  toy-mix & V--T* F--T* &  10 & 3093 & 3195 & 3307 \\ 
  toy-mix & V--T*--F &  10 & 5942 & 6129 & 6296 \\ 
  vaccines & F--T F--T &  10 & 28148 & 30611 & 32884 \\ 
  vaccines & V--T* F--T &  10 & 26536 & 29297 & 31160 \\ 
  vaccines & V--T* F--T* &  10 & 29905 & 31443 & 33514 \\ 
  vaccines & V--T*--F &  10 & 34517 & 35911 & 38177 \\ 
   \bottomrule
\end{tabular}
}
\caption{Quantiles (0.25, 0.5, 0.75) of the ESS for the subset of the runs with KS < 0.1.} 
\label{table:ESS}
\end{table}

\begin{figure}[t]
	\centering
	\includegraphics[width=\textwidth]{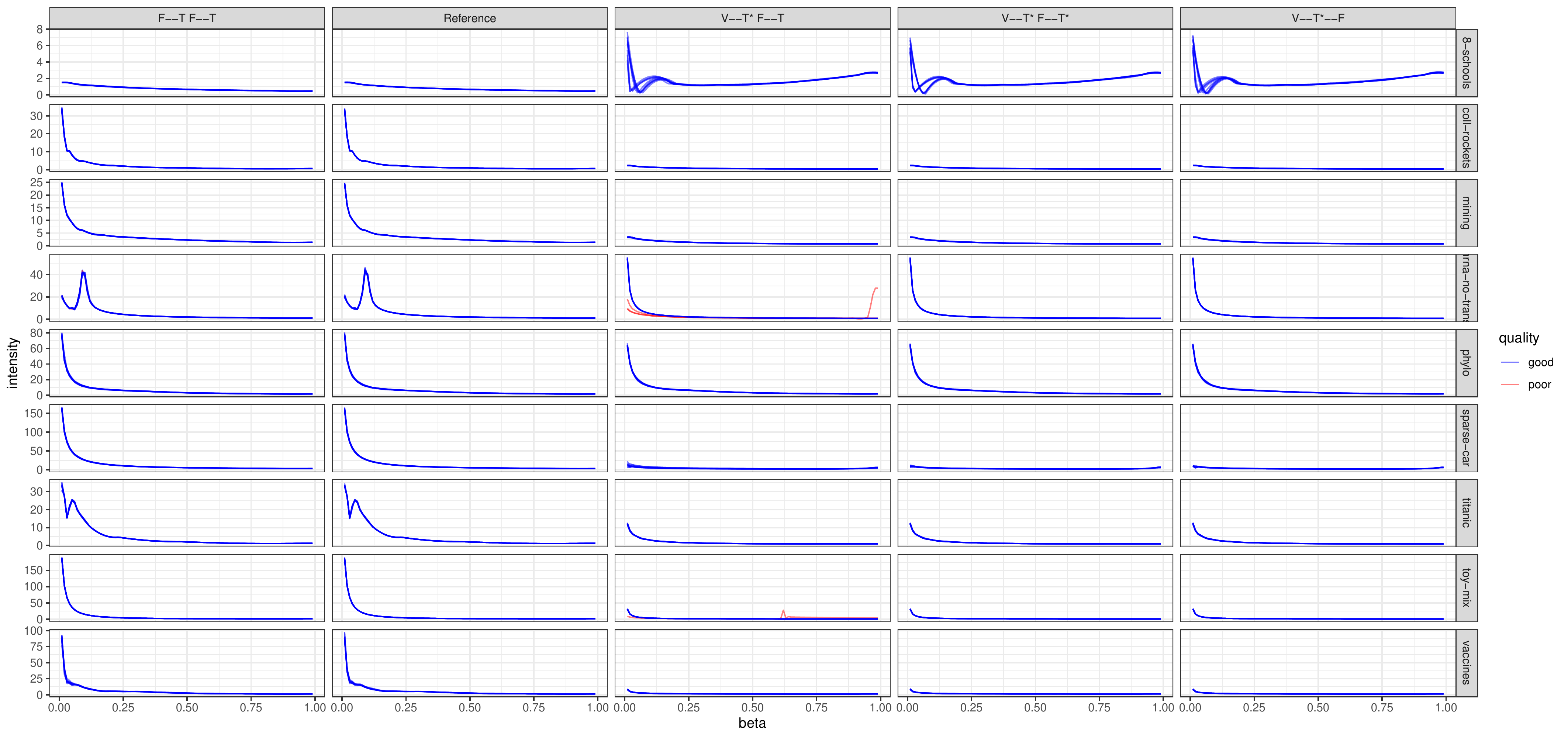}
	\caption{The local communication barrier for nine models (row). For the variational methods, the plot shows the local communication barrier between the variational distribution and the target; for standard PT, the plot shows the local communication barrier between the fixed reference and the target. We show the estimated functions for the five algorithms (columns) and ten random seeds. Red curves indicate ``catastrophic failures'' as described in \cref{fig:demo}.}
	\label{fig:many-lambdas}
\end{figure}

\begin{figure}[t]
	\centering
	\includegraphics[width=\textwidth]{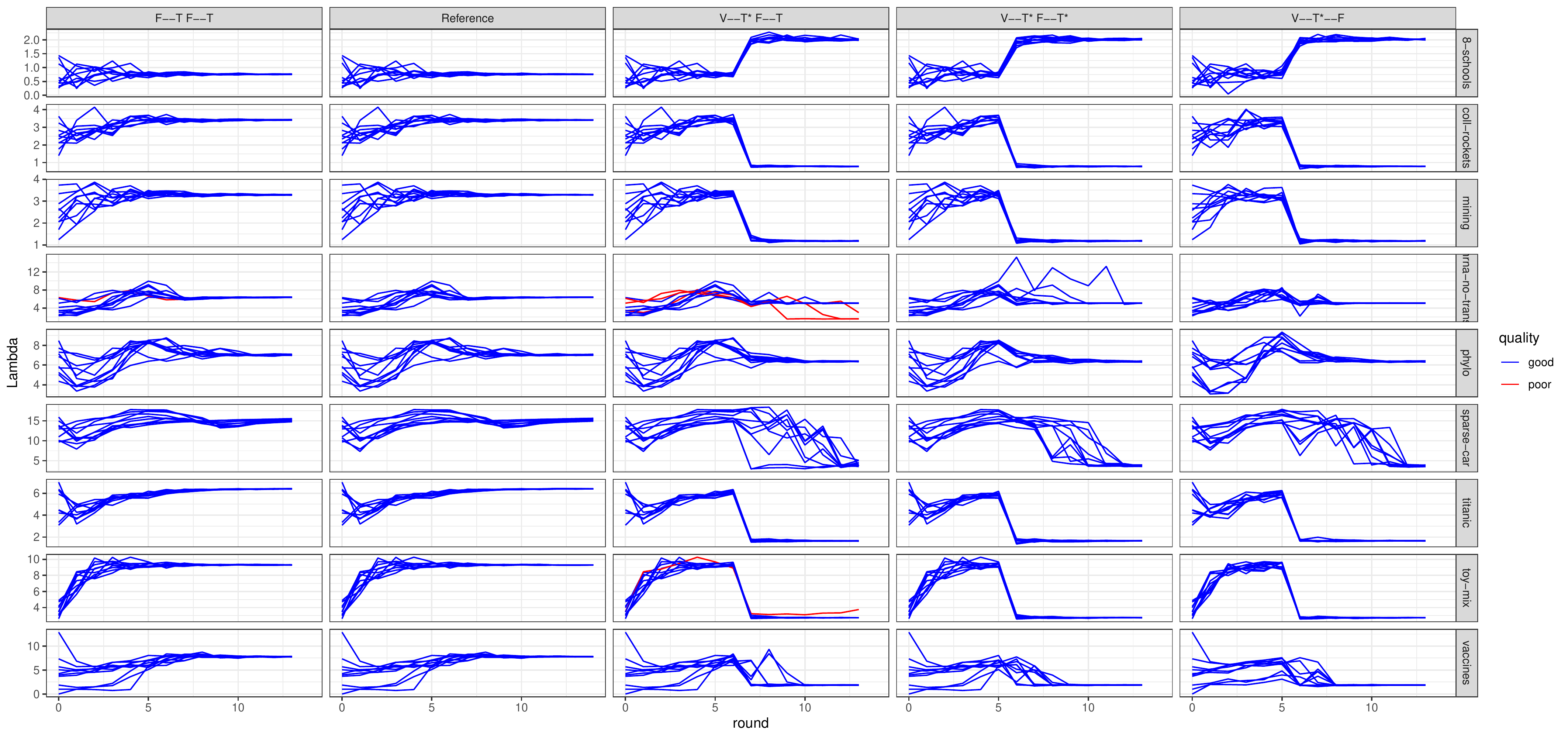}
	\caption{The global communication barrier estimates between the reference and target distributions for nine models (row), each shown as a function of the adaptation round. For the variational methods, the plot shows the global communication barrier between the variational distribution and the target; for standard PT, the global communication barrier is computed between the fixed reference and the target. We show the estimated communication barriers for the five algorithms (columns) and ten random seeds. Red curves indicate ``catastrophic failures'' as described in \cref{fig:demo}. }
	\label{fig:many-GCBs}
\end{figure}

We also show the local and global communication barriers, $\lambda$ and $\Lambda$, for these models and algorithms in \cref{fig:many-lambdas} and \cref{fig:many-GCBs}. Interestingly, in some of the cases, e.g.\ the transfection problem, the high gains in terms of tempered restarts obtained by going from standard PT to variational, are not as large when measured by $\Lambda$. This could be due to the variational $\lambda$ function being generally smoother and hence the optimal schedule easier to approximate in a finite number of rounds. 

To replicate the results in this subsection, run, from the root of the repository \texttt{bl-vpt-nextflow}  
at \url{https://github.com/UBC-Stat-ML/bl-vpt-nextflow} 
the command \texttt{./nextflow run pt\_topologies.nf} and the results will be produced in the directory \texttt{bl-vpt-nextflow/deliverables/pt\_topologies}.

\end{document}